%% file: ClassicThesis.tex
\RequirePackage{silence} 
\documentclass[twoside,openright,titlepage,numbers=noenddot,
                headinclude=true,footinclude=false,cleardoublepage=empty,abstract=on,
                BCOR=5mm,paper=b5,fontsize=11pt]{scrreprt}

\input{classicthesis-config}

\makeatletter \newrobustcmd*{\mknumAlph}[1]{%
\begingroup   \blx@tempcnta=#1\relax   \ifnum\blx@tempcnta>702 %
\else     \ifnum\blx@tempcnta>26 %
\advance\blx@tempcnta\m@ne       \divide\blx@tempcnta26\relax       \blx@numAlph\blx@tempcnta       \multiply\blx@tempcnta26\relax       \blx@tempcnta=\numexpr#1-\blx@tempcnta\relax     \fi   \fi   \blx@numAlph\blx@tempcnta   \endgroup} 

\def\blx@numAlph#1{%
\ifcase#1\relax\blx@warning@entry{Value out of range}\number#1\or   P1\or P2\or P3\or P4\or P5\or P6\or P7\or P8\or P9\or P10\or P11\or P12\or P13\or   P14\or P15\or P16\or P17\or P18\or P19\or P20\or P21\or P22\or P23\or P24\or P25\or P26\else   \blx@warning@entry{Value out of range}\number#1\fi} \makeatother  \DeclareFieldFormat{labelnumber}{\ifkeyword{myPapers}{\footnotesize\mknumAlph{#1}}{#1}}

\DeclareSourcemap{
  \maps[datatype=bibtex]{
    \map{
      \step[fieldsource=keywords, match=\regexp{alphabetical}, final]
      \step[fieldset=usera, fieldvalue={\textcolor{teal}{\scalebox{1.5}{\hspace{-0.4em}*}}}]
    }
  }
}

\AtEveryBibitem{
  \clearfield{issue}
  \clearfield{number}
  \clearfield{month}
}

\begin{document}

\emergencystretch 5em

\interfootnotelinepenalty=10000 

\frenchspacing

\raggedbottom

\selectlanguage{american} 



\pagenumbering{roman}

\pagestyle{plain}



\cleardoublepage\include{FrontBackMatter/Titlepage}	



\cleardoublepage\include{FrontBackMatter/Titleback}

\cleardoublepage\include{FrontBackMatter/Dedication}

\cleardoublepage\include{FrontBackMatter/Agradecimientos}

\cleardoublepage\include{FrontBackMatter/Acknowledgements}

\cleardoublepage\include{FrontBackMatter/Contents}

\cleardoublepage\include{FrontBackMatter/Abstract}

\cleardoublepage\include{FrontBackMatter/Resumen}

\cleardoublepage\include{FrontBackMatter/Publications}


\cleardoublepage

\pagestyle{scrheadings}

\pagenumbering{arabic}


\include{Chapters/Introduction} 


\ctparttext{In the first part of the thesis, we introduce the theoretical background from which the rest of the work is built. We have taken the opportunity to introduce our own notation and conventions, setting the ground for the following parts. Thus, we encourage the reader to have a look at these first pages. This part is divided in the following three chapters.

In \cref{ch:cosmo}, we discuss the description of the Universe provided by the Standard Cosmological Model, together with its problems and extensions, such as inflation, that aim at solving part of them. We therefore introduce the background geometry, and proceed afterward with the quantization of fields in this context in \cref{ch:qftcs}. We also present the spectrum of fluctuations, as well as Gaussian state theory and entanglement. Finally, in \cref{ch:analogs}, we discuss the idea of simulating QFTCS in the laboratory, which is the goal of the field of analog gravity. In particular, we focus on BECs as analog platforms.

Some results from \cite{Entanglement2024} are included in \cref{ch:qftcs}, and of course all chapters are written with the intention of accommodating the framework used in all the publications that constitute this thesis. \\ \bigskip

\emph{Notation.} For this part we set $\hbar = c = k_{\text{B}} = 1$, and use $\mpl=1/\sqrt{G}$ and the metric signature $\left(-, +, ..., +\right)$. Unless explicitly stated, Greek indices~\mbox{$\mu, \nu$} run from $0$ to $D$, while Latin indices $i,j$ run from $1$ to $D$.

} 


\cleardoublepage

\part{Basics of (analog) cosmological particle production}\label{pt:basics}

\include{Chapters/Cosmology}

\include{Chapters/Quantum_field_theory_in_a_curved_background}

\include{Chapters/AnalogGravity}

\cleardoublepage


\ctparttext{In this second part, we discuss cosmological particle production in the early Universe. We study how the expansion of the geometry during the inflation and reheating epochs leads to the production of particles out of the vacuum. In particular, we focus on spectator fields that are only coupled to gravity (non-minimally). This suits the known properties of dark matter, and we consider the viability of this mechanism for reproducing the observed abundance of this elusive component of the Universe.

This part is divided in two chapters. Chapter \ref{ch:singlefield} focuses on both scalar and vector fields in flat FLRW. We consider cosmological production sourced by chaotic, single field inflationary models, analyzing two different potentials: quadratic and Starobinsky's. In \cref{ch:deSitter}, we work with a much simpler inflationary model, so that particle production can be computed analytically. In exchange, we consider scalar fields in FLRW universes with any kind of spatial curvature.

With these two chapters, we cover the contents of the already published articles \cite{ScalarField2023,VectorDM2024}, as well as the works in progress\footnote{These articles were not publicly available on arXiv at the time of defending the thesis. Nevertheless, \cite{Starobinsky2024} is now published in JCAP.    } \cite{Starobinsky2024,CurvedDM2024}. \\ \bigskip 

\emph{Notation.} In this part, we set $\hbar = c = k_{\text{B}} = 1$, and use  $\mpl=1/\sqrt{G}$ and the metric signature $\left(-, +, +, +\right)$. Furthermore, Greek indices $\mu, \nu$ run from $0$ to $3$, while Latin indices $i,j$ run from $1$ to $3$. Most of the time, we work in units of the inflaton mass of the quadratic model $m_{\phi}$.}

\cleardoublepage

\part{Cosmological production of (dark matter) particles in the early Universe}\label{pt:darkmatterproduction}

\def\chapterfolder{figures/DMProductionSlowRoll/}
\include{Chapters/DMProductionSlowRoll}

\def\chapterfolder{figures/DMProductionDeSitter/}
\include{Chapters/DMProductionDeSitter}

\cleardoublepage


\ctparttext{Cosmological particle production is a mechanism that can explain the observed abundance of dark matter, as we have studied in part II. However, the access to scenarios in which this type of phenomenon can be probed is limited. In this part of the thesis, we explore the possibility of using analog systems to simulate cosmological particle production. In particular, we focus on BECs in quasi-$(1+2)$ dimensions and the phononic excitations on top of their ground state, which behave as a massless, minimally coupled scalar field in a curved spacetime whose particular form can be tuned in the laboratory by adjusting the condensate properties.

We begin by developing the mentioned analogy in \cref{ch:becstheory}, and showing how phonons can be understood as moving in an FLRW universe with positive, negative or vanishing curvature, depending on the trapping potential of the BEC. This corresponds to the contents of \cite{CosmologyPaper2022,BECPaper2022}. Moreover, we show the realization of the analogy in an actual experiment, based on the work \cite{Experiment2022}. Then, we present a reinterpretation of cosmological particle production as a one-dimensional scattering problem in Quantum Mechanics, which is part of the reference \cite{ScatteringTheory2024}. The corresponding experimental implementation of these ideas is contained in \cite{ScatteringExp2024}, which is to be published\footnote{This, as well as the rest of the papers related to this chapter, are now published.}. Finally, we also theoretically study entanglement between the cosmologically produced pairs and how to improve its detectability in such BEC setups in \cref{ch:entanglement}, corresponding to the contents of~\cite{Entanglement2024}. \\ \bigskip 

\emph{Notation.} Due to its immediate connection with experiment, in this part of the thesis we work in SI units. The metric signature is still $(-,+,+)$, although we work now in $(1+2)$ dimensions, and therefore Greek indices $\mu, \nu$ run from $0$ to $2$, while Latin indices $i,j$ only run from $1$ to $2$.}

\cleardoublepage

\part{Experimental realization of analog cosmological production}
\label{pt:becs}

\def\chapterfolder{figures/Weakly-interacting_BECs_as_QFT_in_CS_simulators/}
\include{Chapters/Weakly-interacting_BECs_as_QFT_in_CS_Simulators}

\def\chapterfolder{figures/Entanglement/}
\include{Chapters/Entanglement_between_produced_pairs}

\cleardoublepage


\ctparttext{In parts II and III, we have studied the cosmological production of particles in the early Universe, and how this mechanism can be simulated in analog systems, in particular in BECs. In this last part, we aim at transferring insights from one context to the other, and vice versa. This will be very enriching, as viewing the cosmological scenarios from the perspective of analog experiments sometimes allows for a better, or at least different, understanding of the physical processes involved, thus leading to novel ways of approaching arising problems. In fact, \cref{ch:qftcs} in part I is already written in a way that bridges these two scenarios, bringing formalisms and concepts that were born in the context of the BEC analyses carried out to Cosmology.

In \cref{ch:qva}, we make use of the idea that one may eventually measure some occupation number with an apparatus after certain expansion of the geometry to define, characterize and study a family of vacua whose notion of particle can be associated to hypothetical measurements. On the other hand, until \cref{ch:entanglement}, we had considered instantaneous switch-ons and -offs of the analog scale factor in our BEC analyses. We take this idea and explore in \cref{ch:switcheffects} how important the switch-on and -off effects are in the production process, both in a similar BEC experiment, but also in the cosmological context. This is interesting if one understands particle production in the early Universe in terms of a switch-on (the initial state is vacuum, and then inflation starts, smoothly) and a switch-off (which could be understood as a smooth transition to the reheating epoch, leading to a thermal, adiabatic universe). \\ \bigskip

\emph{Notation.} For this part, we go back to setting $\hbar = c = k_{\text{B}} = 1$, keeping the metric signature $\left(-, +, ..., +\right)$. As before, Greek indices are used for all spacetime components, whereas Latin indices run for the spatial part. The working dimensions will depend on the specific section.
}

\cleardoublepage

\part{A bridge between the Universe and the laboratory}\label{pt:beyond}

\def\chapterfolder{figures/QVA/}
\include{Chapters/QVA}

\def\chapterfolder{figures/SwitchEffects/}
\include{Chapters/Switch-on_and_-off_effects_in_particle_production}

\cleardoublepage


\part{Conclusions and coda}

\include{Chapters/Conclusions}

\include{Chapters/What_we_can_learn_from_analog_experiments}


\part*{Appendices}

\appendix

\cleardoublepage

\include{Chapters/AppNumericalParameters}

\include{Chapters/AppCosmoDM}

\def\chapterfolder{figures/Weakly-interacting_BECs_as_QFT_in_CS_simulators/}
\include{Chapters/AppBECTheory}


\cleardoublepage\include{FrontBackMatter/Bibliography}




\end{document}


%% file: classicthesis-config.tex
\PassOptionsToPackage{utf8}{inputenc}
  \usepackage{inputenc}

\PassOptionsToPackage{T1}{fontenc} 
  \usepackage{fontenc}

\PassOptionsToPackage{
  drafting=false,    
  tocaligned=false, 
  dottedtoc=true,  
  eulerchapternumbers=true, 
  linedheaders=false,       
  floatperchapter=true,     
  eulermath=false,  
  beramono=true,    
  palatino=true,    
  style=arsclassica 
}{classicthesis}

\newcommand{\myTitle}{Cosmological production of dark matter in the Universe and in the laboratory\xspace}

\newcommand{\myName}{\'Alvaro Parra L\'opez\xspace}
\newcommand{\myProf}{Luis Javier Garay Elizondo}

\newcommand{\mySupervisor}{José Alberto Ruiz Cembranos\xspace}
\newcommand{\myFaculty}{Faculty of Physical Sciences\xspace}

\newcommand{\myUni}{Complutense University of Madrid\xspace}

\newcommand{\myTime}{December 2024\xspace}
\newcommand{\myVersion}{\xspace}

\newcommand{\mpl}{\text{M}_{\text{P}}}

\newcommand{\amax}{a_{\text{max}}}
\newcommand{\amin}{a_{\text{min}}}

\newcommand{\af}{a_\text{f}}
\newcommand{\ai}{a_\text{i}}
\newcommand{\atwo}{a_{\text{II}}}
\newcommand{\ti}{t_{\text{i}}}
\newcommand{\tf}{t_{\text{f}}}

\newcommand\varpm{\mathbin{\vcenter{\hbox{%
  \oalign{\hfil$\scriptstyle+$\hfil\cr
          \noalign{\kern-.3ex}
          $\scriptscriptstyle({-})$\cr}%
}}}}
\newcommand\varmp{\mathbin{\vcenter{\hbox{%
  \oalign{$\scriptstyle({+})$\cr
          \noalign{\kern-.3ex}
          \hfil$\scriptscriptstyle-$\hfil\cr}%
}}}}

\usepackage{siunitx}
\usepackage{comment}
\usepackage{wrapfig}
\usepackage[dvipsnames]{xcolor}

\newcommand{\tcp}[1]{\mathcheck{\tilde{c}_\tp}}

\newcommand{\ci}{\mathrm{i}}

\newcommand{\tho}{t_{\text{hold}}}

\newcommand{\lr}[1]{\left(#1\right)}

\newcommand{\bs}[1]{\boldsymbol{#1}}
\newcommand{\bS}{\boldsymbol{S}}
\newcommand{\LN}{\text{LN}}

\definecolor{elinor}{rgb}{0.25,0.5,0.55}

\providecommand{\mLyX}{L\kern-.1667em\lower.25em\hbox{Y}\kern-.125emX\@}

\PassOptionsToPackage{spanish,es-lcroman,american}{babel}
    \usepackage{babel}

\usepackage{csquotes}
\PassOptionsToPackage{%
  backend=biber,bibencoding=utf8, 
  language=auto,%
  style=numeric-comp,%
  sorting=none, 
  maxbibnames=10, 
  backref=true,%
  natbib=true, 
  url=false}{biblatex}
    \usepackage[defernumbers=true]{biblatex}
\DeclareFieldFormat[unpublished]{title}{#1} 

\DeclareFieldFormat{usera}{#1}
\renewbibmacro*{begentry}{%
  \printfield{usera}%
}

\usepackage{physics}
\usepackage{amsmath}
\usepackage{braket}      
\usepackage{amsfonts,amsthm, amssymb} 
\usepackage{nicefrac}       
\usepackage{mathtools}

\renewcommand{\vec}[1]{\boldsymbol{#1}} 

\usepackage{bm}
\usepackage{bbm}

\usepackage[normalem]{ulem} 

\newcommand{\diff}{{\text{d}}}

\newcommand{\vx}{\bm{x}}
\newcommand{\vy}{\bm{y}}

\newcommand{\bk}{\bar{\bm{k}}}
\newcommand{\vk}{\bm{k}}
\newcommand{\dchi}{\chi^{\prime}}
\newcommand{\vp}{v^{\prime}}

\newcommand{\asi}{\alpha_{s,\text{i}}}
\newcommand{\asf}{\alpha_{s,\text{f}}}

\newcommand{\cs}{c_{\text{s}}}

\newcommand{\etai}{\eta_{\text{i}}}
\newcommand{\etaf}{\eta_{\text{f}}}
\newcommand{\etaon}{\eta_{\text{on}}}
\newcommand{\etaoff}{\eta_{\text{off}}}

\newcommand{\tA}{\tilde{A}}

\newcommand{\aB}{a_{\text{B}}}

\usepackage{dsfont}

\usepackage{graphicx} %
\usepackage{overpic}
\usepackage{scrhack} 
\usepackage{xspace} 
\usepackage{acronym} 

\usepackage{tabularx} 
\usepackage{caption}
\usepackage{subfig}

\usepackage{enumerate}

\usepackage{float} 

\usepackage[export]{adjustbox}
\newcolumntype{P}[1]{>{\centering\arraybackslash}p{#1}}
\newcolumntype{M}[1]{>{\centering\arraybackslash}m{#1}}

\usepackage{listings}
\usepackage{classicthesis}

\hypersetup{%
  colorlinks=true, linktocpage=true, pdfstartpage=1, pdfstartview=FitV,%
  breaklinks=true, pageanchor=true,%
  pdfpagemode=UseNone, %
  plainpages=false, bookmarksnumbered, bookmarksopen=true, bookmarksopenlevel=1,%
  hypertexnames=true, pdfhighlight=/O,
  urlcolor=CTurl, linkcolor=CTlink, citecolor=CTcitation, 
  pdftitle={\myTitle},%
  pdfauthor={\textcopyright\ \myName, \myUni, \myFaculty},%
  pdfsubject={},%
  pdfkeywords={},%
  pdfcreator={pdfLaTeX},%
  pdfproducer={LaTeX with hyperref and classicthesis},%
}

\makeatletter
\@ifpackageloaded{babel}%
  {%
    \addto\extrasamerican{%
    }%
    \addto\extrasngerman{%
    }%
    }{\relax}
\makeatother

\areaset[current]{400pt}{640pt} 
\usepackage{tikz} 

\usepackage{afterpage} 

\usepackage[bottom]{footmisc} 

\usepackage{lipsum}

\usepackage{cleveref}

\usepackage{overpic}

\usepackage{pdfpages}

%% file: FrontBackMatter/Titlepage.tex

\begin{titlepage}

\begin{addmargin}[-1cm]{-1.6cm}
\null\vspace{\stretch{1}} 
\vspace{0.1cm}
\begin{center}

\large

\textbf{Universidad Complutense de Madrid} \\
Facultad de Ciencias Físicas \\ \bigskip \bigskip

\normalsize

\vspace{0.1cm}

\includegraphics[width=4cm]{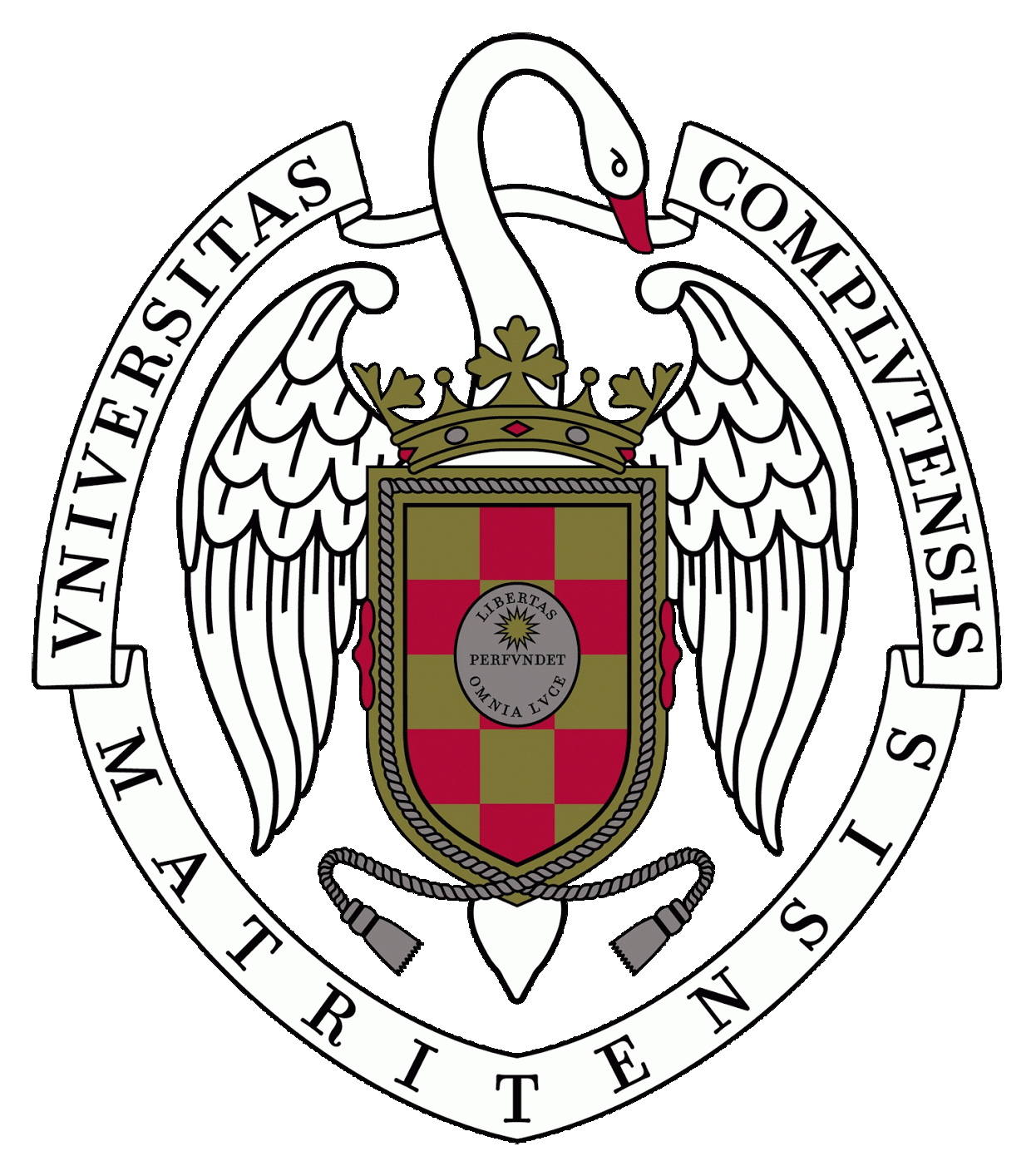} \\ \bigskip \bigskip 

\vspace{0.1cm}

PhD Thesis \\

\bigskip \bigskip


\begingroup
\color{Maroon}\large\spacedallcaps{Cosmological production of dark matter \\ in the universe and in the laboratory} \\ \bigskip 
\endgroup

\par\noindent\rule{\textwidth}{0.4pt}

\vspace{0.6cm}

\begingroup
\color{Maroon}\large\spacedallcaps{Producción cosmológica de materia oscura en el universo y en el laboratorio} \\ \bigskip 
\endgroup

\vspace{0.7cm}

Dissertation submitted for the degree of Doctor of Philosophy in Physics by

\vspace{0.7cm}

\Large\spacedlowsmallcaps{\myName} 

\normalsize

\vspace{0.7cm}


Under the supervision of \\

\vspace{0.7cm}

\Large\spacedlowsmallcaps{ \mySupervisor \\ \bigskip \myProf }\normalsize

\bigskip\bigskip\bigskip 

Madrid, \myTime\  \myVersion 

\vspace*{\stretch{1}} 
    		\null

\end{center}
\end{addmargin}
\vspace*{\fill}
\end{titlepage}

%% file: FrontBackMatter/Titleback.tex

\thispagestyle{empty}

\hfill

\bigskip

\vfill

This thesis was handed in on the 6th of November 2024, and was defended on the 13th of December 2024. After the defense, I kept updating the text to newer versions, listed below:

\begin{itemize}
    \item 10/02/2026 --- First upload to arXiv.
    \begin{enumerate}
        \item Some typos in equations and the text have been fixed, and some sentences have been slightly rewritten in order to improve clarity and readability, without changing their content.
        \item I included some footnotes pointing towards some references for further details, such as the one on page 30 regarding the volume term in \cref{eq:qftcs.NumberOperatorExpectation} and the definition of the number density per mode.  
        \item The bibliography has been updated, as well as the publications directly connected to the content of the thesis. 
        \item I have added a translation of the acknowledgement section to English. 
        \item An additional footnote on page 31 has been added to clarify a point regarding adiabaticity.
        \item In \cref{ch:switcheffects}, I have included a sentence clarifying the effects of switch-on and -off processes in the context of BEC analog experiments (\cref{fig:switch.SpectraSwitch}): \textit{Here, we observe that the effect of slightly decreasing the duration of an already short switch-on and -off is subtle on $A_k$, which tells us that the limit $\delta\to 0$ is achieved fast}. I have slightly modified the conclusions to that end as well: \textit{the total amount of particles produced is not significantly changed by shortening the duration of switch-on and -off processes, as long as they are fast enough not to perturb the process in between (i.e., the limit $\delta \to 0$ is achieved fast)}.
        \item What used to be \cref{fig:ds.SpectraCurvedLargeK} has been split in two for better clarity and presentation. The labels have been corrected, and the curve for positive curvature is depicted in discrete form.
        \item By accident, the old \cref{fig:ds.SpectraCurvedLargeK} featured the results with the ILES vacuum, which has now been corrected to show the results with the adiabatic vacuum.
    \end{enumerate}
\end{itemize}

\vfill

\noindent\myName\textit{\myTitle} 
\textcopyright\ \myTime


%% file: FrontBackMatter/Dedication.tex
\thispagestyle{empty}
\phantomsection
\pdfbookmark[1]{Dedication}{Dedication}

\vspace*{3cm}

\begin{center}
    \textit{How do you prove that you exist...? \\ Maybe we don't exist.} \\ \medskip
    --- Vivi Ornitier
\end{center}


\bigskip

\begin{center}
   \emph{A mis padres, Mari Carmen y Urbano.}  \\ \medskip
   \emph{Pero sobre todo, a Gloria.}
\end{center}

%% file: FrontBackMatter/Agradecimientos.tex

\pdfbookmark[1]{Agradecimientos}{Agradecimientos} 


\begingroup

\let\clearpage\relax
\let\cleardoublepage\relax
\let\cleardoublepage\relax

\chapter*{Agradecimientos}

Sinceramente, creo que tengo mucha suerte.

Tengo suerte porque esto no es una despedida (no por ahora, al menos). Escribir estos agradecimientos es solamente el final de un proyecto, el de mi tesis, pero no el cierre de una etapa. Me siento afortunado porque voy a poder seguir disfrutando de la vida con cada una de las personas que me han acompañado a lo largo de estos años durante unos doce meses más, y esa oportunidad no la tiene todo el mundo después de la defensa. Seguiré viniendo todas las mañanas al claramente mejor despacho del Departamento de Física Teórica de la Universidad Complutense de Madrid, y bajaremos a desayunar juntos, para poder tomarnos ese café que está entre regular y asqueroso, según el día. Y es que durante todo el tiempo que he estado viviendo en Madrid, pero especialmente a lo largo de los últimos meses, he aprendido mucho, no sólo de Física, sino de qué es importante en la vida. Está claro: vosotros; y ahora quiero aprovechar esta oportunidad para contaros a todos ---amigos, familia, Gloria--- cuánto os aprecio. Si estás leyendo esto, o tienes mucha curiosidad, o esperas que te haga mucha ilusión ver tu nombre aquí escrito. En cualquier caso, allá va esta pequeña historia, de la que seguramente formas parte.

Yo no esperaba estar aquí. No esperaba estar en Madrid, ni escribiendo unos agradecimientos en español. Cuando estaba en el último año del Grado (curso~18/19) me fui a Oslo ---¿sabíais que estuve de Erasmus en Noruega? Tuve mucha suerte, porque mi profesor de \textit{Modern Quantum Mechanics}, Joakim Bergli, decidió hacer un proyecto conmigo que supuso mi inicio en la investigación (¡gracias, Joakim!). Le cogí el gusto, así que perseguí mi idea de hacer un máster fuera de España (porque yo \textit{pensaba} que este país no valía nada) y me fui a Heidelberg. Al principio, Alemania pintaba relativamente bien (sólo al cabo de unos meses me di cuenta de que la mayoría de mis amigos eran españoles...), hasta que llegó la pandemia, a los seis meses de empezar el máster. Durante el año y medio restante, todo fue online, estuvimos encerrados prácticamente el segundo curso entero (20/21), y empecé a echar de menos el solecito y las tapitas. Aún así, fue en Heidelberg donde conocí a Gloria, así que eso lo compensa todo. Como desde siempre nos han dicho que es imposible hacer el doctorado en España, mi intención era quedarme en Alemania, ¡y me salió una oferta y todo! ¿Qué hice yo? Rechazarla. Recuerdo perfectamente estar tumbado en la cama mirando al techo, sopesando pros y contras, cuando tomé la decisión de no hacer el doctorado en Alemania. Por si acaso, había pedido la FPU con Jose (Cembranos), así que mi apuesta fue mudarme a Madrid sin saber si me la iban a conceder, y esperar a que el Ministerio decidiese publicar la resolución. Así de grandes eran las ganas que tenía de venir a vivir aquí. Como decía, tengo mucha suerte, así que finalmente me dieron la FPU, y aquí estamos.

El primero al que tengo que agradecer es, precisamente, a Jose, porque confió en mí desde el principio, y me dio la oportunidad de hacer la tesis con él sin conocerme de nada, y viniendo de fuera. Supongo que pensó que se me había ido la olla cuando le conté que había rechazado lo de Alemania, aún sabiendo que no era nada seguro obtener la FPU. Al poco tiempo de empezar la tesis, Luis (Garay) se apuntó. Creo que le caí bien en persona, porque yo le había mandado un correo hacía un año (¡antes de pedir la FPU con Jose!), preguntándole por la posibilidad de hacer la tesis con él, al que me contestó sutilmente que no (por falta de financiación, pienso). No le tengo rencor. Yo lo tengo muy claro: creo que sois los dos mejores directores de tesis que podría haber tenido, ya lo siento por el resto. Además, os complementáis magistralmente. Me habéis dado guía cuando estaba perdido y libertad cuando la necesitaba, y habéis aguantado mis mil preguntas, solicitudes de correcciones y opiniones (soy un poco perfeccionista, lo siento), y me habéis enseñado muchísimo. No sólo de Física, sino de cómo ser una buena persona. Ojalá todos los profesores del mundo fuesen como vosotros. 

Irme de Murcia (sí, es que se me pegan los acentos enseguida, ahora parezco del Barrio del Pilar de toda la vida) significaba alejarme de mi familia, y no siempre es fácil encontrar sustitutos. La verdad es que a veces me arrepiento un poco de irme, pero, siendo sinceros, me arrepentiría de lo contrario igualmente. Así que me hace feliz poder hablaros de mi otra familia, los miembros del despacho-biblioteca, \textit{Biblioprecarios}. Gracias a Dani, por preguntar todos y cada uno de los días: \textit{"¿cómo andamos?"}. Me sorprende y maravilla tu interés genuino por preocuparte de tus amigos, y también que se nos fuese tanto de madre lo de los patos. Gracias a Óscar, por compartir la afición por el mejor deporte del mundo (el pádel) y por mudarte al mejor barrio de Madrid, dándome la razón. Me da pena que no puedas estar en mi defensa, pero unos días después celebramos eso, y tu vuelta de Hamburgo. Gracias a Sara (Piloñeta), porque sabes perfectamente quién dijo la cita del principio de la tesis, y por hacer que me sienta más arropado en el despacho (y no sólo porque ya no hay tanta distancia entre mí y los demás). No sé bien por qué, pero cuando te ríes me haces muy feliz. Gracias a Toni, por ser el mejor Toni del mundo, y por saludarme sólo a mí algunos días cuando entras en el despacho medio en trance. Creo que no puedo contar una anécdota tuya sin reírme a carcajadas.

A veces, más allá de la familia nuclear, con la que normalmente se convive, uno tiene, por ejemplo, primos con los que se lleva especialmente bien, como si fuesen hermanos. En mi caso, en los despachos de alrededor, encontrábamos a Alfredo. Segons la llargada del teu cabell, podem ser aproximadament la mateixa persona/xef. Gràcies per estar literalment sempre aquí, i per atrevir-te a començar una interessant aventura musical amb mi (¡maita!). Espere que no em faltes mai. Gracias a Álvaro, la persona con la que más he disfrutado haciendo Física en mi vida, con diferencia. Trabajar, como concepto, es terrible, pero cuando nos tirábamos tres horas hablando en la pizarra de nuestras movidas se me pasaba el tiempo volando. Si seguimos este camino, espero tenerte siempre a mi lado. Gracias a Isi, por ser el mejor mentor/papá pato que se puede tener. Déjame decirte que, aunque ya no estás tan cerca, nos has enseñado a andar solos. Los seminarios-aperitivos van viento en popa. Gracias a Lucía, por hacerme replantearme si merece la pena la Academia. Es broma. O no, porque es importante que la gente te haga pensar en si lo que haces tiene sentido, a veces el camino es muy complicado. Ánimo con lo de ser influencer. Gracias a Rita, por animarte a tocar la guitarra en un local de mala muerte, y grabar la mejor maqueta de pop español de la historia. Y chicos, ¡a verdade, é portuguesa! Desculpa por revelar o teu segredo. Gracias a Sara (Giordano), por ayudar a crear esta familia, por ser tan buena amiga y cuidar tanto a la gente. Me hace muy feliz haberte podido dar ese empujoncito para que empezases a cantar. Cuando dices que te gustaría vivir en Madrid en el futuro se me alegra el corazón. 
Grazie a Sara (Giordano), per aver contribuito a creare questa famiglia, per essere una così buona amica e prenderti così cura delle persone. Mi rende molto felice averti potuto dare quella piccola spinta per iniziare a cantare. Quando dici che ti piacerebbe vivere a Madrid in futuro, il mio cuore si riempie di gioia. Pero aquí somos muchos primos. También están Álvaro (2000), Darío, Giovanni, Jose (Canario), Juanjo, Mercè, Nacho, Pablo, Patri, Rafa, ...

En un párrafo aparte, porque está unos cuantos pasillos más allá, tenemos a Sara (de Scals). En serio, ¿no había más nombres? Mira que hay muchos Álvaros, pero... En fin, que muchas gracias a Sara por dejarme comerte la oreja (a veces diariamente) en nuestras pausitas de café durante casi dos años ya. Joder, qué rápido pasa el tiempo. Gracias por sorprenderte cada vez que me pongo a hablar en catalán inventado después de unas cervezas. Siento que puedo compartírtelo todo, y eso es muy importante para mí. Menos mal que te encanta Madrid, así sabré dónde encontrarte siempre.
De veritat, no hi havia més noms? Ja hi ha molts Álvaros, però... En fi, moltes gràcies a la Sara per deixar-me donar-te la tabarra (de vegades diàriament) en les nostres pauses de cafè durant gairebé dos anys ja. Com passa de ràpid el temps... Gràcies per sorprendre't cada vegada que em pose a parlar en català inventat després d'unes cerveses. Sent que et puc compartir tot, i això és molt important per a mi. Sort que t'encanta Madrid, així sabré on trobar-te sempre. En mi línea de buena suerte, resulta que tengo en Madrid (también a largo plazo) a Ana y Mercedes (que son de Murcia). Gracias por estar a mi lado durante tantos (tantísimos, Ana, ¿desde los ocho?) años, por acogerme en Madrid cuando aún no vivía aquí, y por mudaros a veinte minutos andando, en el mejor barrio, como decía antes. Por supuesto, también tengo que agradecer a mi familia madrileña adoptiva (otro concepto distinto). Gracias a Gloria (Platero) y a Jose Manuel por cuidarme desde el minuto uno y preocuparse por mí como si fuese uno más del núcleo familiar. Y dejadme que termine con Madrid dando las gracias a unas personitas que, aún siendo amigas de Gloria, me acogieron en su ciudad como si fuese uno más. En especial, tengo en mente a Blanca, Carlos (Reifarth), Eva, Isa, Iván, Jorge, Marta, Miguel (Hernández, sí), Pablo y Richi.

Muchas gracias a Adrià, Bea e Iván, por acogerme y cuidarme en las pantanosas tierras de Louisiana, y por enseñarme tanta Física. Thanks to the Heidelberg and Jena people, for such wonderful science and collaborations, especially to Stefan, who has taught me so much since 2020. Per descomptat, vull donar les gràcies a la Mireia. A més de que estic súper a gust treballant amb tu, sé que sempre puc comptar amb tu per a qualsevol cosa, amiga. To those who are still there since Oslo, despite space and time: Ale, Amalia, Fede and Marianna. 

Como no podía ser de otra forma, siguen ahí aquellos que disfrutamos y sufrimos el Grado en Física en la Universidad de Murcia, especialmente Antonio, Josué, Luke y Quini. Gracias de todo corazón por mantener vivos esos recuerdos, espero que podamos seguir creando nuevos en los próximos años. Además, cada vez que vuelvo a Molina, ahí están Alfonso, Álvaro y Paco, que se han convertido en un refugio, cual manta calentita y estufa en invierno. A veces logro coincidir con Pablo (Carpio), por ejemplo, que igual que yo, regresa a Murcia desde Viena cuando puede. Volver a Molina también significa ensayar con Staring (Into The Woods), que me da la vida. Cómo os eché de menos, Javi, Vicen (y otra vez Álvaro), cuando estaba fuera. Estos años me he dado cuenta de lo necesario que es que mi vida esté de algún modo ligada a la música. Curiosamente, también he encontrado en Madrid a otra personita con la que compartir esta pasión, Mariaje. ¡Dime, por favor, qué posibilidad había de que tú también fueses de Molina! Gracias a todos por hacerme sentir tan bien.

He dejado para el final a los pilares fundamentales de mi vida. Por un lado, mi familia (la de verdad), en especial mis padres, Mari Carmen y Urbano. Mis modelos a seguir en prácticamente todo, con los que pocas veces estoy en desacuerdo, y que me han apoyado con todos mis sueños a lo largo de cada uno de los pasos que he dado en la vida. Gracias por enseñarme a ser una buena persona. Me alegro de que me hayáis sentido más cerca estos años, aunque realmente estaba igual de lejos en coche que antes en avión. Además, quiero dar las gracias a mi hermana, Mari Carmen (Parra), porque sé que estará ahí cuando lo necesite. También quiero mencionar a mi prima Margarita, que siempre ha encontrado el huequito para llamar por teléfono, estuviese en Oslo, Heidelberg, Baton Rouge o Madrid. Por último, quiero dar las gracias a Gloria. Yo no sería la persona que soy hoy si no te hubiese conocido. Quizá suene muy típico, pero lo digo en el sentido más puro y literal posible. Desde que te conocí, no has dejado de enseñarme. Gracias a ti, he aprendido a ser crítico conmigo mismo, a darme cuenta de mis errores, reflexionar y llegar a ser mejor persona. Me has enseñado sobre la vida, sobre política, me has demostrado cuáles son las cosas que realmente importan en este mundo. Me has enseñado a cuidar a las personas a las que quiero, a ser más empático, y a ser más feliz. Gracias por estar a mi lado durante todos estos años. Te quiero muchísimo. 

Ya habéis visto que hay poca Física en estos agradecimientos, pero es que una lección muy importante que he aprendido durante el doctorado es que para hacer ciencia, como para toda actividad humana, lo más importante son las personas. Además, tenéis unas doscientas páginas de tesis para el que las quiera leer, así que tampoco se puede quejar nadie. Yo creo que así ya estaría bien, ¿no? 

\endgroup

%% file: FrontBackMatter/Acknowledgements.tex

\pdfbookmark[1]{Acknowledgements}{Acknowledgements} 


\begingroup

\let\clearpage\relax
\let\cleardoublepage\relax
\let\cleardoublepage\relax

\chapter*{Acknowledgements}

\textit{I translated the acknowledgements to English so that my non-Spanish-speaking friends and colleagues can read them. Have fun.}

\vspace{5mm}

Honestly, I think I am very lucky.

I am lucky because this is not a farewell (not for now, at least). Writing these acknowledgements is only the end of a project, that of my thesis, but not the closing of a stage. I feel fortunate because I'll get to keep enjoying life with everyone who's been with me through these years for about twelve more months, and not everyone has that opportunity after the defense. I will keep coming every morning to what is clearly the best office of the Department of Theoretical Physics at the Complutense University of Madrid, and we will go downstairs to have breakfast together, to drink that coffee that ranges from just okay to disgusting, depending on the day. And it is that during all the time I have been living in Madrid, but especially over the last few months, I have learned a lot, not only about physics but about what is important in life. It is clear: you; and now I want to take this opportunity to tell you all---friends, family, Gloria---how much I appreciate you. If you are reading this, either you are very curious, or hope to be delighted to see your name written here, in any case, here goes this little story, of which you surely are a part.

I did not expect to be here. I did not expect to be in Madrid, nor writing acknowledgements in Spanish. When I was in the last year of my Bachelor's degree (course 18/19) I went to Oslo---did you know I was an Erasmus student in Norway? I was very lucky because my professor of Modern Quantum Mechanics, Joakim Bergli, decided to do a project with me that marked my start in research (thanks, Joakim!). I liked it, so I pursued my idea of doing a master's degree outside Spain (because I thought this country was worthless) and I went to Heidelberg. At first, Germany looked relatively good (only after a few months did I realize most of my friends were Spanish...), until the pandemic arrived six months after starting the master's. During the remaining year and a half, everything was online, we were practically confined for the entire second year (20/21), and I started missing the sunshine and the tapas. Still, it was in Heidelberg where I met Gloria, so that compensates for everything. Since we have always been told it is impossible to do a PhD in Spain, my intention was to stay in Germany, and I even got an offer! What did I do? I rejected it. I remember lying on my bed looking at the ceiling, weighing pros and cons, when I decided not to do the PhD in Germany. Just in case, I had applied for the FPU fellowship with Jose (Cembranos), so my bet was to move to Madrid not knowing if they would grant it, and wait for the Ministry to publish the resolution. That is how badly I wanted to live here. As I said, I am very lucky, so they finally granted me the FPU, and here we are.

The first person I have to thank is precisely Jose, because he trusted me from the beginning and gave me the chance to do the thesis with him without knowing me at all, and coming from abroad. I suppose he thought I was crazy when I told him I had rejected Germany, even knowing it was not certain I would get the FPU. Shortly after starting the thesis, Luis (Garay) joined. I think he liked me in person because I had sent him an email a year ago (before applying for the FPU with Jose!), asking about the possibility of doing my thesis with him, to which he subtly said no (due to lack of funding, I think). I bear no grudge. I have it very clear: I think you are the two best thesis supervisors I could have had, sorry to the rest. Moreover, you complement each other masterfully. You gave me guidance when I was lost and freedom when I needed it, and tolerated my thousand questions, correction requests, and opinions (I am a bit of a perfectionist, sorry), and you have taught me a lot. Not only about physics but about how to be a good person. I wish all professors in the world were like you.

Leaving Murcia (yes, I pick up accents quickly, now I seem like I am from Barrio del Pilar for life) meant distancing myself from my family, and substitutes are not always easy to find. Honestly, sometimes I regret leaving a bit, but to be honest, I would regret the opposite anyway. So I am happy to tell you about my other family, the members of the office-library, Biblioprecarios. Thanks to Dani for asking every single day: "How are we doing?" I am surprised and amazed by your genuine interest in caring for your friends, and also that the duck situation got so out of hand. Thanks to Óscar for sharing the passion for the best sport in the world (pádel) and for moving to the best neighborhood in Madrid, proving me right. I am sad you cannot be at my defense, but a few days later we celebrate that and your return from Hamburg. Thanks to Sara (Piloñeta), because you perfectly know who said the quote at the beginning of the thesis, and for making me feel more supported in the office (not only because there is no longer such a distance between me and others). I don't know why, but when you laugh you make me very happy. Thanks to Toni for being the best Toni in the world and for greeting only me some days when you enter the office half in a trance. I think I cannot tell a single anecdote about you without laughing out loud.

Sometimes, beyond the nuclear family we live with, one has cousins with whom they get especially along with, like brothers. In my case, in the surrounding offices, we found Alfredo. Depending on your hair length, we might roughly be the same person/chef. Thanks for literally always being here, and for daring to start an interesting musical adventure with me (maita!). I hope you never leave me. Thanks to Álvaro, the person with whom I have enjoyed doing physics the most in my life, by far. Work, as a concept, is terrible, but when we would spend three hours talking on the blackboard about our stuff, time flew by. If we continue on this path, I hope to always have you by my side. Thanks to Isi for being the best mentor/duck-dad one can have. Let me tell you that, although you are not so close anymore, you have taught us to walk on our own. The seminar-aperitifs are going great. Thanks to Lucía for making me reconsider if academia is worth it. Just kidding. Or not, because it is important that people make you think if what you do makes sense; sometimes the path is very complicated. Good luck with becoming an influencer. Thanks to Rita for daring to play guitar in a crappy practice room and recording the best demo of Spanish pop history. And guys, she is actually Portuguese! Sorry for revealing your secret. Thanks to Sara (Giordano), for helping create this family, for being such a good friend and caring so much for people. It makes me very happy to have been able to give you that little push to start singing. When you say you would like to live in Madrid in the future, my heart rejoices. But here we are many cousins. Also, Álvaro (2000), Darío, Giovanni, Jose (the Canarian), Juanjo, Mercè, Nacho, Pablo, Patri, Rafa, ...

In a separate paragraph, because she is a few corridors away, we have Sara (de Scals). Seriously, were there no more names? There are many Álvaros, but... Anyway, many thanks to Sara for letting me bug you (sometimes daily) during our coffee breaks for almost two years now. Damn, how fast time flies. Thanks for being surprised every time I start speaking invented Catalan after a few beers. I feel I can share everything with you, and that is very important to me. Luckily you love Madrid; that way, I will always know where to find you. In my streak of good luck, it turns out that I have in Madrid (also long term) Ana and Mercedes (who are from Murcia). Thanks for being by my side for so many (so many, Ana, since we were eight?) years, for welcoming me in Madrid when I still didn't live here, and for moving at a twenty minutes walking distance, to the best neighborhood, as I said before. Of course, I also have to thank my adoptive Madrid family (a different concept). Thanks to Gloria (Platero) and Jose Manuel for taking care of me from minute one and worrying about me as if I were one more of the family. And let me finish with Madrid by thanking some lovely people who, even being friends of Gloria, welcomed me in their city as if I were one more. Especially, I have in mind Blanca, Carlos (Reifarth), Eva, Isa, Iván, Jorge, Marta, Miguel (Hernández, yes), Pablo and Richi.

Many thanks to Adrià, Bea and Iván for welcoming and taking care of me in the swampy lands of Louisiana, and for teaching me so much Physics. Thanks to the Heidelberg and Jena people, for such wonderful science and collaborations, especially to Stefan, who has taught me so much since 2020. Of course, I want to thank Mireia. Besides the fact that I am super comfortable working with you, I know I can always count on you for anything, friend. To those who are still there since Oslo, despite space and time: Ale, Amalia, Fede and Marianna.

As it could not be otherwise, those who enjoyed and suffered the Physics Degree at the University of Murcia are still there, especially Antonio, Josué, Luke and Quini. Thanks from the bottom of my heart for keeping those memories alive; I hope we can keep creating new ones in the coming years. Also, every time I return to Molina, there are Alfonso, Álvaro and Paco, who have become a refuge, like a warm blanket and heater on a winter's day. Sometimes I manage to coincide with Pablo (Carpio), for example, who, like me, returns to Murcia from Vienna whenever he can. Returning to Molina also means rehearsing with Staring (Into The Woods), which gives me life. How I missed you, Javi, Vicen (and Álvaro again), when I was away. These years I have realized how necessary it is that my life is somehow linked to music. Interestingly, I have also found in Madrid another wonderful person with whom to share this passion, Mariaje. Please tell me, what chance was there that you were also from Molina! Thanks to all of you for making me feel so good.

I have saved for last the fundamental pillars of my life. On one hand, my family (the real one), especially my parents, Mari Carmen and Urbano. My role models in practically everything, with whom I rarely disagree, and who have supported me with all my dreams throughout every step I have taken in life. Thanks for teaching me to be a good person. I am glad you have felt closer to me these years, although I was really just as far by car as before by plane. Also, I want to thank my sister, Mari Carmen (Parra), because I know she will be there when I need her. I also want to mention my cousin Margarita, who has always found a little space to call, whether I was in Oslo, Heidelberg, Baton Rouge or Madrid. Finally, I want to thank Gloria. I would not be the person I am today if I hadn't met you. It may sound very typical, but I say it in the purest and most literal sense possible. Since I met you, you have not stopped teaching me. Thanks to you, I have learned to be critical with myself, to realize my mistakes, reflect and become a better person. You have taught me about life, about politics, you have shown me what things really matter in this world. You have taught me to take care of the people I love, to be more empathetic, and to be happier. Thanks for being by my side all these years. I love you very much.

You have seen that there is little Physics in these acknowledgements, but a very important lesson I have learned during the PhD is that to do science, as for all human activity, the most important thing is the people. Besides, you have about two hundred pages of thesis for anyone who wants to read them, so no one can complain. I think this is enough, right?

\endgroup

%% file: FrontBackMatter/Contents.tex
\pagestyle{scrheadings}
\pdfbookmark[1]{\contentsname}{tableofcontents}
\setcounter{tocdepth}{2} 
\setcounter{secnumdepth}{3} 
\manualmark
\markboth{\spacedlowsmallcaps{\contentsname}}{\spacedlowsmallcaps{\contentsname}}
\tableofcontents
\automark[section]{chapter}
\renewcommand{\chaptermark}[1]{\markboth{\spacedlowsmallcaps{#1}}{\spacedlowsmallcaps{#1}}}
\renewcommand{\sectionmark}[1]{\markright{\textsc{\thesection}\enspace\spacedlowsmallcaps{#1}}}
\clearpage
\begingroup
    \let\clearpage\relax
    \let\cleardoublepage\relax
    \phantomsection
    \addcontentsline{toc}{chapter}{\tocEntry{\listfigurename}}
    \listoffigures

    \vspace{8ex}






	\newpage 
    \phantomsection
    \addcontentsline{toc}{chapter}{\tocEntry{Acronyms}}
    \markboth{\spacedlowsmallcaps{Acronyms}}{\spacedlowsmallcaps{Acronyms}}
    \chapter*{Acronyms}
    \begin{acronym}[QFTCS]
        \acro{QFT}{Quantum Field Theory}
        \acro{QFTCS}{Quantum Field Theory in Curved Spacetimes}
        \acro{BH}{Black Hole}
        \acro{CMB}{Cosmic Microwave Background}
        \acro{BEC}{Bose-Einstein condensate}
        \acro{LCDM}[$\Lambda$CDM]{Lambda cold dark matter}
        \acro{FLRW}{Friedmann-Lemaître-Robertson-Walker}
        \acro{WIMP}{Weakly Interacting Massive Particle}
        \acro{KG}{Klein-Gordon}
        \acro{ILES}{Instantaneous lowest energy state}
        \acro{GP}{Gross-Pitaevskii}
    \end{acronym}      
                   
\endgroup

%% file: FrontBackMatter/Abstract.tex

\phantomsection
\addcontentsline{toc}{chapter}{\tocEntry{Abstract}}
\markboth{\spacedlowsmallcaps{Abstract}}{\spacedlowsmallcaps{Abstract}}

\begingroup
\let\clearpage\relax
\let\cleardoublepage\relax
\let\cleardoublepage\relax

\chapter*{Abstract}

In this thesis, we investigate the phenomenon of cosmological particle production within the framework of Quantum Field Theory in Curved Spacetimes, which can be understood as a dark matter production mechanism, and explore its analog simulation using Bose-Einstein condensates. The intersection of General Relativity and Quantum Field Theory has posed significant challenges, particularly in formulating a consistent theory of Quantum Gravity. Nevertheless, the study of quantum fields propagating on a classical, curved spacetime background provides insights into quantum effects in the early Universe and near compact objects.

The primary focus of this work is on cosmological particle production, a process by which particles are generated even out of the vacuum due to the dynamical nature of spacetime, especially during periods of non-adiabatic changes of the geometry such as inflation. This mechanism, crucial in the early Universe, may potentially explain the origin of dark matter particles. With this aim, we investigate particle production from spectator fields, which are not coupled directly to the inflaton or other fields, but influenced solely by the background geometry.

This thesis is structured into four main parts. Part~I lays the theoretical foundation, introducing the Standard Model of Cosmology, inflationary theory, Quantum Field Theory in Curved Spacetimes, quantum vacuum ambiguities, and the concept of analog gravity, focusing on Bose-Einstein condensates as analog platforms. Part~II examines cosmological particle production in realistic and toy inflationary models, analyzing scalar and vector spectator fields in different scenarios, including the impact of spatial curvature. Our findings suggest that cosmological production can naturally account for the observed dark matter abundance without additional physics, and highlight the efficiency of production due to tachyonic instabilities during the oscillatory phase of the Ricci scalar. Part~III delves into analog gravity experiments, establishing a correspondence between phonons in Bose-Einstein condensates and massless scalar fields in various Friedmann-Lemaître-Robertson-Walker universes. We present experimental evidence that demonstrates the ability to reconstruct the expansion history from the spectrum of fluctuations, reinterpret cosmological production in terms of one-dimensional scattering, and propose methods to measure entanglement between produced pairs. These experiments validate the analog gravity approach and provide insights into the quantum nature of particle production. Part~IV synthesizes theoretical and experimental perspectives, addressing quantum vacuum ambiguities in cosmological production and the effects and relevance of the switch-on and -off of the expansion processes on particle creation. We propose a new family of vacua and analyze the relevance of non-adiabatic transitions from and toward static regions in particle production, confirming that production is most important during these intermediate regimes. These studies illustrate how analog simulations allow us to understand both Quantum Field Theory in Curved Spacetimes and cosmological particle production from a new perspective.

Through this work, we aim to highlight the interest of cosmological particle production in the context of dark matter, as well as underscoring the value of analog gravity experiments in enhancing our understanding of Quantum Field Theory in Curved Spacetimes and cosmological particle production. Future research directions include exploring Higgs inflation, the effects of spatial curvature in more realistic models, and further studies of entanglement properties in both cosmological and analog contexts.

\endgroup			

\vfill

%% file: FrontBackMatter/Resumen.tex

\phantomsection

\addcontentsline{toc}{chapter}{\tocEntry{Resumen}}
\markboth{\spacedlowsmallcaps{Resumen}}{\spacedlowsmallcaps{Resumen}}

\begingroup
\let\clearpage\relax
\let\cleardoublepage\relax
\let\cleardoublepage\relax

\chapter*{Resumen}

En esta tesis, investigamos el fenómeno de producción cosmológica de partículas dentro del marco de la Teoría Cuántica de Campos en Espaciotiempos Curvos, un mecanismo que puede ser entendido como una forma de producción de materia oscura, y exploramos su simulación análoga mediante condensados de Bose-Einstein. La intersección entre la Relatividad General y la Teoría Cuántica de Campos ha planteado importantes retos, especialmente en la formulación de una teoría consistente de Gravedad Cuántica. Sin embargo, el estudio de campos cuánticos que se propagan en un espaciotiempo curvo clásico permite comprender efectos cuánticos tanto en el Universo primitivo como cerca de objetos compactos.

Este trabajo se centra principalmente en la producción cosmológica de partículas, un proceso mediante el cual se generan partículas incluso a partir del vacío debido a la naturaleza dinámica del espaciotiempo, particularmente durante periodos de cambios no adiabáticos de la geometría, como ocurre durante inflación. Este mecanismo, crucial en el Universo temprano, podría explicar el origen de las partículas de materia oscura. Con este objetivo, investigamos la producción de partículas a partir de campos espectadores, que no están directamente acoplados al inflatón ni a otros campos, sino que se ven influenciados únicamente por la geometría de fondo.

Esta tesis se estructura en cuatro partes principales. La parte I establece las bases teóricas, introduciendo el Modelo Cosmológico Estándar, la teoría inflacionaria, la Teoría Cuántica de Campos en Espaciotiempos Curvos, las ambigüedades del vacío cuántico y el concepto de gravedad análoga, centrándose en los condensados de Bose-Einstein como plataformas análogas. En la parte II, analizamos la producción cosmológica de partículas en modelos inflacionarios tanto realistas como de juguete, estudiando campos espectadores escalares y vectoriales en diferentes escenarios, teniendo en cuenta el impacto de la curvatura espacial. Nuestros resultados sugieren que la producción cosmológica puede explicar de forma natural la abundancia de materia oscura observada sin física adicional, además de señalar la importancia de las inestabilidades taquiónicas durante la fase de oscilación del escalar de Ricci, que incrementa la eficiencia de la producción. En la parte III, profundizamos en los experimentos de gravedad análoga, estableciendo una correspondencia entre fonones en condensados de Bose-Einstein y campos escalares sin masa en distintos universos de Friedmann-Lemaître-Robertson-Walker. Además, presentamos resultados experimentales que demuestran cómo es posible obtener información de la evolución de la geometría a partir del espectro de fluctuaciones, reinterpretamos la producción cosmológica en términos de un problema de dispersión en potenciales unidimensionales y proponemos métodos para caracterizar el entrelazamiento entre pares producidos. Estos experimentos validan el enfoque de la gravedad análoga y aportan información sobre la naturaleza cuántica de las partículas producidas. La parte IV sintetiza las perspectivas teóricas y experimentales, abordando las ambigüedades del vacío cuántico en la producción cosmológica y los efectos y relevancia del encendido y apagado de los procesos de expansión en la creación de partículas. Proponemos una nueva familia de vacíos y analizamos la importancia de las transiciones no adiabáticas desde y hacia un periodo de estaticidad para la producción de partículas, confirmando que estos regímenes intermedios son más relevantes de cara a la producción. Estos estudios ilustran cómo las simulaciones análogas permiten entender tanto la Teoría de Campos en Espaciotiempos Curvos como la producción cosmológica de partículas desde una nueva perspectiva.

A través de este trabajo, pretendemos destacar el interés de la producción cosmológica de partículas en el contexto de la materia oscura, así como subrayar el valor de los experimentos de gravedad análoga para mejorar nuestra comprensión de la Teoría Cuántica de Campos en Espaciotiempos Curvos y la producción cosmológica de partículas. Las futuras líneas de investigación incluyen la exploración de la inflación de Higgs, el estudio de los efectos de la curvatura espacial en modelos más realistas, así como la continuación del análisis de las propiedades de entrelazamiento tanto en contextos cosmológicos como análogos.

\endgroup			

\vfill

%% file: FrontBackMatter/Publications.tex




\pagestyle{scrheadings}
\cleardoublepage
\phantomsection
\addcontentsline{toc}{chapter}{\tocEntry{Publications}}
\markboth{\spacedlowsmallcaps{Publications}}{\spacedlowsmallcaps{Publications}}

\chapter*{Publications} 

\phantom{
\cite{CosmologyPaper2022}
\cite{BECPaper2022}
\cite{Experiment2022}
\cite{ScalarField2023}
\cite{QVA2023}
\cite{VectorDM2024}
\cite{ScatteringTheory2024}
\cite{ScatteringExp2024}
\cite{Entanglement2024}
\cite{SwitchEffects2024}
\cite{CurvedDM2024}
\cite{Starobinsky2024}
}

Most ideas presented in this thesis belong to the following works:\\


\newrefcontext[sorting=none]
\printbibliography[heading=none, keyword=myPapers]

\noindent The authors are listed in alphabetical order in the publications marked with an asterisk \textcolor{teal}{\scalebox{1.5}{*}}.

%% file: Chapters/Introduction.tex

\addtocontents{toc}{\protect\vspace{\beforebibskip}}

\chapter{Introduction} 

\label{ch:introduction} 




During the last century, we have witnessed a revolution in our understanding of the Universe. General Relativity has successfully passed numerous tests and has become the standard theory of gravity. On the other hand, Quantum Field Theory (QFT), which describes elementary particle interactions in flat spacetime, has been the cornerstone of the Standard Model of Particle Physics, and has been able to explain most of the phenomena observed in the laboratory. However, the combination of these two theories has been a challenge. The formulation of a theory of Quantum Gravity has been a long-standing problem, and the lack of experimental evidence has made it even more difficult.

Below the scales in which quantum effects of gravity become relevant, the Universe seems to be well described by the so-called Quantum Field Theory in Curved Spacetimes (QFTCS). This theory is a generalization of QFT to curved backgrounds, and it has been able to provide a framework for studying the interplay of gravity and quantum fields in the early Universe, or in the vicinity of compact objects. The main idea is to consider spacetime as a classical background, and to quantize the fields propagating on it. When these fields are placed in non-trivial backgrounds, interesting phenomena may appear. The Unruh effect, Hawking emission by Black Holes (BHs), or cosmological pair production are some examples. Moreover, there is a striking and common feature to all QFT (also in flat spacetime, as we will discuss): Starting from a classical theory, there are infinitely many inequivalent quantum theories, each defining a vacuum state. These are the so-called quantum vacuum ambiguities. The choice of vacuum also determines the notion of particle, as excitations on top of it. This ambiguity in the choice of quantum theory is not typically discussed in QFT in flat spacetime because Poincaré invariance provides a unique, \textit{preferred} vacuum. However, in general, the group of symmetries of the classical theory is not large enough to perform this selection. 

One of the phenomena mentioned above, cosmological particle production, will be the main topic explored in this thesis. It is the process by which particles are created, even out of the vacuum, due to the time-dependence of spacetime. The most famous example is the production of particles during inflation, when the Universe undergoes an accelerated expansion. We will discuss how the non-adiabaticity of the changes in the spacetime metric is directly related to fluctuations of the system's particle number. Hence, this production mechanism is particularly important in the early Universe. In fact, it is believed that the fluctuations generated during inflation constitute the seeds of the large-scale structure we observe today, for which the Cosmic Microwave Background (CMB) radiation is strong evidence. Typically, inflation is driven by the so-called inflaton field, which eventually decays into the particles we observe today, transferring all of its energy to the thermal bath of relativistic particles from which the description of the Universe by the Standard Model of Cosmology starts to be valid. We will, nevertheless, adopt a different perspective when looking at early pair production: We will focus on spectator fields, that is, fields that are not directly coupled to the inflaton or any other field, and whose dynamics is only determined by the background geometry. This is a typical scenario for the application of QFTCS, and it is the one we will explore in this thesis. The expansion of the geometry will then produce excitations of these fields, which can be regarded as dark matter particles. This latter type of particle is incorporated into the Standard Model of Cosmology in a phenomenological manner, to account for observations suggesting that most of the gravitational mass in the Universe consists of very weakly interacting, if not entirely non-interacting, matter. Given these necessary weak interactions, it seems natural to expect that dark matter could have been produced through cosmological particle production processes and that it does not interact at all, except gravitationally.

Because directly accessing and measuring the phenomena that arise in QFTCS is a hard task, analog gravity experiments provide a valuable tool to study these effects. Dating back to 1981, Unruh proposed that an analog of Hawking emission could be observed in fluids. This idea has been extended to other systems, with Bose-Einstein condensates (BECs) being one of the most successful, and the main analog platform we will consider in this work. The analogy is based on the fact that the equations of motion of perturbations on top of the ground state in these systems can be written in terms of an effective metric, in such a way that the dynamics of fluctuations coincides with that of a quantum field in a curved spacetime described by said metric, which is characterized by the properties of the system, and in particular its ground state. Experimentally tuning these parameters may allow for the engineering of analogs of BHs and cosmological metrics, and the study of the quantum effects that arise in these scenarios. In particular, one can use these systems to simulate pair production due to a time-dependent geometry. Realizing these simulators will allow us to test the predictions of QFTCS, and to gain valuable insights that can be used to improve our understanding of the actual cosmological scenario.

We have structured this thesis in four distinct parts:

\begin{itemize}
    \item In Part I, we provide a theoretical introduction to the main topics around which this work revolves. In \cref{ch:cosmo}, we briefly introduce the Standard Model of Cosmology, focusing on the concepts most relevant to us, as well as the almost universally accepted inflationary paradigm. We study in \cref{ch:qftcs} how a QFT in such a background is developed. We introduce the concepts of quantum vacuum ambiguities and particle production, with a special mention of the spectrum of fluctuations, one of the most important quantities in the thesis. Moreover, we review the Gaussian states formalism and summarize how entanglement between cosmologically produced pairs can be quantified. Although this chapter also covers textbook material, the reader may find some of the notation unconventional. Finally, in \cref{ch:analogs}, we introduce the concept of analog gravity, and show how the effective metric arises, focusing particularly on BECs. We also briefly discuss the history of analog gravity, and recall the different systems that have been used to simulate QFTCS. This gives a glimpse of the possibilities that these systems offer, and the achievements that have been made in the field. Additionally, we briefly comment on the possibility of simulating cosmological production of massive particles, which could be seen as an analog of dark matter production, discussed in Part II. A reader familiar with these topics can skip this Part, except for notational convenience and some new results regarding entanglement (see \cref{sec:qftcs.entanglement}, corresponding to \cite{Entanglement2024}). 
    \item Part II is devoted to the early Universe, in which cosmological production is most important, with a special focus on the potential of this phenomenon as a dark matter production mechanism. We start with single-field, chaotic, realistic inflationary models in a spatially flat cosmology in \cref{ch:singlefield}. In this context, we study the production of scalar and vector spectator fields in a background set by two typical inflationary potentials: quadratic and Starobinsky's. We compare the relic abundance this process yields with current observations of dark matter in order to restrict the parameter space in which these theories reproduce observations. We introduce spatial curvature in \cref{ch:deSitter}, and consider a toy model of inflation based on a de Sitter geometry. In these chapters, we pay special attention to the choice of vacuum, which is non-trivial in a geometry that is always expanding. This part covers the results of~\mbox{\cite{ScalarField2023,VectorDM2024,CurvedDM2024,Starobinsky2024}}.
    \item In Part III, we turn to analog experiments. We first describe how fluctuations on top of the ground state of a BEC in quasi $(1+2)$ dimensions can be mapped to the motion of massless scalar fields in cosmological geometries of any spatial curvature in \cref{ch:becstheory}, also illustrating the realization of an actual experiment that benchmarks this system as a QFTCS simulator. Additionally, we take another point of view and theoretically study analog cosmological production in the language of one-dimensional scatteringthe corresponding experimental confirmation will appear in \cite{ScatteringExp2024}). Lastly, we theoretically explore the quantification of entanglement between the produced particles in \cref{ch:entanglement}, and show with what precision this can be measured in an actual experiment. The content of these chapters is mainly based on~\mbox{\cite{CosmologyPaper2022,BECPaper2022,Experiment2022,ScatteringTheory2024,ScatteringExp2024,Entanglement2024}}.
    \item Lastly, in Part IV, based on \cite{QVA2023} and \cite{SwitchEffects2024}, we put together the theoretical and experimental insights of previous parts. In \cref{ch:qva} we study quantum vacuum ambiguities in the context of cosmological particle production, and introduce a new family of vacua with an associated physical particle notion in terms of the clicks on some hypothetical detector. In \cref{ch:switcheffects} we consider the effects of switching on and off the expansion in analog systems, and study how fast or slow switches affect the spectrum of produced fluctuations. However, note that the expansion of the actual Universe is very adiabatic at the beginning of inflation and well into reheating, and therefore these transitions can be modeled in terms of switch-on and -off processes as well. We show that, in general, switch-on and -off processes dominate production due to their non-adiabaticity. In the context of analog systems, this poses questions about which particle production is actually probed in experiments such as those in Part III, since starting and ending the experiment may be more relevant for production than what happens in between. In the context of early cosmological production, this tells us that the most relevant periods for particle creation are the transitions between different stages of the Universe, rather than the stages themselves, which coincides with the insights gained from the analyses in Part II. 
\end{itemize}

Finally, we close the thesis by showcasing the main conclusions extracted from all the work in \cref{ch:conclusions}. As a coda, we finish with a reflection on the insights gained from analog experiments, and discuss what we can learn from them in \cref{ch:whatwecanlearn}. Together, these two chapters constitute Part V of the thesis.

We hope that the reader enjoys the journey through the different aspects of QFTCS, and that they find the insights provided by analog experiments useful to understand, perhaps from a fresh perspective, cosmological particle production of (dark matter) particles.

%% file: Chapters/Cosmology.tex

\addtocontents{toc}{\protect\vspace{0.5em}}

\chapter{Standard Cosmology and beyond} 

\label{ch:cosmo} 




Cosmology aims at describing the Universe as a whole. This includes answering the questions of how it originated, how it evolved, what its structure is, and what its fate will be. Its current theoretical formulation can be traced back to the development of Einstein's General Relativity in 1915, which provided a mathematical framework for understanding the gravitational dynamics of the Universe on large scales. During the rest of the 20th century, numerous key observations took place, building step by step what we regard now as the Standard Model of Cosmology~\mbox{\cite{Mukhanov2005,Weinberg2008}}. 

The cosmological description of the Universe relies on homogeneity and isotropy at large scales. This is known as the \textit{Copernican Principle}, and had to be assumed until observations were able to confirm it by the end of the century. Crucially, evidence suggests that most of the content of the Universe is made of the so-called \textit{dark components}. On the one hand, we have dark matter, whose interaction with ordinary matter is (if it exists) very weak \cite{Rubin1970}. On the other hand, there is dark energy, a negative-pressure component which is responsible for the accelerated expansion of the Universe \cite{Hubble1929}. These observations, together with homogeneity and isotropy, are accommodated by the so-called \textit{Lambda Cold Dark Matter} ($\Lambda$CDM) model, which has been established as the Standard Model of Cosmology. This description of the Universe considers dark energy in the form of a cosmological constant $\Lambda$, as well as the existence of \textit{cold} (highly non-relativistic) dark matter.

However, there are still open questions that the $\Lambda$CDM model cannot answer, such as the nature of the dark sector, the problem of baryon asymmetry, or the origin of the large-scale structure of the Universe \cite{Coles2002}. Importantly for this thesis, the Standard Model of Cosmology does not provide a satisfactory explanation for the initial conditions that led to the hot, dense state from which the Universe originated (described by the Hot Big Bang model). As an attempt to address these issues, the concepts of inflation and reheating were introduced \cite{Starobinsky1980,Guth1981}, which are crucial for the understanding the early stages of the Universe and constitute open fields that lie beyond the $\Lambda$CDM model.

In this chapter, we will briefly review the most important elements of the Standard Model of Cosmology concerning this thesis in \cref{sec:cosmo.sm}, along with the thermodynamic description of an expanding universe in \cref{sec:cosmo.thermodynamics}. We will pay special attention to dark matter in \cref{sec:cosmo.dm}, and additionally introduce the concepts of inflation and reheating in \cref{sec:cosmo.inflation} and \cref{sec:cosmo.reheating}, respectively.

 
\section{Standard Model of Cosmology}
\label{sec:cosmo.sm}

In the following section, we will review the elements that constitute the $\Lambda$CDM model, including the description of the geometry of the Universe and its energy content. 

\subsection{Homogeneous and isotropic universe}

The large-scale structure of spacetime is described by the Friedmann-Lemaître-Robertson-Walker (FLRW) metric \cite{Friedman1922, Friedman1924, Lemaitre1931, Robertson1935, Robertson1936a, Robertson1936b, Walker1937}, which corresponds to a homogeneous and isotropic universe. The line element of the FLRW metric, which allows one to measure spacetime distances, is given in $(1+D)$-dimensions\footnote{We will write everything for general $D$ spatial dimensions, so that we incorporate already the several scenarios discussed in the thesis.} by
\begin{equation}
    \dd s^2 = g_{\mu\nu}\dd x^{\mu}\dd x^{\nu} = -\dd t^2 + a^2(t)q_{ij}\dd x^{i}\dd x^{j},
\label{eq:cosmo.FLRWLineElementGeneral}    
\end{equation}
where $g_{\mu\nu}$ is the spacetime metric, $a(t)$ is the scale factor that governs the expansion of spacetime and $q_{ij}$ is the metric of the spatial sections without accounting for the scale factor. Using spherical coordinates, the line element takes the form
\begin{equation}
\dd s^2 = -\dd t^2 + a^2(t)\left(\frac{\dd r^2}{1-\kappa r^2} + r^2 \dd\Omega_{\text{D}}^2\right),
\label{eq:cosmo.FLRWLineElementSpherical}
\end{equation}
where $\kappa$ is the curvature of the spatial sections (which can be either positive, negative or zero), and $\dd\Omega_{\text{D}}^2$ is the differential solid angle of the $D$-dimensional unit sphere. It is often useful to work in conformal time $\eta$, which is defined as $\dd \eta = \dd t/a(t)$, so that
\begin{equation}
    \dd s^2 =  a^2(\eta)\left(-\dd \eta^2 +\frac{\dd r^2}{1-\kappa r^2} + r^2 \dd\Omega_{\text{D}}^2\right).
\label{eq:cosmo.FLRWLineElementConformal}
\end{equation}

For a metric of the form \eqref{eq:cosmo.FLRWLineElementConformal}, the Ricci scalar can be written as
\begin{equation}
    R(\eta) = \frac{D}{a^{2}(\eta)} \left[(D-3) \left( \frac{a'(\eta)}{a(\eta)} \right)^2+2 \frac{a''(\eta)}{a(\eta)} + (D-1)\kappa \right],
\label{eq:cosmo.RicciScalar}
\end{equation}
where the prime denotes derivative with respect to conformal time $\eta$.

The dynamics of spacetime follows from Einstein's field equations~\mbox{$G_{\mu\nu} = 8\pi G T_{\mu\nu}$}, which relate the geometry of the universe with its energy-matter content, in the form of the Einstein tensor $G_{\mu\nu}$ (completely determined by the metric $g_{\mu\nu}$) and the stress-energy tensor $T_{\mu\nu}$, respectively. As is well known, Einstein's equations can be derived via the variational principle from the action
\begin{equation}
    S = S_{\text{EH}} + S_{\text{m}},
\end{equation}
where $S_{\text{EH}}$ is the Einstein-Hilbert action,
\begin{equation}
    S_{\text{EH}} = \int \dd^D x \sqrt{-g} R,
\label{eq:cosmo.EHAction}
\end{equation}
and $S_{\text{m}}$ is the action of the matter and radiation fields,
\begin{equation}
    S_{\text{m}} = \int \dd^D x \sqrt{-g} \mathcal{L}_{\text{m}},
\label{eq:cosmo.MatterAction}
\end{equation} 
with $g$ denoting the determinant of the spacetime metric.

It remains to specify what is the energy-momentum tensor $T_{\mu\nu}$, which is defined through 
\begin{equation}
    T^{\mu\nu} = -\frac{2}{\sqrt{-g}}\frac{\delta S_{\text{m}}}{\delta g_{\mu\nu}}.
\end{equation}
The components of the Universe are well described by barotropic, perfect fluids following the equation of state $p=\omega \rho$, so that $T^{\mu}_{\nu}$ can be written in the frame comoving with the fluid as
\begin{equation}
    T^{\mu}_{\nu} = \text{diag}(-\rho, p, p, p).
\label{eq:cosmo.PerfectFluidTensor}
\end{equation}
We see that \cref{eq:cosmo.PerfectFluidTensor} only depends on the density $\rho$ and the pressure $p$ of the fluid, which furthermore must be only function of $t$ due to the aforementioned symmetries.

We are now ready to introduce both the FLRW metric \eqref{eq:cosmo.FLRWLineElementGeneral} and the stress-energy tensor \eqref{eq:cosmo.PerfectFluidTensor} into the field equations. This yields the well-known Friedmann equation,
\begin{equation}
   H^2 = \frac{8\pi}{3\mpl^2} \rho - \frac{\kappa}{a^2},
\label{eq:cosmo.FriedmannEquation}
\end{equation}
where $H=\dot{a}/a$ is the Hubble parameter (namely the rate of expansion of the universe) and the dot denotes derivative with respect to cosmological time $t$, as well as the acceleration equation
\begin{equation}
    \frac{\ddot{a}}{a}=-\frac{4\pi}{3\mpl^2}\left(\rho + 3p\right).
\label{eq:cosmo.AccelerationEquation}
\end{equation}
Note that the sign of the Hubble parameter $H$ tells us about the expansion or contraction rate of the universe. It is customary to define the deceleration parameter
\begin{equation}
    q = -\frac{\ddot{a}}{aH^2},
\label{eq:cosmo.DecelerationParameter}
\end{equation}
which is positive for decelerating universes, negative for accelerating ones, and zero for a universe with a constant expansion rate $H$. Dividing Friedmann equation~\eqref{eq:cosmo.FriedmannEquation} by $H^2$ leads to
\begin{equation}
    1 = \frac{\rho}{\rho_{\text{c}}} - \frac{\kappa}{a^2H^2} = \Omega - \frac{\kappa}{a^2H^2},
\label{eq:cosmo.FriedmannDivided}
\end{equation}
where
\begin{equation}
\Omega = \frac{\rho}{\rho_{\text{c}}} = \frac{8\pi}{3\mpl^2H^2} \rho
\label{eq:cosmo.AbundanceDefinition}
\end{equation}
is the abundance. The quantity $\rho_{\text{c}} = 3\mpl^2H^2/(8\pi)$ is called the critical density because~\mbox{$\rho = \rho_{\text{c}}$} implies that the curvature term in \cref{eq:cosmo.FriedmannDivided} above must vanish. If one considers $\rho$ as the sum of the energy density of all the components of the universe, and furthermore interprets the curvature term as a contribution to the energy density, in the sense that one defines
\begin{equation}
\rho_{\kappa} = -\frac{3\mpl^2\kappa}{8\pi a^2},
\label{eq:cosmo.CurvatureDensity}
\end{equation} 
then Friedmann \cref{eq:cosmo.FriedmannEquation} can be rewritten as
\begin{equation}    
    1 = \Omega_M + \Omega_R + \Omega_{\Lambda}  + \Omega_{\kappa}, 
\label{eq:cosmo.AbundanceSumRule}
\end{equation}
where we have put forward the assumed components of the Universe, which we discuss in the following: non-relativistic matter ($\Omega_M$), radiation ($\Omega_R$), dark energy~\mbox{($\Omega_{\Lambda}$)} and curvature ($\Omega_{\kappa}$). 

By invoking the conservation of the stress-energy tensor, $\nabla_{\mu}T^{\mu\nu}=0$, one can derive the continuity equation
\begin{equation}
    \dot{\rho} + 3H(\rho + p) = 0.
\label{eq:cosmo.ContinuityEquation}
\end{equation}
Note that since we are considering barotropic fluids, it only suffices to know one of the three \cref{eq:cosmo.FriedmannEquation,eq:cosmo.AccelerationEquation,eq:cosmo.ContinuityEquation} to determine the dynamics of the universe. In this case, \cref{eq:cosmo.ContinuityEquation} can be integrated to obtain
\begin{equation}
    \rho \propto a^{-3(1+\omega)},
\label{eq:cosmo.DensityEvolution}
\end{equation}
which tells us how the density evolves with the expansion.

Let us now discuss the different components that form part of the $\Lambda$CDM model:
\begin{enumerate}
    \item \textbf{Non-relativistic matter} behaves as a pressureless fluid, and therefore~\mbox{$\omega = 0$}. This implies that the energy density of matter scales as $a^{-3}$, that is, it dilutes with the expansion as expected. If one considers this to be the only content of the Universe, the Friedmann equation \eqref{eq:cosmo.FriedmannEquation} yields for the scale factor~\mbox{$a \propto t^{2/3} \propto \eta^2$}. Importantly, observations are compatible with the existence of dark matter, which is a matter component that interacts very weakly with ordinary matter, and will be discussed in more detail in \cref{sec:cosmo.dm}. In particular, the Standard Cosmological Model assumes the existence of \textit{cold} dark matter, which behaves as non-relativistic with respect to structure formation. It is estimated to dominate the matter sector, constituting around $85\%$ of the matter content of the Universe as measured today.
    \item \textbf{Radiation} is characterized by $\omega = 1/3$, and therefore its energy density scales as $a^{-4}$. If this component dominates, the scale factor evolves as $a \propto t^{1/2} \propto \eta$. Since it dilutes faster than matter, radiation is expected to dominate the energy content of the Universe at early times, and become negligible at late times.
    \item \textbf{Curvature} behaves as a fluid with $\omega = -1/3$, and therefore its energy density scales as $a^{-2}$, which agrees with our definition~\eqref{eq:cosmo.CurvatureDensity}. Current observations set the actual value of $\Omega_{\kappa}$ to be very close to (and compatible with) zero. We will discuss the implications of this fact in subsection \ref{subsec:cosmo.curvature}.
    \item \textbf{Dark energy} is an exotic component, which in the $\Lambda$CDM model contributes as a cosmological constant $\Lambda$ in the right-hand side of Einstein's equations, in accordance with the fact that \mbox{$\omega = -1$}, leading to a constant energy density for dark energy. The scale factor when dark energy dominates acquires the form \mbox{$a \propto e^{Ht} = -1/(H\eta)$}. Crucially, the pressure of dark energy is negative, which contributes in \cref{eq:cosmo.AccelerationEquation} with the sign opposite to that of matter and radiation, and allows for an accelerated expansion of the Universe.
\end{enumerate}

From the evolution of the densities of the different components, one can infer that the early Universe was radiation-dominated. Then, after radiation-matter equality, the Universe became matter-dominated. Finally, at late times, the Universe is dominated by dark energy, which is consistent with the accelerated expansion of the Universe that has been observed (see subsection \ref{subsec:cosmo.de}). The $\Lambda$CDM model accommodates numerous observations that are compatible with the existence of the components discussed above, which are detailed in the following (sub)sections.

\subsection{Dark energy, accelerated expansion and Hubble tension}
\label{subsec:cosmo.de}

The observation of the redshift of the light of distant galaxies by Hubble in 1929 \cite{Hubble1929} was crucial in establishing the idea that the Universe is expanding, leading to the formulation of Hubble's law. In this way, the very same prediction stemming from the dynamics of an FLRW was confirmed. Until the end of the 20th century, the expansion of the Universe was thought to be decelerating, which is precisely what FLRW dynamics predicts in the absence of dark energy. However, the observation of supernovae of type Ia in the late 90s \cite{Riess1998,Perlmutter1999} provided the first evidence that the expansion of the Universe is actually accelerating, which requires the existence and domination of the negative pressure component that we call dark energy. Most observations, including those of the recent Pantheon+ \cite{Pantheon2022} and Sh$0$es \cite{Shoes2022} collaborations, are compatible with the existence of dark energy in the form of a cosmological constant $\Lambda$, which comprises $70\%$ of the energy budget of the Universe today.

Another important observation corresponds to the discovery of the Cosmic Microwave Background (CMB) radiation in 1965 \cite{Penzias1965}. In the early Universe the temperature was so high that space was filled with a plasma of charged particles, which prevented radiation from propagating freely. As the Universe expanded and cooled down, the plasma recombined into neutral atoms, which allowed radiation to decouple from matter and propagate freely. Photons that escaped the plasma at that instant began traveling through the Universe, suffering redshift due to its expansion, and constituting one of the most powerful probes of the early Universe. In agreement with the Standard Cosmological Model, the CMB has been measured to be isotropic to a very high degree, for example by the COBE collaboration \cite{Mather1994} and more recently by the Planck satellite \cite{Planck2018-1}, providing further evidence of the validity of the $\Lambda$CDM model (whose theoretical predictions are used to fit the background). In fact, its spectrum is very well described by blackbody radiation with a temperature of \mbox{$T_{\text{CMB}} = 2.7255 \pm 0.0006$ K}, which tells us that photons were in equilibrium before they last scattered. One finds, however, very small anisotropies in the CMB, corresponding to temperature fluctuations of the order of $\Delta T/T \sim 10^{-5}$. These are thought to be the seeds of the large-scale structure of the Universe, and will be further discussed in \cref{sec:cosmo.inflation}.

Let us mention that the value of the Hubble constant today is found to be~\mbox{$H_0 = 73.5 \pm 1.1 \, \text{km}\,\text{s}^{-1}\,\text{Mpc}^{-1}$} according to the Pantheon+~\cite{Pantheon2022} and Sh$0$es \cite{Shoes2022} collaborations (in concordance with other \textit{local measurement} analyses), whereas the Planck collaboration obtains \mbox{$H_0 = 67.4 \pm 0.5 \, \text{km}\,\text{s}^{-1}\,\text{Mpc}^{-1}$} \cite{Planck2018-1}. This discrepancy is known as the Hubble tension, and is one of the most important open questions in modern Cosmology \cite{DiValentino2021, Abdalla2022}. The origin of this discrepancy is still unknown, and it is not clear whether it is due to systematic errors in the measurements or if it is a sign of new physics beyond the $\Lambda$CDM model.

\subsection{Spatial curvature and the flatness problem}
\label{subsec:cosmo.curvature}

Let us now discuss spatial curvature. Observations tell us that our Universe is essentially flat. Spatial curvature, if it exists, must be negligible. More concretely, Planck 2018 data \cite{Planck2018-1} sets the value of the curvature abundance today to be~\mbox{$\Omega_{\kappa} = -0.0096 \pm 0.0061$}, which is not compatible with $0$. On the other hand, the study of Baryon Acoustic Oscillations (see \cite{Chen2024} for details on the recent DESI collaboration) sets an independent bound compatible with $\kappa=0$ \cite{DESI2024}. When put together, these two measurements constrain the abundance of curvature in the $\Lambda$CDM model to~\mbox{$\Omega_{\kappa} = 0.0024 \pm 0.0016$}, in favor of a positively curved universe. Therefore, we cannot discard a small but nonzero curvature with current observations. At the same time, the Friedmann equation \eqref{eq:cosmo.FriedmannEquation} tells us that the curvature abundance has been growing since the radiation-dominated epoch. This means that at that point,~$\Omega_{\kappa}$ must have been incredibly small. That is, the initial conditions of the Universe must have been set with an incredible precision in order to have $\Omega_{\kappa}$ so close to zero today, since even a small perturbation of the curvature abundance in the early Universe would lead to a large positive or negative abundance nowadays. This is known as the flatness problem, and it is one of the main motivations for the introduction of the inflationary paradigm, which we will discuss further below.

In this thesis, we will study the implications of the existence of spatial curvature, be it positive or negative, in the gravitational production of dark matter particles during the early Universe, in particular in \cref{ch:deSitter}.

\section{Equilibrium        Thermodynamics}
\label{sec:cosmo.thermodynamics}

As one can deduce from \cref{eq:cosmo.DensityEvolution}, the Universe was hotter and denser in the past. In the early times, all particles that filled spacetime were relativistic and highly interacting, in such a way that the state of the Universe can be well described by the notion of \textit{local equilibrium}, in the sense that the entropy is maximal. That is, the rate of interaction of particles was much larger than the rate of expansion, and therefore the entropy is maximized before any substantial change in the size of the Universe takes place. Thus, we discuss in the following some notions of thermodynamics for species in equilibrium (see e.g. \cite{Mukhanov2005,Weinberg2008}).

The number density of relativistic bosons\footnote{The chemical potential for bosons must satisfy $\mu \leq m$.}, i.e. $T \gg m, \mu$, can be written as
\begin{equation}
    n = \frac{\zeta(3)}{\pi^2}sT^3, \quad \rho = \frac{\pi^2}{30}sT^4,
\label{eq:cosmo.RelBosonEnergyDensity}
\end{equation}
whereas for fermions\footnote{The chemical potential for fermions can exceed the mass. However, we are assuming here approximately non-degenerate fermions, and we can safely neglect the chemical potential \cite{Mukhanov2005}.} one has
\begin{equation}
    n = \frac{3}{4}\frac{\zeta(3)}{\pi^2}sT^3, \quad \rho = \frac{7}{8}\frac{\pi^2}{30}sT^4,
\label{eq:cosmo.RelFermionEnergyDensity}
\end{equation}
where $\zeta(3) \approx 1.202$ is the Riemann zeta function. In all cases, $p\simeq\rho/3$, as expected for relativistic particles. In the non-relativistic case, in which $T \ll m$, these relations turn out to be
\begin{equation}
    n = \frac{g}{(2\pi)^{3/2}} \left(\frac{mT}{2\pi}\right)^{3/2} e^{(\mu - m)/T}, \quad \rho = m n, \quad p = nT.
\label{eq:cosmo.NonRelEnergyDensity}
\end{equation}
Because the temperature is very small as compared to the mass in this case, all quantities above are exponentially suppressed. Moreover, the equation of state can be written as $\omega = p/\rho = T/m \ll 1$, therefore justifying the assumption of $\omega=0$ in the Standard Cosmological Model.

Now, we have said that in the early Universe, all species were in (local) equilibrium, and therefore one can associate a temperature with the corresponding thermal bath. The energy density entering in e.g. the Friedmann equation \eqref{eq:cosmo.FriedmannEquation} in this stage of the history of the Universe has contributions of the form of \cref{eq:cosmo.RelBosonEnergyDensity} from bosons and \cref{eq:cosmo.RelFermionEnergyDensity} from fermions. As the Universe expands, it cools down, the interaction rate $\Gamma$ of particles decreases, and they decouple from the bath when it becomes of the order of the Hubble parameter, $\Gamma \sim H$. At this point, the interaction stops being efficient enough to maintain local equilibrium. Decoupling will happen at different temperatures for different particles, and these can behave relativistically or non-relativistically at this point, turning into relics and maintaining that same density distribution, since interactions are no longer efficient. However, note that the energy density of non-relativistic species \eqref{eq:cosmo.NonRelEnergyDensity} is much smaller than that of relativistic species \eqref{eq:cosmo.RelBosonEnergyDensity}, and therefore the energy density of the Universe can be approximately written in terms of the latter, as
\begin{equation}
    \rho =  \frac{\pi^2}{30}g_*(T)T^4,
\label{eq:cosmo.DensityBath}
\end{equation}
where $g_*(T)$ is the effective number of relativistic degrees of freedom at temperature~$T$,
\begin{equation}
    g_*(T) = \sum_{i} \left(\frac{7}{8}\right)^{\alpha_i} g_i \left(\frac{T_i}{T}\right)^4,
\label{eq:cosmo.reldof}
\end{equation}
with $\alpha_i = 0$ for bosons and $\alpha_i = 1$ for fermions. Note that this includes also the possible contribution to the density of some species that decoupled at a temperature~$T_i$ different from $T$ while being relativistic.

Let us finally address entropy density. As long as particles can be considered as part of a thermal bath, the entropy $S$ in a comoving volume $V$ is conserved,
\begin{equation}
    S = sa^3 = a^3\frac{\rho + p - \mu n}{T} \simeq  a^3\frac{\rho + p}{T},
\end{equation}
where we used that the quantity $na^3$ is conserved in equilibrium and in the last step we assumed $T\gg \mu$. Here, $s$ can be interpreted as an entropy density, and using \cref{eq:cosmo.DensityBath} (and $p=\rho/3$), it can be written to a good approximation in terms of the relativistic degrees of freedom as
\begin{equation}
    s = \frac{2\pi^2}{45}g_{*S}(T)T^3,
\label{eq:cosmo.RelEntropyDensity}
\end{equation}
where $g_{*S}(T)$ denotes now the effective number of relativistic degrees of freedom that contribute to the entropy density,
\begin{equation}
    g_{*S}(T) = \sum_{i} \left(\frac{7}{8}\right)^{\alpha_i} g_i \left(\frac{T_i}{T}\right)^3.
\end{equation}
Note that $g_{*S}(T)$ will coincide with $g_{*}(T)$ as long as no relativistic species has decoupled from the bath. Indeed, if all species share the same temperature, the corresponding factors above become one. On the other hand, if there exists a decoupled relativistic species with a thermal distribution at a temperature $T_i$,~$g_{*S}(T)$ will be different from $g_{*}(T)$ due to the different power in the temperature factors.

Crucially, the fact that $sa^3$ is conserved together with \cref{eq:cosmo.RelEntropyDensity} implies that
\begin{equation}
    T \propto a^{-1}g_{*S}^{-1/3}(T),
\end{equation} 
and therefore if the entropic degrees of freedom are constant, one may use the relation $T\propto a^{-1}$ for the bath temperature.

\section{Dark matter and structure formation}
\label{sec:cosmo.dm} 

The first observation regarding dark matter concerns the study of the Coma Cluster by Zwicky in 1933 \cite{Zwicky1933}, who realized that the velocity of the galaxies in the cluster was much larger than the velocity expected if only visible matter was considered. It was therefore suggested the existence of some kind of \textit{invisible} matter\footnote{Zwicky called this \textit{Dunkle Materie}, which means indeed dark matter.}, in the sense that it does not interact with light. Another crucial observation was that of the galactic rotation curves by Rubin in 1970 \cite{Rubin1970}, who noted that the rotation curves of galaxies are mostly flat, instead of decreasing with the distance of the observed stars to the center, compatible with the existence of a large dark matter halo surrounding galaxies (see~\cite{Sofue2001} for a more recent analysis). More evidence of the existence of dark matter comes from analyses of gravitational lensing~\mbox{\cite{Bartelmann2001, Clowe2004}}, studies regarding the large-scale structure of the Universe \cite{SDSS2000, SDSS2023} (see also \cite{Springel2005} for numerical simulations), or the already discussed CMB observations \cite{Planck2018-1}. The most recent measurements of the abundance of dark matter in the Universe suggest that it constitutes around $85\%$ of its matter content, and around $27\%$ of the total energy content today. In other words, most of the matter in the Universe is dark, and therefore dark matter is crucial in understanding the formation of large-scale structures. Although there are some baryonic candidates (as for example massive compact halo objects or MACHOs \cite{MACHO2000}), most of the dark matter particles considered in the literature are non-baryonic. Indeed, the $\Lambda$CDM model assumes the existence of \textit{cold} dark matter, which is a highly non-relativistic component that interacts very weakly with ordinary, baryonic matter. Weakly Interacting Massive Particles (WIMPs) are perhaps the most studied cold dark matter candidates~\cite{Jungman1996}. However, the nature of dark matter is still unknown.

Since dark matter interacts very weakly with the Standard Model fields (if at all), it is natural to ask how it was produced in the first place (see \cite{Arbey2021} for a review). If the interaction with the primordial plasma is strong enough, and dark matter particles are in equilibrium in the early Universe, one can talk about \textit{thermal production}. In this scenario the dark matter particles decouple from the thermal bath as discussed in \cref{sec:cosmo.thermodynamics}, becoming a thermal relic. When the decoupling temperature is much larger than the mass of the dark matter particle, the relic is said to be \textit{hot}. In this case, dark matter cannot cluster on galaxy scales until it becomes non-relativistic, and structure formation results from fragmentation of primordial structures \cite{Coles2002}. The classical example of hot dark matter is neutrinos, which are relativistic at the time of decoupling. On the other hand, if the decoupling temperature is much smaller than the mass of the dark matter particle, the relic is said to be \textit{cold}. As we mentioned above, WIMPs are good candidates for cold dark matter, such as the neutralino, which is a candidate for the dark matter particle in supersymmetric theories \cite{Salam1974}. Since they are non-relativistic, cold dark matter candidates lead to structure formation by gravitational collapse and clustering. If the interaction with the thermal bath is weak, dark matter particles are said to be \textit{non-thermally produced}. Some exhaustively studied examples include that of a homogeneous condensate of a weakly interacting scalar field, such as the axion \cite{Peccei1977}, or \textit{fuzzy dark matter}, which consists of free, very light ($m \sim 10^{-22} \text{eV}$) scalar particles \cite{Hu2000}. The way in which cold dark matter is produced depends on the particular model, but there is a very interesting mechanism for producing dark matter in a non-thermal way that is universal to all dark matter types and has received much attention in the last years, which is gravitational production. This is the main topic of this thesis, and we will review it in detail in the following subsection.

\subsection{Gravitationally produced dark matter candidates}
\label{subsec:cosmo.cosmopp}

As originally discussed by Parker \cite{Parker1968,Parker1969,Parker1971}, and also a few years later by Ford~\cite{Ford1987}, the mere expansion of spacetime is able to produce particles out of the vacuum, depending on the characteristics of the expansion itself. This mechanism is particularly important during the early stages of the Universe, namely during inflation and reheating, since the geometry changes most violently during these periods. Cosmological\footnote{In the following, we will use the word \textit{cosmological} to denote production due to the time-dependence of spacetime, whereas we understand the term \textit{gravitational particle production} as more general, including all particle production mechanisms that come from the framework of QFTCS, such as Hawking radiation, the Unruh effect, etc.} particle production (see \cite{Ford2021,Kolb2023a} for recent reviews on the field) is particularly interesting as a dark matter production mechanism because it does not require any coupling to other fields, besides gravitational interaction: One only needs that the corresponding field is non-conformally coupled to the geometry. Typically, it is assumed that interactions with the Standard Model fields are weak enough that one can regard the dark matter field as a spectator \cite{Markkanen2018, Herring2020}, in such a way that the produced density after the initial expansion of the Universe remains as a relic abundance and dilutes until today. Furthermore, the spectator field does not backreact on the geometry, that is, its corresponding energy density is subdominant, at least in the early Universe.

The first works on gravitational production of dark matter were done in the context of supermassive spectator scalar fields, called \mbox{\textit{WIMPZillas}} \cite{Chung1998a,Chung1998b,Kolb1998,Kuzmin1999,Chung2001,Chung2003}, in which the mass of the dark matter field $m_{\varphi}$ is of the order of the inflaton mass $m_{\phi}$ (see \cref{sec:cosmo.inflation}), with production of particles above this scale being strongly suppressed. Cosmological production for lower masses is also efficient, but the resulting abundance does not fall within the observational constraints \cite{Planck2018-1} on isocurvature perturbation in the case of a minimal coupling to curvature \cite{Chung2005}. However, a non-minimal coupling suppresses these constraints, allowing for light dark matter abundances compatible with observations in different inflationary scenarios \cite{Markkanen2017,Fairbairn2019,Tenkanen2019,Garcia2023a,Garcia2023b,Kolb2023c,Garcia2024b}. 
 
Recently there have been many works exploring cosmological production of scalar fields in the context of different inflationary models and couplings to gravity~\mbox{\cite{Ema2016, Markkanen2018, Fairbairn2019, Hashiba2019,Herring2020,Kainulainen2023,Garcia2023a}}. Of special importance are the works \cite{Ema2016, Markkanen2017, Ema2018, Chung2019, Cembranos2020,Yu2023}, which incorporated analyses of the relevance of the oscillatory behavior of the background geometry in the production. Most studied cases concern fields that are coupled to the geometry through a term of the form $\xi R \varphi^2$ in the action, with $\varphi$ denoting the spectator field, although other options have been explored, such as those including derivative couplings~\cite{Borrajo2020}. These oscillations induce tachyonic instabilities during reheating, which greatly enhance particle production. Importantly, the type of dark matter produced in these tachyonic regimes is adiabatic \cite{Markkanen2017,Fairbairn2019}, and therefore isocurvature perturbations do not have to be considered.

In recent years, the production of nonzero spin fields has also been studied in the context of gravitational production, primarily of vector fields~\mbox{\cite{Graham2015,Ema2019,Bastero2019,Ahmed2020,Kolb2021,Bastero2022,Sato2022,Ozsoy2024,VectorDM2024,Capanelli2024a,Capanelli2024b,Capanelli2024c}}. Crucially, in this case, the isocurvature constraints are always fulfilled \cite{Graham2015}. Until recently, only a minimal coupling to gravity was considered (that is, $\xi=0$ above). However, it has been found that a non-minimal coupling could enormously influence particle production in these scenarios, although it requires a careful treatment of the arising instabilities of the theory \cite{Capanelli2024a,Capanelli2024b,Capanelli2024c,Ozsoy2024,VectorDM2024}. These complications are also present in the interesting case of spin-2 fields, which have been examined in \cite{Kolb2023b}, whereas production of higher spin, bosonic fields was treated in the work~\cite{Alexander2020}. Fermionic fields have also been investigated \cite{Ema2019}, although Pauli exclusion principle leads to a suppression of cosmological production in these cases. 

In this thesis, we will discuss cosmological production of particles in the early Universe due to the expansion of spacetime, focusing on its application as a dark matter production mechanism. We will apply these ideas mainly to non-minimally coupled scalar fields (see next chapter), improving on the description of the background geometry. In particular, we will consider the exact equations of motion of the background, and solve them numerically when an analytic solution is not possible, instead of considering approximations. Moreover, we will pay special attention to the role of the oscillatory behavior of the background geometry, which will prove crucial for production, although it is often neglected. We will adopt a more fundamental rather than phenomenological point of view when discussing the particularities of the choice of vacuum in our quantum field theory. In this sense, our work also differs from previous references. Additionally, we consider the effects of spatial curvature in this context for the first time, and more importantly, our results are compatible with the bounds in non-minimally coupled scalar production outlined in reference \cite{Garcia2024b}, coming from several observational constraints, including isocurvature perturbations. Furthermore, our studies on cosmological production of dark matter vector fields allowing for non-minimal couplings to the geometry with a complete treatment of the background inflaton equations constitute one of the first works in this direction.

\section{Inflation}
\label{sec:cosmo.inflation}

We have seen that the $\Lambda$CDM model provides a very well tested description of the evolution of the Universe since the early moments in which it was a hot, dense plasma of relativistic particles. However, there are some problems that the model does not address, mainly related to its initial conditions (besides the fact that the dark sector implementation is purely phenomenological, and in this sense it is an incomplete framework as well). One of these problems is the so-called flatness problem, which we have already discussed in subsection \ref{subsec:cosmo.curvature}. There is also the fact that CMB is isotropic to a very high degree on scales far larger than the causal horizon at the time photons began their journey. That is, these different regions of the sky could not have been causally connected, and yet they share the same temperature. This is known as the horizon problem. Inflation (see e.g. \cite{Linde1990, Liddle2000, Ellis2023} for more details) aims at solving these problems, while also providing explanation for the origin of the deviations beyond homogeneity that gave rise to the large-scale structure of the Universe and that can be observed in the CMB.

The inflationary paradigm was introduced by Guth in 1981 \cite{Guth1981} as a mechanism able to address the problems mentioned above. Inflation is characterized by a period of accelerated expansion ($q < 0$) that would have taken place before the radiation-dominated era in the early, hot Universe. It is easy to see that the accelerated expansion condition can be rewritten as
\begin{equation}
\frac{\dd}{\dd t} \frac{1}{H a} < 0,
\label{eq:cosmo.InflationCondition}
\end{equation} 
which states that the comoving Hubble length, namely the size of the observable Universe as viewed by a comoving observer, must decrease during this period. In other words, this allows for the currently observable Universe to originate from a very small, causally connected region in the past. On the other hand, taking into account \cref{eq:cosmo.CurvatureDensity}, condition \eqref{eq:cosmo.InflationCondition} implies that the curvature abundance decreases during an accelerated expansion, and therefore a sufficiently long inflationary period would be able to smear out any initial trace of curvature. As mentioned in \cref{sec:cosmo.sm}, one needs a negative-pressure component in the Universe (with~\mbox{$\omega<-1/3$}) in order to have an accelerated expansion. We will discuss how this is achieved below.

These first models of inflation were based on the existence of a scalar field $\phi$, the inflaton, which undergoes a phase transition from a high temperature, metastable state to a stable state. These models were not free of problems, and the original proposal by Guth had to be modified \cite{Linde1982, Albrecht1982}, in what was called the \textit{new inflationary universe scenarios} \cite{Linde1990}, in contrast to the original proposal\footnote{Nowadays, all inflationary models involving phase transitions from such high temperature metastable states are called \textit{old inflationary scenarios} \cite{Linde2007}.}. However, the latter still presented some caveats, which were solved by Linde in 1983 with the introduction of chaotic inflation \cite{Linde1983}, perhaps the most popular today, and the one we will be focusing on in this thesis. 

This scenario considers that the inflaton can take any initial value in different regions of the early Universe, hence the name chaotic. However, in the infinite Universe, there will exist regions in which the inflaton field is nearly homogeneous, which will lead to our observable Universe. Such a scalar field is considered to be the dominant contribution to the energy-momentum tensor of the Universe during the inflationary epoch, in the form of a homogeneous, perfect fluid with density and pressure given by
\begin{equation}
    \rho = \frac{1}{2}\dot{\phi}^2 + V(\phi), \quad p = \frac{1}{2}\dot{\phi}^2 - V(\phi),
\label{eq:cosmo.InflatonDensityAndPressure}
\end{equation}
where $V(\phi)$ corresponds to the inflaton potential. The Friedmann equation \eqref{eq:cosmo.FriedmannEquation} in this situation reads
\begin{equation}
    H^2 = \frac{4\pi}{3\mpl^2}\left[\dot{\phi}^2 + 2V(\phi)\right] - \frac{\kappa}{a^2},
\label{eq:cosmo.InflatonFriedmann}
\end{equation}
whereas the equation of motion for the inflaton is (as it corresponds to a Klein-Gordon field in an FLRW background with such a potential, as we will discuss in \cref{ch:qftcs})
\begin{equation}
    \ddot{\phi} + 3H\dot{\phi} + V^{\prime}(\phi) = 0,
\label{eq:cosmo.InflatonEoM}
\end{equation}
where the prime denotes total derivative (with respect to $\phi$). From \cref{eq:cosmo.InflatonDensityAndPressure}, one can see that the condition for an accelerated expansion \mbox{$\rho + 3 p < 0$} (cf. \cref{eq:cosmo.DecelerationParameter}) implies~\mbox{$\dot{\phi}^2 < V(\phi)$} (this could be realized with a sufficiently flat potential, for example). Additionally, one typically assumes the so-called slow-roll approximation~\mbox{\cite{Mukhanov2005, Weinberg2008}}, which implies the stronger condition $\dot{\phi}^2 \ll V(\phi)$, together with \mbox{$\ddot{\phi} \ll H\dot{\phi}, \, V^{\prime}(\phi)$}. These conditions can be reformulated as
\begin{equation}
    \epsilon = \frac{\mpl^2}{16\pi}\left(\frac{V^{\prime}}{V}\right)^2 \simeq - \frac{\dot{H}}{H} \ll 1, \quad \eta = \frac{\mpl^2}{8\pi}\frac{V^{\prime\prime}}{V} \ll 1.
\label{eq:cosmo.SlowRollConditions}
\end{equation}
Note that the above equations are nothing but requiring that the expansion rate is sufficiently constant, which leads to a quasi-exponential expansion of spacetime. If these conditions are maintained for a sufficiently long number $\mathcal{N}$ of $e$-folds, defined~as
\begin{equation}
    \mathcal{N} = \log{\frac{a_{\text{end}}}{a_{\text{start}}}},
\end{equation}
with $a_{\text{end}}$ and $a_{\text{start}}$ being the scale factors at the end and start of inflation, respectively, one can solve the horizon and flatness problems  (approximately $\mathcal{N}=50$ is sufficient), and also provide a mechanism for the generation of primordial fluctuations that will give rise to the large-scale structure of the Universe. We will discuss the latter in the next subsection. Eventually, the inflaton rolls down the potential, and starts oscillating around its minimum, while the slow-roll approximation ($\epsilon, \eta \ll 1$) breaks down. The energy density of the inflaton is then transferred to the rest of the fields that permeate space through a process called reheating, and the Universe finally enters the radiation-dominated era. From this point on, the $\Lambda$CDM model is able to explain the cosmic evolution. The reheating process, however, is crucial in order to explain the hot Universe we observe today, and we will discuss it in \cref{sec:cosmo.reheating}. 

We will concentrate on two particular chaotic inflation scenarios in this thesis: quadratic inflation \cite{Mukhanov2005,Weinberg2008} and Starobinsky inflation \cite{Starobinsky1980}, which will be described in detail in \cref{ch:singlefield}. Interestingly, the latter was introduced by Starobinsky in 1980, even before Guth's original work, for a different purpose. However, it is one of the most successful models of inflation today. Other popular chaotic inflationary models constitute \textit{Higgs inflation} \cite{Bezrukov2008, Rubio2019}, in which the inflaton is identified with the Higgs field, therefore avoiding the need for introducing an extra degree of freedom.

For completeness, let us mention before closing this section some models other than chaotic inflation that have been proposed in the literature. These include, for example, \textit{hilltop inflation} \cite{Boubekeur2005}, in which inflation happens near the maximum of a very flat potential, or \textit{natural inflation} \cite{Freese1994}, which combines the idea of a hilltop and a quadratic minimum from standard chaotic inflation with a quadratic potential. On the other hand, \textit{hybrid inflation} \cite{Linde1994} tries to avoid super-Planckian values for the inflaton field, which occur in chaotic inflationary models, by introducing an extra field. Lastly, a very interesting proposal concerns \textit{quintessential inflation} \cite{Wetterich1987,Spokoiny1993,Wetterich1994,Peebles1999,Brax2005,Rubio2017,Bettoni2021}, in which the role of the inflaton and the dark energy field are played by the same scalar field. We refer the reader to references such as \cite{Lyth2009,Dimopoulos2020,Ellis2023} for more details. 

\subsection{Primordial fluctuations}
\label{subsec:cosmo.fluctuations}

Although the inflationary paradigm was originally introduced to solve the horizon and flatness problems, it also provides a mechanism for the generation of primordial fluctuations that will give rise to the large-scale structure of the Universe. These fluctuations are imprinted in the CMB, and lead to the small anisotropies that can be observed in its temperature. The idea is that quantum fluctuations of the inflaton field are stretched to cosmological scales during inflation, and then become classical. These fluctuations are responsible for the density perturbations that will give rise to the large-scale structure of the Universe. 

So far we have regarded the inflaton as a homogeneous, classical field. Nevertheless, one may account for small fluctuations $\delta\phi$,
\begin{equation}
    \phi(\vec{x}, t) = \phi_0(t) + \delta\phi(\vec{x}, t),
\end{equation} 
which will obviously break homogeneity, and lead to deviations of the metric on top of the FLRW background solution so that Einstein's equations are perturbed as~\mbox{$\delta G_{\mu\nu} = 8\pi G \delta T_{\mu\nu}$}. One can then obtain the equations of motion for the perturbations (conveniently written in terms of the Bardeen's variables \cite{Bardeen1980}) and quantize the perturbation field. As for any quantum field in the presence of an external, time-dependent agent (in this case the metric), gravitational particle production will lead to the creation of a scalar spectrum of fluctuations. The power spectrum of these fluctuations is defined as (see e.g. \cite{Liddle2000,Mukhanov2005,})
\begin{equation}
    \mathcal{P}_{\phi} = \frac{k^3}{2\pi^2} \braket{\delta\phi_{\vk} \delta\phi_{-\vk}}. 
\end{equation}
Similarly, a tensor power spectrum is generated by the quantum fluctuations of the metric, and is responsible for primordial gravitational waves (perhaps indirectly measurable from CMB B-modes \cite{Kamionkowski2016}, for example).

\section{Reheating}
\label{sec:cosmo.reheating}

We have anticipated that inflationary models must have a mechanism for stopping the accelerated expansion and transitioning into a radiation-dominated era, connecting to the Hot Big Bang scenario, and somehow transferring all the energy density of the inflaton to the rest of the particle fields. This is the goal of reheating~\mbox{\cite{Kofman1997,Basset1998}}, which we discuss as the next stage after chaotic inflation.

When the slow-roll regime ends, the inflaton field starts oscillating around the minimum of the potential, converting all the stored potential energy into kinetic energy (see \cite{Linde1994} for implementation in hybrid models). In this regime, the Hubble rate, which was approximately constant during slow-roll, drops below the inflaton mass $H<m_{\phi}$, and many oscillations happen over a Hubble time (for details on the explicit inflaton dynamics during slow-roll we refer to subsection \ref{subsubsec:sf.slowroll}). When averaged over many oscillations\footnote{Although here averaged, oscillations will prove crucial for cosmological particle production, as we will see in part II of the thesis.}, the field pressure vanishes, and the energy density reads
\begin{equation}
    \braket{\rho} \simeq \frac{\mpl^2}{6\pi t^2},
\end{equation}
as corresponds to a non-relativistic matter field. Thus, the inflaton behaves as a condensate of non-relativistic particles, leading to a scale factor of the form~\mbox{$a(t) \propto t^{2/3}$}.

Typically, it is assumed that the inflaton is coupled to other scalar and fermionic fields, which will eventually give rise to the Standard Model particles. Reheating ends when the Hubble parameter $H$ is of the order of the total decay rate of the inflaton $\Gamma_{\phi}$. At this point, one can write the temperature of the thermal bath at the end of reheating as\footnote{One must ensure that these decay products of the inflaton thermalize before the formation of structures.} \cite{Bassett2006}
\begin{equation}
    T_{\text{rh}} \simeq \left(\frac{90}{\pi^2 g_{*, \text{rh}}}\right)^{1/4}\sqrt{\Gamma_{\phi} \mpl},
\end{equation}
where $g_{*,\text{rh}}$ is the corresponding effective number of relativistic degrees of freedom~\eqref{eq:cosmo.reldof} at the end of reheating.

More importantly, the inflaton may decay through non-perturbative mechanisms. In the literature, this process is sometimes referred to as \textit{preheating} \cite{Traschen1990,Linde1994,Kofman1997}, since this decay happens in a very short amount of time, to distinguish it from the perturbative and long-lasting decay mechanism discussed above. Preheating may produce a large number of particles, which makes it a viable route only for bosonic decays, since fermionic decays are suppressed by the Pauli exclusion principle. However, the exact details of this process depend on the particular theoretical model under consideration, and it is beyond the scope of this thesis. Nevertheless, the main idea is that, when the effects of Bose condensation are taken into account, the inflaton field may explosively decay into a large number of particles. This is because the inflaton oscillations cause the appearance of parametric resonances in the equation of motion of the bosonic fields it is coupled to, leading to an exponential instability and production of particles. 

Other alternatives to the standard reheating mechanisms have been proposed in the literature, such as the so-called \textit{gravitational reheating}~\mbox{\cite{Ford1987,Spokoiny1993,Chun2009,Haque2023}}, in which particles are produced gravitationally via the same mechanism discussed in subsection \ref{subsec:cosmo.cosmopp}. This is useful when considering, for example, quintessential models of inflation, since the oscillating phase is absent, and instead one has a \textit{kination} period. Since the inflaton does not oscillate, the standard reheating mechanisms are not applicable here. Note that the mechanism via which reheating occurs is often related to the particular inflationary model in consideration. In fact, a somewhat different inflationary model is \textit{warm inflation}, in which it is assumed that the inflaton couples to other degrees of freedom already during inflation, with the slow-roll conditions being determined by this coupling \cite{Berera1995,Bastero2009,Bastero2016}. This would avoid the need for a \textit{graceful exit} from inflation, and eliminate the reheating phase.

In this thesis, we will only consider the background dynamics of the inflaton field during the reheating process, since we will consider cosmological particle production of spectator fields that are not coupled to the Standard Model particles. Thus, how exactly reheating is realized is not relevant for our purposes, as long as the inflaton field undergoes an oscillating phase and decays into a thermal bath that will eventually lead to the radiation-dominated Universe.

%% file: Chapters/Quantum_field_theory_in_a_curved_background.tex

\chapter{Quantum field theory in a cosmological background} 

\label{ch:qftcs} 




This thesis revolves around cosmological particle production \cite{Parker1968,Parker1969}, understood as the production of pairs due to the time-dependence of the geometry, i.e. the expansion (or contraction) of spacetime. The requirements of homogeneity and isotropy---as the Copernican principle implies in Cosmology---lead to the description of the Universe in terms of an FLRW metric, and the spacetime under consideration will be, at any point in this thesis, a particular realization of this geometry\footnote{There will be only one exception, which concerns an example of production due to the presence of an external electric potential, namely the Schwinger effect, which we will discuss in \cref{sec:qftcs.schwinger}.}.

At the same time, most of the problems examined in this work concern real scalar field theories, with a few exceptions involving complex scalar and vector fields. Therefore, we will focus on this particular case throughout this chapter, in which we provide a general introduction to cosmological particle production and the set of tools and concepts that will be required in this thesis. In particular, we will study the quantization of a scalar field in an arbitrary background and examine the ambiguities associated with the choice of vacuum~\mbox{\cite{Birrell1982,Mukhanov2007}}. Later, we will describe the process of particle production in the early Universe in terms of equal-time two-point functions, and define the spectrum of fluctuations, which is the equivalent of the inflaton power spectrum in the cosmological literature, discussed in subsection~\ref{subsec:cosmo.fluctuations}. Moreover, we will show how the production process can be understood in terms of Gaussian states, and how the entanglement between the produced particles can be quantified \cite{Serafini2017,Brady2022}. 

While this is an introductory chapter, we take the opportunity to introduce some of the notation and conventions that will be used throughout this work, as well as some non-standard definitions, derivations and results from \cite{CosmologyPaper2022}, \cite{Entanglement2024}, and  \cite{SwitchEffects2024}. 

\section{Dynamics of a scalar field in a \mbox{D-dimensional} FLRW universe}
\label{sec:qftcs.scalarfielddynamics}

The action of a free, massive scalar field in a $(1+D)$-dimensional spacetime can be written as
\begin{equation}
S=-\frac{1}{2}\int \diff t \diff ^D x \sqrt{-g} \left[g^{\mu\nu}\partial_{\mu}\varphi\partial_{\nu}\varphi + \left(m^2 + \xi R\right)\varphi^2\right],
\label{eq:qftcs.CurvedActionNM} 
\end{equation}
where we allowed for a coupling to the Ricci scalar $R$ of the form~\mbox{$\xi R \varphi^2$}, with $\xi$ being a dimensionless coupling constant. This is the only possible local coupling of this type with the correct dimensions. Note that we do not consider here any coupling of the derivatives of the scalar field (see \cite{Borrajo2020}). The dynamics of the scalar field, obtained from the variation of the action \eqref{eq:qftcs.CurvedActionNM}, is described by the Klein-Gordon (KG) equation of motion
\begin{equation}
    \partial_{\mu}\left(\sqrt{-g} g^{\mu\nu} \partial_{\nu}\varphi\right) - \sqrt{-g}\left(m^2+\xi R\right)\varphi = 0.
\label{eq:qftcs.KGEquation}
\end{equation} 
The KG equation \eqref{eq:qftcs.KGEquation} has an associated inner product that is invariant under the action of the isometries of spacetime, which is given by
\begin{equation}
(\varphi_1, \varphi_2) = i \int \diff^{D}x \left[\varphi_1^*(t, \vx)\pi_{\varphi_2}(t, \vx) - \varphi_2(t, \vx)\pi_{\varphi_1}^*(t, \vx)\right],
\label{eq:qftcs.KGInnerProduct}
\end{equation}
where $\pi_{\varphi} = \delta S/\delta\dot{\varphi}$ is the conjugate momentum to the field $\varphi$. Note that we have extended \cref{eq:qftcs.KGInnerProduct} to the complex domain, which will be useful in the following. The non-positive inner product \eqref{eq:qftcs.KGInnerProduct} is conserved in time, and is the so-called Klein-Gordon product.

As we mentioned above, we will restrict ourselves (for the most part) to metrics of the FLRW form (cf. \cref{eq:cosmo.FLRWLineElementGeneral}). In this case, the equation of motion \eqref{eq:qftcs.KGEquation} takes the form
\begin{equation}
\ddot{\varphi} + D \frac{\dot{a}}{a}\dot{\varphi} - \left[\frac{\Delta}{a^2} - m^2 - \xi R\right]\varphi = 0,
\label{eq:qftcs.ScalarFieldEOM}
\end{equation}
where the Ricci scalar is given by \cref{eq:cosmo.RicciScalar} and $\Delta$ is the $D$-dimensional Laplace-Beltrami operator for the spatial metric $q_{ij}$, defined as
\begin{equation}
\Delta = \frac{1}{\sqrt{q}}\partial_{i}\left(\sqrt{q}q^{ij}\partial_{j}\right), \quad q = \text{det}\,q_{ij}.
\label{eq:qftcs.LaplaceBeltrami}
\end{equation}
The KG product reduces in this case to
\begin{equation}
(\varphi_1, \varphi_2) = i a^D \int \diff^{D}x \sqrt{q} \left(\varphi_1^*\dot{\varphi}_2 - \varphi_2\dot{\varphi}_1^*\right),
\label{eq:qftcs.KGInnerProductFLRW}
\end{equation}
where we used that in this case $\pi_{\varphi} = a^D\sqrt{q}\dot{\varphi}$.


It will be convenient to work with the rescaled field written in conformal time\footnote{We will switch between cosmological and conformal time depending on what is most convenient in each situation.} $\eta$, 
\begin{equation}
\chi(\eta, \vx) =  a^{(D-1)/2}(\eta) \varphi(\eta, \vx),
\label{eq:qftcs.FieldsRelation}
\end{equation}
which satisfies the equation of motion
\begin{equation}
    \chi^{\prime\prime}(\eta, \vx) - \left\{\Delta - \kappa - a^2(\eta)\left[m^2 + \left(\xi - \frac{D-1}{4D}\right) R\right]\right\}\chi(\eta, \vx)=0.
    \label{eq:qftcs.ChiEOM}
\end{equation}
The field $\chi$ can be expanded in terms of the eigenfunctions of the Laplace-Beltrami operator, which we denote by $\mathcal{H}_{\vk}$, and satisfy 
\begin{equation}
-\Delta \mathcal{H}_{\vk}(\vx) = h^2(\vec{k}) \mathcal{H}_{\vk}(\vx),
\label{eq:qftcs.EigenvalueEquation}
\end{equation}
where $\bm{k}$ is the set of labels that characterize the eigenfunctions, and can be either continuous or discrete, depending on the particular spatial sections. For instance, for flat spatial sections in Cartesian coordinates,~$\vk$ would correspond to the wavevector $(k_{x_1}, k_{x_2}, k_{x_3}, ...)$, whereas in spherical coordinates it denotes the radial and angular wavenumbers $(k, l, q, ...)$. The corresponding eigenvalues of the operator $-\Delta$ are denoted by $h^2(\vk)$, and can be degenerate in general, that is, different sets of labels~$\vk$ can lead to the same eigenvalue $h^2(\vk)$. The eigenvalues are real, and therefore equation \eqref{eq:qftcs.EigenvalueEquation} is satisfied also by the conjugate eigenfunction $\mathcal{H}_{\vk}^*$. This means that~\mbox{$\mathcal{H}_{\vk}^*=\mathcal{H}_{\bk}$} for some $\bk$ such that $h^{2}(\vec{k})=h^{2}(\bk)$. For example, in the case of Euclidean spatial sections, the eigenvalues are given by $h^2(\vec{k}) = k^2$, where~\mbox{$k = \abs{\vk}$}. In this situation, complex conjugation on the eigenfunctions (which are plane waves) is equivalent to performing the operation $\bk = -\vk$, which leaves the eigenvalues invariant. Furthermore, the eigenfunctions $\mathcal{H}_{\vk}$ are normalized such that
\begin{equation}
\int \dd^{D}x \sqrt{q} \, \mathcal{H}_{\vk}(\vx)\mathcal{H}^*_{{\vk}^{\prime}}(\vx) = \delta^{(D)}({\vk}, {\vk}^{\prime}),
\label{eq:qftcs.EigenfunctionsOrthogonality}
\end{equation}
where we introduced the $\delta$-function with respect to the measure $\dd\mu_{\vk}$, which will be specified in each particular case and fulfills \cite{Birrell1982}
\begin{equation}
\int \dd\mu_{{\vk}^{\prime}} f(\vk^{\prime}) \delta^{(D)}({\vk}, {\vk}^{\prime}) = f({\vk}).
\end{equation}

With this, we can write the most general solution to the KG equation~\eqref{eq:qftcs.ChiEOM} as the following linear combination of particular solutions labeled by $\vec{k}$,
\begin{equation}
\begin{split}
\chi(\eta,\vx) &= \int \dd\mu_{\vk} \mathcal{H}_{\vk}(\vx) \Tilde{\chi}_{\vk}(\eta) \\ 
&= \int \dd\mu_{\vk} \mathcal{H}_{\vk}(\vx) \left[a_{\vk} v_h(\eta) + a^*_{\bk} v_h^*(\eta)\right],
\label{eq:qftcs.ChiFieldExpansion}
\end{split}
\end{equation}
where the coefficients of the linear combination $a_{\vk}$ and $a_{\vk}^*$ are related by complex conjugation to ensure the reality of the field $\chi$. Note that we used the notation~$v_h$ on the temporal modes since, as we will see below, they only depend on the eigenvalues of the Laplace-Beltrami operator, and not on the specific set of labels $\vk$.

Introducing the expansion \eqref{eq:qftcs.ChiFieldExpansion} into the equation of motion \eqref{eq:qftcs.ChiEOM}, one finds that the time-dependent mode functions $v_h(\eta)$ and $v_h^*(\eta)$ satisfy a decoupled system of ordinary differential equations of the form 
\begin{equation}
    v^{\prime\prime}_h(\eta) + \omega_{h}^2(\eta) v_h(\eta) = 0.
    \label{eq:qftcs.ModeEquation}
\end{equation}
These are harmonic oscillator equations with time-dependent frequencies, given by
\begin{equation}
\begin{split}
    \omega_h^2(\eta) =& h^2({\vk}) + m^2a^2(\eta) \\
    &+ \frac{1+(4\xi-1)D}{4} \left[(D-3) \left( \frac{a'(\eta)}{a(\eta)} \right)^2+2 \frac{a''(\eta)}{a(\eta)} \right].
\label{eq:qftcs.MasterFrequency}
\end{split}
\end{equation}
These frequencies capture all the information about the gravitational background, which we consider external, classical, and not affected by the dynamics of the field~$\chi$. It is homogeneity that simplifies the treatment, decoupling the dynamics of each mode $v_h$ from the others. Moreover, isotropy ensures that the modes only depend on the eigenvalue $h^2({\vk})$, and not on the complete set ${\vk}$, hence the labelling of the modes. Let us note that if the coupling to the Ricci scalar is
\begin{equation}
\xi = \frac{D-1}{4D},
\label{eq:qftcs.ConformalCoupling}
\end{equation}
the term containing the Ricci scalar, and therefore the derivatives of the scale factor $a(\eta)$ vanish. This is called conformal coupling, and will be a very important particular case throughout this work.

Note that the KG normalization condition together with the expansion \eqref{eq:qftcs.ChiFieldExpansion} imply that two particular solutions $\varphi_{\vk}$ and $\varphi_{{\vk}^{\prime}}$ must obey 
\begin{equation}
\begin{split}
   (\varphi_{\vk}, \varphi_{{\vk}^{\prime}}) &= i a^{D} \int \diff^{D}x \sqrt{q}\left[\varphi_{\vk}^*\dot{\varphi}_{{\vk}^{\prime}} - \varphi_{{\vk}^{\prime}}\dot{\varphi}_{\vk}^*\right] \\
    &= i a \, \left[\Tilde{\chi}_{\vk}^*\dot{\Tilde{\chi}}_{\vk^{\prime}} - \Tilde{\chi}_{\vk^{\prime}}\dot{\Tilde{\chi}}_{\vk}^*\right] \int \diff^{D}x \sqrt{q} \mathcal{H}_{\vk}(\vec{u})\mathcal{H}^*_{{\vk}^{\prime}}(\vec{u}) \\
    &= i \, \left[\Tilde{\chi}_{\vk}^*\Tilde{\chi}^{\prime}_{\vk^{\prime}} - \Tilde{\chi}_{\vk^{\prime}}\Tilde{\chi}_{\vk}^{\prime\,*}\right] \delta^{(D)}({\vk}, {\vk}^{\prime}) \\
    &=\delta^{(D)}({\vk}, {\vk}^{\prime}), 
\end{split}
\end{equation}
where we used the orthogonality relation of the eigenfunctions, \cref{eq:qftcs.EigenfunctionsOrthogonality}, and the relation between fields \eqref{eq:qftcs.FieldsRelation}. The last step requires that the temporal part of particular solutions to \cref{eq:qftcs.ModeEquation} satisfy
\begin{equation}
    \Tilde{\chi}_{\vk}^*\Tilde{\chi}_{\vk^{\prime}}^{\prime} - \Tilde{\chi}_{\vk^{\prime}}\Tilde{\chi}_{\vk}^{\prime \, *}= -i,
\label{eq:qftcs.WronskianFullMode}
\end{equation}
which written in terms of the Wronskian of particular solutions to the mode equation~$v_h$ reads
\begin{equation}
    \text{Wr}\left[v_h, v_h^*\right] =v_h v^{\prime\, *}_h - v^{\prime}_h v_h^* = i.
\label{eq:qftcs.WronskianModes}
\end{equation}
Similarly, one can show that, as long as the Wronskian condition \eqref{eq:qftcs.WronskianModes} is satisfied, $(\varphi_{\vk}^*, \varphi_{{\vk}^{\prime}}^*) = -\delta^{(D)}(\vk,\vk^{\prime})$ and $(\varphi_{\vk}, \varphi_{{\vk}^{\prime}}^*) = 0$. Note that the variables $a$ and~$a^*$ can be used as phase space coordinates, satisfying the Poisson bracket condition~\mbox{$\left\{a_{\vk}, a_{{\vk}^{\prime}}^*\right\} = i\delta^{(D)}({\vk}, {\vk}^{\prime})$}.

The modes~$v_h$ and $v^*_h$ constitute one basis of the space of solutions of the mode equation \eqref{eq:qftcs.ModeEquation}, and lead to an expansion in terms of the coefficients $a_{\vk}$ and $a^*_{\vk}$, as we have written in \cref{eq:qftcs.ChiFieldExpansion}. Nevertheless, one may choose a different basis, for example, the modes $u_h$ and $u^*_h$. The corresponding expansion coefficients will of course be different,~$b_{\vk}$ and $b^*_{\vk}$. Writing the $u_h$ modes in terms of the $v_h$ modes allows one to find a relation between the two sets of solutions to the mode equation,
\begin{equation}
    u_h =\alpha_h v_h + \beta_{h} v_h^{*}, \quad\quad\quad v_h = \alpha^*_h u_h - \beta_h u^*_h, 
    \label{eq:qftcs.BogoliubovTransformationModes}    
\end{equation}
where $\alpha_h$ and $\beta_h$ are the so-called Bogoliubov coefficients \cite{Birrell1982}, which satisfy the relation 
\begin{equation}
    \abs{\alpha_h}^2 - \abs{\beta_h}^2 = 1
\label{eq:qftcs.BogoliubovCoefficientsNorm}
\end{equation} 
in light of \cref{eq:qftcs.WronskianModes}. These coefficients can be written in terms of the Wronskian of the modes as 
\begin{equation}
    \alpha_h = -i\text{Wr}[u_h, v_h^*], \quad \beta_h = i\text{Wr}[u_h, v_h].
\label{eq:qftcs.BogoliubovCoefficientsWronskian}
\end{equation}
Since the field can be expanded in either basis, one must have that
\begin{equation}
\begin{split}
    \chi(\eta, \vx) =& \int \dd \mu_{\vk} \mathcal{H}_{\vk}(\vx) \left[a_{\vk} v_h(\eta) + a_{{\bk}}^* v_h^*(\eta)\right] \\
    =& \int \dd \mu_{\vk} \mathcal{H}_{\vk}(\vx) \left[b_{\vk} u_h(\eta) + b_{{\bk}}^* u_h^*(\eta)\right].
\end{split}
\end{equation}
Introducing the Bogoliubov transformation between the modes \eqref{eq:qftcs.BogoliubovTransformationModes} in the expression above, one can find the relation between the coefficients $a_{\vk}$ and $b_{\vk}$, which reads
\begin{equation}
    b_{\vk} = \alpha_h^* a_{\vk} - \beta_h^*a_{{\bk}}^* \quad \text{and} \quad a_{\vk} = \alpha_h b_{\vk} + \beta_h^*b_{{\bk}}^*.
\label{eq:qftcs.BogoliubovTransformationCoefficients}
\end{equation}

The form of the frequency in \cref{eq:qftcs.MasterFrequency} is completely general for a real scalar field in a $D$-dimensional FLRW universe. By taking different values for the mass~$m$, the coupling $\xi$, the spatial curvature $\kappa$, or the dimension $D$, one can particularize to the different scenarios that will be studied in this thesis, with a few exceptions. In part~II, we will consider early Universe scenarios of massive, non-minimally coupled fields, and therefore we will take $D=3$. We will also study vector fields, but the corresponding analysis follows closely that of scalar fields. On the other hand, part~III will be devoted to the study of the analog realization of cosmological production in a spin-$0$, $(1+2)$-dimensional BEC, which corresponds to setting~\mbox{$\xi=m=0$} and~\mbox{$D=2$}. Lastly, in part IV we will also comment on the Schwinger effect, in which the role of the background field is played by an external electric potential, instead of the spacetime metric. In all cases, one arrives at an equation of motion for the modes of the form \eqref{eq:qftcs.ModeEquation}, with a time-dependent frequency whose explicit form depends on the particularities of the problem.

In the next section, we will proceed with the quantization of the field $\varphi$ following the canonical approach.

\section{From classical to quantum fields}
\label{sec:qftcs.quantization}

In order to canonically quantize the field $\chi$, one promotes the coefficients of the expansion \cref{eq:qftcs.ChiFieldExpansion} to creation and annihilation operators, fulfilling standard commutation relations, which are explicitly given~by
\begin{equation}
\big[\hat{a}_{\vk}^{\dagger}, \hat{a}_{{\vk}^{\prime}}^{\dagger}\big] = \big[\hat{a}_{\vk}, \hat{a}_{{\vk}^{\prime}}\big]=0,\quad \big[\hat{a}_{\vk}, \hat{a}_{{\vk}^{\prime}}^{\dagger}\big] = \delta^{(D)}({\vk}, {\vk}^{\prime}).
\label{eq:CreationAnnihilationCommutationRels}
\end{equation}
These operators define the vacuum state
\begin{equation}
\hat{a}_{\vk}\ket{0^a} = 0, \quad \forall\, {\vk}.
\end{equation}
By acting with the creation operator on the vacuum state, one can build a basis of the corresponding Hilbert space (so-called Fock space~\cite{Birrell1982}). Note that, in general, each set of coefficients, $a_{\vk}$ and $b_{\vk}$ (and their complex conjugates), associated with the basis $v_h$ and $u_h$, respectively, defines two different notions of vacuum \cite{Mukhanov2007}. Then, the Bogoliubov transformation between the coefficients \eqref{eq:qftcs.BogoliubovTransformationCoefficients} becomes a relation between the two sets of creation and annihilation operators,
\begin{equation}
    \hat{b}_{\vk} = \alpha_h^* \hat{a}_{\vk} - \beta_h^*\hat{a}_{{\bk}}^{\dagger} \quad \text{and} \quad \hat{a}_{\vk} = \alpha_h \hat{b}_{\vk} + \beta_h^*\hat{b}_{{\bk}}^{\dagger}.
\label{eq:qftcs.BogoliubovTransformationOperators}
\end{equation}
Importantly, the Bogoliubov transformation only mixes creation and annihilation operators for the $\vk$ and $\bk$ modes, for which the mode equation coincides, as both sets of labels lead to the same Laplace-Beltrami eigenvalue $h^2$.

The number of $b$-particles in the $a$-vacuum, which will be, in general, a non-vacuum state according to the $\hat{b}_{\vk}$ operators, is given by the expectation value of the number operator $\hat{N}_{\vk}^b = \hat{b}_{\vk}^{\dagger}\hat{b}_{\vk}$ in the $a$-vacuum,
\begin{equation}
\bra{0^a}\hat{N}_{\vk}^b\ket{0^a} = \abs{\beta_h}^2 \mathcal{V},
\label{eq:qftcs.NumberOperatorExpectation}
\end{equation}
where $\mathcal{V}$ is related to the volume of the spatial sections of the metric, which could be infinite in general. Thus, one usually works with the mean number density, which is simply given by\footnote{More details on how the volume term arises and how to regularize it can be found in \cite{CurvedDM2024}, together with a more detailed discussion on the definition of the number density.}
\begin{equation}
    n_h^b = \abs{\beta_h}^2/(2\pi^2).
\label{eq:qftcs.MeanNumberDensity}
\end{equation}
Integrating \cref{eq:qftcs.NumberOperatorExpectation} over all modes, we find the total mean density,
\begin{equation}
   n^b = \frac{1}{V} \int \dd \mu_{\vk} \bra{0^a}\hat{N}_{\vk}^b\ket{0^a},
\label{eq:qftcs.TotalNumberDensity}
\end{equation}
where $V$ is the volume of the spatial sections. As long as \cref{eq:qftcs.TotalNumberDensity} remains finite, the Bogoliubov transformation~\eqref{eq:qftcs.BogoliubovTransformationOperators} is well-defined, and the dynamics is implemented unitarily~\cite{Alvarez2022}.

We have showcased one of the most striking features of QFT, which is the ambiguity in the choice of quantum theory starting from a given classical theory. Let us remark that this is natural to all QFT, even in Minkowski spacetime. In the latter scenario, however, one typically asks the quantum theory to respect the Poincaré symmetry of the classical theory. This selects a unique basis of solutions, leading to the Minkowski vacuum. However, when time-translational invariance is broken, as may be the case when the geometry changes with time, the group of symmetries of the classical theory is not enough to restrict the possible choices of quantization. Therefore, the notions of vacuum and particle become ambiguous, and quantum vacuum ambiguities are inherent to quantum field theories in curved spacetime \cite{QVA2023}. 

In the following section, we will introduce the concept of cosmological production of particles, which is intimately related with the choice of vacuum and particle notions.

\subsection{Particle production as an \textit{in}-\textit{out} process}
\label{subsec:qftcs.inout}

We have seen that when quantizing a classical field theory, there are in general infinitely many possibilities of choosing a vacuum state, each leading to a different quantum field theory. As a result, the notion of particle becomes ambiguous, since it depends on the particular quantum theory chosen, reflected in the basis with which the field operator is expanded. However, in some cases, a preferred quantum theory can be identified, leading to a well-defined notion of particle. Typically, the quantum theory is required to preserve the same symmetries as the classical theory. If the spacetime under consideration is maximally symmetric, it may be possible to select a unique, preferred vacuum state. This is the case of Minkowski spacetime.

Let us now consider the situation in which the frequency of the mode equation~\eqref{eq:qftcs.ModeEquation} acquires a constant value for $\eta\leq\eta_{\text{i}}$ and \mbox{$\eta\geq\eta_{\text{f}}$},~$\omega_{h, \text{i}}$ and $\omega_{h, \text{f}}$, respectively, in such a way that there is an intermediate period $\eta_{\text{i}} < \eta < \eta_{\text{f}}$ in which the frequency is time-dependent. In the \textit{in} ($\eta \leq \eta_{\text{i}}$) and \textit{out} ($\eta \geq \eta_{\text{f}}$) regions, the field behaves locally as in Minkowski spacetime. The solution to the mode equation in these regimes can be written as the linear combination 
\begin{equation}
    v_h = A_h e^{-i\omega_h \eta} + B_h e^{i\omega_h \eta},
\label{eq:qftcs.MinkowskiModesGeneral}
\end{equation}
with the corresponding \textit{in} and \textit{out} frequencies in each case. In particular, the natural notion of vacuum for an observer living in the \textit{in} region will correspond to modes that behave as
\begin{equation}
    v_h = \frac{1}{\sqrt{2\omega_{h, \text{i}}} }e^{-i\omega_{h, \text{i}} \eta}, \quad \text{for} \quad \eta \leq \eta_{\text{i}},
\label{eq:qftcs.InModeIC}
\end{equation}
and similarly for an observer living in the \textit{out} region,
\begin{equation}
    u_h = \frac{1}{\sqrt{2\omega_{h, \text{f}}} }e^{-i\omega_{h, \text{f}} \eta},\quad \text{for} \quad \eta \geq \eta_{\text{f}},
\label{eq:qftcs.OutModeIC}
\end{equation}
where the modes are normalized according to the Wronskian \eqref{eq:qftcs.WronskianModes}. This choice of modes is the same that leads to the Minkowski vacuum in that spacetime. Expanding the field in terms of $v_{h}$ or $u_{h}$ and its complex conjugates leads to two particular sets of creation and annihilation operators defining the \textit{in} and \textit{out} vacua, respectively. It is important to note that \eqref{eq:qftcs.InModeIC} and \eqref{eq:qftcs.OutModeIC} are two different criteria that select particular solutions of the mode equation \eqref{eq:qftcs.ModeEquation}, which is defined for all times. Thus, each set of modes will evolve in the rest of the regions as dictated by the mode equation. That is, the \textit{in} mode $v_h(\eta)$, for example, is the solution, at all times~$\eta$, to the mode equation \eqref{eq:qftcs.ModeEquation} with \textit{time-dependent} frequency \eqref{eq:qftcs.MasterFrequency}, that for $\eta \leq \eta_{\text{i}}$ behaves as a positive frequency plane wave (i.e. as a Minkowski mode). However, the behavior for~\mbox{$\eta > \eta_{\text{i}}$} will not be of this form. In particular, for~\mbox{$\eta \geq \eta_{\text{f}}$},~$v_h$ will not correspond to the Minkowski notion of vacuum in that region, which is in fact associated with the mode $u_h$, but will behave instead as some linear combination of positive and negative frequency plane waves, as \cref{eq:qftcs.ModeEquation} dictates in that regime.

Since each mode, together with its conjugate, is a basis of solutions, the modes~$v_h$ and $u_h$ are related by a Bogoliubov transformation of the form \eqref{eq:qftcs.BogoliubovTransformationModes}, which in general will involve a non-vanishing $\beta_h$ coefficient. This indicates that the notion of vacuum before and after the expansion is not the same, or, put differently, that the expansion of the geometry produces particles---measured by an observer after this process---out of the vacuum, as understood by an observer before the change of spacetime. This is the essence of the \textit{in}-\textit{out} formalism, and will be the typical situation in analog cosmological production, in which there will be exact \textit{in} and \textit{out} regions before and after the expansion of the analog geometry, between which we compute the number density of produced particles.

However, in Cosmology, spacetime is always expanding. Nevertheless, at times the expansion can be very adiabatic. The effect non-adiabaticity of spacetime has in particle production can be quantified with the use of the dimensionless function that we call the \textit{adiabatic coefficient} $\mathcal{C}_h$,
\begin{equation} 
    \mathcal{C}_h(\eta)\equiv\abs{\omega_h^{\prime}(\eta)/\omega_h^2(\eta)}.
    \label{eq:qftcs.AdiabaticCoefficient}
\end{equation}
If the expansion of the geometry is very adiabatic, $\abs{\mathcal{C}_h(\eta)} \ll 1$, and therefore particle production becomes negligible\footnote{Characterizing adiabaticity, or relating it to the absence of production is not straightforward. However, $\mathcal{C}_h$ can be used as an indicator of the amount of production in most typical scenarios, as discussed in \cite{SwitchEffects2024}.}, at the same time as the notion of vacuum becomes stable. We will often consider scenarios in which spacetime expansion is very adiabatic at early and late times, such as at the beginning of inflation and well into the reheating epoch. In these regimes, which can be thought of as approximately \textit{in} and \textit{out} regions, we can define the so-called zeroth-order adiabatic vacuum via the choice of modes\footnote{This is related to approximate solutions to the mode equation in the WKB formalism \cite{Bender1999}.}
\begin{equation}
\begin{split}
    v_h(\eta_{\text{ad}}) &= \frac{1}{\sqrt{2\omega_h(\eta_{\text{ad}})}}e^{-i\omega_h(\eta_{\text{ad}}) \eta_{\text{ad}}}, \\ 
    v_h^{\prime}(\eta_{\text{ad}}) &= -\frac{1}{\sqrt{2\omega_h(\eta_{\text{ad}})}}\left(i\omega_h(\eta_{\text{ad}}) + \frac{1}{2}\frac{\omega_h^{\prime}(\eta_{\text{ad}})}{\omega_h(\eta_{\text{ad}})}\right)e^{-i\omega_h(\eta_{\text{ad}}) \eta_{\text{ad}}}.
\label{eq:qftcs.AdiabaticVacuum}
\end{split}
\end{equation}
This is a suitable vacuum prescription \textit{at each time $\eta_{\text{ad}}$} as long as the adiabaticity condition is satisfied. In other words, it will correspond to the notion of vacuum of an observer living at, for example, $\eta \geq \eta_{\text{ad}}$, if $\abs{\mathcal{C}_h(\eta)} \ll 1$ in this region.

Let us stress that the fact that particles are produced is a consequence of the energy input into the system coming from the expansion of the geometry, and not a consequence of a somewhat arbitrary choice of vacuum. The Bogoliubov formalism is one way of describing the fact that the Hamiltonian of our system has evolved during the expansion, and with it the notion of vacuum. Another way of looking at this problem is to carry the time-evolution together with states (i.e. working in the Schrödinger picture), in such a way that one starts with vacuum before the expansion, and the state of the system evolves into something that is populated due to the time-dependent metric.

\section{Gaussian states}
\label{sec:qftcs.gaussian}

We have seen that evolution only mixes ${\vk}$ and ${\bk}$ modes, such that~\mbox{$h(\vk) = h(\bk)$}, which is expected in a homogeneous and isotropic universe (recall that, for Euclidean spatial sections, these would be modes propagating in opposite directions). Hence, in order to describe our system, it is sufficient to focus on subsystems of two degrees of freedom described by the vector of canonical operators
\begin{equation}
    \hat{\bs{A}}_{\vk}=\lr{\hat{a}_{\vk},\hat{a}_{\vk}^{\dagger},\hat{a}_{\bk},\hat{a}_{\bk}^{\dagger}}^\top,
    \label{eq:qftcs.PhaseSpaceVec}
\end{equation}
where the superscript $\top$ denotes the transpose. With this notation, we can write Heisenberg evolution as
\begin{equation}
    \hat{\bs{B}}_{\vk}=\bS_h \cdot\hat{\bs{A}}_{\vk}\,,
    \label{eq:qftcs.QuantEvol}
\end{equation}
where $\hat{\bs{B}}_{\vk}$ is formed by the \textit{out} operators (cf. \cref{eq:qftcs.BogoliubovTransformationOperators}) and
\begin{equation}
    \bS_h=
    \begin{pmatrix}
        \alpha_h^* & 0 & 0 & -\beta_h^*\\
        0 &  \alpha_h & - \beta_h & 0\\
        0 & -\beta_h^* & \alpha_h^* & 0\\
        -\beta_h & 0 & 0 & \alpha_h\\
    \end{pmatrix}.
    \label{eq:qftcs.SmatrixTh}
\end{equation}
This matrix is always symplectic, in the sense that $\bS\cdot\bs{\Omega}\cdot\bS^\top=\bs{\Omega}$, where $\Omega$ is the symplectic matrix
\begin{equation}
    \bs{\Omega}=\begin{pmatrix}
    0 & 1\\
    -1 & 0
    \end{pmatrix}
    \oplus
    \begin{pmatrix}
    0 & 1\\
    -1 & 0
    \end{pmatrix}.
\end{equation}    
The matrix $\bS_h$ is non-unitary if $\beta_h\neq0$, as can be verified using the constraints satisfied by the Bogoliubov coefficients \eqref{eq:qftcs.BogoliubovCoefficientsNorm}.

In order to find the specific values of the elements of $\bS_h$, one simply needs to evaluate \cref{eq:qftcs.BogoliubovCoefficientsWronskian} at any time $\eta$. Since this expression compares two particular solutions of the mode equation \eqref{eq:qftcs.ModeEquation}, the Bogoliubov coefficients must be the same regardless of the time $\eta$ in which the evaluation is made. In practice, and taking into account the \textit{in-out} scenario we just described, we will solve the mode equation~\eqref{eq:qftcs.ModeEquation} with initial conditions at $\eta\leq \etai$ of the form \eqref{eq:qftcs.InModeIC}. Then, we will evaluate this solution at $\etaf$ and compare it to the particular solution defined by initial conditions~\eqref{eq:qftcs.OutModeIC} at $\eta \geq \etaf$ through \cref{eq:qftcs.BogoliubovCoefficientsWronskian} in order to extract the two Bogoliubov coefficients. We would like to remark that, especially in cosmological contexts, the \textit{in} and \textit{out} regions may be reached asymptotically (or approximately realized). Later, we will use this procedure to compute several observables, including entanglement, for particular examples of dynamical universes that are relevant for current or future experiments. 

General quantum states are characterized by all $n$-point functions of their canonical variables. In our case, these correspond to quantum averages of all the possible products of components of the vector $\hat{\bs{A}}$ defined in \eqref{eq:qftcs.PhaseSpaceVec}, in all possible orderings. However, we will consider that the state of our system is Gaussian (which is a sensible assumption both in Cosmology and in the analog experiments we will discuss). Gaussian states (see \cite{Brady2022} for a review) are a subset of quantum states which are completely specified by the one- and two-point functions of the components of $\hat{\bs{B}}$, called mean $\bs{\mu}$ and covariance matrix $\bs{\sigma}$, which are defined by
\begin{equation}
\begin{split}
    &{\mu}^i_{\vk} = \braket{\hat{A}^i_{\vk}}/\,\mathcal{V},\\
    &{\sigma}_{\vk}^{ij} = \braket{\{\hat{A}^i_{\vk}-\mu^i_{\vk},\hat{A}^j_{\vk}-\mu^j_{\vk}\}}/\,\mathcal{V},
\end{split}
\label{eq:qftcs.MeanAndCov}
\end{equation}
where $i$ and $j$ run here over the four components of $\hat{\bs{A}}_{\vk}$, and $\{\cdot,\cdot\}$ stands for anticommutator. The antisymmetric part of the two-point function contains no information on the state, since it just encodes the canonical commutation relations. Note that the expectation values of the operators are proportional to the volume~$\mathcal{V}$, as discussed in \cref{eq:qftcs.NumberOperatorExpectation}. The components of $\hat{\bs{A}}$ can be expressed in any other basis of canonical operators. For instance, we could write $\hat{\bs{A}}_{\vk}^{(\chi)}=(\hat{{\Tilde{\chi}}}_{\vk},\hat{\Tilde{\pi}}_{\vk},\hat{\Tilde{\chi}}_{\bk},\hat{\Tilde{\pi}}_{\bk})^\top$, where the canonical momentum conjugate to $\chi$ is given by $\pi_{\chi} = \delta S/\delta \chi^{\prime} = \chi^{\prime}$. The change of canonical basis from creation/annihilation to field/momentum operator basis is given by the matrix\footnote{These operators comprise the time-dependent part of the field decomposition in modes in \cref{eq:qftcs.ChiFieldExpansion}.}
\begin{equation}
    \bs{P}_h=
    \begin{pmatrix}
   v_h & 0 & 0 & v_h^* \\
  v^{\prime}_h & 0 & 0 & v^{\prime *}_h \\
    0 & v_h^* & v_h & 0 \\
    0 & v_h^{\prime *} & v^{\prime}_h & 0
    \end{pmatrix},
\end{equation}
with $\hat{\bs{A}}_{\vk}^{(\chi)}=\bs{P}_h \cdot\hat{\bs{A}}_{\vk}$ so that also $\bs{\sigma}^{(\chi)}_{\vk}=\bs{P}_h \cdot\bs{\sigma}_{\vk}\cdot\bs{P}_h^\top$.

In Gaussian states, all higher-order $n$-point functions can be expressed in terms of the mean and the covariance matrix. This property greatly simplifies computations involving Gaussian states as opposed to generic quantum states. Furthermore, Gaussian states include vacuum, thermal, or squeezed states, which describe well the outcomes of the experiments that we will analyze.

Linear evolution preserves Gaussianity. Hence, from \eqref{eq:qftcs.QuantEvol}, a quantum system prepared in a Gaussian state characterized by $\bs{\mu}(\eta_0)$ and~$\bs{\sigma}(\eta_0)$ will evolve into another Gaussian state given by
\begin{equation}
\begin{split}
    &\bs{\mu}(\eta)=\bS(\eta-\eta_0)\cdot\bs{\mu}(\eta_0),\\
    &\bs{\sigma}(\eta)=\bS(\eta-\eta_0)\cdot\bs{\sigma}(\eta_0)\cdot\bS^\top(\eta-\eta_0).
\end{split}
\end{equation}

In a scenario such as the one described in subsection \ref{subsec:qftcs.inout}, assuming that we start with a system initially prepared in a Gaussian state~\mbox{$(\bs{\mu}_{\text{in}}, \bs{\sigma}_{\text{in}})$}, the system at time $\eta_{\textrm f}$ (right when the expansion of the geometry ends) will be in the Gaussian state
\begin{equation}
    \bs{\mu}_{\text{out}}=\bS\cdot\bs{\mu}_{\text{in}}\qquad\text{and}\qquad\bs{\sigma}_{\text{out}}=\bS\cdot\bs{\mu}_{\text{in}}\cdot\bS^\top,
\end{equation}
and evolve freely afterward. If we consider a single $({\vk},\bk)$ subsystem,~$\bS$ is given by \eqref{eq:qftcs.SmatrixTh} when $\bs{\mu}$ and $\bs{\sigma}$ are written in the creation/annihilation operator basis. These formulas allow us to predict any observable provided that the initial state is Gaussian. We will use the above machinery later, in section \ref{sec:qftcs.entanglement}, to discuss entanglement between the~\mbox{$({\vk}, \bk)$} produced modes. 

Note that the covariance matrix for the scalar field $\chi$ described in \cref{sec:qftcs.scalarfielddynamics} can be completely determined by the symmetrized two-point functions of the field and its conjugate momentum,
\begin{equation}
    \bs{\sigma}^{(\chi)}_{\vk} =\frac{1}{\mathcal{V}}
    \begin{bmatrix}
    0 & 0 &  \braket{\{\hat{\Tilde{\chi}}_{\vec{k}}, \hat{\Tilde{\chi}}_{\bk}\}} &  \braket{\{\hat{\Tilde{\chi}}_{\vec{k}}, \hat{\Tilde{\pi}}_{\bk}\}}\\
    0 & 0 &  \braket{\{\hat{\Tilde{\pi}}_{\vec{k}}, \hat{\Tilde{\chi}}_{\bk}\}} &  \braket{\{\hat{\Tilde{\pi}}_{\vec{k}}, \hat{\Tilde{\pi}}_{\bk}\}} \\
     \braket{\{\hat{\Tilde{\chi}}_{\bk}, \hat{\Tilde{\chi}}_{\vec{k}}\}} & \braket{\{\hat{\Tilde{\chi}}_{\bk}, \hat{\Tilde{\pi}}_{\vec{k}}\}} & 0 & 0 \\
    \braket{\{\hat{\Tilde{\pi}}_{\bk}, \hat{\Tilde{\chi}}_{\vec{k}}\}} & \braket{\{\hat{\Tilde{\pi}}_{\bk}, \hat{\Tilde{\pi}}_{\vec{k}}\}} & 0 & 0
    \end{bmatrix}.
\label{eq:CovarianceMatrixFields}
\end{equation} 

The three linearly independent correlations that one needs to access are $\braket{\{\hat{\Tilde{\chi}}_{\vk}(\eta), \hat{\Tilde{\chi}}_{\bk}(\eta)\}}$, $\braket{\{\hat{\Tilde{\chi}}_{\vk}(\eta)\, \hat{\Tilde{\pi}}_{\bk}(\eta)\}}$, and $\braket{\{\hat{\Tilde{\pi}}_{\vk}(\eta), \hat{\Tilde{\pi}}_{\bk}(\eta)\}}$. However, note that if we invoke the relation between field and conjugate momentum, as well as the equation of motion \eqref{eq:qftcs.ScalarFieldEOM}, we can write all correlations in terms of one of them and its derivatives. Let us exemplify this with the momentum-momentum correlator, from which the other can be extracted as
\begin{equation}
\begin{split}
    \frac{\dd}{\dd \eta} \braket{\{\hat{\Tilde{\pi}}_{\vk}(\eta), \hat{\Tilde{\pi}}_{\bk}(\eta)\}} &= -2\omega_h^2\braket{\{\hat{\Tilde{\chi}}_{\vk}(\eta), \hat{\Tilde{\pi}}_{\bk}(\eta)\}}, \\
    \frac{\dd^2}{\dd \eta^2} \braket{\{\hat{\Tilde{\pi}}_{\vk}(\eta), \hat{\Tilde{\pi}}_{\bk}(\eta)\}} &= -4\omega_h\omega_h^{\prime} \braket{\{\hat{\Tilde{\chi}}_{\vk}(\eta), \hat{\Tilde{\pi}}_{\bk}(\eta)\}} \\
   &\hspace{0.8cm}-2\omega_h^2 \braket{\{\hat{\Tilde{\pi}}_{\vk}(\eta), \hat{\Tilde{\pi}}_{\bk}(\eta)\}} \\
    &\hspace{0.8cm} + 2\omega_h^4\braket{\{\hat{\Tilde{\chi}}_{\vk}(\eta), \hat{\Tilde{\chi}}_{\bk}(\eta)\}}.
\label{eq:qftcs.DerivativesPiCorrelator}
\end{split}
\end{equation}
The above relations hold at any time, and allow us to characterize the state at the time~$\eta$. We will discuss these correlations further in the next section, along with their relation to cosmological particle production and the power spectrum. 

\section{Power spectrum of produced particles}
\label{sec:qftcs.powerspectrum}

We have seen that Gaussian states are determined by the first and second order moments, $\vec{\mu}$ and $\vec{\sigma}$, respectively. For all the quantum states we will consider, we will have that $\vec{\mu}=0$. Therefore, in order to completely determine the state of the field, we only need to specify the momenta, i.e. the two-point correlation functions.

In particular, let us write the symmetrized, equal-time two-point function of the field $\chi$, which is given by
\begin{equation}
\begin{split}
    \mathcal{G}_{\chi\chi} (\eta, \bm{x}, \vec{y}) &= \frac{1}{2}\braket{\{\hat{\chi}(\eta,\vec{x}), \hat{\chi}(\eta, \vec{y})\}}\\
    &= \frac{1}{2}\int \dd \mu_{\vk} \dd \mu_{\vk^{\prime}} \mathcal{H}_{\vk} (\vec{x}) \mathcal{H}_{\vk^{\prime}} (\vec{y}) \braket{\{\hat{\Tilde{\chi}}_{\vk}(\eta), \hat{\Tilde{\chi}}_{{\vk}^{\prime}}(\eta)\}} \\
    &= \int \dd \mu_{\vk} \dd \mu_{\vk^{\prime}}\mathcal{H}_{\vk} (\vec{x}) \mathcal{H}_{\vk^{\prime}} (\vec{y}) \Big[v_h v_{h^{\prime}}\braket{\hat{a}_{\vk}\hat{a}_{{\vk}^{\prime}}}\\
    &\hspace{0.5cm}+v_h v_{h^{\prime}}^*\braket{\hat{a}_{\vk}\hat{a}_{\vk^{\prime}}^{\dagger}}+v_h^*v_{h^{\prime}}\braket{\hat{a}_{\vk}^{\dagger}\hat{a}_{{\vk}^{\prime}}}+v_h^*v_{h^{\prime}}^*\braket{\hat{a}_{\vk}^{\dagger}\hat{a}_{\vk^{\prime}}^{\dagger}}\Big]
\label{eq:qftcs.PhiPhiCorrelatorGeneral}
\end{split}
\end{equation}
where we have decomposed the field as in \eqref{eq:qftcs.ChiFieldExpansion}. If the state of the system $\ket{\Psi}$ is homogeneous and isotropic, then the two-point function will depend only on the spatial distance between $\vx$ and $\vy$, as given by the line element \eqref{eq:cosmo.FLRWLineElementGeneral}. Let us consider a thermal state, for which the only expectation values that survive\footnote{Note that in a thermal state $\braket{\hat{a}_{\vk}^{\dagger}\hat{a}^{\dagger}_{{\vk}^{\prime}}}=\braket{\hat{a}_{\vk}\hat{a}_{{\vk}^{\prime}}}=0$, since after acting with the operators the resulting states are orthogonal.} are $\braket{\hat{a}_{\vk}\hat{a}_{{\vk}^{\prime}}^{\dagger}}=n_{h}^{\text{in}} \delta^{(D)}({\vk}, \vk^{\prime})$ and $\braket{\hat{a}_{\vk}^{\dagger}\hat{a}_{{\vk}^{\prime}}}=(1+n_{h}^{\text{in}}) \delta^{(D)}({\vk}, \vk^{\prime})$, where $n_{h}^{\text{in}}$ follows a thermal distribution. The proportionality to $\delta^{(D)}(\vk, \vk^{\prime})$ allows one to perform one of the two integrals, so that the two-point function can be written as
\begin{equation}
    \mathcal{G}_{\chi\chi} (\eta, \vx, \vy) = \int \dd \mu_{\vk} \mathcal{H}_{\vk}(\vx)\mathcal{H}_{\vk}(\vy) \left(1+2n_{h}^{\text{in}}\right)\abs{v_h(\eta)}^2.
\label{eq:qftcs.ChiChiCorrelator}
\end{equation} 
Similarly, the two-point function of the derivative of the field can be written as
\begin{equation}
    \mathcal{G}_{\dchi\dchi} (\eta, \vx, \vy) = \int \dd \mu_{\vk} \mathcal{H}_{\vk}(\vx)\mathcal{H}_{\vk}(\vy) \left(1+2n_{h}^{\text{in}}\right)\abs{\vp_h(\eta)}^2.
\label{eq:qftcs.ChiDotChiDotCorrelator}
\end{equation}

Let us now define the spectrum of fluctuations as the combination
\begin{equation}
    S_h(\eta) = \left(1+2n_{h}^{\text{in}}\right) \frac{\abs{\vp_h(\eta)}^2}{\omega_h(\eta)},
\label{eq:qftcs.SpectrumDefinition}
\end{equation} 
so that the two-point correlator \eqref{eq:qftcs.ChiDotChiDotCorrelator} can be expressed as
\begin{equation}
    \mathcal{G}_{\dchi\dchi} (\eta, \vx, \vy) = \int \dd \mu_{\vk} \mathcal{H}_{\vk}( \vx)\mathcal{H}_{\vk}(\vy) \omega_h(\eta) S_h(\eta).
\end{equation}
The quantity \eqref{eq:qftcs.SpectrumDefinition} is analogous to the power spectrum\footnote{Note that we chose the convention of calling spectrum to the momentum transform of the two-point function of $\dchi$ instead of that of $\chi$.} defined in subsection~\ref{subsec:cosmo.fluctuations}. When considering an \textit{in}-\textit{out} process such as the one described in the previous section, we can think of $v_h$ as the modes associated with the notion of vacuum of an observer before certain expansion of the geometry, living at $\eta\leq \eta_{\text{i}}$. However, we can also write \eqref{eq:qftcs.ChiDotChiDotCorrelator} in terms of the $u_h$ modes. Making use of the Bogoliubov transformation \eqref{eq:qftcs.BogoliubovTransformationModes} one obtains the spectrum of fluctuations in terms of the Bogoliubov coefficients between $v_h$ and $u_h$ modes. For $\eta\geq \eta_{\text{f}}$, this reads
\begin{equation}
    \!S_h(\eta)\!=\!\left(1\!+\!2n_h^{\text{in}}\right)\!\Bigg\{\frac{1}{2}\!+\!\abs{\beta_h}^2\!+\!\abs{\alpha_h\beta_h}\!\cos{\Big[2\omega_{k,\text{f}}\eta\!+\!\text{arg}(\alpha_h\beta_h)\Big]}\Bigg\},
\label{eq:qftcs.SpectrumBogoliubov}
\end{equation}
where we have used that the $u_h$ modes behave as plane waves \eqref{eq:qftcs.OutModeIC} in this regime, with $\omega_{k, \text{f}}$ being the frequency of the mode equation \eqref{eq:qftcs.ModeEquation} for $\eta \geq \eta_{\text{f}}$. The spectrum~\eqref{eq:qftcs.SpectrumBogoliubov} is non-vanishing even for $n_h^{\text{in}}=0$, i.e. in vacuum. This corresponds to the so-called spontaneous production, and is independent of the initial occupations of the state. Here, one recognizes the vacuum contribution in the first term, the mean number density defined in~\eqref{eq:qftcs.MeanNumberDensity} in the second, and a term that oscillates with twice the frequency after the expansion due to the interference between the produced quanta. The factor in front, corresponding to the initial thermal occupations, will enhance the quantum production in what is known as the stimulated production. For completeness, let us mention that one could also write \cref{eq:qftcs.SpectrumBogoliubov} at an instant~$\eta$ before the end of the expansion, if one uses the same vacuum prescription, i.e. \cref{eq:qftcs.OutModeIC}, making the substitution $\omega_{h,\text{f}} \to \omega_h(\eta)$ (this is what is called the instantaneous lowest energy state (ILES) \cite{Alvarez2022}, and we will discuss it further in this thesis). However, we will focus on the \textit{in}-\textit{out} scenario for the moment. 

A very convenient way of writing \eqref{eq:qftcs.SpectrumBogoliubov} is in terms of an offset $S_{h,0}$, an amplitude~$A_h$ and a phase $\vartheta_h$, as in
\begin{equation}
    S_h(\eta) = S_{h,0} + A_h\cos{\left[2\omega_{k,\text{f}}\left(\eta-\eta_{\text{f}}\right)+\vartheta_h\right]},
\label{eq:qftcs.SpectrumOffsetAmpPhase}
\end{equation}
where
\begin{equation}
\begin{split}
    S_{h,0} &= \left(1+2n_h^{\text{in}}\right)\left[\frac{1}{2}+\abs{\beta_h}^2\right],\\
    A_h &= \left(1+2n_h^{\text{in}}\right)\abs{\alpha_h\beta_h},\\
    \vartheta_h &= \text{arg}(\alpha_h\beta_h) + 2\omega_{k, \text{f}}\eta_{\text{f}}.
\label{eq:qftcs.OffsAmpPhaseDef}
\end{split}
\end{equation}
Note that by means of the Bogoliubov condition, $\abs{\alpha_h} = \sqrt{1+\abs{\beta_h}^2}$, we can write
\begin{equation}
\begin{split} 
A_h &= \sqrt{S_{h,0}^2-\frac{\left(1+2n_h^{\text{in}}\right)^2}{4}}\,.
\label{eq:qftcs.AmpOfOff}
\end{split}
\end{equation}
Thus, both $S_{h,0}$ and $A_h$ are determined solely by the mean occupation number (as long as the initial spectral density $n_h^{\text{in}}$ is known). On the other hand, in order to specify the phase, one needs to know the argument of both $\alpha_h$ and $\beta_h$. With the definition of the spectrum of fluctuations \eqref{eq:qftcs.SpectrumDefinition}, it is easy to rewrite the two-point functions of the field mode $\Tilde{\chi}_{\vk}$ and its conjugate momentum $\Tilde{\pi}_{\vk}$ as
\begin{equation}
\begin{split}
        \expval{\{\hat{\Tilde{\pi}}_{\vk},\hat{\Tilde{\pi}}_{\bk}\}}&=2\omega_h S_h\,,\\
        \expval{\{\hat{\Tilde{\chi}}_{\vk},\hat{\Tilde{\pi}}_{\bk}\}}&=-\frac{\omega_h^{\prime}}{\omega_h^2}S_h-\frac{1}{\omega_h} S^{\prime}_h\,, \\
        \expval{\{\hat{\Tilde{\chi}}_{\vk},\hat{\Tilde{\chi}}_{\bk}\}}&=\frac{1}{2\omega_h^4}\left[2\left(\omega_h^{\prime\prime} - 2\frac{\omega_h^{\prime \, 2}}{\omega_h} + 2\omega_h^3\right)S_h + 2\omega_hS_h^{\prime\prime}\right],
\label{eq:qftcs.CorrelationsSkRelations}    
\end{split}
\end{equation}
which is consistent with \cref{eq:qftcs.DerivativesPiCorrelator}. These relations are valid at all times, and simplify in a static regime in which the scale factor is constant. With the above equations we can write any element of the mean and covariance matrix of the subsystem $({\vk},\bk)$ in terms of the spectrum~$S_h$, as well as their first and second time derivatives. If the state is Gaussian, the three correlations \eqref{eq:qftcs.CorrelationsSkRelations} suffice to characterize the state of the field.

Note that to arrive at this result we have only made use of three assumptions: 1) the homogeneity and isotropy of the system, 2) the specific dynamics of the system, and 3) the Gaussianity of the state. The role of each assumption consists on the following. Assumption 1) guarantees that the dynamical equations can only mix ${\vk}$ and $\bk$ modes and depend only on $h$ (see the mode equation \eqref{eq:qftcs.ModeEquation}). This is sufficient to ensure that the quantum state of the system---and in particular the covariance matrix---can be reconstructed from the spectrum as defined above, only in terms of $h$. Assumption 2) provides the specific dependence of the quantum state---and in particular the covariance matrix---in terms of $h$. Assumption 3) guarantees that knowledge of the covariance matrix (and first moments) is enough to reconstruct the full quantum state. Changing assumption 2)---i.e. changing the specific dynamics of the system---would lead to different relations \eqref{eq:qftcs.CorrelationsSkRelations}. This will lead to different coefficients of the covariant matrix, but would still allow to fully determine the quantum state (that is, to perform \textit{state tomography}) in the same way.

It is straightforward to check that knowledge of the offset $S_{h, 0}$, amplitude $A_h$, and phase $\vartheta_h$ are equivalent to knowing $S_h$ as a function of time, and therefore these parameters also encode complete information about the state of the subsystem of modes $({\vk},\bk)$. This will be very useful when discussing analog experiments. For completeness, we give the form of the covariance matrix in terms of these parameters (in the creation/annihilation operator basis),
\begin{equation}
   \bs{\sigma}_{\vk}^{(\text{out})} = 
   2
    \begin{pmatrix}
    0 & S_{h, 0} & -A_h e^{-\ci\vartheta_h} & 0\\
    S_{h, 0} &  0 & 0 & -A_h e^{\ci\vartheta_h} \\
   -A_h e^{-\ci\vartheta_h} & 0 & 0 & S_{h, 0} \\
    0 & -A_h e^{\ci\vartheta_h} & S_{h, 0} & 0
    \end{pmatrix}.
\label{eq:qftcs.CovarianceMatrixOffAmpPhase}
\end{equation}

Once the state of the system is completely determined, one can proceed to study entanglement between the modes $({\vk},\bk)$. 

\section{Entanglement}
\label{sec:qftcs.entanglement}

Entanglement is a definitive indicator of the quantum nature of a composite system, since classical subsystems cannot be entangled. However, identifying and measuring this quantity can be challenging in many scenarios. To detect the presence of entanglement, one can use a so-called entanglement witness, which is a measure that distinguishes between separable and non-separable (i.e. entangled) states in a quantum system. An example is the Cauchy-Schwarz inequality~\cite{Busch2014a,Busch2014b,Wasak2014}. In particular, for a two-mode quantum system $(A,B)$ we will consider
\begin{equation}
    \Delta = \abs{\langle \hat{a}_{A}\hat{a}_{B}\rangle}^2 - \langle\hat{a}_{A}^\dagger\hat{a}_{A}\rangle\langle\hat{a}_{B}^\dagger\hat{a}_{B}\rangle > 0.
    \label{eq:qftcs.CS}
\end{equation}
This inequality is satisfied if and only if the two modes are entangled, though its violation does not guarantee the absence of entanglement. Therefore, Cauchy-Schwarz provides a sufficient condition to confirm entanglement between $A$ and~$B$, and thus it is called an unfaithful witness\footnote{For the witness to be faithful, it must provide a sufficient and necessary condition for entanglement.}. Moreover, the value of $\Delta$ does not correlate with the amount of entanglement: Higher (positive) $\Delta$ does not necessarily imply a greater degree of entanglement. To go beyond detection and actually quantify the amount of entanglement, we turn to an entanglement quantifier. This is a quantity that behaves as an entanglement monotone: A higher value indicates a greater amount of entanglement (see \cite{Plenio2007} for further details). For systems in a pure state, a common entanglement quantifier is the von Neumann entropy of each subsystem. However, for mixed states, this measure also includes classical correlations and thus does not exclusively quantify entanglement. This is precisely our case, since we are interested in quantifying the entanglement of two-mode subsystems $(\vk, \bk)$, which are in a mixed state if the initial state is thermal.

Logarithmic negativity is a faithful entanglement quantifier for bipartite systems $(A,B)$ as long as the state $\hat{\bm{\rho}}_{AB}$ is Gaussian and one of the subsystems has only one degree of freedom, regardless of whether the state is pure or mixed \cite{Peres1996,Plenio2005}. Under these conditions, which are met by our system, logarithmic negativity is positive if and only if entanglement exists, and a higher logarithmic negativity corresponds to more entanglement. Logarithmic negativity is defined from the density matrix of the system as
\begin{equation}
    \LN(\hat{\bm{\rho}}_{AB}) = \log_2\left(\text{tr}\,\hat{\bm{\rho}}_{AB}^{\top_{A}}\right)
\end{equation}
where $\hat{\bm{\rho}}_{AB}^{\top_{A}}$ is the partial transpose of the density matrix with respect to subsystem~$A$~\cite{Plenio2005}. For a bipartite Gaussian state with covariance matrix $\bm{\sigma}_{AB}$, LN can be computed from the symplectic eigenvalues~$\tilde{\nu}_I$ of the partially transposed (with respect to any of the subsystems) covariance matrix $\tilde{\bm{\sigma}}_{AB}$. The symplectic eigenvalues of a matrix~$\bm{\sigma}_{AB}$ are defined as the absolute value of the eigenvalues of the matrix~\mbox{$i\bm{\Omega}\bm{\sigma}_{AB}$ \cite{Serafini2017}}, whereas the partially transposed covariance matrix~$\tilde{\bm{\sigma}}_{AB}$ is obtained from~$\bm{\sigma}_{AB}$ by reversing the sign of all the components in~$\bm{\sigma}_{AB}$ involving momenta of the subsystem with respect to which the partial transpose is performed---or equivalently by switching rows and columns corresponding to creation and annihilation operators. LN can then be expressed as
\begin{equation}
    \LN(\bm{\sigma}_{AB}) = \sum_I \text{Max}\left[0, -\log_2 \Tilde{\nu}_I\right]\,,
    \label{eq:qftcs.LN}
\end{equation}
where the eigenvalues $\tilde{\nu}_I$ are independent of the subsystem chosen for the partial transpose. From the above, it is clear that entanglement between $A$ and $B$ will exist if and only if at least one symplectic eigenvalue is smaller than $1$. In our case, the two symplectic eigenvalues of the partially transposed covariance matrix associated to~\eqref{eq:qftcs.CovarianceMatrixOffAmpPhase} are~\mbox{$\Tilde{\nu}_h=2\abs{A_h \pm S_{h,0}}$}. Using the definition of offset, amplitude and phase in \eqref{eq:qftcs.OffsAmpPhaseDef} and the identities satisfied by the Bogoliubov coefficients \eqref{eq:qftcs.BogoliubovCoefficientsNorm}, we can demonstrate that $A_h \geq 0$ and $S_{h,0}\geq 1/2$, and therefore $2(A_h+S_{h, 0})\geq1$, for any possible value of $\beta_h$. Hence, only one symplectic eigenvalue of the partially transposed covariance matrix contributes to LN, with the value
\begin{equation}
    \Tilde{\nu}_h^{\text{min}}=2(S_{h, 0} - A_h)\,,
    \label{eq:qftcs.EigenvaluesOffAmp}
\end{equation}
so that one can write
\begin{equation}
    \LN_h = \text{Max}\left\{0, -\log_2 \left[2(S_{h, 0} - A_h)\right]\right\}.
\label{eq:qftcs.LNOffAmp}
\end{equation}
Thus, LN$_h$ indicates entanglement in our system if and only if \mbox{$S_{h, 0}-A_h<1/2$}. Moreover, for a covariance matrix in the form \eqref{eq:qftcs.CovarianceMatrixOffAmpPhase} the Cauchy-Schwarz inequality~\eqref{eq:qftcs.CS} reads
\begin{equation}
    \Delta_h =  A_h^2 - \left(S_{h, 0}-1/2\right)^2  \,.
    \label{eq:qftcs.CSOffAmp}
\end{equation}
Since $A_h\geq0$ and $S_{h, 0}\geq1/2$, Cauchy-Schwarz inequality detects entanglement if~\mbox{$S_{h, 0}-A_h < 1/2$}. This is the same condition derived for LN, which tells us that $\Delta_h$ behaves as a faithful witness in this cosmological scenario. Furthermore, the degree of entanglement of the subsystem only depends on the eigenvalue $h^2$.

To determine whether $\Delta_h$ is also an entanglement monotone, note that both $A_h$ and $S_{h, 0}$ increase monotonically with $\beta_h$, while \mbox{$S_{h, 0}-A_h$} decreases monotonically (while remaining  positive), or vanishes in the limit $|\beta_h|\to\infty$. This implies that, when nonzero, LN increases monotonically with $|\beta_h|$. It is also straightforward to verify that LN decreases monotonically with $n_{h}^{\text{in}}$. On the other hand, using \eqref{eq:qftcs.AmpOfOff} we find
\begin{equation}
    \Delta_h=S_{h, 0}-\frac{(1+2n_h^{\text{in}})^2+1}{4} = \left(1+2n_h^{\text{in}}\right)\abs{\beta_h}^2 - (n_h^{\text{in}})^2.
\end{equation} 
While $\Delta_h$ also increases monotonically with $|\beta_h|^2$, given that $S_{h, 0}$ does, its behavior with respect to $n_h^{\text{in}}$ depends on the magnitude of the number density produced. If $|\beta_h|^2>n_h^{\text{in}}$, then $\Delta_h$ acts as a quantifier. However, in general, $\Delta_h$ can only be considered a faithful witness (assuming that the conditions 1 to 3 stated before eq.~\eqref{eq:qftcs.CovarianceMatrixOffAmpPhase} are met).

Although all these tools apply in general to all scenarios studied in the thesis, we will make use of them to analyze in particular entanglement in analog pair production in BECs and determine its detectability in \cref{ch:entanglement}.

\section{The Schwinger effect}
\label{sec:qftcs.schwinger}

Spontaneous particle production occurs in the presence of an external, time-dependent agent, independently of the latter being an expanding metric or any other field. Let us now introduce the Schwinger effect, to which we will dedicate some chapters and sections in part IV of this thesis.

The Schwinger effect \cite{Sauter1931,Schwinger1951} is a particle creation phenomenon generated by intense electromagnetic fields. This occurs for electric strengths larger than the so-called Schwinger limit \cite{Heisenberg1936}, \mbox{$1.3 \cdot 10^{18} \, \text{V/m}$}. Below this regime, production is exponentially suppressed. This effect is non-perturbative in the case of a constant electric field. However, for rapidly-varying contributions, mostly high-frequencies are involved, and pair production becomes perturbative. This phenomenon is also called multiphoton Breit-Wheeler pair production~\cite{Breit1934,Reiss1962}. We will refer, as it is usual in the literature, to the Schwinger effect including both contributions, as the non-perturbative treatment also incorporates the perturbative effects.

In the Schwinger effect, a strong electromagnetic field excites a charged matter field in $(1+D)$-Minkowski spacetime, generating particle pairs in the same way as a time-dependent geometry does in Cosmology\footnote{In fact, recent works have considered Schwinger effect as a dark matter production mechanism~\cite{Bastero2023a,Bastero2023b}.}. While in the latter case the time-dependent external agent is the spacetime metric, now this role is played by an electromagnetic potential~$A_{\mu}(t,\vx)$. The action of a complex scalar field $\psi(t,\vx)$ coupled to such potential is
\begin{equation}
S = -\int \dd t \dd ^Dx \left[\left(\partial^{\mu}-iqA^{\mu}\right)\psi^*\left( \partial_{\mu}+iqA_{\mu}\right)\psi + m^2\psi^*\psi\right],
\label{eq:qftcs.SchwingerAction}
\end{equation}
where $q$ and $m$ are its charge and mass, respectively. The dynamics of the field is determined by the Klein-Gordon equation
\begin{equation} 
    \left[\left(\partial_{\mu}+iqA_{\mu}\right)\left(\partial^{\mu}+iqA^{\mu}\right)+m^2\right]\psi(t, \vx) = 0.
\label{eq:qva.FieldEOM}
\end{equation}
If we consider a homogeneous but time-dependent electric field $\textbf{E}(t)$, the temporal gauge, defined by $A_{\mu}(t,\vx)=(0,\textbf{A}(t))$, makes the equations of motion explicitly homogeneous. In this gauge, $\textbf{E}(t)=-\dot{\textbf{A}}(t)$. Without loss of generality, we can fix the electric field to lie in the $z$-direction.

Unlike in the cosmological case, the Fourier modes~$\psi_{\vk}(t)$ directly satisfy decoupled harmonic oscillator equations of the type~\eqref{eq:qftcs.ModeEquation}, with no need of rescaling the field as in \eqref{eq:qftcs.FieldsRelation}. In addition, in contrast to the previous cases, the privileged direction of the electric field breaks isotropy. This translates into time-dependent frequencies that depend on the angle $\theta$ between the wavevector $\vk$ and the direction of the vector potential $\textbf{A}(t)$ through an additional term linear on $k$,
\begin{equation}
    \Omega_{\vk}^2(t) = k^2 + 2qkA(t)\cos{\theta} + q^2A^2(t) + m^2.
    \label{eq:qftcs.SchwingerFrequency}
\end{equation}
Other than that, quantization and particle production follow similarly. As in cosmological production, we will neglect backreaction effects, although this should be taken into consideration for supercritical electric field intensities and large particle densities.

Let us mention that no derivatives of the electric potential $\vec{A}(t)$ appear in the frequency \eqref{eq:qftcs.SchwingerFrequency}, and therefore this phenomenon is close to the conformally coupled case in Cosmology (cf. \cref{eq:qftcs.ConformalCoupling}). This will be of particular importance when discussing the effects of the switch-on and -off of the electric potential or the scale factor in \cref{ch:switcheffects}.

%% file: Chapters/AnalogGravity.tex


\chapter{Analog gravity experiments} 

\label{ch:analogs} 




QFTCS aims to describe the quantum behavior of fields in a fully classical background. However, probing strong curvature effects directly is a challenging task. The vast expansion of the early Universe or Black Hole scenarios are obviously hard to reach, and this is where analog gravity experiments come into play.

Analog gravity is based on the principle of analogical reasoning~\cite{Bartha2022}, which states that if two systems share the same mathematical structure, they are likely to exhibit the same physical behavior. Typically, one of these systems is inaccessible, and the other can be realized in the laboratory. Although some works are interested in reproducing Einstein's field equations (see for example refs.~\mbox{\cite{Girelli2008,Sindoni2009}}), we will consider analog systems which are described by, in our case, a scalar field whose effective action acquires the form \eqref{eq:qftcs.CurvedActionNM} for a certain number of spatial dimensions~$D$, mass of the field $m$, and coupling to the Ricci scalar $\xi$, i.e., moving in a fixed gravitational field. In other words, we aim at establishing an analogy at the level of the equations of motion. Analogs of other fields can also be achieved, and we will briefly discuss them in \cref{sec:analogs.dm}, although the scalar field is the most commonly reproduced scenario. 

In this chapter, we will introduce the reader to the main ideas of analog gravity and briefly review some of the experiments that have been carried out in recent years. We will pay special attention to Bose-Einstein condensates, as they are one of the most common platforms for analog gravity experiments and the main analog system in this thesis. We have mainly based this chapter on references \cite{Visser2002,Barcelo2011,Jacquet2020,Almeida2023}, and the references therein.

\section{A dynamical analogy}

Analogies have been used historically to understand complex systems, and in the context of General Relativity, they have been used to explain concepts such as BHs. It was precisely after the description of BH radiation by Hawking in~1975~\mbox{\cite{Hawking1975,Gibbons1977}} that Unruh proposed in 1981 that sound waves encountering a so-called \textit{acoustic horizon} in a fluid could be seen as an analogy of the Hawking effect that could be realized in the laboratory \cite{Unruh1981}. The additional ingredient included in Unruh's proposal, in contrast to the traditional use of analogical reasoning, was the consideration of the analogy as a means to demonstrate the existence of the inaccessible phenomena. This marked the beginning of the analog gravity program as it is known today\footnote{Some authors \cite{Almeida2023} consider previous analog developments as an earlier period of the history of analog gravity, with Unruh's work as the starting point of the modern analog gravity program.}.

The idea behind the fluid analogy for BHs is very simple, and we will try to illustrate it with an example. Imagine a river in which there is a region where the water flows faster than the maximum speed at which a fish can swim. As long as the fish is in the slow region, it will be able to swim against the current. However, as soon as it reaches the fast regime, it will be dragged by the flow, and it will not be able to escape. Sound waves on top of a background fluid behave very much like our fish. The limit between the two flow regimes is called the \textit{acoustic horizon}, and it is the point where the flow reaches the speed of sound. No sound signals can escape the trapped, supersonic region. Photons and BHs behave similarly. Upon quantization of the fluctuating field, exactly in the same way as we described in \cref{ch:qftcs}, phonons (the quasiparticle associated with sound) will be created in pairs at the horizon, analogously to Hawking radiation. 

The feature that sound in fluids and the rest of the analog gravity systems share is the form of the equations of motion of small excitations on top of a classical background. All of them can be rewritten or interpreted in terms of an effective metric, the so-called \textit{acoustic metric}, characterized by the properties of the ground state of the system, such that the action for these fluctuations acquires the typical form of that of a field in a curved spacetime (for scalar fields, that would be \cref{eq:qftcs.CurvedActionNM}). In this way, by tuning the experimental system, one can modify the acoustic metric, thus probing different analog spacetime geometries for the fluctuating field. We have discussed BHs for historical reasons, but we are interested in cosmological scenarios, to which we devote part of this thesis. Indeed, the following discussion should be thought of as applied to Cosmology. In particular, if one is able to reproduce an FLRW metric, the analog of cosmological particle production could be studied. 

\subsection{The acoustic metric}

Fluid dynamics is governed by the continuity equation,
\begin{equation}
    \partial_t \rho + \nabla \cdot (\rho \mathbf{v}) = 0,
\label{eq:analogs.Continuity}
\end{equation}
and the Euler equation
\begin{equation}
    \rho \left[ \partial_t \mathbf{v} + (\mathbf{v} \cdot \nabla) \mathbf{v} \right] = - \nabla p,
\label{eq:analogs.Euler}
\end{equation}
where $\rho$ is the fluid density, $\mathbf{v}$ is the velocity field, and we have assumed an inviscid fluid so that the only force in the picture is the pressure~$p$. Assuming a locally irrotational, barotropic fluid, \cref{eq:analogs.Euler} can be rewritten as\footnote{Note that this is Bernoulli's equation.}
\begin{equation}
    -\partial_t \phi + h + \frac{1}{2}(\nabla \phi)^2 = 0,
\label{eq:analogs.Bernoulli}
\end{equation}
where we introduced the velocity potential $\mathbf{v} = -\nabla \phi$ and the specific enthalpy $h$ is a function of $p$ only,
\begin{equation}
    h(p) = \int_0^p \frac{\dd p^{\prime}}{\rho(p^{\prime})}.
\end{equation}
The next step is to linearize the above equations of motion~\eqref{eq:analogs.Continuity} and~\eqref{eq:analogs.Bernoulli} around a background flow characterized by a velocity potential $\phi_0$, pressure~$p_0$, and density~$\rho_0$,
\begin{equation}
    \phi = \phi_0 + \phi_1, \quad \rho = \rho_0 + \rho_1, \quad p = p_0 + p_1,
\label{eq:analogs.Linearization}
\end{equation}
and keep only the linear terms in perturbations. This leads to the wave equation\footnote{Barotropicity implies that to linear order in perturbations $h = h_0 + p_1/\rho_0$, which simplifies the equations.}
\begin{equation}
\begin{split}
    \hspace{-2em}&-\partial_t\left[c_{\text{s}}^{-2}\rho_0 \left(\partial_t \phi_1 + \vec{v}_0 \cdot \nabla \phi_1 \right)\right] \\
    &\hspace{1.5cm}+ \nabla \left[\rho_0 \nabla \phi_1 - c_{\text{s}}^{-2} \rho_0 \vec{v}_0 \left(\partial_t\phi_1 + \vec{v}_0 \cdot \nabla \phi_1 \right)\right] = 0,
\label{eq:analogs.WaveEquation}
\end{split}
\end{equation}
for the fluctuating field\footnote{For clarity, we have omitted the linearized equations for the density and pressure perturbations, which can be directly obtained from $\phi_1$.} $\phi_1$, where the speed of sound $c_{\text{s}}$ is defined as
\begin{equation}
    c_{\text{s}}^2 = \frac{\dd p}{\dd \rho}.
\label{eq:analogs.SpeedOfSoundFluid}
\end{equation}
Crucially, \cref{eq:analogs.WaveEquation} can be rewritten in terms of an effective metric $g_{\mu\nu}$ (for $D\geq 2$) as
\begin{equation}
    \frac{1}{\sqrt{g}}\partial_{\mu}\left(\sqrt{g}g^{\mu\nu}\partial_{\nu}\phi_1\right),
\label{eq:analogs.EffectiveWaveEquation}
\end{equation}
with
\begin{equation}
    g_{\mu\nu} = \left(\frac{\rho_0}{c_{\text{s}}}\right)^{2/(D-1)} \begin{bmatrix}
    -(c_{\text{s}}^2-v_0^2) & -\vec{v}_0^{\top} \\
    -\vec{v}_0 & \mathbb{1}_{D}
    \end{bmatrix}.
\label{eq:analogs.AcousticMetricGeneral}
\end{equation}
This is precisely the KG equation for a massless, minimally coupled scalar field in a curved spacetime described by the acoustic metric, whose corresponding line element is given by
\begin{equation}
    \dd s^2 = \left(\frac{\rho_0}{c_{\text{s}}}\right)^{2/(D-1)}\left[-\left(c_{\text{s}}^2 - v_0^2\right)\dd t^2 - 2\vec{v}_0 \cdot \dd t \dd\vec{x}  + \dd\vec{x}^2\right].
\end{equation}
Note that the acoustic metric \eqref{eq:analogs.AcousticMetricGeneral} is not well-defined for the case $D=1$. This is related to the fact that conformal symmetries always allow a mapping to Minkowski spacetime. However, this will not be a problem for us, because we will be mainly interested in the cases $D=2$ and $D=3$. Importantly, it is the background parameters that determine the form of the metric; therefore, by tuning the laboratory system, one can achieve different effective geometries, such as BH-like or cosmological.

We have seen how the motion of sound in a fluid is ruled by the same dynamics as a scalar field in a curved spacetime. All analog gravity systems aim at reproducing this analogy, not necessarily realizing the same exact form of the acoustic metric, although this is indeed the case of Bose-Einstein condensates.

\section{Brief history of analog gravity}
\label{sec:analogs.history}

Unruh's proposal was largely unnoticed until the $90$s, when the first works by Jacobson further pursued the idea of the BH analogy~\mbox{\cite{Jacobson1991, Jacobson1993}}, and many theoretical works followed, not only by his group, but also by Comer \cite{Comer1992}, Visser \cite{Visser1993,Visser1998}, Volovik \cite{Volovik1995,Volovik1996} and Unruh~\cite{Unruh1995}, to cite a few. By the start of the new century, there was a focus on condensed matter systems through the works \cite{Eltsov1998, Baldovin2000, Garay2000,Barcelo2001}. More theoretical developments on other platforms include references~\mbox{\cite{Volovik2003,Barcelo2003,Barcelo2003b,Barcelo2003c,Volovik2009,Barcelo2011,Delhom2024}}.

Despite the theoretical boom in the field, it was not until 2008 that the first experiments took place. The first one was carried out by Philbin \textit{et al.} \cite{Philbin2008}, who observed the formation of a horizon in a fiber-optical analog. In the following years, a myriad of experiments in different analog systems would come, probing QFTCS in different scenarios, such as Hawking effect \cite{Philbin2008,Weinfurtner2011,Euve2016,MunozDeNova2019,Drori2019,Shi2023}, superradiance~\mbox{\cite{Torres2017,Giacomelli2021,Braidotti2022}}, pair creation in expanding universes\footnote{Pay special attention to the $1+1$ cases, which are special, as we mentioned above.} \cite{Eckel2018,Wittemer2019,Banik2021,Steinhauer2022,Experiment2022,ScatteringExp2024}, or the dynamical Casimir effect \cite{Jaskula2012}. These experiments have been carried out in different physical platforms, confirming the universality of these phenomena \cite{Barcelo2011}. New proposals of analog gravity systems have been put forward in the last years, such as microcavity-polaritons \cite{Jacquet2022,Jacquet2023,Falque2024}, Dirac materials \cite{Tolosa2023,Haller2023} or superfluid helium-4 \cite{Svancara2024}.

Nevertheless, there are still many challenges to overcome in the field. The most important one is the detection of unequivocal quantum signals from these phenomena \cite{Robertson2017a,Alonso2018,Berges2018a,Berges2018b,Kunkel2022}, which would establish their quantum origin. Indeed, there have been recent claims of the detection of entanglement in the context of analog BHs \cite{Steinhauer2016}, but we are still lacking the corresponding experimental confirmation in cosmological analog systems\footnote{There have been some success in $1+1$ systems such as in \cite{Chen2021}, but note that in one spatial dimension the acoustic FLRW metric cannot be realized.} \cite{Jacquet2020,Almeida2023}. The reason is the necessity of stimulating these processes to attain their observability, as well as the non-vanishing temperature of the environment. As a consequence of both, quantum signals have a poor signal-to-noise ratio in the performed experiments, if present at all. Though this might be seen as a minor aspect, the observation of unambiguous quantum signals is crucial to ensuring the quantum field theoretic origin of the observed phenomena, which could otherwise be fully reproduced by classical theories \cite{Martin2015,Green2020,Ashtekar2020,Agullo2022b}. Notably, QFT is our most successful physical theory, unifying a vast range of observed phenomena under a single framework. However, most of the confirmed predictions lie in the realm of perturbative phenomena. As such, efforts should be directed to the confirmation of basic non-perturbative predictions from the theory, such as quantum pair creation in presence of dynamical backgrounds. Detecting entanglement in this context would provide unmistakable proof of pair creation by QFT mechanisms. Hence, the relevance of experiments with this goal extends beyond analog gravity or QFTCS into the realm of fundamental physics. In chapter~\ref{ch:entanglement}, we analyze the feasibility of detecting entanglement in pair production in a $(1+2)$ BEC \cite{Entanglement2024}, similar to the one used in the experiments \cite{Experiment2022,ScatteringExp2024}, in order to find the optimal regime of experimentally tunable parameters to maximize its detection.

As we will comment in \cref{ch:whatwecanlearn}, one may ask whether these analogies possess confirmatory power in the sense that they can prove phenomena in the target system that they are trying to reproduce, and this has triggered many interesting discussions in the context of Philosophy (see e.g. \cite{Dardashti2017,Crowther2021}). However, we would like to stress that analog experiments do not need to validate inaccessible phenomena to be interesting. They allow us to probe the behavior of fields in non-trivial backgrounds, which may or may not be mapped to some curved spacetime, and to transfer insights between different scenarios (cosmological or not). In this sense, we believe that analog gravity experiments should be regarded as QFT in non-trivial backgrounds simulations, rather than evidence for Hawking radiation or other phenomena.

\section{Bose-Einstein condensates}
\label{sec:analogs.becs}

Let us now particularize to Bose-Einstein condensates and discuss the emergence of the acoustic metric in these systems, since they will be the main analog platform in this thesis. 

BECs (for an exhaustive introduction, see \cite{Pitaevskii2016}) were predicted by Einstein in~1925~\cite{Einstein1925}, based on work by Bose in~1924~\cite{Bose1924}, as a phase transition occurring at low temperatures in a gas of bosons. The atoms in this gas are cooled down to temperatures close to absolute zero, where the de Broglie wavelength of the atoms becomes comparable to the interatomic distance. In this regime, the atoms start to behave as a single quantum entity, and they all occupy the same quantum state, characterized by a quantum field describing what is called a \textit{condensate}. Its realization needed an enormous experimental effort, and it was not until 1995 that it was achieved~\cite{Anderson1995, Davis1995, Bradley1995}.

In the context of analog gravity, BECs were first proposed as an analog system by Garay \textit{et al.}~\mbox{\cite{Garay2000, Garay2001}}, and subsequently developed by Barceló \textit{et al.}~\mbox{\cite{Barcelo2001,Barcelo2003,Barcelo2003b,Barcelo2003c}} and Fedichev \textit{et al.} \cite{Fedichev2003,Fedichev2004}. From there, further theoretical work has been done during the first two decades of the 21st century (some of the most important references are \cite{Novello2002,Fischer2004,Fischer2004b,Uhlmann2005,Calzetta2005,Liberati2006a,Liberati2006b,Weinfurtner2006,Weinfurtner2007,Weinfurtner2009,Schuetzhold2009,Prain2010,Bilic2013}). Thanks to the large experimental efforts in the last decade, BECs and other cold atom platforms have also been extensively explored experimentally, both in the context of analog BHs~\mbox{\cite{Carusotto2008,Lahav2010,Horstmann2010,Weinfurtner2011,Steinhauer2014,Steinhauer2016,MunozDeNova2019}}, Unruh effect~\mbox{\cite{Rodriguez-Laguna2017,Hu2019,Gooding2020}}, analog cosmologies~\mbox{\cite{Eckel2018,Banik2021,Chen2021,Experiment2022,Tajik2023,ScatteringExp2024}}, false vacuum decay~\mbox{\cite{Zenesini2024,Jenkins2024}} or even the dynamical Casimir effect \cite{Jaskula2012}. For a glimpse of two-component BECs, which allow for an additional massive phononic mode, we refer the reader to~\mbox{\cite{Liberati2006a,Liberati2006b}}.

For later convenience, we will recover units in this section, and work in spherical coordinates. 

\subsection{Dynamics of the ground state}

The dynamics of a BEC is described by the Gross-Pitaevskii (GP) equation \cite{Gross1961, Pitaevskii1961,Lifshitz1980},
\begin{equation}
    i \hbar \partial_t \Psi = \left(-\frac{\hbar^2}{2m}\nabla^2 + V + \lambda|\Psi|^2\right)\Psi,
\label{eq:analogs.GrossPitaevskii}
\end{equation}
where $\Psi$ is the wavefunction of the condensate, $m$ is the mass of the atoms, $V$ is the external trapping potential, and $\lambda$ is the interaction strength. The GP equation can be obtained from the effective action
\begin{equation}
    \!\Gamma [\Psi]\!=\!\int \text{d}t \, \text{d}^Dx \left\{i\hbar \Psi^* \partial_t \Psi - \frac{\hbar^2}{2m} \nabla \Psi^* \nabla \Psi - V |\Psi|^2\!- \frac{\lambda}{2}|\Psi|^4\right\}.
\label{eq:analogs.QuantumEffectiveAction}
\end{equation}
This is an accurate description for the dynamics of the BEC, which is a weakly-coupled Bose gas, where most atoms occupy the ground state.

The condensate behaves as a superfluid \cite{Bogoliubov1946}, which can easily be seen by writing~$\Psi$ in the Madelung representation \cite{Madelung1927},
\begin{equation}
    \Psi (t, \vec{r}) = \sqrt{n (t, \vec{r})} e^{i S (t, \vec{r})},
\label{eq:analogs.MadelungRepresentation}
\end{equation}
with $n (t, \vec{r}) = \abs{\Psi (t, \vec{r})}^2$ denoting the background particle number density and~$S (t, \vec{r})$ being the background phase of the condensate's mean field. Using the parametrization \eqref{eq:analogs.MadelungRepresentation} in the GP eq.\ \eqref{eq:analogs.GrossPitaevskii} leads to hydrodynamic equations almost (see below) equivalent to \cref{eq:analogs.Continuity,eq:analogs.Euler}. Namely, one obtains the local conservation law or continuity equation
\begin{equation}
    0 = \partial_t n + \nabla(n \vec{v}),
\label{eq:analogs.ContinuityBEC}
\end{equation}
and the Bernoulli equation (whose divergence yields the Euler equation)
\begin{equation}
    0 = \hbar \partial_t S + V + \frac{\hbar^2}{2 m} (\vec{\nabla} S)^2 + \lambda n - \frac{\hbar^2}{2 m} \frac{\nabla^2 \sqrt{n}}{\sqrt{n}},
\label{eq:analogs.EulerBEC}
\end{equation}
where the superfluid velocity is given by
\begin{equation}
    \vec{v} = \frac{\hbar}{m} \vec{\nabla} S.
\label{eq:analogs.SuperfluidVelocity}
\end{equation}
The last term in \eqref{eq:analogs.EulerBEC} is called the quantum potential, and is expected to be subleading for a sufficiently smooth density. It is precisely this term that differentiates the background dynamics from those of an irrotational, inviscid fluid, and can be neglected for wavenumbers larger than the so-called healing length, 
\begin{equation}
    \xi = \frac{\hbar}{\sqrt{2 m \lambda n}},
\label{eq:analogs.HealingLength}
\end{equation}
which is the length scale over which the condensate density changes significantly. This is the typical assumption leading to the acoustic approximation \cite{Volovik2009,Barcelo2011}.

\subsection{Deriving the acoustic metric}
\label{subsec:bec.AcousticMetric}

Let us introduce small fluctuations on top of the ground state as 
\begin{equation}
    \Psi = \psi_0 + \frac{e^{iS_0}}{\sqrt{2}}\left[\phi + i 2m \varphi\right],
\label{eq:analogs.BackgroundSplit} 
\end{equation}
where $\psi_0 = \sqrt{n_0}e^{iS_0}$ is the condensate mean field, and fulfills the GP equation~\eqref{eq:analogs.GrossPitaevskii},~\mbox{$\phi, \varphi$} are real, and the factor $2m$ has been introduced for convenience so that the field $\varphi$ has the standard dimensions. We do not consider any backreaction of fluctuations to the form of the action. Fluctuations are assumed to be small enough and will be kept only to linear order in the equations of motion corresponding to quadratic order in the action. Conceptually, this corresponds to a background field which is well described by mean field equations. We will also not consider any renormalization of the couplings in the action. In this sense, the action \eqref{eq:analogs.QuantumEffectiveAction} can actually be identified with (an approximation of) the quantum \textit{effective} action, which is renormalized already.

As a next step, we consider the dynamics of the fluctuations parametrized by the two real fields $\phi$ and $\varphi$ introduced in eq.~\eqref{eq:analogs.BackgroundSplit}. To that end, we expand the effective action \eqref{eq:analogs.QuantumEffectiveAction} around the background solution $\psi_0$ to quadratic order in the fluctuating fields $\phi$ and $\varphi$, which yields
\begin{equation}
    \Gamma [\Psi] = \Gamma [\psi_0] + \text{terms linear in $\phi, \varphi$} + \Gamma_2 [\phi, \varphi],
\end{equation}
with 
\begin{equation}
    \begin{split}
        \Gamma_2 [\phi, \varphi] =& \int \text{d} t \, \text{d}^2 r \Big\{\hbar 2m\varphi \partial_t \phi - \frac{\hbar^2}{4m}\left[(\vec{\nabla} \phi)^2 +4m^2(\vec{\nabla} \varphi)^2  \right]  \\
        & -\frac{1}{2}\left(V+\hbar\partial_t S_0 + \hbar^2\frac{(\vec{\nabla}S_0)^2}{2m}\right)(\phi^2 + 4m^2\varphi^2)\\
		&- \hbar^2 (\vec{\nabla}S_0)(\phi \vec{\nabla}\varphi -\varphi \vec{\nabla}\phi) -\frac{\lambda n_0}{2}(3\phi^2+4m^2\varphi^2)\Big\}.
\end{split}
\end{equation}
Terms linear in the fluctuating fields $\phi$ and $\varphi$ cancel out because the background satisfies the GP equation of motion. Therefore, we only have to consider the quadratic part $\Gamma_2 [\phi, \varphi]$. The latter can be simplified using Bernouilli equation \eqref{eq:analogs.EulerBEC},
\begin{equation}
    \begin{split}
        \Gamma_2 [\phi, \varphi] = \int &\text{d} t \, \text{d}^2 r \Big\{ - \hbar^2m(\vec{\nabla} \varphi)^2  -\frac{1}{2}\phi\left(2\lambda n_0-\hbar^2\frac{\vec{\nabla}^2}{2m} \right)\phi  \\
		&  + \phi\left[-\hbar2m\partial_t \varphi -2\hbar^2 (\vec{\nabla}S_0) \vec{\nabla}\varphi -  \hbar^2   (\vec{\nabla}^2 S_0)\varphi\right] \Big\}.
    \end{split}
\label{eq:analogs.FullFluctAction}    
\end{equation}
We can also neglect the terms containing $-\hbar^2\vec{\nabla}^2/2m$ in light of the acoustic approximation. Moreover, we assume a vanishing background velocity $\vec{v}_0$, which is indeed fulfilled for the scenarios described in this work. This implies that $\vec{\nabla}^2 S_0 = 0$, and allows us to write the field $\phi$ in terms of $\varphi$ in a very simple way, 
\begin{equation}
    \phi = -\frac{\hbar}{\lambda (t) n_0 (r)}\partial_t \varphi,
    \label{eq:analogs.FluctuationFieldsRelation}
\end{equation}
leading to a quadratic effective action for $\varphi$ only. We find
\begin{equation}
        \Gamma_2 [\varphi] = \frac{\hbar^2}{2} \int \text{d} t \, \text{d}^2 r \left\{\frac{1}{c_{\text{s}}^2}  (\partial_t \varphi)^2-  (\vec{\nabla} \varphi)^2 \right\},
\label{eq:analogs.QEAPhiTwo}
\end{equation}
where we introduced the time- and space-dependent speed of sound
\begin{equation}
    c_{\text{s}}^2 (t, \vec{r}) = \frac{\lambda (t) \, n_0 (\vec{r})}{m}.
    \label{eq:analogs.SpeedOfSoundBEC}
\end{equation}
Finally, the latter effective action can be rewritten as an action for a free, minimally coupled, massless scalar field in a curved spacetime determined by the acoustic metric $g_{\mu\nu}$,
\begin{equation}
    \Gamma_2 [\varphi] = -\frac{\hbar^2}{2} \int \text{d} t \, \text{d}^2 r \, \sqrt{-g} \, g^{\mu\nu} \partial_\mu \varphi \partial_\nu \varphi.
    \label{eq:analogs.QEACurvedSpacetime}
\end{equation}
That is, under the mentioned conditions, the dynamics of the fluctuations on top of the ground state of the condensate is the same as that of the scalar field described generally in~\cref{eq:qftcs.CurvedActionNM}, namely given by the KG equation \eqref{eq:qftcs.KGEquation}, for the particular case in which~\mbox{$D=2$}, $m=\xi=0$, and the geometry is described by the acoustic metric 
\begin{equation}
    g_{\mu\nu} = c_{\text{s}}^{-2} \begin{bmatrix}
        -c_{\text{s}}^2 & 0 \\
        0 & \mathbb{1}_{2}
    \end{bmatrix}.
\label{eq:analogs.AcousticMetricBEC}
\end{equation}
Except for the background density term missing in front, this has the same form as the fluid acoustic metric \eqref{eq:analogs.AcousticMetricGeneral} for the case of a stationary flow, with the differences being due to the splitting of the fluctuations into real and imaginary parts (cf. \cref{eq:analogs.BackgroundSplit}, see e.g.~\mbox{\cite{Floerchinger2008,Heyen2020}} for discussions regarding this parametrization) instead of directly perturbing the density and the phase in the Madelung form \eqref{eq:analogs.MadelungRepresentation}. The relation between the two parametrizations is easily found, since at linear order in perturbations
\begin{equation}
    \Psi = \sqrt{n_0+n_1}e^{i(S_0+S_1)} = \sqrt{n_0}e^{iS_0} + \frac{e^{iS_0}}{\sqrt{2}}\left(\frac{n_1}{2n_0} + i\sqrt{2n_0}S_1\right),
\end{equation}
from which we can identify the perturbations in the density and the phase as
\begin{equation}
    n_1 = 2n_0 \phi, \quad S_1 = \sqrt{2}m\varphi.
\label{eq:analogs.ParametrizationsRelation}
\end{equation}
In this case, the corresponding acoustic metric takes the form \eqref{eq:analogs.AcousticMetricGeneral}. For a constant background density profile $n_0$, the two parametrizations lead to acoustic metrics related by a constant, a power of $n_0$. However, if the density depends on the spatial coordinates, this factor in front of the acoustic metric \eqref{eq:analogs.AcousticMetricGeneral} spoils the mapping to FLRW universes. Crucially, the parametrization \eqref{eq:analogs.BackgroundSplit} allows for a mapping to FLRW metrics with any spatial curvature, depending on the specific background density profile, in contrast with previous works, which typically focused on the flat case. The specific dynamics of the background variables required to mimic FLRW universes for the fluctuating variables will be discussed in \cref{ch:becstheory}.

\subsection{Beyond the acoustic approximation}
\label{subsec:analogs.beyondacoustic}

Although in this thesis we will focus mainly on the acoustic regime of the BEC, let us briefly comment on the possibility of going beyond the acoustic approximation for completeness. 

The effects of the quantum potential can be included through the \textit{eikonal} approximation \cite{Barcelo2001,Visser2002,Liberati2006a,Liberati2006b,Weinfurtner2006,Weinfurtner2009,Barcelo2010}, which leads to the Bogoliubov dispersion relation \cite{Bogoliubov1946} for the fluctuations,
\begin{equation}
    \omega_k = \frac{\hbar}{2m}\sqrt{k^2(k^2+2/\xi^2)}.
\label{eq:analogs.FullBogoliubovDispersion}
\end{equation}
The above becomes phononic for low momenta, $k \ll \xi^{-1}$, so that 
\begin{equation}
    \omega_k \simeq \frac{\hbar\sqrt{2}}{m\xi}k = c_{\text{s}} k
\end{equation}
At the level of the acoustic metric, this results in an effective, $k$-dependent speed of sound in \cref{eq:analogs.AcousticMetricBEC}, in such a way that the geometry experienced by each mode depends on its wavenumber (what has been called \textit{rainbow} metrics in e.g. \cite{Weinfurtner2009}). Crucially, if the density profile is flat, the eikonal approximation becomes an exact extension to the ultraviolet of the cosmological analogy.

\section{Analog systems for dark matter fields}
\label{sec:analogs.dm}

We have described above the analogy emerging in spin-$0$ BECs, whose excitations can be mapped to a massless, real scalar field. Therefore, these systems cannot reproduce cosmological production of dark matter fields, which must be massive. Nevertheless, let us stress that the insights obtained from the exploration of spin-$0$ BECs will be very useful in cosmological analyses (as we will explicitly show throughout the thesis). However, in order to provide a greater degree of completeness, we will briefly discuss in this section the possibility of analog cosmological dark matter production in several systems, including the natural extension to~\mbox{spin-$1$ BECs}.

\subsection{Spin-1 Bose-Einstein condensates}

Let us consider a spin-$1$ BEC as a generalization of the previous section. The elementary results concerning spinor condensates come mainly from \cite{Kawaguchi2012}, and the derivation of the acoustic metric in these systems corresponds to the upcoming work of Brunner, Flörchinger, and Schmidt\footnote{A preprint of this work is now publicly available, see \cite{Schmidt2025}}, whom we thank for sharing these preliminary results in private communications.
 
The effective action for the three-component spinor $\Phi_m$ describing a spin-$1$ BEC in $(1+2)$ dimensions is given by \cite{Ohmi1998, Ho1998}
\begin{equation}
\begin{split}
    \Gamma &= \int \dd t \dd^2 x \Bigg\{ \sum_{s=-1}^1 \Psi^{\dagger}_s \left(i\hbar \partial_t + \frac{\hbar}{2m}\Delta - V + z_1s - z_2s^2\right) \Psi_s \\
    & - \frac{c_0}{2}n^2 - \frac{c_1}{2}\Bigg[\left(\abs{\Psi_+}^2 - \abs{\Psi_-}^2\right)^2 \\
    &- 2 \abs{\Psi_0}^2\left(\abs{\Psi_+}^2 + \abs{\Psi_-}^2\right) -2\Psi_0^2\Psi_+^*\Psi_-^* + 2(\Psi_0^*)^2\Psi_+\Psi_-\Bigg]\Bigg\},
\label{eq:spin1.GeneralAction}
\end{split}
\end{equation}
where in addition to the terms for the spin-$0$ case (cf. \cref{eq:analogs.FullFluctAction}) one has that~\mbox{$s=\pm 1, 0$} denotes the three spin components; $z_1$ and $z_2$ are the linear and quadratic Zeeman coefficients; $n = \sum_s \abs{\Phi_s}^2$ is the condensate density; and $c_0$ and~$c_1$ are the density and spin interaction strengths, respectively.

One can follow the same steps as in the spin-$0$ case to obtain the effective metric for the spin-$1$ BEC, namely allow for small fluctuations on top of the ground state of the form
\begin{equation}
    \Psi_s = \bar{\psi}_{s} + \delta\psi_{s}.
\end{equation}
In order to obtain the quadratic action for the fluctuations, one must specify the ground state around which the mean field expansion takes place. There are four possible ground states of spin-$1$ BECs: the polar, the ferromagnetic, the antiferromagnetic, and the broken axisymmetric ground state. Since our intention is to exemplify how an effective metric for a possible dark matter model can be constructed, we will concentrate on the polar case.

\subsubsection*{Acoustic metric in the polar ground state}

Neglecting the quantum pressure, that is, in the acoustic approximation, and in the case of vanishing flow ($\bm{v}_0=0$), the polar ground state is given by
\begin{equation}
    \bar{\psi}(\vx) = (0, \sqrt{n_0(\vx)}, 0)^{\top},
\end{equation}
provided the potential is engineered to be $V = -c_0 n_0$, $z_1=0$, and the relation~\mbox{$z_2>2n_0\abs{c_1}$} holds. Brunner \textit{et al.} showed that the effective action for the fluctuations in this case reads
\begin{equation}
\begin{split}
    \Gamma_2 = -\frac{\hbar^2}{2}\int \dd t \dd^2r \Big[\sqrt{g_0}g_0^{\mu\nu}&\partial_\mu \varphi\partial_{\nu}\varphi \\
    &+ \sqrt{g_1}\left(g_1^{\mu\nu}\partial_{\mu}\phi\partial_{\nu}\phi^* - M^2\phi\phi^*\right)\Big],
\end{split}
\end{equation}
where $\varphi = \delta \psi_0$ is analog to the phononic fluctuating field in \cref{sec:analogs.becs}, and we introduced the massive, complex field $\phi =\delta\psi_{+} + i\delta\psi_{-} $ with spacetime-dependent mass
\begin{equation}
    M^2(t, \vx) = \frac{1}{\hbar^2}\left[z_2^2(t) + 2z_2(t)c_1(t) n_0(\vx)\right].
\end{equation}
The effective metrics determining the dynamics of the fields are
\begin{align}
    g_{0, \mu\nu} = \text{diag}[-1, a_0^2(t, \vx),a_0^2(t,\vx)], \\
    g_{1, \mu\nu} = \text{diag}[-1, a_1^2(t, \vx),a_1^2(t,\vx)],
\end{align}
where the scale factors read
\begin{equation}
    a_0^2 (t, \vx) = \frac{m}{c_0(t)n_0(\vx)}, \quad a_1^2 (t, \vx) = \frac{2m}{q(t) +c_1(t)n_0(\vx)}.
\label{eq:analogs.Spin1ScaleFactors}
\end{equation}
In other words, the same massless mode that we found in the spin-$0$ case is present in the spin-$1$ scenario, but now there are two additional massive modes that behave as a complex scalar field. The mass of these modes is determined by the density of the condensate, and the interaction strengths $c_0$ and $c_1$. Therefore, by tuning these experimental parameters, we can change both the acoustic metric and the mass of the $\phi$ field, which could be used to mimic the dynamics of a dark matter field in the early Universe. Let us briefly explore this in the next subsection.

\subsubsection*{Tuning the effective metric}

Note that the effective metric describing the dynamics of each field is different in principle, although they could be tuned to be the same. Indeed, looking at the scale factors \eqref{eq:analogs.Spin1ScaleFactors}, we see that if
\begin{equation}
    c_0(t)n(\vx)=\frac{q(t)+c_1(t)n_0(\vx)}{2},
\end{equation}
the two metrics coincide. This implies that the background density must be constant, and that
\begin{equation}
    2c_0(t)-c_1(t)=\frac{q(t)}{n_0}.
\end{equation}

Let us now allow for different acoustic metrics for each field. If we consider a constant value for the background density, the spatial dependence of the scale factors $a_0$ and $a_1$ disappears, as well as that of the mass $M$. However, in order to map the dynamics of the field~$\phi$ to that of a dark matter field, the mass has to be constant as well. We find that one can do so by proper tuning of the experimental parameters, namely the Zeeman coefficient $z_2$ and the interaction strength $c_1$. Explicitly, given certain scale factor $a_1$, in order to achieve a constant mass $M$ one must solve the system
\begin{align}
    M(t) &= \hbar^{-2}\left[z_2^2(t) + 2z_2(t)c_1(t)n_0\right], \\
    a_1^2(t) &= \frac{2m}{z_2(t) + c_1(t)n_0}.
\end{align}
From here, we derive 
\begin{equation}
    z_2(t) = \frac{2m}{a^2_1(t)}\left(1\pm\sqrt{1-\hbar^2M(t)\frac{a_1^4(t)}{4m^2}}\right),
\end{equation}
which allows us to solve the above system for $c_1(t)$ as well. In case the density is not constant, one must engineer a space- and time-dependent Zeeman coefficient~$z_2(t, \vx)$ through the external magnetic field. More concretely, in order to implement a particular scale factor $a_1$, one must engineer
\begin{equation}
    z_2(t, \vx) = -c_1(t)n_0(\vx) + \frac{2m}{a_1^2(t)}. 
\end{equation} 
This is a more challenging task, but theoretically possible.

\subsection{Other analog dark matter models}

We have briefly showcased in these preliminary comments how spin-$1$ BECs can be used to mimic the dynamics of massive fields in cosmological backgrounds, and therefore be used to study the cosmological production of spectator dark matter. However, there are other systems that can be considered beyond BECs.

There have been recent developments regarding the study of analog systems for fermionic fields. Proposals such as those in \cite{Kennes2021,Tolosa2023} and \cite{Fulgado2023} have shown the emergence of analog, massive fermionic fields in bilayer graphene and Fermi gases, respectively. However, let us note that gravitational production is suppressed for fermions via Pauli blocking as soon as the initial state occupation number is non-vanishing \cite{Parker1971} (which, although commonly assumed in theory, is not the case in an experiment).

Building analog systems capable of simulating dark matter production in the early Universe is an exciting task that we would like to pursue in the future. As we mentioned above, in this thesis we will study spin-$0$ BECs, which allow for the study of cosmological production of massless scalar particles. Even though one would require an analog massive field for simulating cosmological production of dark matter, we will see that these systems can be very useful in understanding this production mechanism in general.

%% file: Chapters/DMProductionSlowRoll.tex

\addtocontents{toc}{\protect\vspace{0.5em}}

\chapter{Particle production sourced by single-field inflation} 

\label{ch:singlefield} 



In order to study particle production in the early Universe, one needs to specify the background dynamics, which determines the explicit shape of the mode equation frequency \eqref{eq:qftcs.MasterFrequency}. In this chapter, we study cosmological production under realistic inflationary scenarios, which characterize the rapid expansion of the Universe in its early stages, and the subsequent reheating period, which, as discussed in \cref{sec:cosmo.reheating}, is responsible for the transition from the inflationary epoch to the radiation-dominated era. We focus on single-field chaotic inflationary models (cf. \cref{sec:cosmo.inflation}), for which we study the dynamics of the inflaton field $\phi$ under two different potentials: a quadratic potential and the Starobinsky potential. The former is a simple model that has been widely studied, and although its simplest realization is ruled out by CMB observations, it allows us to compare our results with previous literature (see e.g.~\cite{Markkanen2017a, Cembranos2020, Markkanen2018, Fairbairn2019, Kainulainen2023}). On the other hand, the Starobinsky potential was the first proposed inflationary potential and, although more complex, it fits within the current observational constraints (for recent works regarding cosmological production in this context, see \cite{Dorsch2024,Racco2024}). We also study the late-time dynamics of the inflaton field, which is relevant for the discussion of particle production and the choice of vacuum. We focus on spatially flat sections in this chapter, and study both massive scalar and vector fields ($\varphi$ and~$A_{\mu}$, respectively) with non-minimal couplings to the curvature scalar and the Ricci tensor, which behave as spectator fields. Once the background geometry is known, we can proceed to solve the corresponding mode equations. The goal is to extract the number of particles produced during the evolution of the early Universe due to the spacetime dynamics, which can be seen as a dark matter production mechanism. If the field is non-interacting after this period, this abundance will be related to that measured today only by dilution due to expansion. In this way, we set constraints on the possible values for each of the coupling strengths and the mass of the produced quanta so that it does not lead to overproduction.

\section{Single-field slow-roll inflation}
\label{sec:sf.chaoticinf}

We describe the early epoch of the Universe with a chaotic inflationary model consisting of a single, homogeneous, and isotropic scalar field~$\phi$ with a potential~$V(\phi)$. The equation of motion for the inflaton is therefore the already discussed \cref{eq:cosmo.InflatonEoM}, which, together with the Friedmann equation \eqref{eq:cosmo.InflatonFriedmann} for vanishing spatial curvature, provides the background dynamics. Through Einstein's equations it is easy to write the Ricci curvature scalar in terms of the inflaton, which reads
\begin{equation}
R =\frac{8\pi}{\mpl^2}\left[4V(\phi) - \dot{\phi}^2\right] = \frac{8\pi}{\mpl^2}\left[4V(\phi) - a^{-2}\phi^{\prime 2}\right],
\label{eq:sf.RicciStressEnergy}
\end{equation}
as well as the traceless Ricci tensor~\mbox{$\Tilde{R}_{\mu\nu} = R_{\mu\nu} - g_{\mu\nu}R/4$}, whose components are given by 
\begin{equation}
    \Tilde{R}^{00}=\frac{6\pi}{\mpl^2}a^{-4}\phi^{\prime \, 2}, \qquad \Tilde{R}^{ii}=\frac{2\pi}{\mpl^2}a^{-4}\phi^{\prime \, 2},
    \label{eq:sf.RicciTensor}
\end{equation} 
and will be useful when studying a vector field $A_{\mu}$ with a coupling to curvature through these terms (more details on its dynamics are given below).

\begin{figure}[t!]
    \includegraphics[width=\textwidth]{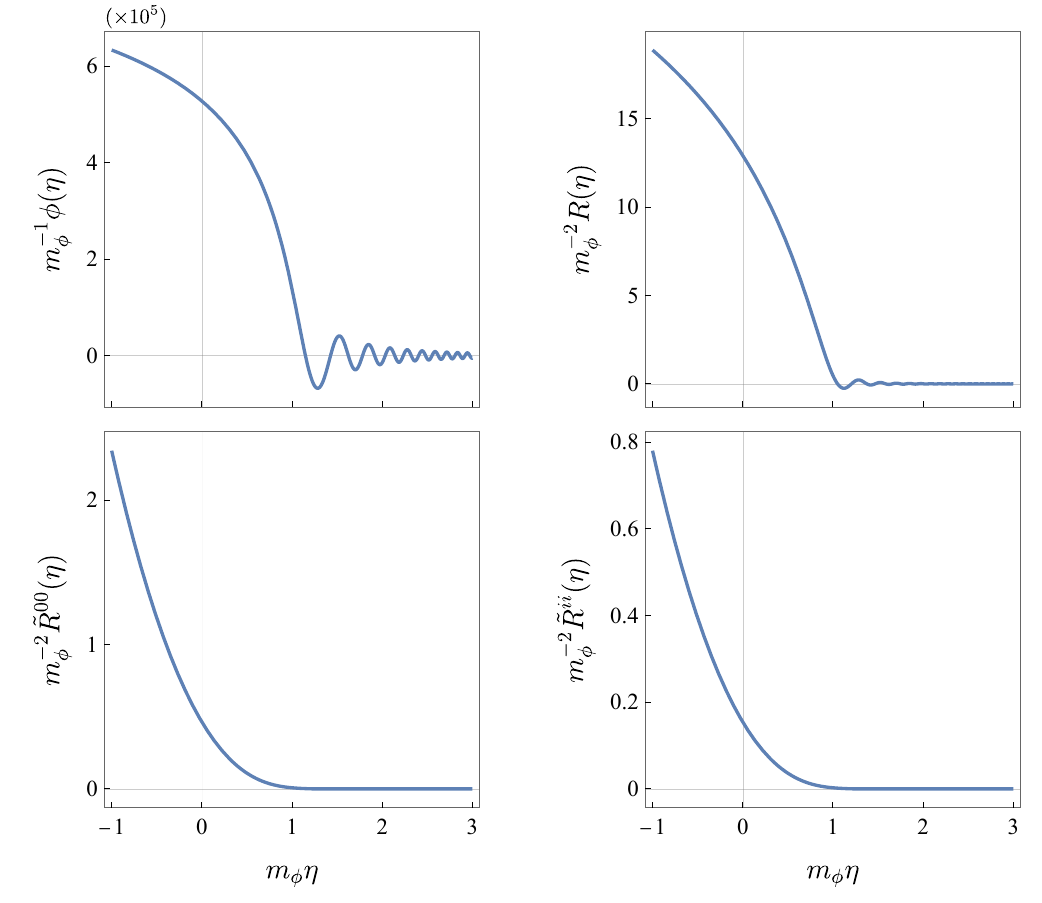}
    \begin{picture}(0,0)
    \put(298, 265){\includegraphics[width=0.23\textwidth]{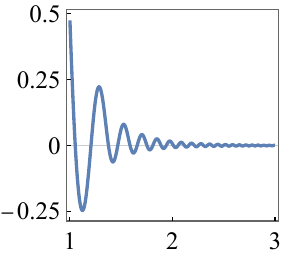}}
    \end{picture}
    \begin{picture}(0,0)
    \put(93, 101){\includegraphics[width=0.23\textwidth]{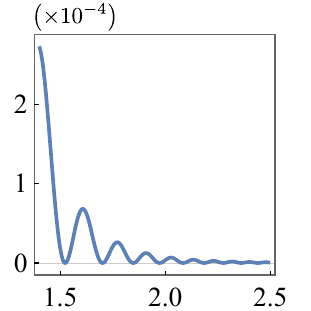}}
    \end{picture}
    \begin{picture}(0,0)
    \put(298, 101){\includegraphics[width=0.23\textwidth]{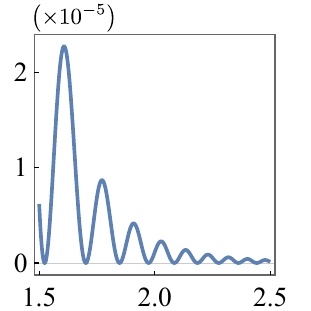}}
    \end{picture}
    \caption[Inflaton field, curvature scalar, and traceless Ricci components for a quadratic inflationary potential as a function of conformal time.]{Inflaton field $\phi(\eta)$ (upper-left panel), curvature scalar $R(\eta)$ (upper-right panel), and components of the traceless Ricci tensor (bottom row), as functions of conformal time. The range of time corresponds to the end of inflation and the beginning of reheating. The parameters used for all figures in this chapter are given in \cref{sec:param.sf}. Figure from \cite{VectorDM2024}.}
    \label{fig:sf.InflatonAndRicciConformal}
\end{figure}

As we discussed in \cref{sec:cosmo.inflation}, although the inflaton equation \eqref{eq:cosmo.InflatonEoM}, together with the Friedmann equation \eqref{eq:cosmo.InflatonFriedmann}, has no analytic solution in general, one can find approximations for certain regimes. When this is not possible, we must rely on numerical computation. For most of the inflationary period, we can use the slow-roll approximation introduced in \cref{sec:cosmo.inflation} to obtain a solution to the inflaton equation of motion (see subsection \ref{subsubsec:sf.slowroll} below). However, during the transition between inflation and reheating, the dynamics of the inflaton has to be obtained numerically. Both the inflaton field $\phi$ and the Ricci scalar $R$ start to oscillate with decreasing amplitude, as can be observed in figure~\ref{fig:sf.InflatonAndRicciConformal}, where $\phi(\eta)$ and $R(\eta)$ are depicted for an interval of time during the transition phase. This is the epoch in which most of the particles are produced, and the inflaton dynamics is solved up to a numerically accessible time~$\eta_{\text{f}}$, when production becomes negligible; that is, when the expansion becomes adiabatic and some kind of approximate \textit{out} regime is reached (cf. subsection \ref{subsec:qftcs.inout}). We therefore analyze the two following regions in conformal time,
\begin{equation}
\eta = \begin{cases}
\eta_{\text{i}} \leq \eta < \eta_*, \quad &\text{slow-roll approximation},\\
\eta_* \leq \eta \leq \eta_{\text{f}}, \quad &\text{numerical solution}.
\end{cases}
\end{equation}
For late times, deep in the reheating era, we can also use an analytic approximation for the solution of the inflaton equation of motion, given in subsection \ref{subsubsec:sf.latereheating}, which---although not used in our calculations---will be useful to make some remarks in \cref{sec:sf.abundance}.

Let us now particularize to the quadratic and Starobinsky potentials.

\subsection{Quadratic potential}
\label{subsec:sf.quadratic}

We first consider the case of a quadratic potential, $V(\phi) = \frac{1}{2}m_{\phi}^2\phi^2$, where $m_{\phi}$ denotes the inflaton mass.

\subsubsection*{Inflationary era: slow-roll approximation}
\label{subsubsec:sf.slowroll}

We consider that the inflationary period starts at the negative initial time $t_{\text{i}}$. Within the slow-roll approximation, we can neglect the derivative of the field in favor of the potential (cf. \cref{eq:cosmo.SlowRollConditions}), and the field slowly rolls over until it falls to a minimum and starts oscillating. With this assumption, we can approximately write the Friedmann equation~\eqref{eq:cosmo.InflatonFriedmann} during slow roll as
\begin{equation}
H\simeq \sqrt{\frac{8\pi}{3\mpl^2}V(\phi)}.
\label{eq:cosmo.HubbleSlowRoll}
\end{equation}
A slowly varying inflaton implies that $H \simeq \text{constant}$ in this regime. Hence, the expansion of spacetime is said to be {quasi}-exponential, as it resembles the pure de Sitter solution. Slow-roll conditions are maintained long enough to solve the flatness and horizon problems, and in this regime the inflaton equation \eqref{eq:cosmo.InflatonEoM} becomes easily solvable,
\begin{equation}
\dot{\phi}\simeq -\frac{V^{\prime}(\phi)}{3H}\simeq -V^{\prime}(\phi)\frac{\mpl}{\sqrt{24\pi V(\phi)}}.
\label{eq:sf.SlowRollInflatonEOM}
\end{equation}
For the particular potential $V(\phi)=\frac{1}{2}m_{\phi}^2\phi^2$, the solution to the above equation~\eqref{eq:sf.SlowRollInflatonEOM}~is
\begin{equation}
\phi_{\text{SR}}(t) = \phi_0 - \frac{\mpl}{\sqrt{12\pi}}m_{\phi} t,
\end{equation}
where $t<0$ corresponds to the inflationary period. Note that $t=0$ and $\phi_0$ are the ending time of inflation and the value of the field at this instant, respectively. From here, it is straightforward to obtain an explicit expression for the Ricci scalar, introducing the solution into~\eqref{eq:sf.RicciStressEnergy}.

The scale factor is obtained by integrating the Hubble rate. In the slow-roll approximation it reads
\begin{equation}
a_{\text{SR}}(t) = a_0 \exp{-\int_{\phi_0}^{\phi(t)}\, \dd \phi \frac{8\pi}{\mpl^2}\frac{V(\phi)}{V^{\prime}(\phi)}},
\end{equation}
which for the quadratic potential becomes
\begin{equation}
a_{\text{SR}}(t)=a_0 \exp{-\frac{2\pi}{\mpl^2}\left[\phi_{\text{SR}}^2(t) - \phi_0^2\right]}.
\end{equation}

We also need the relation between cosmological and conformal time in order to write both $a(\eta)$ and $R(\eta)$. This relation can be obtained numerically from~\mbox{$\eta = \eta_0 + \int_{0}^t \dd t/a(t)$}. These are the necessary ingredients for determining the frequency of the mode equation in this region, under the slow-roll approximation.

This regime is valid as long as the slow-roll conditions \eqref{eq:cosmo.SlowRollConditions} are fulfilled (which in fact are the same in the case of a quadratic potential). When these no longer hold, at say, $t>t_{*}$ with $t_{*}<0$, the inflaton equation of motion~\eqref{eq:cosmo.InflatonEoM} has to be solved numerically. The field begins to exit the inflationary regime and $t=\eta=0$ marks both the end of inflation and the beginning of reheating. At this point, the scale factor reaches the value $a_0$, which merely sets the scale and hence we normalize it to $a_0=1$.

\subsubsection*{Late reheating}
\label{subsubsec:sf.latereheating}
For late times, well into the reheating epoch ($\eta_*\ll \eta \lesssim \eta_{\text{f}}$), and assuming\footnote{For sufficiently small spectator field masses, this is not the case. We will deal with this situation at the end of subsection~\ref{subsec:sf.vacuumchoice}.} $\eta_{\text{f}} < \eta_{\text{rh}}$, where $\eta_{\text{rh}}$ denotes the end of reheating, one can find an approximate solution to the inflaton equation \eqref{eq:cosmo.InflatonEoM} \cite{Mukhanov2005,Weinberg2008}. We do not use it for obtaining our results, but it will be important for the discussion in subsection~\ref{subsec:sf.vacuumchoice}. In this approximation, the Hubble rate reads
\begin{equation}
H(t)\simeq \frac{2}{3t}\left[1 - \frac{\sin{(2m_{\phi}t - 2\theta)}}{2m_{\phi}t}+ \mathcal{O}(m_{\phi}^{-2}t^{-2})\right]^{-1},
\label{eq:sf.HubbleRateInflaton}
\end{equation}
where $\theta$ is an arbitrary phase, whereas the inflaton field is given by the expression
\begin{equation}
\phi(t) = \frac{\Phi_0}{t}\sin{(m_{\phi}t)}\left[1 -\frac{\cos{(2m_{\phi}t)}}{2m_{\phi}t} + \mathcal{O}(m_{\phi}^{-2}t^{-2})\right],
\label{eq:sf.LateReheatingInflaton}
\end{equation}
with $\Phi_0 \equiv {\mpl}/{(\sqrt{3\pi}m_{\phi})}$.
This solution is valid as long as $m_{\phi}t \gg 1$, a condition that is fulfilled during reheating, since, as we will see, the scale factor behaves as that of a matter-dominated universe. Indeed, we can integrate $H(t)$ in order to approximately obtain the scale factor~$a(t)$, 
\begin{equation}
a(t) = \mathcal{C} t^{2/3}\left[1 +  \mathcal{O}(m_{\phi}^{-2}t^{-2}) \right].
\end{equation}
The constant $\mathcal{C}$ is determined by requiring that the value of the scale factor at late times coincides with the one obtained from the numerical simulation in the previous region. One can now integrate the scale factor in order to obtain $t(\eta) = \left({\mathcal{C}}\eta/3\right)^3$.

Now that we have a solution for the inflaton field and the scale factor valid for late times, we can obtain the Ricci scalar from  \eqref{eq:sf.RicciStressEnergy} by taking the solution for $\phi(t)$ to first order in $(m_{\phi}t)^{-1}$. We end up with
\begin{equation}
R(t)\!=\!\frac{8}{3t^2}\left[2\sin^2{(m_{\phi}t)}\!-\!\left(\cos{(m_{\phi}t)} - \frac{\sin{(m_{\phi}t)}}{m_{\phi}t}\right)^2\!+\!\mathcal{O}(m_{\phi}^{-3}t^{-3})\right].
\label{eq:sf.LateRicci}
\end{equation}

With this, we are able to describe the frequency of the mode equation until very late times, for which the approximations derived in this subsection become even better. The density of produced particles will be calculated at a sufficiently large time $\eta_{\text{f}}$, such that particle production is negligible from that point in time onwards.

\subsection{Starobinsky potential}
\label{subsec:sf.starobinsky}

\begin{figure} 
    \centering
    \includegraphics[width=0.5\textwidth]{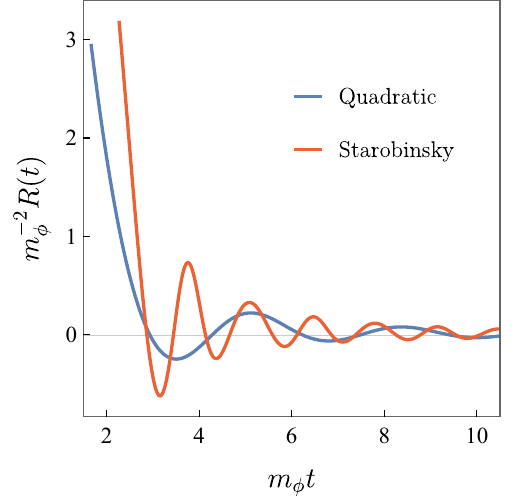}
    \caption[Ricci scalar for quadratic and Starobinsky inflationary potentials]{Ricci scalar during the transition from inflation to reheating for the quadratic (blue) and the Starobinsky (red) potential, as a function of cosmological time.}
    \label{fig:sf.RicciStarobinsky}
\end{figure}

Let us now introduce the Starobinsky potential, which is given by~\mbox{\cite{Starobinsky1980,Maeda1987,Lust2024}}
\begin{equation} 
    V(\phi)=L^4\left[1-\exp\left(-4\sqrt{\frac{\pi}{3}}\frac{\phi}{\mpl}\right)\right]^2,
\label{eq:sf.StarobinskyPotential}
\end{equation}
where $L$ is a free parameter. In this case, one also assumes that inflation starts in the slow-roll regime. Hence, the derivations and arguments in subsection \ref{subsubsec:sf.slowroll} follow exactly in the same way, simply replacing the quadratic potential by \cref{eq:sf.StarobinskyPotential}. Also, we can obtain similar expressions for the late-time solutions for the inflaton field and the curvature scalar (\cref{eq:sf.LateReheatingInflaton} and \cref{eq:sf.LateRicci}). Note that at late times, when~\mbox{$\phi/\mpl\ll1$}, this potential also behaves quadratically,
\begin{equation}
    V(\phi) \simeq L^4 \frac{16\pi}{3}\frac{\phi^2}{\mpl^2},
\end{equation}  
which allows us to identify an inflaton mass for the Starobinsky potential as
\begin{equation}
\bar{m}_{\phi}^2 = \frac{32\pi}{3}\frac{L^4}{\mpl^2}. 
\end{equation}
In figure \ref{fig:sf.RicciStarobinsky} we show the Ricci scalar corresponding to both potentials for comparison (note that we use cosmological time with this in mind), in the region of transition from inflation to reheating. We observe that, at a qualitative level, both potentials behave similarly. However, the Starobinsky curvature scalar oscillates faster, which can have an impact on the number of particles produced during the transition from inflation to reheating, since, as we will study in the following, they play an important role in this process.

\section{Field dynamics in a (1+3) FLRW background}
\label{sec:sf.fielddynamics}

In this section we describe the dynamics of scalar and vector fields in a $(1+3)$ FLRW universe. We consider both massive scalar and vector fields non-minimally coupled to gravity. The dynamics of these fields will be encoded in the corresponding actions, which will be written in terms of the scale factor and the Ricci scalar of the FLRW metric (see \cref{ch:qftcs}). We then solve the corresponding mode equations in order to extract the number of particles produced during the evolution of the universe.

\subsection{Scalar field in a flat FLRW universe}
\label{subsec:sf.scalarfield}

We have extensively discussed the dynamics of a scalar field in a~\mbox{$(1+D)$}-dimensional FLRW in \cref{ch:qftcs}. Here, we focus only on the~$\kappa=0$ scenario, work with three spatial dimensions and Cartesian coordinates, and expand the field $\varphi$ in Fourier modes, which are the eigenfunctions of the Laplace-Beltrami operator in those coordinates, namely $\mathcal{H}_{\vk}(\vec{x}) = e^{i\vec{k}\vec{x}}/(2\pi)^{3/2}$, with $\vk$ denoting the usual wavevector. Note that in this case $\bk = -\vk$. Then, the field can be expanded as (cf. \cref{eq:qftcs.ChiFieldExpansion})
\begin{equation}
    \varphi(t, \vec{x}) = \int \frac{\dd^3\vec{k}}{(2\pi)^{3/2}} e^{i\vec{k}\vec{x}} \Tilde{\varphi}_{\vk}(t),
\end{equation}
with $\Tilde{\varphi}_{\vk}(t) = \Tilde{\varphi}^*_{-\vk}(t)$ in order to ensure that $\varphi$ is real. The frequency of the mode equation \eqref{eq:qftcs.ModeEquation} for the rescaled field $\chi=a\varphi$ (cf. \cref{eq:qftcs.FieldsRelation}) is given by
\begin{equation}
    \omega_k^2(\eta) = k^2 + a^2(\eta)\left[m^2 + \left(\xi- 1/6\right)R(\eta)\right],
\label{eq:sf.ScalarFrequency}
\end{equation}
where we used that the eigenvalues of the Laplace-Beltrami operator are given by~\mbox{$h^2(\vk) = k^2$}, with~\mbox{$k=+\sqrt{\vec{k}^2}$} being the norm of the wavevector, and therefore we can use this as label for the modes $v_k$. From here, quantization follows as outlined in \cref{sec:qftcs.quantization}, where we have discussed the ambiguity in the choice of the vacuum state. However, we can make use of the fact that our background expands very adiabatically at sufficiently early and late times before and after inflation in order to make sensible choices of \textit{in} and \textit{out} vacua. In particular, we will assume the system is initially in the Bunch-Davies vacuum, which is the preferred choice in a de Sitter geometry (for a massive scalar field~\mbox{\cite{Allen1985}}), to which our background approaches as we go back in time. For the notion of vacuum of an observer in the late reheating regime, we will consider the so-called adiabatic vacuum prescription \cite{Birrell1982}. We will give more details on these choices in \cref{sec:sf.scalarmodes}.

Note that a value of $\xi$ smaller than $1/6$ would lead to tachyonic instabilities in the asymptotic past. On the other hand, although for a general scalar field it is possible to have $\xi>1$ (in contrast to what happens in the case of the Higgs field~\cite{Markkanen2017b}), the behavior of particle production is in this case qualitatively the same as for $\xi \sim 1$. Nevertheless, interesting results have been obtained for the regime of very large~$\xi$ when the backreaction of the spectator field is taken into account~\cite{Lebedev2022}. For these reasons, we restrict ourselves to the range~\mbox{$1/6 \leq \xi \leq 1$}.

Before calculating particle production of such scalar field in the background described in section \ref{sec:sf.chaoticinf}, let us discuss the dynamics of a massive vector field, which has not been analyzed so far in the thesis.

\subsection{Vector field in a flat FLRW universe}
\label{subsec:sf.vectorfield}

In this subsection, we discuss the dynamics of a massive vector field $A_{\mu}$ in a flat FLRW universe in $(1+3)$ dimensions which is non-minimally coupled to the geometry, extending therefore the minimally coupled model presented in references~\cite{Ema2019, Bastero2019, Ahmed2020}. The idea is similar to what has been done with the scalar field, but the equations of motion become more involved. Since this case is not included either in the general introduction of chapter \ref{ch:qftcs}, we derive the corresponding dynamics in the following. 

The interaction with gravity is taken into account through the action
\begin{equation}
\!S\!=\! -\frac{1}{2} \int \text{d}^4x \sqrt{-g} \Bigg\{\frac{1}{2}F_{\mu\nu}F^{\mu\nu} \!+\! \left[\left(m^2 + \gamma R\right) g^{\mu\nu} \!+ \sigma \Tilde{R}^{\mu\nu}\right]\!A_{\mu}A_{\nu}\Bigg\},
\label{eq:sf.VectorGeneralAction}
\end{equation}
where $m$ is now the mass of the spin-1 boson, $F_{\mu\nu}=\partial_{\mu}A_{\nu} - \partial_{\nu}A_{\mu}$ is the field strength, and $\gamma$ (similar to $\xi$ in the scalar field) and $\sigma$ are the couplings to the Ricci scalar $R$ and the traceless Ricci tensor~\mbox{$\Tilde{R}^{\mu\nu} = R^{\mu\nu} - g^{\mu\nu}R/4$}, respectively. For simplicity, we use conformal time $\eta$ and Cartesian coordinates, such that the metric can be written in a conformally flat form, $g_{\mu\nu} =  a^2(\eta) \eta_{\mu\nu}$, where $\eta_{\mu\nu}$ is the Minkowski metric, which will be used from now on for raising and lowering indices. Introducing this metric in the action \eqref{eq:sf.VectorGeneralAction}, we can explicitly write
\begin{equation}
S = -\frac{1}{2}\int \dd^4x \left[\frac{1}{2}F^{\mu\nu}F_{\mu\nu} + M^{\mu\nu}A_{\mu}A_{\nu}\right],
\label{eq:sf.VectorAction}
\end{equation}
where we have defined the mass tensor $M^{\mu\nu}$ as
\begin{equation}
M^{\mu\nu} \equiv a^2\eta^{\mu\nu}\left(m^2 + \gamma R\right) + a^4\sigma\Tilde{R}^{\mu\nu}.
\end{equation}
This action leads to the equation of motion
\begin{equation}
\partial_{\mu}\partial^{\mu}A^{\nu} - \partial_{\mu}\partial^{\nu}A^{\mu} =M^{\mu\nu}A_{\mu},
\label{eq:sf.ProcaEquation}
\end{equation}
which is a generalization of the \textit{Proca equation} for massive spin-1 bosons~\cite{Peskin1995} (the latter being recovered in the case of vanishing couplings $\gamma$ and $\sigma$).

We expand the field $A_{\mu}$ in momentum space using Fourier modes, so that we can write
\begin{equation}
A_{\mu}(\eta, \vec{x}) = \int \frac{\dd^3k}{(2\pi)^{3/2}} e^{i\vec{k}\cdot\vec{x}} \Tilde{A}_{\vk,\mu}(\eta).
\end{equation}
Additionally, spatial components are divided in \textit{longitudinal} and \textit{transverse} modes, defined as
\begin{equation}
    \vec{k} \cdot \vec{\tA}_{\vk} = k \tA_{\vk,\text{L}}, \qquad \text{and} \qquad \vec{k} \cdot  \vec{\tA}_{\vk,\text{T}} = 0,
\end{equation}
so that
\begin{equation}
\vec{\tA}_{\vk} = \vec{\tA}_{\vk,\text{T}} + \tA_{\vk,\text{L}} \frac{\vk}{k}.
\end{equation}

Writing the equation of motion~\eqref{eq:sf.ProcaEquation} in momentum space, we observe that $\tA_{\vk,0}$ is not dynamical, but can actually be expressed in terms of the longitudinal part,
\begin{equation}
\left[k^2 - M^{00}\right]\tA_{\vk,0}(\eta) = -ik\tA_{\vk,\text{L}}^{\prime}(\eta).
\end{equation}
Moreover, as long as~$k^2 - M^{00}$ is different from $0$, one can write
\begin{equation}
\tA_{\vk,0}(\eta) = -\frac{ik\tA_{\vk, \text{L}}^{\prime}(\eta)}{k^2 - M^{00}}.
\label{eq:sf.0Component}
\end{equation}

Expanding the action \eqref{eq:sf.VectorAction} in Fourier modes and making use of \eqref{eq:sf.0Component}, we arrive~at
\begin{equation}
\begin{split}
    S = \frac{1}{2}\int \frac{\dd\eta \dd^3k}{(2\pi)^{3/2}}&\Bigg[\abs{\vec{\tA}_{\vk,\text{T}}^{\prime}}^2- \left(k^2 + M^{\text{TT}}\right)\abs{\vec{\tA}_{\vk,\text{T}}}^2\\ 
    &\hspace{0.5cm}-\frac{M^{00}}{k^2 -M^{00}}\abs{\tA_{\vk,\text{L}}^{\prime}}^2 -M^{\text{LL}}\abs{\tA_{\vk,\text{L}}}^2\Bigg],
    \label{eq:sf.MomentumAction}
\end{split}
\end{equation}
where
\begin{equation}
    M^{\text{TT}} = a^2\left(m^2 + \gamma R\right) + a^4\sigma\Tilde{R}^{\text{TT}},
\end{equation}
with $\Tilde{R}^{\text{TT}}$ denoting the traceless Ricci tensor in the direction transverse to the wavevector. However, due to the isotropy of the background, in our case all spatial components coincide. We note that the dynamics of transverse and longitudinal degrees of freedom are independent of each other, and as such can be treated separately. Crucially, from \eqref{eq:sf.MomentumAction} it is clear that, for a positive value of $M^{00}$, the kinetic term of the longitudinal mode becomes negative, leading to ghost instabilities for momenta below $\abs{M^{00}}$. In this case, the quantum theory is ill-defined and exhibits divergences even at the linearized level (see e.g. \cite{Himmetoglu2009} for more details). In order to avoid this situation, we require that $M^{00}$ be negative for all values of $m, \gamma$, and $\sigma$, which restricts the allowed parameter space of our theory. Note that this is indeed a fundamental limitation of the theory, and not an operational requirement. We will analyze the allowed region of parameters at the end of this subsection (see \cite{Capanelli2024a} for an exhaustive analysis of the parameter space in this kind of theories).

\subsubsection*{Dynamics of transverse modes}

The action for the transverse modes is given by
\begin{equation}
S_{\text{T}} = \frac{1}{2}\int \frac{\dd\eta \, \dd^3k}{(2\pi)^{3/2}}\left[\abs{\vec{\tA}_{\vk,\text{T}}^{\prime}}^2 - \left(k^2 + M^{\text{TT}}\right)\abs{\vec{\tA}_{\vk,\text{T}}}^2\right],
\label{eq:sf.TransverseAction}
\end{equation}
and leads to the equation of motion
\begin{equation}
\vec{\tA}_{\vk,\text{T}}^{\prime\prime}(\eta, \vec{k}) + \left[k^2 + M^{\text{TT}}(\eta)\right]\vec{\tA}_{\vk,\text{T}}(\eta, \vec{k}) = 0.
\end{equation}
The most general solution can be written as
\begin{equation}
\vec{\tA}_{\vk,\text{T}}(\eta, \vec{k}) = \sum_{s=\pm}\left[a_{\vk, s} v_{k,\text{T}}(\eta)\vec{\epsilon}_s  +  a^*_{-\vk, s}v_{k,\text{T}}^*(\eta) \vec{\epsilon}^*_s\right],
\label{eq:sf.TransverseExpansion}
\end{equation}
where $\vec{\epsilon}_s$ is a polarization vector ($s=+1$ or $-1$). Note that this expression fulfills the condition $\vec{\tA}^*_{-\vk,\text{T}}(\eta) = \vec{\tA}_{\vk,\text{T}}(\eta)$, as required by the reality of $A_{\mu}$. In turn, this implies that~$v_{k,\text{T}}(\eta)$ only depends on the norm $k$ of the $3$-momentum, as expected from homogeneity and isotropy. Therefore, the mode equation for $\vec{\tA}_{\vk,\text{T}}$ boils down~to
\begin{equation}
v_{k,\text{T}}^{\prime\prime}(\eta) + \omega_{k,\text{T}}^2(\eta)v_{k,\text{T}}(\eta) = 0,
\label{eq:sf.TransverseModeEquation}
\end{equation}
where
\begin{equation}
   \omega_{k,\text{T}}^2(\eta)=k^2 + M^{\text{TT}}(\eta).
\label{eq:sf.TransverseFrequency}
\end{equation}
This has the shape of a harmonic oscillator equation with a time-dependent frequency, and for $\sigma=0$, it has exactly the form of the equation of the non-minimally coupled scalar field discussed in subsection~\ref{subsec:sf.scalarfield}.

\subsubsection*{Dynamics of longitudinal modes}

On the other hand, the action for the longitudinal modes reads
\begin{equation}
S_{\text{L}} = \frac{1}{2}\int \frac{\dd\eta \dd^3k}{(2\pi)^{3/2}}\left[-\frac{M^{00}}{\vec{k}^2 - M^{00}}\abs{\tA_{\vk,\text{L}}^{\prime}}^2 - M^{\text{LL}}\abs{\tA_{\vk,\text{L}}}^2\right],
\label{eq:sf.LongitudinalAction}
\end{equation}
which can be written in a form similar to \eqref{eq:sf.TransverseAction}, provided we introduce the auxiliary field $\mathcal{A}_{\vk,\text{L}}$ as
\begin{equation}
    \mathcal{A}_{\vk,\text{L}}(\eta) = f_k(\eta) \tA_{\vk,\text{L}}(\eta), \qquad f_k(\eta)=\frac{\sqrt{-M^{00}(\eta)}}{\sqrt{k^2 - M^{00}(\eta)}},
\end{equation}
where the corresponding element of the mass tensor reads
\begin{equation}
     M^{00}(\eta)=-a^2(\eta)\left[m^2+\gamma R(\eta)\right] + a^4(\eta)\sigma \Tilde{R}^{00}(\eta).
\label{eq:sf.M00}
\end{equation}
Then, equation \eqref{eq:sf.LongitudinalAction} can be expressed as
\begin{equation}
S_{\text{L}} = \frac{1}{2}\int\frac{\dd\eta \dd^3k}{(2\pi)^{3/2}}\left\{\abs{\mathcal{A}^{\prime}_{\vk,\text{L}}}^{2} - \omega_{k,\text{L}}^2(\eta)\abs{\mathcal{\tA}_{\vk,\text{L}}}^2\right\},
\end{equation}
where the frequency is given by
\begin{equation}
\begin{split}
\omega^2_{k,\text{L}}(\eta)&=\frac{M^{\text{LL}}(\eta)}{f_k(\eta)^2} - \frac{f_k^{\prime\prime}(\eta)}{f_k(\eta)}, \\ M^{\text{LL}}(\eta) &= a^2(\eta)\left[m^2+\gamma R(\eta)\right] + a^4(\eta)\sigma \Tilde{R}^{\text{LL}}(\eta),
\label{eq:sf.LongitudinalFrequency}
\end{split}
\end{equation}
with $\Tilde{R}^{\text{LL}}$ denoting the traceless Ricci tensor in the direction of the wavevector, which coincides with $\Tilde{R}^{\text{TT}}$. As before, the field can generally be written as the linear combination
\begin{equation}
\mathcal{A}_{\vk,\text{L}}(\eta) = a_{\vk,\text{L}}v_{k,\text{L}}(\eta) + a_{-\vk,\text{L}}^*v_{k,\text{L}}^*(\eta),
\label{eq:sf.LongitudinalExpansion}
\end{equation}
leading to the mode equation
\begin{equation}
v^{\prime\prime}_{k,\text{L}}(\eta) + \omega^2_{k,\text{L}}(\eta)v_{k,\text{L}}(\eta) = 0.
\label{eq:sf.LongitudinalModeEquation}
\end{equation}
We see here that, contrary to the transverse mode and the scalar field case, the form of the longitudinal frequency is not simply $k^2$ plus an $\eta$-dependent term. It is more involved: Temporal and wavenumber dependencies mix. 

Canonical quantization of the transverse and longitudinal fields, $\vec{\tA}_{\vk,\text{T}}$ and $\mathcal{A}_{\vk,\text{L}}$, respectively, is completely analogous to the quantization of a real scalar field discussed in \cref{sec:qftcs.quantization}. Each field can be written as a linear combination of any basis of solutions of its corresponding mode equation, \eqref{eq:sf.TransverseModeEquation} or \eqref{eq:sf.LongitudinalModeEquation}, and depending on the choice of basis one arrives at a different quantum theory. We will select the initial state for the field $A_{\mu}$ in analogy to the scalar case, and compute the abundance of dark matter resulting from the expansion of the geometry during inflation and reheating by obtaining the mean number of particles an observer would measure after spacetime expansion becomes irrelevant (for particle production) in the state of the system. From this point on, particle production will be negligible, provided the expansion is adiabatic enough. Therefore, one can again extrapolate this abundance to the present and compare with observations, as long as the dark matter field is non-interacting.
In order to solve the mode equations~\eqref{eq:sf.TransverseModeEquation} and~\eqref{eq:sf.LongitudinalModeEquation} for obtaining the number of particles that are produced in each mode, we only consider the case of the single-field inflationary model with quadratic potential for the vector field, an analysis that corresponds to reference \cite{VectorDM2024}.

\subsubsection*{Parameter space}

Let us now discuss in which region of parameter space our theory is well-defined, starting with the coupling $\gamma$. Stability of initial conditions (which will be analyzed further below) requires $\gamma \geq 1/6$, as for the scalar field. Any value of $\gamma$ within this region is valid, as long as $m$ and~$\sigma$ are such that $M^{00}<0$. However, backreaction has to be taken into account for very large values of~$\gamma$, for which the energy density of the vector field becomes significant and starts to affect the geometry (again, see ref. \cite{Lebedev2022} for such an analysis in the context of non-minimally coupled scalar fields). Since we assume that the energy density of the field is small, and production of particles for~\mbox{$\gamma \gtrsim 1$} is qualitatively similar to that of $\gamma = 1$, we consider the range~\mbox{$1/6\leq \gamma\leq 1$}. At the same time, we want to allow for the possibility of vanishing $\sigma$, so that we can compare with scalar field production, for example. This, together with the range of $\gamma$ discussed above, sets a limit for the smallest mass we can consider,~$m=0.5m_{\phi}$. This is clear when looking at the explicit form of $M^{00}$, given in \eqref{eq:sf.M00}. The Ricci scalar oscillates during reheating around $0$, and therefore the mass term in $M^{00}$ has to compensate for this behavior so that $M^{00}$ remains negative. On the other hand, the traceless Ricci tensor component is always positive or $0$ (see figure \ref{fig:sf.InflatonAndRicciConformal}). This means that a positive value of $\sigma$ is going to contribute always against the mass term. Therefore,~$\sigma$ has to be negative if we want to keep the chosen range in~$\gamma$ while having~\mbox{$m\geq 0.5m_{\phi}$}. Because of all the previous considerations, we restrict ourselves to the following region of parameter space:
\begin{equation}
m \geq 0.5m_{\phi}, \qquad \gamma \in [1/6, 1], \qquad \sigma \leq 0.
\label{eq:sf.ParameterSpace}
\end{equation}
One can consider smaller values of the mass while keeping the same allowed range in $\gamma$, but then the coupling $\sigma$ has to become negative and $\sigma=0$ is not allowed. At the same time, positive values of $\sigma$ are possible, but in this case it is required that the mass of the test field becomes greater than $0.5m_{\phi}$.

\section{Scalar mode equation and vacuum choice}
\label{sec:sf.scalarmodes}

Let us first consider the case of the scalar field. We solve the mode equation~\eqref{eq:qftcs.ModeEquation} in this case, and discuss the possible vacuum choices in order to extract particle production. Since the background is the same, most of the ideas presented can be extrapolated to the vector field case, which is discussed in the following section. Note that this section applies to both quadratic and Starobinsky inflationary potentials, since the key elements needed for the derivations are common, and the differences enter only in the numerical calculations.

In order to compute the gravitational production during inflation and reheating, we need to extract the $\beta_k$ Bogoliubov coefficient (see \cref{eq:qftcs.MeanNumberDensity}) relating the notions of vacuum of observers in the \textit{in} and \textit{out} regions. Since spacetime is always expanding, this task is not obvious, as we will see. In practice, we will choose some initial conditions for the \textit{in} modes $v_k$ at $\etai$ and evolve them following the mode equation~\eqref{eq:qftcs.ModeEquation} until a time $\etaf$, when particle production becomes negligible (see below for details). Then, it is sufficient to know the value of the \textit{out} modes at $\etaf$ in order to extract the mean number of produced particles via the Wronskian \eqref{eq:qftcs.BogoliubovCoefficientsWronskian}. 

\subsection{\textit{In} mode solution to the scalar mode equation}
\label{subsec:sf.scalarmodeequation}

The task is to solve the mode equation \eqref{eq:qftcs.ModeEquation} from the onset of inflation at $\etai$ until a time $\etaf$ well inside the adiabatic regime at the end of reheating, with the frequency of the oscillator determined by the background geometry described in section \ref{sec:sf.chaoticinf}. In a similar way as we did for the background dynamics, the mode equation is solved in the regions
\begin{equation}
\eta = \begin{cases}
\eta_{\text{i}} \leq \eta \leq \eta_*, \quad \text{slow-roll approximation},\\
\eta_*\leq \eta \leq \eta_{\text{f}}, \quad \text{numerical solution}.
\end{cases}
\end{equation}
In a strict de Sitter geometry, the Hubble rate is exactly constant,~\mbox{$H=H_0$}, the Ricci scalar is $R = 12 H_0^2$, and the scale factor reads~\mbox{$a_{\text{dS}}(\eta) = -1/\left[H_0(\eta-\tilde{\eta})\right]$}, where $\tilde{\eta}$ is an arbitrary time shift. Therefore, the frequency \eqref{eq:qftcs.MasterFrequency} takes the form
\begin{equation}
    \omega_{k, \text{dS}}^2 = k^2 + \frac{\mu^2}{\left(\eta-\tilde{\eta}\right)^2}, \qquad \mu^2 = m^2/H_0^2 + 12(\xi-1/6),
\label{eq:sf.deSitterFrequency}
\end{equation}
where $\mu$ plays the role of a dimensionless, effective mass. We can now identify~$H_0$ with the Hubble rate at the beginning of inflation $H(\eta_{\text{i}})$ and take~\mbox{$\tilde{\eta} = \eta + 1/\left[a(\eta_i)H_0\right]$}, so that the exact frequency~\eqref{eq:sf.ScalarFrequency} approaches the above \cref{eq:sf.deSitterFrequency} in the limit $\eta \to \eta_{\text{i}}$. The solution to the mode equation \eqref{eq:qftcs.ModeEquation} in this simplified scenario that behaves as a positive frequency plane wave in the limit~\mbox{$k\abs{\eta} \to \infty$} is given by
\begin{equation}
\begin{split}
v_{k, \text{dS}}(\eta) &= \sqrt{\pi\abs{\eta - \tilde{\eta}}/2}\, e^{i\pi \nu / 2} H_{\nu}^{(1)}\left(k\abs{\eta - \tilde{\eta}}\right),  \\ 
\nu &= \sqrt{{1}/{4}- \mu^2}.
\label{eq:sf.DeSitterSolution}
\end{split}
\end{equation}
These modes are associated with the Bunch-Davies vacuum discussed above \cite{Birrell1982}. Note that there is a critical value $\mu^2=1/4$ for which $\nu=0$, which separates the regimes of real and imaginary $\nu$. In particular, for~$m^2/H_0^2 \ll 1$, we can approximately write $\mu^2 \approx 12\left(\xi-1/6\right)$, and therefore $\mu^2 = 1/4$ for~$\xi = 3/16$. At this point, the behavior of the modes qualitatively changes~\cite{Borrajo2020}, and this fact will be important for the analysis in subsection \ref{subsec:sf.quadraticabundance}. 

However, our background geometry is \emph{not} exactly de Sitter, but given by the inflaton dynamics derived in section \ref{sec:sf.chaoticinf}. Within the slow-roll approximation, valid from the start of inflation at $\eta_{\text{i}}$ until $\eta_*$, the mode equation to solve is
\begin{equation}
v^{\prime\prime}_k(\eta) + \omega_{k, \text{SR}}^2(\eta) v_k(\eta) = 0,
\label{eq:ModeEquationSR}
\end{equation}
where the scale factor and the Ricci scalar in $\omega_{k, \text{SR}}(\eta)$ correspond to the slow-roll analysis in section \ref{sec:sf.chaoticinf}. Nevertheless, in the slow-roll regime, and for a certain range in $k, m$, and $\xi$, we can approximate the solution satisfying Bunch-Davies initial conditions by (see next subsection for details)
\begin{equation}
\begin{split}
    v_{k, \text{SR}}(\eta) &\simeq \sqrt{\pi\abs{\tau_k}/2} e^{i\pi\nu/2} H_{\nu}^{(1)}\left(k\abs{\tau_k}\right), \\ 
\tau_k &= \frac{\omega_{k, \text{SR}}(\eta)}{\omega_{k, \text{dS}}(\eta)}(\eta-\eta_{*, k}) + \eta_{*, k}-\eta_0,
\label{eq:sf.ApproximateSRSolution}
\end{split}
\end{equation}
where $\eta_{*, k}$ marks the limit of validity of the approximation, and we have chosen the constant\footnote{The reasons behind this choice of time shift will be clear after the next subsection.} $\tilde{\eta} = \eta_0 = 1/H_0$. From this point on, the mode equation~\eqref{eq:qftcs.ModeEquation} has to be solved numerically, independently of the background dynamics being numerical or analytical, taking as initial condition solution \eqref{eq:sf.ApproximateSRSolution} and its derivative at $\eta_{*, k}$. The frequency one has to use in this regime is the one with the scale factor and Ricci scalar given by the full inflaton dynamics. In this way, we obtain the modes~$v_k$ that characterize the \textit{in} vacuum. Before going on, let us see how to obtain expression~\eqref{eq:sf.ApproximateSRSolution}.

\subsection{Slow-roll approximation for the solution to the mode equation}
\label{subsec:sf.approximations}
Because the mode equation cannot be solved analytically, even considering a slowly rolling inflaton field, one would need to use numerical methods in order to find a solution. However, the large amount of $e$-folds to cover (see \cref{sec:cosmo.inflation}) makes it more interesting and feasible to rely on an analytic approximation, such as~\eqref{eq:sf.ApproximateSRSolution}. We dedicate this subsection to formally develop the approximation and to test its validity. For notational convenience, in the calculations that follow we will write~\mbox{$\eta-\tilde{\eta}$} as $\eta$, and drop the mode index $k$. Let us start by defining the following small parameters for given values of $k, m$ and~$\xi$ which will be useful in the following. 
\begin{itemize}
\item First, we have
\begin{equation}
    \epsilon(m, \xi) = \underset{ \eta\in I_1}{\text{max}}\Bigg|1-\frac{\omega_{{\text{SR}}}(  \eta; m, \xi)}{\omega_{{\text{dS}}}( \eta; m, \xi)}\Bigg|, \quad \text{with} \quad I_1 = (-\infty, \eta_1),
\end{equation}
where $\eta_1$ is chosen such that $\epsilon \ll 1$. Then, we can define~\mbox{$f(\eta; m, \xi)$} by 
\begin{equation}
    \frac{\omega_{\text{SR}}}{\omega_{\text{dS}}} = 1 + \epsilon f.
\end{equation}
By construction, $\abs{f(\eta)} \leq 1$ for $\eta \in I_1$.

\item It will also be convenient to define
\begin{equation}
    \sigma(m, \xi) = \underset{\eta\in I_2}{\text{max}}\Big|f^{\prime}(\eta; m, \xi)\eta\Big|, \quad \text{with} \quad I_2 = (-\infty, \eta_2),
\end{equation}
and choose $\eta_2$ such that $\sigma \leq \epsilon$. Then, we introduce $g(\eta; m, \xi)$ as
\begin{equation}
    f^{\prime}(\eta) = \frac{\sigma g(\eta)}{\eta},
\end{equation}
for which again we have that $\abs{g(\eta)}\leq 1$ for $\eta \in I_2$.

\item Similarly, we define
\begin{equation}
    \rho(m, \xi) = \underset{\eta \in I_3}{\text{max}} \Bigg|\frac{\omega^{\prime}_{\text{dS}}(\eta)}{\omega_{\text{dS}}(\eta)}\eta \Bigg|, \quad \text{with} \quad I_3 = (-\infty, \eta_3),
\end{equation}
and choose $\eta_3$ such that $\rho \leq \epsilon$.

Now, we take $\eta_*=\text{min}(\eta_1, \eta_2, \eta_3)$ and $I=(-\infty, \eta_*)$, where $I$ is the interval for which the three parameters $\epsilon, \sigma, \rho$ are small. Note that $\eta_*<0$ since inflation ends at $\eta=0$. 
\end{itemize}

The task is to solve equation \eqref{eq:ModeEquationSR}, for which we define a new time coordinate $\zeta$ within the interval $I$,
\begin{equation}
    \dd\zeta = \frac{\omega_{\text{SR}}(\eta)}{\omega_{\text{dS}}(\eta)} \dd\eta= \left[1+\epsilon f(\eta)\right] \dd\eta.
\end{equation}
After integration until $\eta \in I$ and taking the absolute value, this becomes
\begin{equation}
    \abs{(\zeta - \zeta_*) - (\eta - \eta_*)} = \epsilon \Bigg|\int^{\eta_*}_{\eta} \dd t f(t)\Bigg| = \mathcal{O}(\epsilon)(\eta - \eta_*).
\end{equation}
Then, choosing $\zeta_*=\eta_*$, this can be expressed as
\begin{equation}
    \zeta = \eta\left[1 + \mathcal{O}(\epsilon)\right].
\end{equation}
We change time coordinates $\eta\to \zeta$ in the mode equation, which takes the form
\begin{equation}
    \ddot{w}( \zeta) + \omega_{\text{dS}}^2\left[\eta( \zeta)\right] w( \zeta ) + \epsilon f^{\prime}\left[\eta( \zeta)\right]\frac{\omega_{\text{dS}}^2\left[\eta(\zeta)\right]}{\omega_{\text{SR}}^2\left[\eta(\zeta)\right]} \dot{w}( \zeta)=0,
\label{eq:sf.ModeEquationZeta}
\end{equation}
where $w( \zeta) = v\left[\eta( \zeta)\right]$ and the dot denotes here derivative with respect to $\zeta$. 

Let us analyze the last term. With this aim, we introduce the dimensionless time~$\bar \zeta=\zeta/\eta_*$. Then, in terms of $\bar \zeta$, the equation above has the same form except for the last term, that acquires an extra factor. Using the definition of $f'$ and $\sigma$ above, the coefficient of this term is
\begin{equation}
\epsilon f'\frac{\omega_{\text{dS}}^2 }{\omega_{\text{SR}}^2 }\eta_*=
\epsilon\sigma g(1+\epsilon f)\frac{\eta_*}{\eta}=\mathcal{O}(\epsilon^2)\frac{\eta_*}{\eta}
\end{equation}
Since by definition $\abs{\eta_*} \leq \abs{\eta}, \, \forall \eta \in I$, this coefficient is of order $\mathcal{O}(\epsilon^2)$. Furthermore, the frequency in the second term of \eqref{eq:sf.ModeEquationZeta} is
\begin{align}
    \omega_{\text{dS}}^2(\eta(\zeta)) &= \omega_{\text{dS}}^2\left( \zeta\left[1+\mathcal{O}(\epsilon)\right]\right)\\
    &= \omega_{\text{dS}}^2(\zeta)\left[1 + 2\frac{\omega_{\text{dS}}^{\prime}}{\omega_{\text{dS}}}\Bigg|_{\zeta}\cdot\zeta\,\mathcal{O}(\epsilon)\right] \\
    & = \omega_{\text{dS}}^2\left(\zeta\right)\left[1 + \mathcal{O}(\epsilon^2)\right],
\end{align}
provided that $|\zeta\,\omega_{\text{dS}}^{\prime}(\zeta)/\omega_{\text{dS}}(\zeta)| \leq \rho = \mathcal{O}(\epsilon)$. 
This is satisfied for~\mbox{$\zeta=\eta\left[1+\mathcal{O}(\epsilon)\right]<\eta_*$}, i.e. for $\eta<\eta_*$.
Thus, the equation for $w$ can finally be written as
\begin{equation}
    \ddot{w}(\zeta) + \omega_{\text{dS}}^2(\zeta) w(\zeta) = \mathcal{O}(\epsilon_k^2).
\end{equation}

We can perturbatively solve the differential equation above by writing~\mbox{$w = w_{0} + \epsilon w_{1} + \mathcal{O}(\epsilon^2)$}. The solution to order $\epsilon^0$ is nothing but the de Sitter modes \eqref{eq:sf.DeSitterSolution},
\begin{equation}
    w_{0}(\zeta) = \sqrt{\pi\abs{\zeta}}\, e^{i\pi \nu}/2 H_{\nu}^{(1)}\left(k\abs{\zeta}\right),  \qquad \nu = \sqrt{{1}/{4}- \mu^2},
\end{equation}
and as a consequence, $w_{k, 0}$ behaves as a plane wave for $k\abs{\zeta} \to \infty$. On the other hand, the coefficients of the solution to order $\epsilon^1$ will satisfy the same original equation but with the initial conditions that~\mbox{$w_1=0 \quad \text{for} \quad k\abs{\zeta} \to \infty$}  and therefore~$w_{1}$ is identically zero. We can then write $w$ as
\begin{equation}
\begin{split}
    w(\zeta) &= w_{0}(\zeta )\left[1 + \mathcal{O}(\epsilon^2)\right] \\
    &=  \sqrt{\pi\abs{\zeta}}\, e^{i\pi \nu}/2 H_{\nu}^{(1)}\left(k\abs{\zeta}\right)\left[1 + \mathcal{O}(\epsilon^2)\right].
\end{split}
\end{equation}

In order to undo the coordinate transformation $\zeta \to \eta$ while keeping the error up to~$\mathcal{O}(\epsilon^2)$, we need to consider the $\mathcal{O}(\epsilon^1)$ terms in~\mbox{$\zeta = \eta\left[1+\mathcal{O}(\epsilon)\right]$}. For this, we note that
\begin{equation}
\begin{split}
    \Bigg| \left(\zeta - \eta_*\right) - &\frac{\omega_{\text{SR}}(\eta)}{\omega_{\text{dS}}(\eta)} (\eta - \eta_*) \Bigg| \\
    &= \Bigg| \left(\eta - \eta_*\right) + \epsilon \int_{\eta_*}^{\eta} \dd t f(t) - \left[1 + \epsilon f(\eta)\right] \left(\eta - \eta_* \right) \Bigg|\\
    &=\epsilon\Bigg| \int_{\eta_*}^{\eta} \dd t f(t) - \int_{\eta^*}^{\eta} \dd t f(\eta) \Bigg| \\
    &\leq \epsilon \int_{\eta_*}^{\eta} \dd t \abs{f(t) - f(\eta)} \\
    &=\epsilon \int^{\eta}_{\eta_*} \dd t \left| f^{\prime}(\eta)(t - \eta) + \frac{1}{2!}f^{\prime\prime}(\eta)(t - \eta)^2 + \cdots\right|  \\
    &\leq \epsilon \left\{\left|\frac{1}{2} f^{\prime}(\eta)(\eta - \eta_*)^2\right| +\left| \frac{1}{3!}f^{\prime\prime}\left(\eta - \eta_*\right)^3 \right| + \cdots\right\}.
\end{split}
\end{equation}
This means that, as long as the  terms in curly brackets are of order~$\mathcal{O}(\epsilon)$, we can write
\begin{equation}
\begin{split}
    \zeta &= \eta_* + \left[\frac{\omega_{\text{SR}}(\eta)}{\omega_{\text{dS}}(\eta)} + \mathcal{O}(\epsilon^2) \right] (\eta - \eta_*) \\
    &= \eta_* + \frac{\omega_{\text{SR}}(\eta)}{\omega_{\text{dS}}(\eta)} (\eta - \eta_*)\left[1 + \mathcal{O}(\epsilon^2) \right].
\end{split}
\end{equation}
The first term is equal to
\begin{equation}
 \frac{1}{2}\sigma \left|g(\eta)\frac{\eta - \eta_*}{\eta} \right|=\mathcal{O}(\epsilon).
\end{equation}
The next terms are of the form $f^{(n)}\left(\eta - \eta_*\right)^{n+1}/n!$, which numerically can be seen to be smaller than the first one.

Therefore, undoing the translation of $\eta$ to $\eta-\tilde{\eta}$ that we did at the beginning of this calculation, the solution to the mode equation can be written as \eqref{eq:sf.ApproximateSRSolution} up to terms of order~$\mathcal{O}(\epsilon^2)$, with the choice $\tilde{\eta}=\eta_0$.
With fixed $\xi$, and choosing~$\eta_*$ independent of $k$, the error $\epsilon_k$ increases with increasing $m$ and decreasing $k$.

\begin{figure}[t!]
    \centering
    \includegraphics[width=1\textwidth]{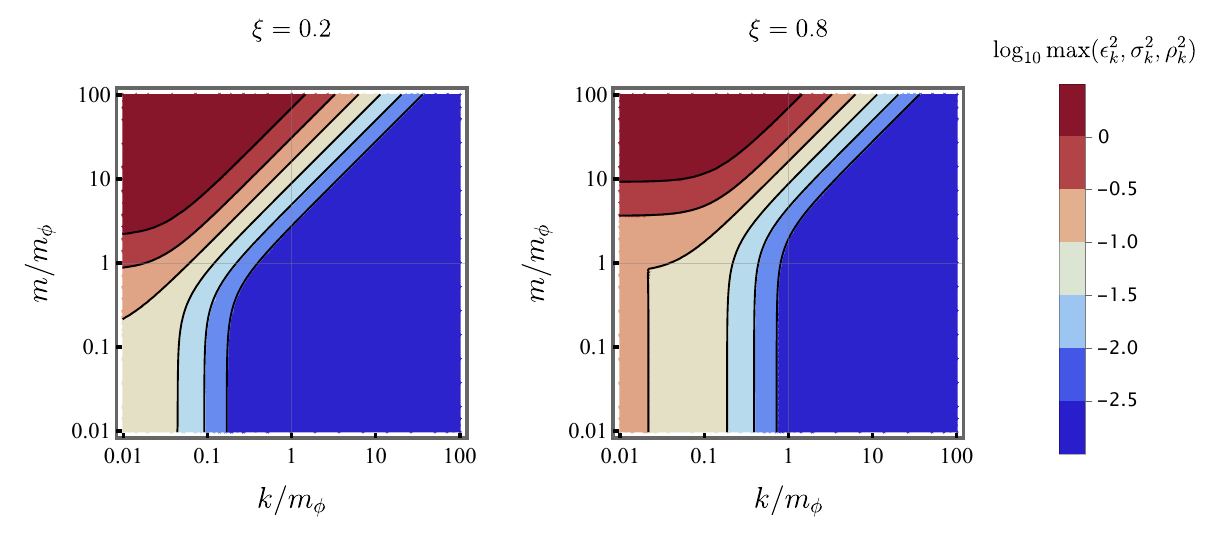}
    \caption[Slow-roll approximation errors for the scalar field]{Maximum of the errors squared as a function of the wavenumber~$k$ and the field mass $m$, for $\xi=0.2$ (left) and $\xi=0.8$ (right). We take $\eta_*=-500m_{\phi}$ for all values of $k$, $m$ and $\xi$. Figure from \cite{ScalarField2023}.}
    \label{fig:sf.error}
    \end{figure}

\begin{figure}[t!]
    \centering
    \includegraphics[width=1\textwidth]{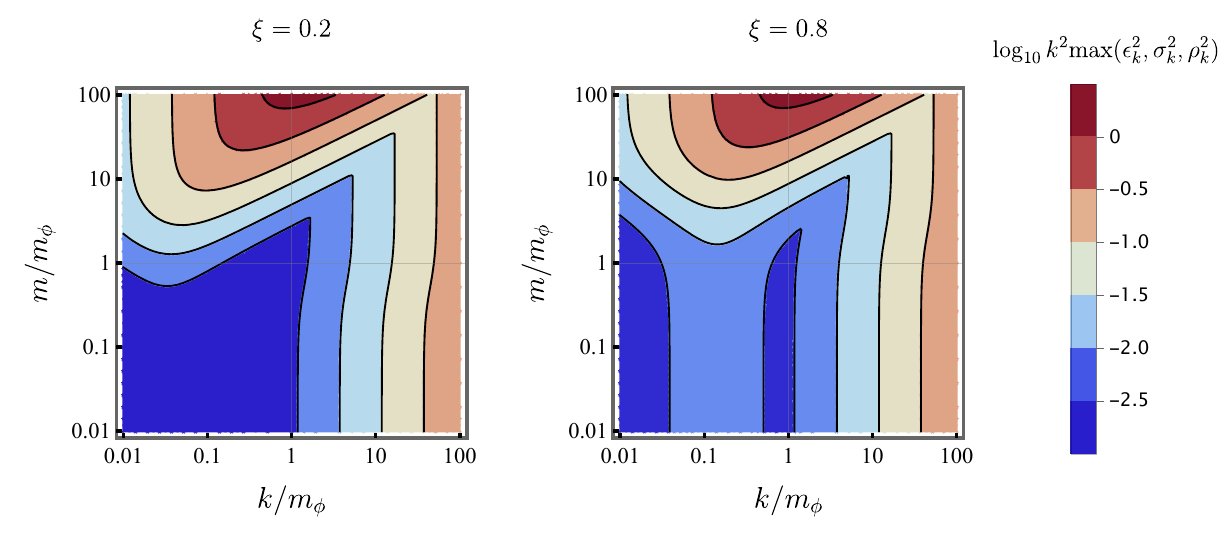}
    \caption[Slow-roll approximation errors times $k^2$ for the scalar field]{Maximum of the errors squared times $k^2$ as a function of the wavenumber $k$ and the field mass $m$, for $\xi=0.2$ (left) and $\xi=0.8$ (right). We take $\eta_*=-500m_{\phi}$ for all values of $k$, $m$ and $\xi$. Figure from \cite{ScalarField2023}.}
\label{fig:sf.k2error}
\end{figure}

When we numerically solve the mode equation \eqref{eq:qftcs.ModeEquation} from $\eta_*$, the error in the initial condition coming from the slow-roll solution \eqref{eq:sf.ApproximateSRSolution} carries through as
\begin{equation}
    v_k(\eta) = v_{k, 0}(\eta) \left[1 + \mathcal{O}(\epsilon_k^2)\right],
\end{equation}
such that $v_k(\eta) \to v_{k, \text{SR}}(\eta)$ as $\eta \to \eta_*$. Therefore, we have for the total density defined in~\eqref{eq:qftcs.TotalNumberDensity} that
\begin{equation}
    \!n(m, \xi) = \int_0^{\infty} \frac{\dd k}{2\pi^2} k^2 \abs{\beta_k}^2 = n_0 \left[1 + \frac{1}{n_0}\int_{0}^{\infty} \frac{\dd k}{2\pi^2}k^2\abs{\beta_{k, 0}}^2\mathcal{O}(\epsilon_k^2)\right],
    \label{eq:sf.DensityOfParticles}
\end{equation}
where $n_0 = \int_0^{\infty} \frac{\dd k}{2\pi^2} k^2 \abs{\beta_{k, 0}}^2$. Although the error $\epsilon_k$ increases as $k$ decreases, the factor~$k^2$ compensates this increase for low $k$. Essentially, although $\epsilon_k^2$ increases for~\mbox{$k<m_{\phi}$}, the quantity~$k^2 \epsilon_k^2$ remains small, whereas $\abs{\beta_{k, 0}}^2$ is roughly of the same order. More explicitly, for our calculations, we take $\eta_* = -500 m_{\phi}$, for which the maximum of the three small parameters squared, $\epsilon_k^2, \sigma_k^2, \rho_k^2$, as a function of mass and wavenumber, for two different choices of coupling $\xi$, is shown in figure~\ref{fig:sf.error}. For~\mbox{$m\leq m_{\phi}$} and~\mbox{$k\geq 0.1m_{\phi}$}, the error is of order~$\mathcal{O}(0.01)$ or smaller for the various values of $\xi$ considered, and thus the approximation is controlled in this regime. At the same time, we can observe in figure \ref{fig:sf.k2error} that $k^2\epsilon_k^2$ decreases as we move to the low-part of the momentum range. This guarantees that this region of the spectrum is robust against errors in the mode equation approximation we used. On the other hand, from figure \ref{fig:sf.k2error} we observe that the quantity $k^2\epsilon_k^2$ grows with~$k$ for~$k>m_{\phi}$, since the decrease in $\epsilon_k^2$ (cf. figure \ref{fig:sf.error}) can not compensate the power~$k^2$. However, gravitational production for high-momentum particles is very small, namely~\mbox{$\abs{\beta_k}^2 \approx 0$} for~$k\gg m_{\phi}$. As a consequence, $n(m, \xi) \approx n_0$ approximates well the total number density of particles produced, since the weight of wavenumbers~\mbox{$k\gg m_{\phi}$} is very small when compared to the rest of the spectrum.


\begin{figure}[t]
\centering
\includegraphics[width=\textwidth]{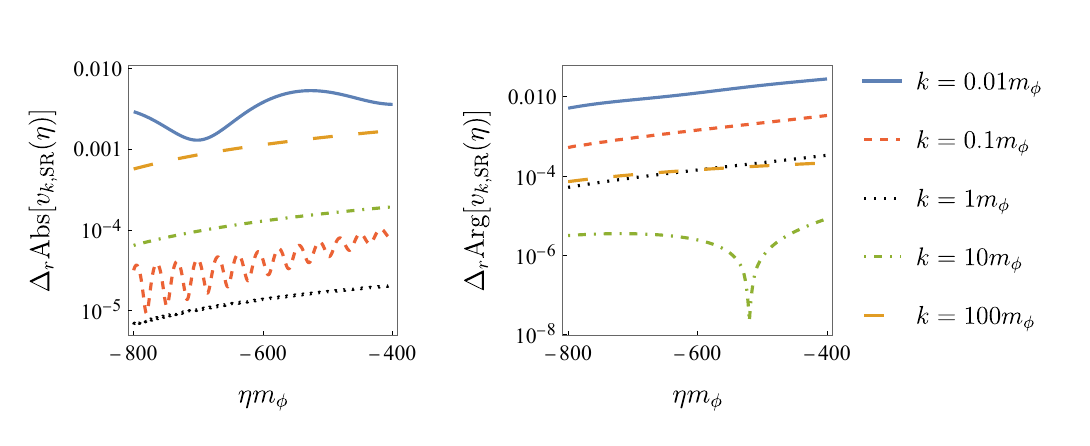}
\caption[Error of the analytic slow-roll approximation with respect to the numerical solution]{Relative error in the absolute value (left panel) and the phase (right panel) of the numerical solution to the exact mode equation \eqref{eq:qftcs.ModeEquation} compared to the analytical approximation~\eqref{eq:sf.ApproximateSRSolution}, for wavenumbers ranging from $k=0.01m_{\phi}$ to $k=100 m_{\phi}$, and \mbox{$m=m_{\phi}, \xi=0.5$}. Here, we take~$\eta_{\text{dS}} = -1000/m_{\phi}$ and \mbox{$\eta_* = -500/m_{\phi}$}. Figure from \cite{ScalarField2023}.}
\label{fig:sf.SRvsWOSR}
\end{figure}

Furthermore, we can test the validity of \eqref{eq:sf.ApproximateSRSolution} when compared to the numerical solution of \eqref{eq:qftcs.ModeEquation} by putting ourselves in the following scenario: Let us assume that the geometry can be approximated by a de Sitter spacetime during the early stages of inflation, such that the solution \eqref{eq:sf.DeSitterSolution} is valid in a region $\eta_{\text{i}}\leq\eta<\eta_{\text{dS}}$. At $\eta_{\text{dS}}$, slow-roll starts to matter, and deviations from the de Sitter solution $v_{k, \text{dS}}(\eta)$ occur. In this scenario, we explore two different paths to continue solving the equation:
\begin{enumerate}
    \item We assume slow-roll inflation is a good description for the background dynamics in the region $\eta_{\text{dS}}\leq\eta<\eta_*$, and take as solution the approximation~\eqref{eq:sf.ApproximateSRSolution}.
    \item We solve numerically the exact equation of motion for the inflaton, eq. \eqref{eq:cosmo.InflatonEoM}, obtaining the frequency corresponding to \eqref{eq:qftcs.ModeEquation}, equation which we again solve numerically. This solution, $v_k(\eta)$, will be valid even for $\eta \geq \eta_*$.
\end{enumerate}
In figure \ref{fig:sf.SRvsWOSR}, we compare the analytical slow-roll solution with the exact numerical solution by plotting the relative difference between their absolute values, 
\begin{equation}
     \Delta_r \text{Abs}\left[v_{k, \text{SR}}(\eta)\right] \equiv \Bigg|\frac{\text{Abs}\left[v_k(\eta)\right] - \text{Abs}\left[v_{k, \text{SR}}(\eta)\right]}{\text{Abs}\left[v_k(\eta)\right]}\Bigg|,
\end{equation}
as well as their phase difference,
\begin{equation}
     \Delta_r \text{Arg}\left[v_{k, \text{SR}}(\eta)\right] \equiv \Bigg|\frac{\text{Arg}\left[v_k(\eta)\right] - \text{Arg}\left[v_{k, \text{SR}}(\eta)\right]}{\pi}\Bigg|.
\end{equation}
We do so for different wavenumbers, ranging from $k=0.01m_{\phi}$ to~\mbox{$k=100m_{\phi}$}, denoted by the different shapes in figure \ref{fig:sf.SRvsWOSR}. We have taken $\eta_{\text{dS}}=-1000/m_{\phi}$ as the start of slow-roll and~$\eta_* = -500/m_{\phi}$ as the time when the slow-roll approximation breaks down. For $k=m_{\phi}$, the relative error is very small, of order $\sim 10^{-4}$ at $\eta_*$. For wavenumbers larger than the mass of the inflaton, $k>m_{\phi}$, the approximation is still good, although it worsens. On the other hand, the error for~\mbox{$k=0.01 m_{\phi}$} starts becoming significant, and gets worse for $k<0.01 m_{\phi}$. However, the corresponding region of the spectrum of produced particles is highly suppressed, as discussed above, and therefore the contribution to the total density of particles is negligible. Similarly, particle production is very small for wavenumbers larger than~\mbox{$k>100m_{\phi}$}, and therefore the range of interest in $k$ is under control. Hence, we can assume the approximation is valid in the region $\eta_{\text{dS}}\leq\eta<\eta_*$. 

Note that if this solution behaves well in this region, it has to become an even better approximation before $\eta_{\text{dS}}$, since the further toward the past we go, the more de Sitter-like is the geometry. Thus, eq.~\eqref{eq:sf.ApproximateSRSolution} can be taken as well as a solution to the mode equation in the region~\mbox{$\eta_{\text{i}}\leq \eta <\eta_{\text{dS}}$}. Under these approximations, the mode equation~\eqref{eq:qftcs.ModeEquation} can be solved analytically from the start of inflation, $\eta_{\text{i}}$, until $\eta_*$, for which the slow-roll approximation starts to fail. From there, the mode equation is solved numerically.

\subsection{Choice of \textit{out} vacuum}
\label{subsec:sf.vacuumchoice}
The solution $v_k(\eta)$ to the mode equation is associated with the Bunch-Davies vacuum, which is a reasonable notion of vacuum at the onset of inflation, since the geometry approaches de Sitter. The procedure in subsection \ref{subsec:sf.scalarmodeequation} allows us to evaluate $v_k(\eta_{\text{f}})$. However, in order to obtain the Bogoliubov coefficient $\beta_k$, we also need~$u_k(\eta_{\text{f}})$ (cf. \cref{eq:qftcs.BogoliubovCoefficientsWronskian}), which is the solution to the mode equation associated with the vacuum notion at this point in time. Then, from the Bogoliubov coefficients~$\beta_k$, we will be able to extract the number density of produced particles at $\eta_{\text{f}}$ using \cref{eq:qftcs.MeanNumberDensity}, but note that we have not defined $\eta_{\text{f}}$ itself. This time is defined by the condition that particle production becomes negligible, which is fulfilled in the adiabatic regime (cf. subsection \ref{subsec:qftcs.inout}), i.e. when
\begin{equation}
\mathcal{C}_k(\eta) = \Bigg | \frac{\omega_k^{\prime}(\eta_{\text{f}})}{\omega_k^2(\eta_{\text{f}})} \Bigg| \ll 1.
\label{eq:sf.AdiabaticCondition}
\end{equation}
The value of $\eta_{\text{f}}$ highly depends on the parameters of the scalar field, and in particular, it becomes larger as the mass $m$ decreases. This is why, for certain regions in parameter space, it may be convenient to use the late-time approximation for the background dynamics, instead of solving numerically the equation of motion of the inflaton field. It is worth mentioning that at the same time, a smaller coupling~$\xi$ to the curvature implies that the Ricci scalar oscillations, which are the main source of non-adiabaticity, are less important, therefore resulting in an earlier~$\eta_{\text{f}}$ at which~\eqref{eq:sf.AdiabaticCondition} holds true.

As long as the background is not static, the meaning of vacuum will change in time. Nevertheless, if the evolution is adiabatic enough, namely condition \eqref{eq:sf.AdiabaticCondition} is fulfilled, one can use the adiabatic prescription \eqref{eq:qftcs.AdiabaticVacuum} to define the \textit{out} vacuum at a given instant~$\eta_{\text{f}}$,
\begin{equation}
\begin{split}
u_k(\eta_{\text{f}}) &= \frac{1}{\sqrt{2\omega_k(\eta_{\text{f}})}}, \\ u_k^{\prime}(\eta_{\text{f}}) &= -\frac{1}{\sqrt{2\omega_k(\eta_{\text{f}})}}\left(i\omega_k(\eta_{\text{f}}) + \frac{1}{2}\frac{\omega_k^{\prime}(\eta_{\text{f}})}{\omega_k(\eta_{\text{f}})}\right).
\label{eq:sf.AdiabaticVacuum}
\end{split}
\end{equation}
In fact, it is this feature that allows us to extrapolate the results obtained at $\eta_{\text{f}}$ to the present when considering fields that interact only gravitationally \cite{Ema2018, Cembranos2020}, because after reheating the background expands more and more adiabatically, and so condition~\eqref{eq:sf.AdiabaticCondition} is always fulfilled.

From the Ricci expression at late times \eqref{eq:sf.LateRicci} we see that $R \sim 1/t^2$. Introducing this in the frequency \eqref{eq:sf.ScalarFrequency}, one finds that, as long as $mt \gg 1$, particle production will be governed by the mass term of the frequency~\eqref{eq:sf.ScalarFrequency},
\begin{equation}
    \omega_k^2(\eta) \simeq k^2 + a^2(\eta)m^2.
\end{equation}
Since the scale factor at late times behaves as $a(\eta) \sim \eta^2$, condition~\eqref{eq:sf.AdiabaticCondition} is fulfilled soon after the Ricci scalar oscillations become unimportant. For masses of the order of the inflaton, this happens at a time $\eta_{\text{f}}$ small enough that we do not need to invoke the late-time solution for the background, since everything can be calculated numerically in an efficient way. This is not the case for masses smaller than the inflaton, for which production stabilizes after many, many oscillations, given that~\mbox{$mt \gg 1$} is fulfilled at later times. As a consequence, if we want to use the adiabatic vacuum prescription, we need to go up to a very large $\eta_{\text{f}}$, and therefore we need to use the analytic approximation for the inflaton dynamics described by~\cref{eq:sf.LateReheatingInflaton}.

Alternatively, we can take a different definition for the vacuum that allows us to calculate the number density of produced particles at some $\bar{\eta}\ll\eta_{\text{f}}$, even for~\mbox{$m\ll m_{\phi}$}. Although it will be still important in terms of adiabaticity, the oscillating term in \eqref{eq:sf.ScalarFrequency} does not affect particle production at sufficiently large (numerically accessible)~$\bar{\eta}$, and therefore we can define the frequency
\begin{equation}
    \omega_k^{(\text{avg})}(\eta)=\sqrt{k^2 + a^2(\eta)\left[m^2 + \left(\xi-1/6\right)\braket{R}(\eta)\right]},
\end{equation}
where the Ricci scalar oscillations are averaged. We can take this frequency to calculate the \emph{averaged} vacuum
\begin{equation}
\begin{split}
u_k^{(\text{avg})}(\bar{\eta}) &= \frac{1}{\sqrt{2\omega_k^{(\text{avg})}(\bar{\eta})}}, \\
u_k^{(\text{avg})\,\prime}(\bar{\eta}) &= -\frac{1}{\sqrt{2\omega_k^{(\text{avg})}(\eta)}}\left(i\omega_k^{(\text{avg})}(\bar{\eta}) + \frac{1}{2}\frac{\omega_k^{(\text{avg})\,\prime}(\bar{\eta})}{\omega_k^{(\text{avg})}(\bar{\eta})}\right).
\label{eq:sf.AveragedVacuum}
\end{split}
\end{equation}
This prescription of vacuum is such that the spectrum of produced particles obtained at $\bar{\eta}$ essentially coincides with the one given by the adiabatic vacuum at the time when we reach the adiabatic regime, $\eta_{\text{f}}$, namely
\begin{equation}
n_k^{(\text{avg})}\Big|_{\eta = \bar{\eta}} \simeq n_k^{(\text{ad})}\Big|_{\eta = \eta_{\text{f}}}.
\label{eq:sf.DensityVacua}
\end{equation}
The larger discrepancies will reside in low wavenumbers, for which~\mbox{$k^2 \sim a^2(\eta)\braket{R}$}, but this region of momentum space is suppressed in the total density of produced particles by a factor~$k^2$ (see \cref{eq:sf.DensityOfParticles} in previous subsection). As a consequence, no differences are appreciated at the chosen $\bar{\eta}$. 

This procedure has a limitation: It is valid up to the smallest mass~$m$ for which the dynamics presented here remain the same until~$\eta_{\text{f}}$. If reheating ends\footnote{Recall from \cref{sec:cosmo.reheating} that the inflaton decays while oscillating, giving rise to the thermal Universe. This is unnoticed by the spectator field, which only feels the background.} before~$\eta_{\text{f}}$ for a particular mass, in principle, the result provided by the averaged vacuum is not strictly correct. However, one can argue that production after reheating will be negligible when compared to the number density of particles that have already been produced. In fact, the dynamics of the Ricci scalar will be the same after the end of reheating, $\eta_{\text{rh}}$, since radiation does not contribute to the stress-energy tensor, and the scale factor will behave as $\eta$ instead of as $\eta^2$. Therefore, the comoving spectra obtained once the adiabaticity regime is reached can be regarded the same independently of $\eta_{\text{rh}}$ being before or after $\eta_{\text{f}}$. 

On the other hand, if this mechanism aims at explaining the observed abundance of dark matter, we have to require that production ends before the time when structures start to form, around $\sim 10^{12}\, \text{s}$ after inflation. If this is not the case, the dark matter abundance observed nowadays will not correspond to the one obtained in this analysis. Nevertheless, a simple estimation using the scale factor of a radiation-dominated universe shows that masses above the order of~$m\sim10^{-30} \, \text{eV}$ would reach adiabaticity early enough (i.e., the condition $mt \gg 1$ is fulfilled before $10^{12}\,\text{s}$). This is many orders of magnitude below the mass of fuzzy cold dark matter, and hence all the interesting range of masses lie within the regime of validity of our method.

\subsection{Adiabaticity and oscillations}
\label{subsec:sf.adiabaticity}

\begin{figure}[t!]
    \centering
    \includegraphics[width=0.5\textwidth]{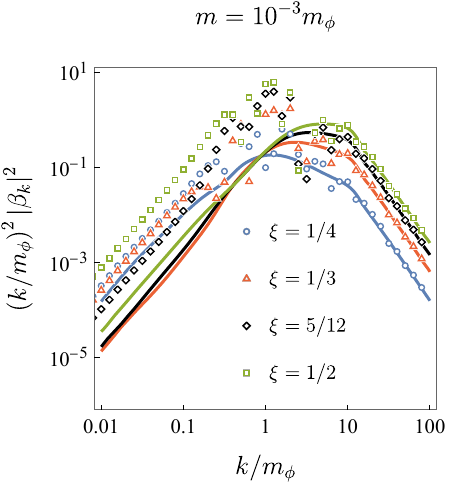}
    \caption[Stability of the adiabatic scalar spectra for mass $m=10^{-3}m_{\phi}$]{Spectra of produced particles of mass $m=10^{-3}m_{\phi}$ and different values of $\xi$, obtained with the adiabatic prescription of the vacuum. The dots correspond to $\eta = 16.33/m_{\phi}$, before the adiabatic regime has been reached for this value of the mass. The solid lines correspond to $\eta = \eta_{\text{f}} = 100/m_{\phi}$, when most of the particles have been produced. Figure from \cite{ScalarField2023}.} 
    \label{fig:sf.AdiabaticityLog}
\end{figure}

\begin{figure}[t!]
    \centering
    \includegraphics[width=0.5\textwidth]{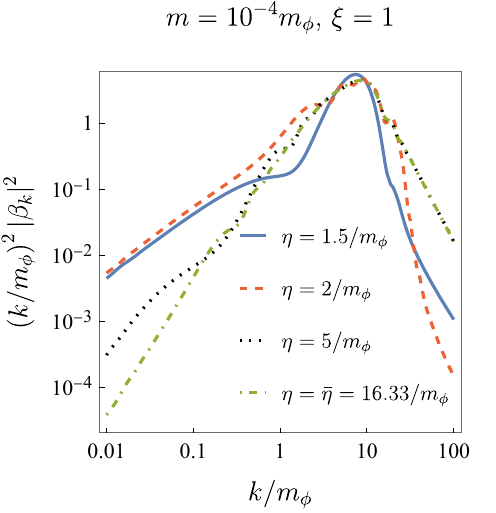}
    \caption[Scalar spectra for $m=10^{-4}m_{\phi}$ and $\xi=1$]{Spectra for $m=10^{-4}m_{\phi}$ and $\xi=1$, obtained with the averaged vacuum prescription, for different instants of time. The spectrum stabilizes after very many oscillations of the curvature scalar. Figure from \cite{ScalarField2023}.} 
    \label{fig:sf.OscillationsInfluence}
\end{figure}

In order to illustrate the importance of the choice of vacuum, we studied the evolution of spectra when calculated using prescription~\eqref{eq:sf.AdiabaticVacuum} before the dynamics has entered the adiabatic regime. As an example, we plotted in figure \ref{fig:sf.AdiabaticityLog} the spectra of particles with mass~$m = 10^{-3}m_{\phi}$ obtained at two different times. The dots correspond to~$\eta = 16.33/m_{\phi}$, whereas the solid lines denote $\eta = \eta_{\text{f}} = 100/m_{\phi}$. For this particular choice of mass, the latter time lies within the adiabatic regime, and this is the reason why the non-adiabatic dots relax to their final value as we approach this limit. As expected, the effect is less noticeable the lower the coupling to the geometry is, as it is the main source of non-adiabaticity in the frequency.

At the same time, we also characterized the importance of the first oscillations of the curvature scalar in the final spectrum of produced particles, obtained with the averaged vacuum defined in eq. \eqref{eq:sf.AveragedVacuum}. As can be seen in figure \ref{fig:sf.OscillationsInfluence}, even after several oscillations of~$R(\eta)$ (for example, at $\eta = 2/m_{\phi}$), the production changes greatly if one compares with the obtained spectra at $\bar{\eta}$. Even when looking only at the total number density of produced particles in eq.~\eqref{eq:qftcs.MeanNumberDensity}, differences are still significant. We observe that the spectrum does not stabilize until~\mbox{$\eta \simeq 5/m_{\phi}$}, which for our model means after hundreds of oscillations of the curvature scalar $R(\eta)$. With this, we want to stress that obtaining the particle production after one or two oscillations does not account for the whole process.

\section{Vector mode equation and vacuum choice}
\label{sec:sf.vectormodes}

Let us now apply the same logic and background to the massive vector field described in subsection \ref{subsec:sf.vectorfield}. The transverse modes behave similarly to a scalar field, and thus we closely follow the analysis done previously in section \ref{sec:sf.scalarmodes}, where now we have an additional coupling to take into account. On the other hand, the longitudinal modes behave somewhat differently, and the corresponding mode equation \eqref{subsec:sf.longitudinalmodeeq} becomes more involved. However, we will see that most of the ideas discussed for scalars can be completely extrapolated here.

\subsection{Transverse modes}
\label{subsec:sf.transversemodeeq}

Similarly to what has been done in section \ref{sec:sf.scalarmodes}, we calculate the number density of particles \eqref{eq:qftcs.MeanNumberDensity} for the transverse modes by comparing at the same time the two solutions corresponding to the (Bunch-Davies) vacuum at the beginning of inflation and to the one at the time when the process is over, for which we take the adiabatic vacuum prescription defined in \cref{eq:sf.AdiabaticVacuum}. As before, making this evaluation requires solving equation \eqref{eq:sf.TransverseModeEquation} from the beginning of inflation, $\eta_{\text{i}}$, until the time for which the adiabatic regime is reached, $\eta_{\text{f}}$. Note that the background evolution stays the same, given by the inflaton equation of motion \eqref{eq:cosmo.InflatonEoM}. Therefore, we again make use of the slow-roll approximation derived in subsection \ref{subsec:sf.approximations}, which for the transverse mode reads\footnote{Note that we have directly introduced the full, numerical frequency $\omega_{k,\text{T}}$ instead of the slow-roll frequency $\omega_{k,\text{T}}^{\text{SR}}$ in the definition of $\tau_k(\eta)$. Nevertheless, subsection \ref{subsec:sf.approximations} follows similarly by making this substitution.}
\begin{equation}
\begin{split}
    v_{k, \text{T}}^{\text{SR}}(\eta) &\simeq \sqrt{\pi\tau_k(\eta)/2} e^{i\pi\nu/2} H_{\nu}^{(1)}\left(k\tau_k(\eta)\right), \\ 
\tau_k(\eta)&= \Bigg|\frac{\omega_{k,\text{T}}(\eta)}{\omega_{k,\text{T}}^{\text{dS}}(\eta)}(\eta-\eta_*) + \eta_*-\eta_0\Bigg|.
\label{eq:sf.ApproximateTSolution}
\end{split}
\end{equation}
Recall that the approximate solution \eqref{eq:sf.ApproximateTSolution} is compatible with Bunch-Davies initial conditions and valid until~$\eta_*$, whose particular value depends on the wavenumber~$k$, the mass of the field $m$, and the couplings $\gamma$ and $\sigma$ (see subsection~\ref{subsec:sf.approximations}). Here, $\omega_{k,\text{T}}^{\text{dS}}$ is the frequency in a de Sitter geometry\footnote{Let us recall that the de Sitter frequency used in the approximation is shifted with respect to the one obtained by taking the limit $\omega_{k,\text{T}}(\eta \to \eta_{\text{i}})$}, 
\begin{equation}
    \omega_{k,\text{T}}^{\text{dS}}(\eta)^2 = k^2 + \frac{\mu^2}{(\eta-\eta_0)^2}, \qquad \text{with} \qquad \mu^2 = m^2/H_0^2 + 12\gamma,
\end{equation}
where $H_0$ and $\nu$ are defined as for the scalar field. Note that if we write~\mbox{$\gamma=\xi-1/6$} we recover the equations for the case of gravitational production of scalar particles in the same background, with~$\xi$ being the coupling to $R$. Thus, the initial vacuum state becomes unstable for~\mbox{$\gamma<0$} in the case of transverse mode production, being $\gamma=0$ the conformal point (equivalent to $\xi=1/6$ in the scalar field case). Equation~\eqref{eq:sf.TransverseModeEquation} is solved in the two regions,
\begin{equation}
\eta = \begin{cases}
\eta_{\text{i}} \leq \eta \leq \eta_*, \quad \text{slow-roll approximation},\\
\eta_*\leq \eta \leq \eta_{\text{f}}, \quad \text{numerical solution},
\end{cases}
\end{equation}
similar to what we did with the scalar field, and where numerical computation is needed once the slow-roll approximation described above is no longer valid (that is, for $\eta > \eta_*$). Let us remark that the use of this analytical solution is crucial for alleviating the numerical workload, especially given that the space of parameters is now three-dimensional, in contrast to the non-minimally coupled scalar case.

Then, it remains only to specify the vacuum of the observer living at $\eta_{\text{f}}$, for which we use again the customary zeroth-order adiabatic prescription \eqref{eq:sf.AdiabaticVacuum}, which for the transverse modes takes the form
\begin{equation}
\begin{split}
      u_{k,\text{T}}(\eta_{\text{f}})&=\frac{1}{\sqrt{2\omega_{k,\text{T}}(\eta_{\text{f}})}}, \\
      u^{\prime}_{k,\text{T}}(\eta_{\text{f}})&=-\frac{1}{\sqrt{2\omega_{k,\text{T}}(\eta_{\text{f}})}}\left(i\omega_{k,\text{T}}(\eta_{\text{f}}) + \frac{1}{2}\frac{\omega_{k,\text{T}}^{\prime}(\eta_{\text{f}})}{\omega_{k,\text{T}}(\eta_{\text{f}})}\right).
\end{split}
\end{equation}
This prescription is a good notion of vacuum as long as the dynamics are adiabatic at~$\eta_{\text{f}}$, which is reached at a different point in time depending on the parameters~\mbox{$k, m, \gamma$} and $\sigma$, and therefore one has to choose an end point $\eta_{\text{f}}$ that fulfills the adiabaticity condition \eqref{eq:sf.AdiabaticCondition} for all the parameter space considered. 

\begin{figure}[t!]
    \centering
    \includegraphics[width=0.9\textwidth]{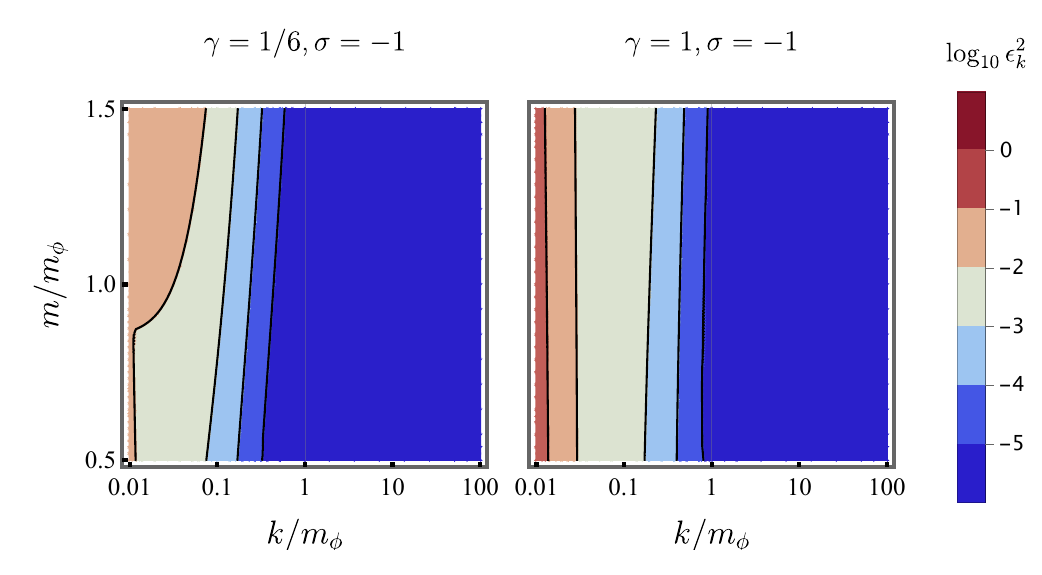}
    \caption[Slow-roll approximation error squared for the transverse modes]{Slow-roll approximation error squared $\epsilon_k^2$ for the transverse modes as a function of the wavenumber $k$ and the field mass $m$, for $\gamma=1/6$ (left) and $\gamma=1$ (right), and $\sigma=-1$, for~\mbox{$\eta_*=-500m_{\phi}$}. The behavior for $\sigma=0$ is qualitatively similar. Figure from~\cite{VectorDM2024}.}
    \label{fig:sf.errorT}
    \end{figure}

With this, we can calculate the number of produced particles, 
\begin{equation}
\begin{split}
   n_{\text{T}}(m, \gamma, \sigma) &= \int \frac{\dd k}{2\pi^2} k^2\abs{\beta_{k,\text{T}}}^2(m, \gamma, \sigma) \\
   &= n_{\text{T}}^{\text{SR}} + \int_0^{\infty}\frac{\dd k}{2\pi^2}k^2\abs{\beta_{k,\text{T}}^{\text{SR}}}^{2}\mathcal{O}(\epsilon_k^2),
   \label{eq:sf.particleproductionT}
\end{split}   
\end{equation}
with $n_{\text{T}}^{\text{SR}} = \int_0^{\infty}\frac{\dd k}{2\pi^2}k^2\abs{\beta_{k,\text{T}}^{\text{SR}}}^2$ being the result obtained when using the approximation~$v_{k,\text{T}}^{\text{SR}}$. The error in \eqref{eq:sf.particleproductionT} due to the use of the approximate solution \eqref{eq:sf.ApproximateTSolution} is shown in figure \ref{fig:sf.errorT} for $\eta_*=-500m_{\phi}$ on specific but representative values of the parameters for illustrative purposes. We note that $\epsilon_k^2$ is essentially independent of the value of the mass $m$ within this range, and is small for $k>m_{\phi}$. For wavenumbers smaller than the mass of the inflaton, the error increases. However, the density of produced particles is suppressed in this range by the $k^2$ factor in~\eqref{eq:sf.particleproductionT}, and therefore $n_{\text{T}}(m, \gamma, \sigma) \approx n_{\text{T,SR}}$ is a good approximation.

\subsection{Longitudinal mode equation}
\label{subsec:sf.longitudinalmodeeq}

Production of longitudinal modes is more involved than that of transverse modes or a scalar field since the form of the frequency \eqref{eq:sf.LongitudinalFrequency} is qualitatively different. However, in the limit when we approach the beginning of inflation, it can be written as a de Sitter-like frequency, which we will use to develop a slow-roll approximation similar to~\eqref{eq:sf.ApproximateTSolution}.

Let us again use the fact that the geometry approaches de Sitter at the onset of inflation, as we did in subsection \ref{subsec:sf.transversemodeeq}. In this limit, \mbox{$M^{\text{LL}}=-M^{00}=\mu^2H_0^2$}. The longitudinal frequency \eqref{eq:sf.LongitudinalFrequency} becomes
\begin{equation}
    \omega_{k,\text{L}}^2(\eta) = k^2 + \frac{\mu^2}{\left(\eta-\tilde{\eta}\right)^2} - \frac{k^2\left[2\left(\eta-\tilde{\eta}\right)^2k^2 - \mu^2\right]}{\left[\left(\eta-\tilde{\eta}\right)^2k^2+\mu^2\right]^2}.
\label{eq:sf.LFrequencyDeSitter}
\end{equation}
Note that $\omega_{k, \text{L}}^2(\eta)/k^2$ only depends on $k\eta$ and $\mu$. If we expand it in powers of $\mu/(k\eta)$ for fixed $\mu$ to second order we obtain
\begin{equation}
    \omega_{k,\text{L}}^2(\eta) \simeq k^2 + \frac{\mu^2-2}{\left(\eta-\tilde{\eta}\right)^2},
\label{eq:sf.LFrequencyApproximatedDeSitter}
\end{equation}
which is a good approximation for $\eta \to \eta_{\text{i}}$. Note that in this limit, the conformal case is, similarly to the scalar field, $\gamma=1/6$, value below which the initial vacuum becomes unstable. This is precisely the reason why we required $\gamma \geq 1/6$.

\begin{figure}[t!]
    \centering
    \includegraphics[width=0.9\textwidth]{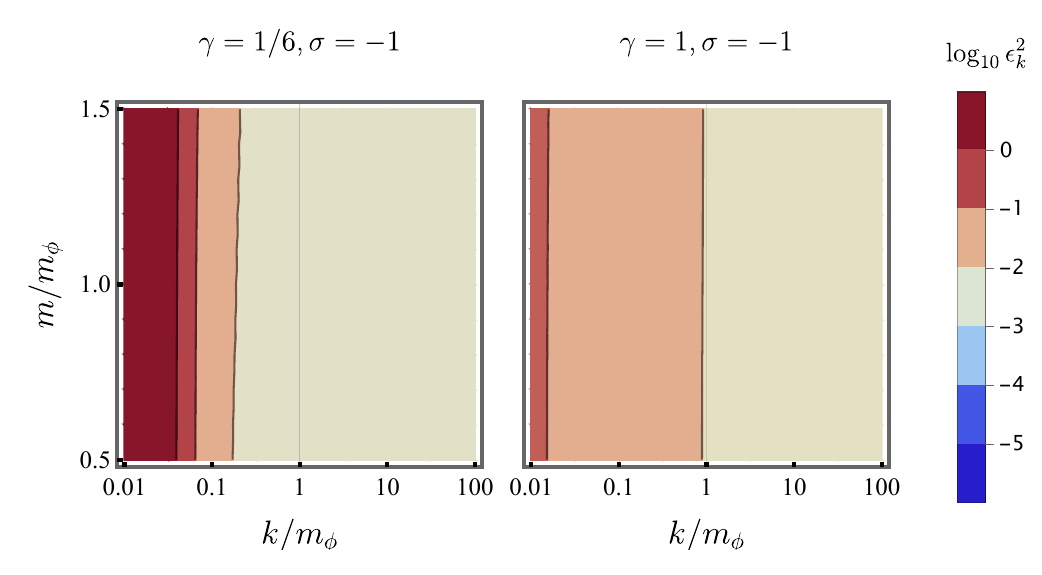}
    \caption[Slow-roll approximation error squared for the longitudinal modes]{Slow-roll approximation error squared $\epsilon_k^2$ for the longitudinal modes as a function of the wavenumber $k$ and the field mass $m$, for $\gamma=1/6$ (left) and~\mbox{$\gamma=1$} (right), and $\sigma=-1$, with $\eta_*$ fixed at $-500m_{\phi}$. The behavior for $\sigma=0$ is qualitatively similar. Figure from \cite{VectorDM2024}.}
    \label{fig:sf.errorL}
\end{figure}

Since the early-time behavior is the same as for the transverse modes, we can write an approximation to the solution during slow-roll for the longitudinal mode equation as
\begin{equation}
\begin{split}
    v_{k,\text{L}}^{\text{SR}}(\eta) &\simeq \sqrt{\pi\tau(\eta, k)/2} e^{i\pi\nu/2} H_{\nu}^{(1)}\left(k\tau_k(\eta)\right),\\
\tau_k(\eta)&= \Bigg| \frac{\omega_{k,\text{L}}(\eta)}{\omega_{k,\text{L}}^{\text{dS}}(\eta)}(\eta-\eta_*) + \eta_*-\eta_0\Bigg |,
\label{eq:sf.ApproximateLSolution}
\end{split}
\end{equation}
where the de Sitter frequency is now given by
\begin{equation}
    \omega_{k,\text{L}}^{\text{dS}}(\eta)^2 = k^2 + \frac{\mu^2-2}{(\eta-\eta_0)^2}, \qquad \text{with} \qquad \mu^2 = m^2/H_0^2 + 12\gamma.
\end{equation}
From this point on, the analysis follows closely what has been discussed in subsection~\ref{subsec:sf.transversemodeeq}. In particular, the error in the number density of produced longitudinal modes due to the use of~\eqref{eq:sf.ApproximateLSolution} is shown in figure~\ref{fig:sf.errorL} for the same value~\mbox{$\eta_*=-500m_{\phi}$}. Note that for these modes, the slow-roll approximation is slightly worse, although the behavior is qualitatively similar to that of the transverse modes. However, the error becomes important again for low values of $k$, which are suppressed in the density, and the integral of the spectra is not significantly affected. 

\section{Relic abundance of dark matter}
\label{sec:sf.abundance}

Let us finally give the results for the spectra of produced particles as a function of the parameters of the field: the mass $m$ and the couplings to the curvature $\xi$ in the case of the scalar field, and $\gamma$ and $\sigma$ in the case of the vector field. Most of the results correspond to the quadratic inflationary potential, but we also present some preliminary results on the Starobinsky potential for the case of the scalar field in subsection \ref{subsec:sf.starobinskyabundance}, which will appear in the upcoming work \cite{Starobinsky2024}.

\subsection{Scalar spectra and abundance of dark matter: quadratic potential}
\label{subsec:sf.quadraticabundance}

\begin{figure}[t!]
    \centering
    \includegraphics[width=0.5\textwidth]{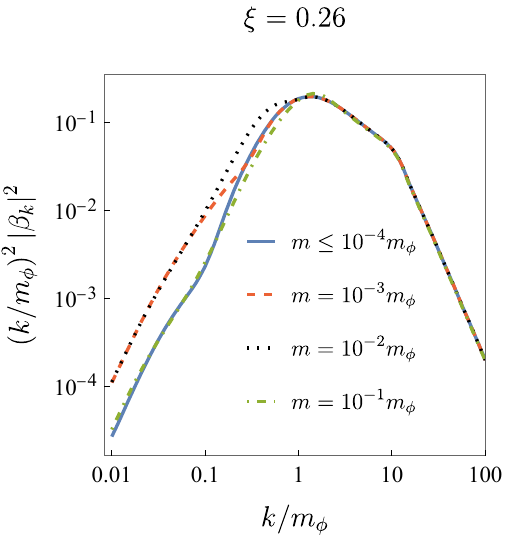}
    \caption[Scalar spectra for masses below the mass of the inflaton and~\mbox{$\xi=0.26$}]{Spectrum of particles for masses below the mass of the inflaton, with $\xi=0.26$. For very small masses ($m\leq 10^{-4}m_{\phi}$), production is dominated by curvature. In the region $10^{-3}\leq m \leq 10^{-1}m_{\phi}$, differences in production due to the mass can be noticed, especially for low values of $k/m_{\phi} \simeq 0.1-1$. Figure from \cite{ScalarField2023}.} 
    \label{fig:sf.SpectrumMassesBelow}
\end{figure}

In this case, calculations have been performed using the averaged vacuum prescription at $\bar{\eta}=16.33/m_{\phi}$. 

\begin{figure}[t!]
    \centering
    \includegraphics[width=0.9\textwidth]{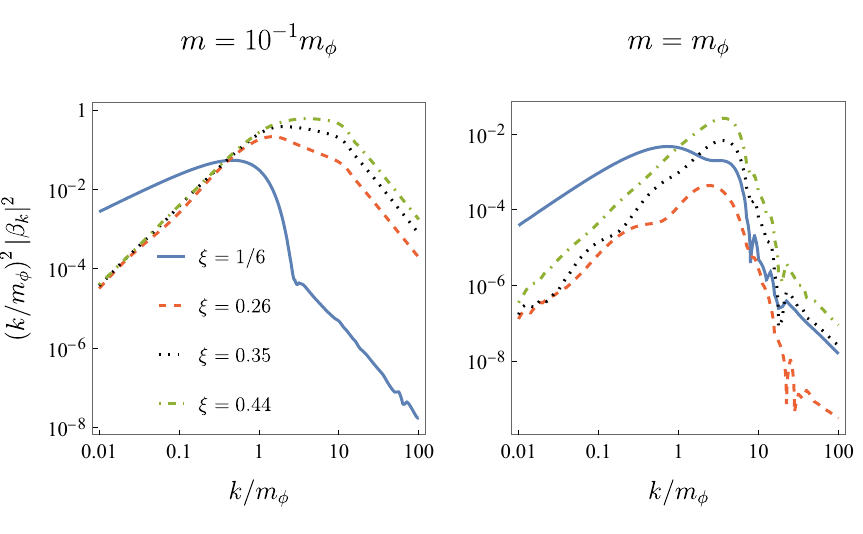}
    \caption[Scalar spectra for particles with $m=10^{-1}m_{\phi}$ and $m = m_{\phi}$, for different values of the coupling]{Spectra of particles with $m=10^{-1}m_{\phi}$ (left) and $m = m_{\phi}$ (right), for different values of the coupling. In general, particle production increases when the curvature term becomes more important, and the maximum of the spectrum is shifted toward higher values of $k$. However, when $m=m_{\phi}$, increasing the coupling does not translate directly into an increase of particle production (compare $\xi=1/6$ and $\xi=0.26$ in the right panel). This can be more clearly seen by examining the total density of particles (see \cref{fig:sf.TotalDensity}). Figure from \cite{ScalarField2023}.} 
    \label{fig:sf.Spectrum01AndInflatonMass}
\end{figure}

We explore first the regime of masses below the inflaton mass. Represented by the solid line in figure \ref{fig:sf.SpectrumMassesBelow}, we have masses~\mbox{$m\leq 10^{-4}m_{\phi}$}. For these values, the mass contribution to the frequency becomes negligible, and the dynamics is entirely given by the coupling to the geometry. As a consequence, the corresponding spectra (blue line in \cref{fig:sf.SpectrumMassesBelow}) lie on top of each other. We do observe larger differences in the shape of the spectrum when increasing the mass, especially for small wavenumbers, as the rest of the curves in figure~\ref{fig:sf.SpectrumMassesBelow} show. We can choose a mass in this regime,~\mbox{$m=10^{-1}m_{\phi}$}, and explore the influence of the coupling $\xi$ in the final result. This is shown in the left panel of figure \ref{fig:sf.Spectrum01AndInflatonMass}, where one observes increasing production of particles with larger values of the coupling.

Lastly, let us come to the scenarios in which the mass of the dark matter field coincides with that of the inflaton. The corresponding spectra are shown in the right panel of figure \ref{fig:sf.Spectrum01AndInflatonMass}. In such cases, it is harder to characterize the behavior with $\xi$. It is clear, nevertheless, that particle production decreases as the mass of particles becomes larger.

\begin{figure}[t!]
    \centering
    \includegraphics[width=\textwidth]{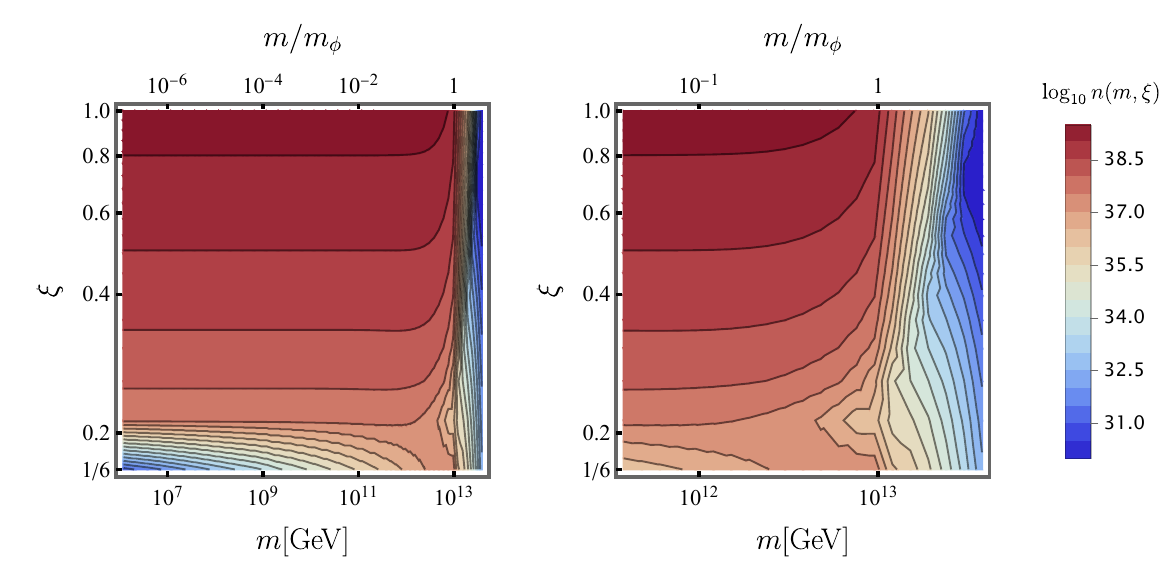}
    \caption[Total scalar density of produced particles]{Logarithm of the total density of produced particles for different values of~$m$ and $\xi$. In order to give the mass and density in units of GeV, we took~\mbox{$m_{\phi}=1.2 \times 10^{13} \, \text{GeV}$} for the mass of the inflaton. We explore a wide range of masses in the left panel while we focus on a smaller region close to the mass of the inflaton on the right panel in order to appreciate the dependence of the total density with the coupling $\xi$. Figure from \cite{ScalarField2023}.} 
    \label{fig:sf.TotalDensity}
\end{figure}

It is easier to characterize particle production in this regime using the total number density of particles \eqref{eq:qftcs.MeanNumberDensity}, which we show in figure~\ref{fig:sf.TotalDensity} as a function of the two parameters of the field, $m$ and~$\xi$. Here, one clearly sees that the prediction is independent of the value of the mass as long as it is below $m \sim 10^{-2}m_{\phi}$, in particular for a sufficiently high value of the coupling, $\xi \gtrsim 0.2$. In this case, the mass is completely negligible when compared to the dynamics of the curvature scalar. Only when the coupling to the curvature is close to $\xi \sim 1/6$, the production of particles is still sensitive to $m$, up to~\mbox{$m \sim 10^{-7}m_{\phi}$}. Below this value, the relevant wavenumbers~\mbox{$k \sim a m \ll m_{\phi}$} are very small, and particle production is therefore suppressed (cf. \cref{eq:sf.DensityOfParticles}). In all this regime of low masses, the number of produced particles increases with larger coupling $\xi$. Closer to the mass of the inflaton,~\mbox{$10^{-2}m_{\phi}<m<m_{\phi}$}, the fact that a heavier particle translates into a lower production becomes apparent. Lastly, in the region around the mass of the inflaton, $m\sim m_{\phi}$, the behavior with the coupling is different, and production may even decrease when raising the value of $\xi$. In fact, there appears to exist a critical value $\xi_{\text{c}} \simeq 0.22$ which separates two qualitatively different regimes. 
As we commented previously, this value is related to the parameter $\mu^2=1/4$ of the Hankel functions, which were a good approximation of the mode functions of our problem. For $m<m_{\phi}$, the number density drops very rapidly if $\xi<\xi_{\text{c}}$. For~\mbox{$m\sim m_{\phi}$},~$\xi_{\text{c}}$ is the value below which production decreases with $\xi$, and above which it increases. This is also illustrated in figure~\ref{fig:sf.Spectrum01AndInflatonMass}, where production for $\xi=1/6$ is larger than for~\mbox{$\xi=0.26$}, and from there it increases again with the coupling. Moreover, we observe the expected strong suppression in the number density of produced particles for masses above the mass of the inflaton. We can confirm this behavior by calculating the spectra for even higher masses, provided we select a negative enough~$\eta_*$---and therefore leading to a longer computation---in this case, as explained in subsection \ref{subsec:sf.approximations}. Note that we took~\mbox{$m_{\phi}=1.2 \times 10^{13} \, \text{GeV}$} for the mass of the inflaton, and as a consequence, the density in figure \ref{fig:sf.TotalDensity} is given in units of $\text{GeV}^{3}$.

Finally, one can consider these gravitationally produced scalar particles as dark matter. In this case, it is  necessary to compare the resulting abundance with observations. The physical density of produced particles is related to the comoving density shown in figure \ref{fig:sf.TotalDensity} only by the scale factor. Assuming that the scalar field is non-interacting, which is mandatory for the gravitational production to be important, as it cannot reach thermal equilibrium, the evolution of the density of created particles from $\eta_{\text{rh}}$ until today will be dictated solely by the dilution due to the isentropic expansion of the background. The predicted abundance can be written in terms of the background radiation temperature \cite{Cembranos2020} as
\begin{equation}
    \Omega (m, \xi) = \frac{8\pi}{3\mpl^2H^2_{\text{today}}}\frac{g_{*S}^{\text{today}}}{g_{*S}^{\text{rh}}}\left(\frac{T_{\text{today}}}{T_{\text{rh}}}\right)^3 m \, \frac{n(m, \xi)}{a_{\text{rh}}^3},
\label{eq:sf.Abundance}
\end{equation}
where $T_{\text{today}}$ and $T_{\text{rh}}$ are the radiation temperature today and at the end of reheating, respectively, and $g^{\text{today}}_{*S}$ and $g^{\text{rh}}_{*S}$ are the corresponding relativistic degrees of freedom (see \cref{sec:cosmo.thermodynamics}). The scale factor at the end of reheating~$a_{\text{rh}}$ is obtained using that, when radiation dominates, at~$\eta_{\text{rh}}$, the Hubble rate can be written as
\begin{equation}
    H^2_{\text{rh}} = \frac{8\pi}{3\mpl^2}\frac{\pi^2}{30}g_{*}^{\text{rh}}T_{\text{rh}}^4,
\end{equation}
which allows one to obtain $\eta_{\text{rh}}$ as a function of the reheating temperature. The equation above results from simply inserting the radiation density, given by \cref{eq:cosmo.DensityBath} into the Friedmann equation \eqref{eq:cosmo.FriedmannEquation}. This sets an upper limit on the reheating temperature, since $\bar{\eta} < \eta_{\text{rh}}$ only for~\mbox{$T_{\text{rh}} \lesssim 10^{13} \, \text{GeV}$}. Let us remark that for the region of parameter space considered, the comoving energy density of the spectator field, namely~\mbox{$mn$}, is many orders of magnitude lower than that of the inflaton, and therefore neglecting backreaction is a well-justified assumption.

\begin{figure}[t!]
    \centering
    \includegraphics[width=\textwidth]{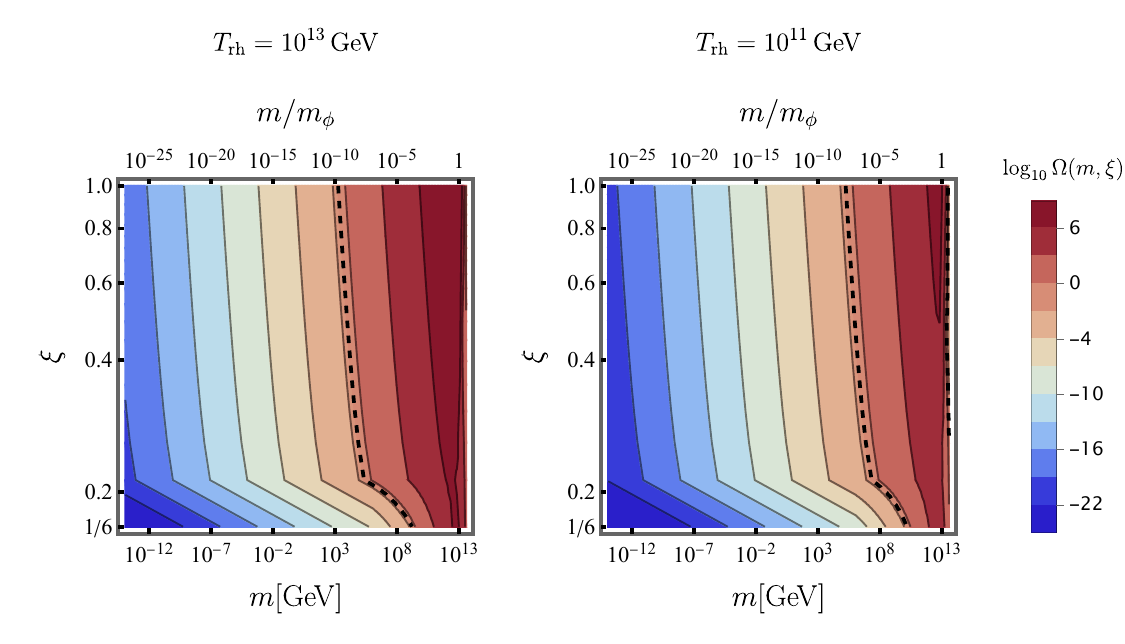}
    \caption[Abundance of produced scalar particles]{Logarithm of the predicted abundance of dark matter today for different values of $m$ and~$\xi$, and a reheating temperature of~\mbox{$T_{\text{rh}}=10^{13} \, \text{GeV}$} (left) and~\mbox{$T_{\text{rh}}=10^{11} \, \text{GeV}$} (right). The observed abundance corresponds to the dashed line. In order to give the mass and density in units of GeV, we took~\mbox{$m_{\phi}=1.2 \times 10^{13} \, \text{GeV}$} for the mass of the inflaton. Figure from~\cite{ScalarField2023}.} 
    \label{fig:sf.Abundance}
\end{figure}

The abundance is represented in figure \ref{fig:sf.Abundance} for different reheating temperatures, together with the observed dark matter abundance, given by the dashed line. We observe that the proposed mechanism can explain observations if the dark matter candidate is light enough ($m \lesssim 10^8 \, \text{GeV}$ for $T_{\text{rh}}=10^{13}\, \text{GeV}$), independently of the value of the coupling $\xi$ for the range that we considered. In addition, heavier particles can also reach the observed dark matter abundance since their production is strongly suppressed above the inflaton mass. 

\subsection{Vector spectra and abundance of dark matter: quadratic potential}

Let us now present the results for the vector field, which in this case have been obtained using the adiabatic vacuum prescription, since the range of masses considered due to the instability of the initial vacuum is such that the adiabatic regime is reached at numerically feasible times $\eta_{\text{f}}$.

\subsubsection*{Spectra of produced transverse modes}
\label{subsubsec:transversalspectra}

\begin{figure}[t!]
\centering
\includegraphics[width=0.9\textwidth]{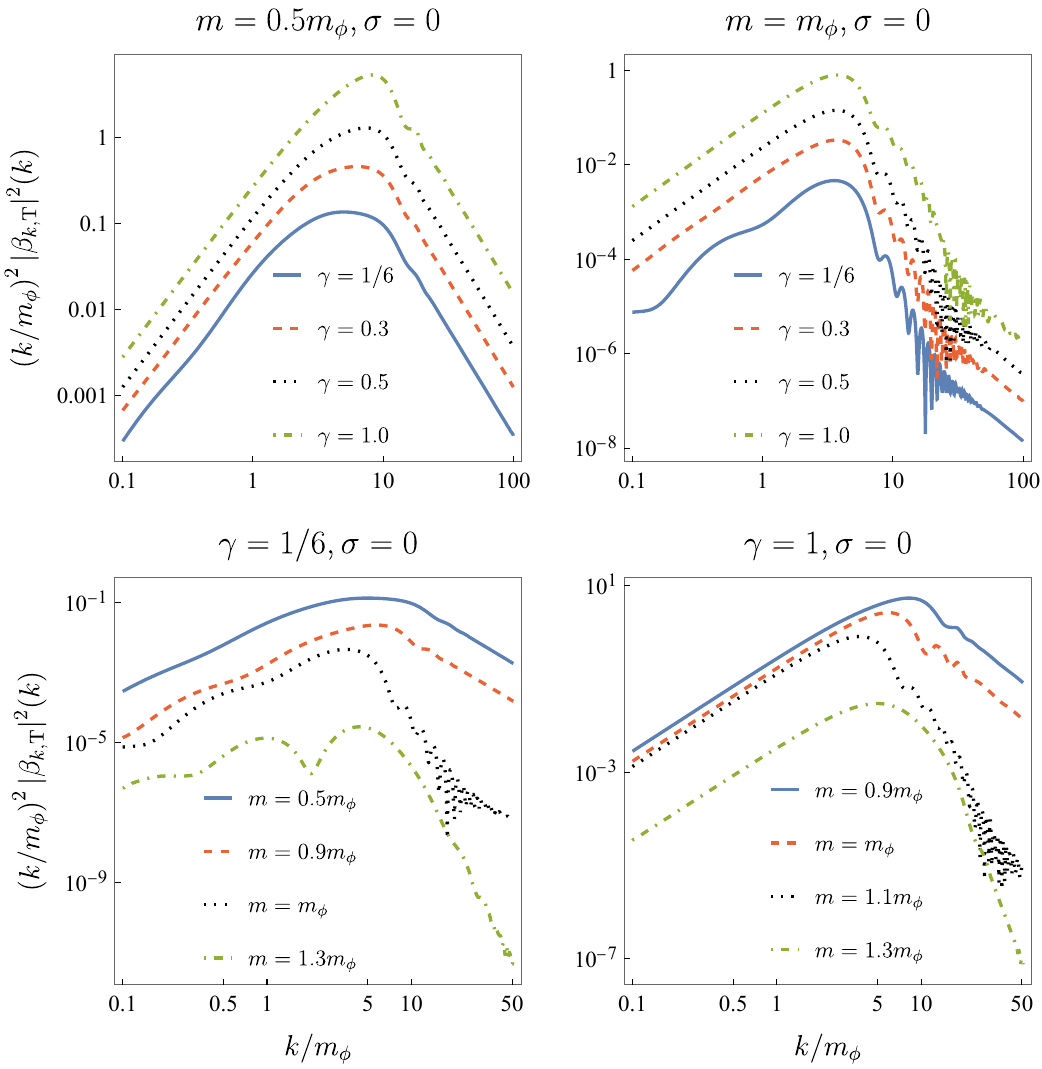}
\caption[Transverse spectra for vanishing $\sigma$]{Spectra of produced transverse modes, in logarithmic scale, for a vanishing coupling $\sigma$. The upper panels show the spectra for several values of $\gamma$, given $m=0.5m_{\phi}$ (left) or $m=m_{\phi}$ (right). On the other hand, the lower panels show the spectra for several values of $m$, given $\gamma=1/6$ (left) or $\gamma=1$ (right). Figure from \cite{VectorDM2024}.}
\label{fig:sf.SpectraT1}
\end{figure}

As for the scalar field, we focus on the study of $k^2 \abs{\beta_{k,\text{T}}}^2$ as it is the interesting quantity for particle density production. 

We first consider a vanishing value of the coupling $\sigma$, that is to say, no coupling to the traceless Ricci tensor. The resulting spectra can be found in figure \ref{fig:sf.SpectraT1}, where the top panels correspond to a fixed value of $m$, whereas the bottom panels are for a fixed value of $\gamma$. A logarithmic scale was used so that the ultraviolet (i.e. $km_{\phi}\gg 1$) and infrared (i.e. $km_{\phi}\ll 1$) behavior can be observed. Note that increasing the mass of the dark matter particle leads to less production in all scales. In particular, spectra for $m=m_{\phi}$ feature oscillations for large values of $k$, and fall off much faster. This falling is enhanced when the mass goes beyond this value. On the other hand, increasing the value of $\gamma$ strengthens the coupling to the geometry, and therefore particle production increases. Interestingly, the maximum of the spectra depends on both the value of~$m$ and $\gamma$.

\begin{figure}[t!]
\centering
\includegraphics[width=0.9\textwidth]{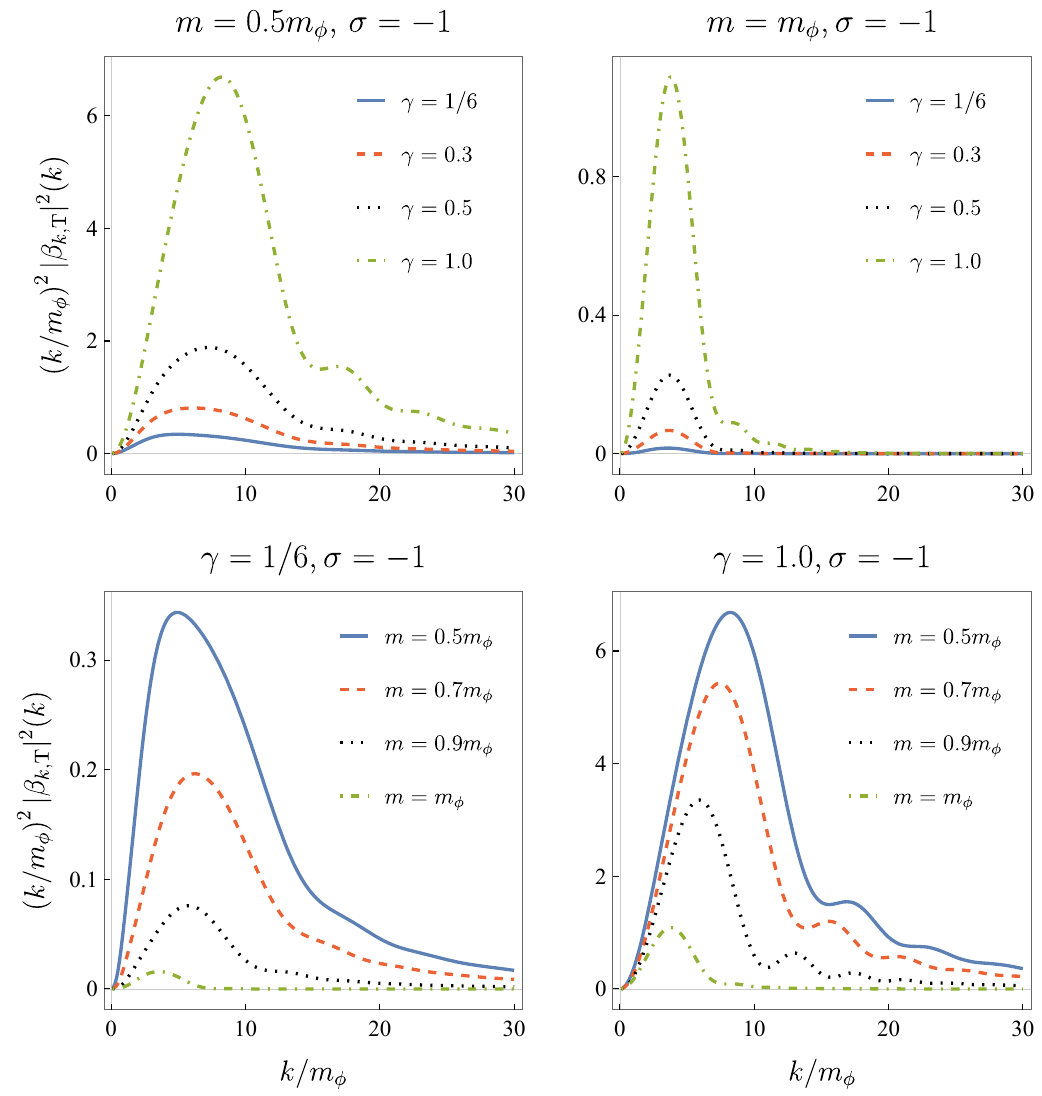}
\caption[Transverse spectra for $\sigma=-1$]{Spectra of produced transverse modes, in linear scale, for a fixed value $\sigma=-1$. The upper panels show the spectra for several values of $\gamma$, given $m=0.5m_{\phi}$ (left) or $m=m_{\phi}$ (right). On the other hand, the lower panels show the spectra for several values of $m$, given $\gamma=1/6$ (left) or $\gamma=1$ (right). Figure from \cite{VectorDM2024}.}
\label{fig:sf.SpectraT2}
\end{figure}

Let us consider now a negative value of $\sigma$. In particular, resulting spectra for~\mbox{$\sigma=-1$} are depicted in figure \ref{fig:sf.SpectraT2}. Note that a linear scale is used in this case for illustrating the magnitude of $n_{\text{T}}$. The behavior is similar to the case $\sigma=0$, but particle production is enhanced by the extra term in the frequency  \eqref{eq:sf.TransverseFrequency}. Indeed, a negative value of $\sigma$ implies that the traceless Ricci tensor term in the frequency always contributes as a negative term (see figure \ref{fig:sf.InflatonAndRicciConformal}). This means that it causes tachyonic instabilities as well as enhances those coming from the oscillations of the curvature scalar around $0$, which are precisely the main source of non-adiabaticity, as analyzed in \cite{Ema2016,Markkanen2017a,Ema2018,Cembranos2020, ScalarField2023}. Similarly, one also finds that a positive value of~$\sigma$ yields less production, since the corresponding term always contributes against tachyonic instabilities. Note that this is not in contradiction with the requirement~\mbox{$M^{00}<0$}. A large, positive value of $\sigma$ has to be compensated with the values of~$m$ and $\gamma$. However, once within the validity of the theory, a negative value of $\sigma$ will contribute toward the frequency becoming imaginary.

\subsubsection*{Spectra of produced longitudinal modes}
\label{subsec:longitudinalspectra}

Let us now show the particle production spectra for longitudinal modes, according to the background dynamics discussed in section~\ref{sec:sf.chaoticinf}. As before, we analyze the quantity $k^2 \abs{\beta_{k,\text{T}}}^2$ for different values of the parameters. 

\begin{figure}[t!]
    \centering
    \includegraphics[width=0.9\textwidth]{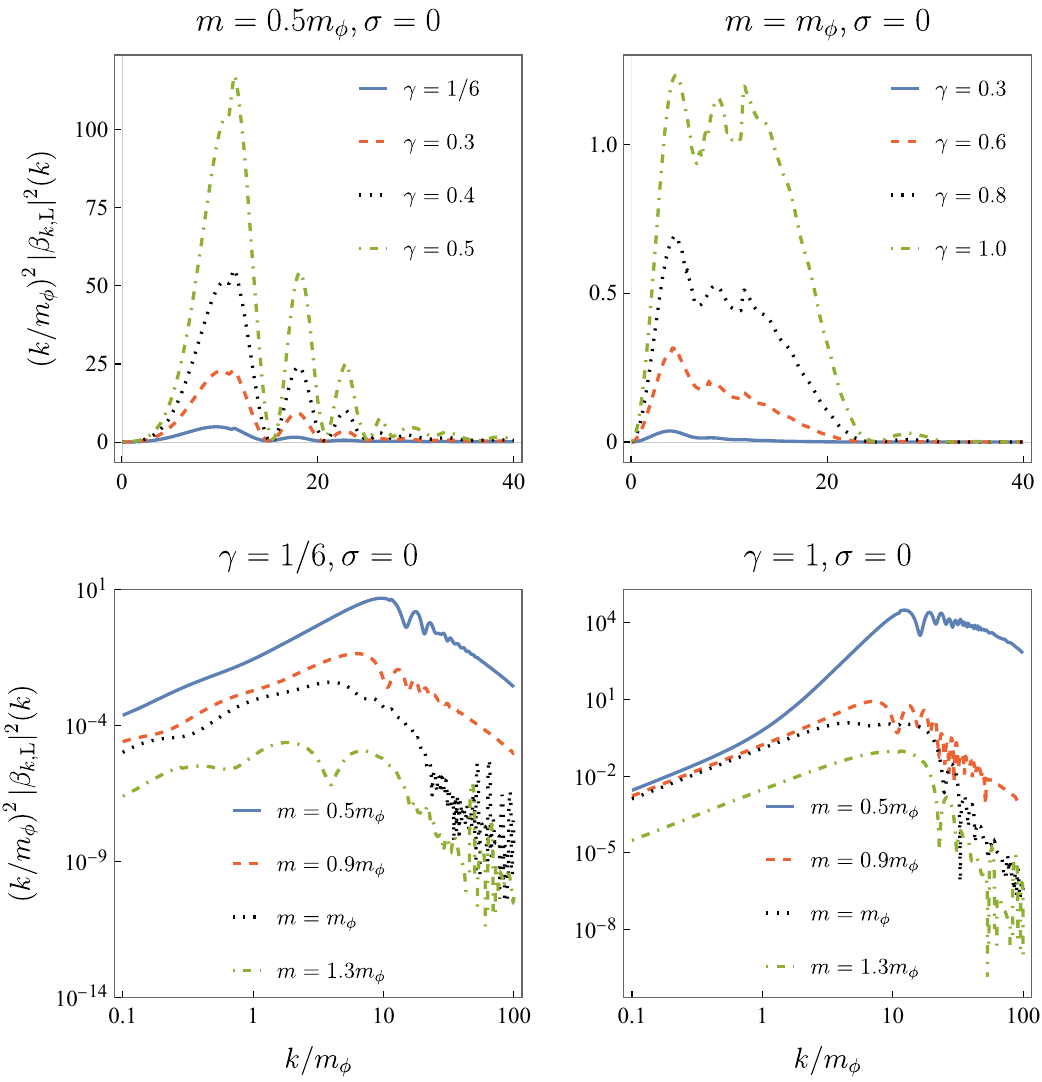}
    \caption[Longitudinal spectra for vanishing $\sigma$]{Spectra of produced longitudinal modes for a vanishing coupling $\sigma$. The upper panels, in linear scale, show the spectra for several values of $\gamma$, given~\mbox{$m=0.5m_{\phi}$} (left) or $m=m_{\phi}$ (right). On the other hand, the lower panels, in log scale, show the spectra for several values of $m$, given $\gamma=1/6$ (left) or~\mbox{$\gamma=1$} (right). Figure from \cite{VectorDM2024}.}
    \label{fig:sf.SpectraL1}
\end{figure}

Because the shape of the frequency is different, the spectra of produced particles differ qualitatively from those obtained in the case of transverse modes. Let us analyze the same cases in the same order. We first concentrate in the situation in which the $\sigma$ coupling vanishes, represented in figure \ref{fig:sf.SpectraL1}. Contrary to the transverse case, all spectra feature significant oscillations in $k$, not only for $m=m_{\phi}$. As before, for fixed mass (top panels), particle production increases with the coupling $\gamma$, whereas it increases when decreasing the dark matter mass (bottom panels) given a fixed value of $\gamma$. Crucially, spectra for~\mbox{$m=m_{\phi}$} fall off faster, as it was observed before. Note that, in general, production of particles is much more important in the longitudinal modes than in the transverse ones: Spectra are broader and maxima can be several orders of magnitude larger. These effects are particularly noticeable for small masses and large values of the coupling $\gamma$, i.e. in scenarios in which production is enhanced (see top left and bottom right panels of figure \ref{fig:sf.SpectraL1}).

\begin{figure}[t!]
    \centering
    \includegraphics[width=0.9\textwidth]{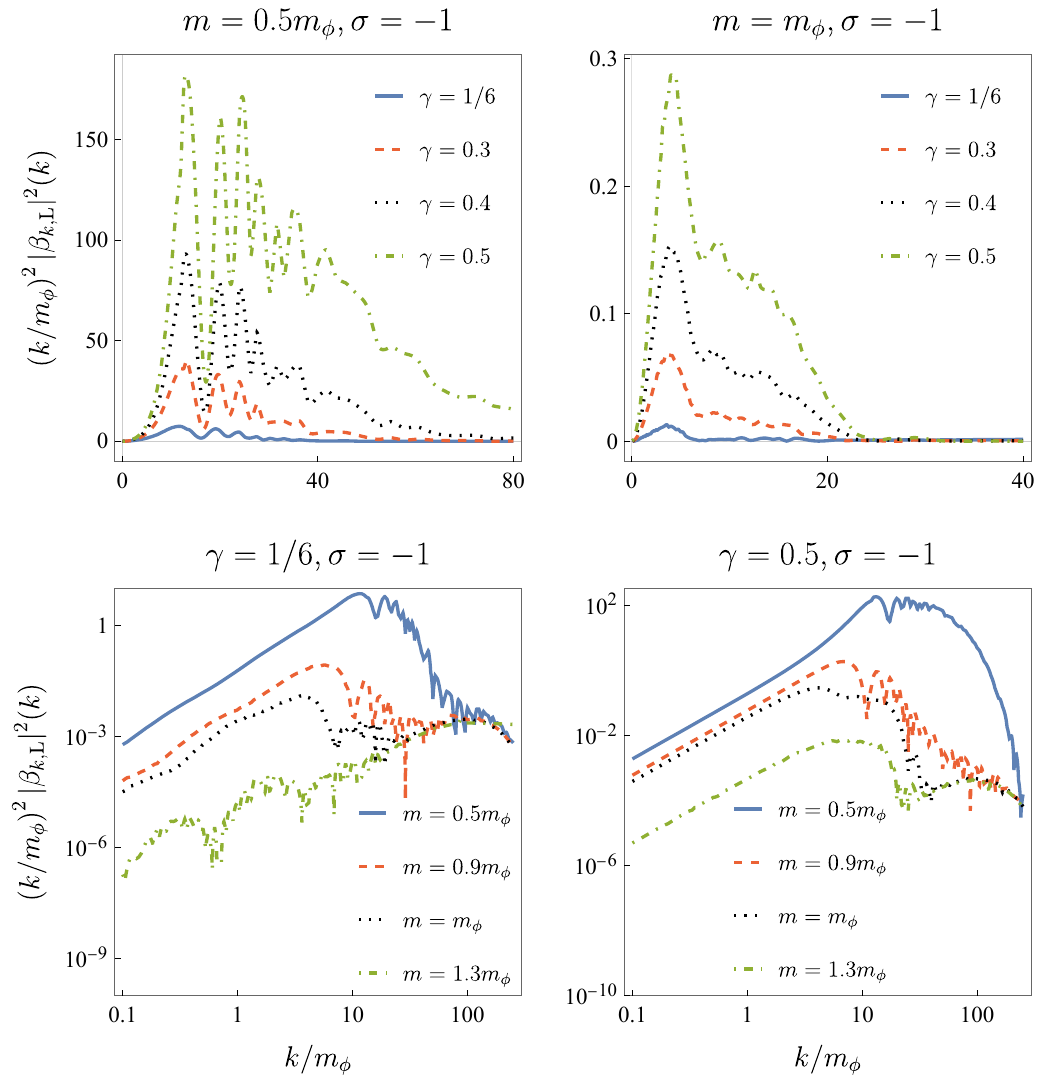}
    \caption[Longitudinal spectra for $\sigma=-1$]{Spectra of produced longitudinal modes for a fixed value \mbox{$\sigma=-1$}. The upper panels, in linear scale, show the spectra for several values of $\gamma$, given~\mbox{$m=0.5m_{\phi}$} (left) or $m=m_{\phi}$ (right). On the other hand, the lower panels, in log scale, show the spectra for several values of $m$, given $\gamma=1/6$ (left) or $\gamma=0.5$ (right). Figure from \cite{VectorDM2024}.}
    \label{fig:sf.SpectraL2}
\end{figure}

As anticipated, non-vanishing, negative values of $\sigma$ will induce tachyonic instabilities, yielding larger production. This is illustrated in figure \ref{fig:sf.SpectraL2}, where one can observe that oscillations in the spectra are much more violent as well, as compared to the $\sigma=0$ case. Because spectra are wider, we resolve them up to $k\approx250m_{\phi}$ and restrict ourselves to $\gamma\leq 0.5$ in order to calculate the total density of particles produced. Similarly to the transverse modes, a positive value of $\sigma$ would lead to less production, since instabilities would become less important. This amplification of the tachyonic behavior of the frequency through the coupling $\sigma$ is absent in the non-minimally coupled scalar field case for obvious reasons, but it is an important mechanism for producing a large abundance of dark matter.

As we have seen, the mechanism of cosmological particle production results in a non-vanishing density of particles also for the test field~$A_{\mu}$. As we have done in subsection \eqref{subsec:sf.quadraticabundance}, we consider the dark matter field as non-interacting (i.e. a spectator field), and assume that particle production becomes negligible after~$\eta_{\text{f}}$. Then, this comoving density will be related to the physical density only by a factor that takes into account the dilution due to the expansion of the universe,~\mbox{$n_{\text{T,phys}}(\eta_{\text{today}})a^3(\eta_{\text{today}})=n_{\text{T}}$} (and similarly for the longitudinal mode). The abundance can be written in terms of the background radiation temperature as in~\eqref{eq:sf.Abundance}, where now $n = n_{\text{L}}+2n_{\text{T}}$, in order to include in the abundance the production for the three modes of the vector field. Similar considerations regarding the upper limit for the reheating temperature to those made in \cref{sec:sf.abundance} apply here, namely that in the whole parameter space considered adiabaticity is reached way before the end of reheating (i.e., ~$\eta_{\text{f}} < \eta_{\text{rh}}$), as long as $T_{\text{rh}} \lesssim 10^{13} \, \text{GeV}$.

\begin{figure}[t!]
    \centering
    \includegraphics[width=0.85\textwidth]{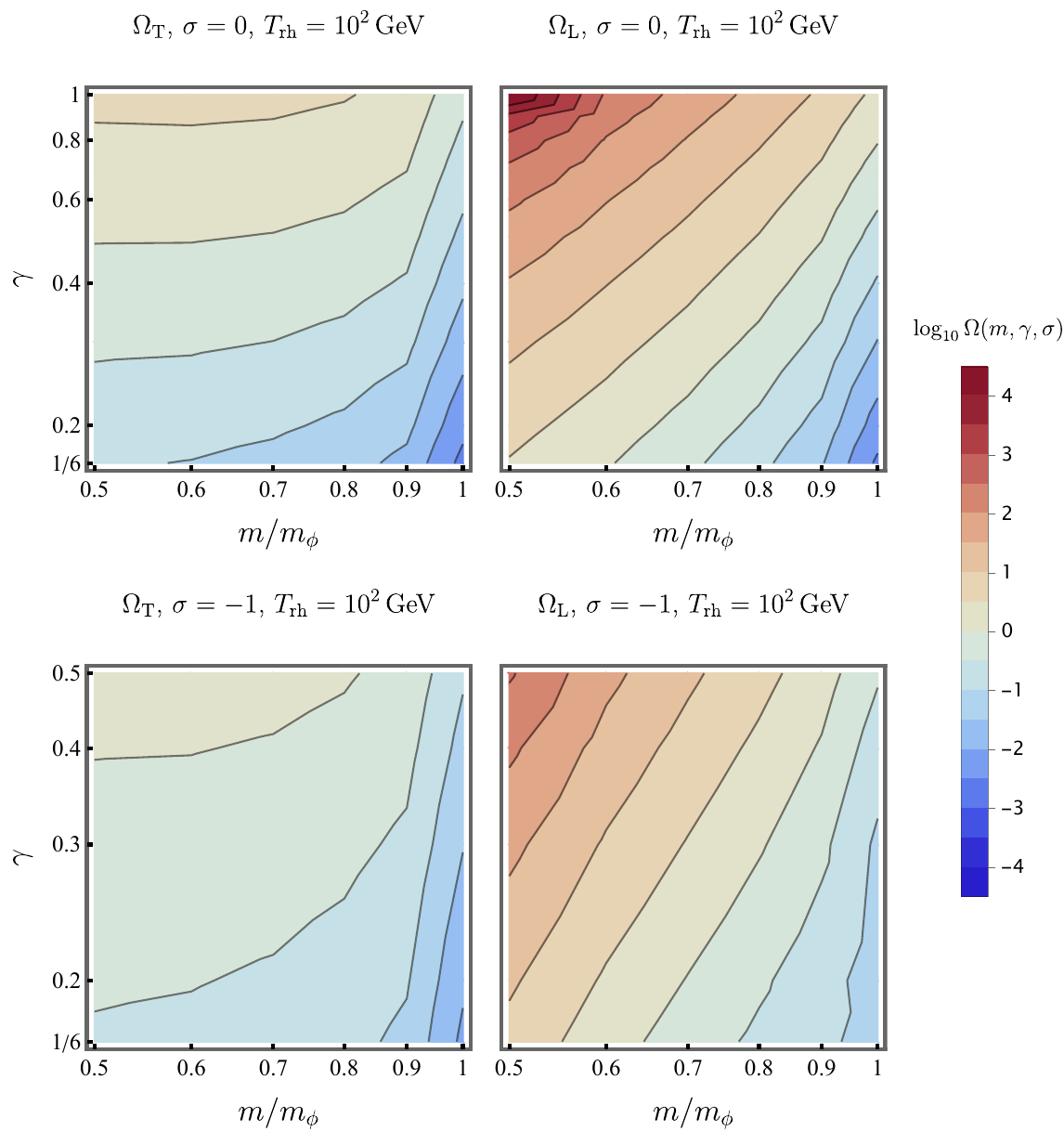}
    \caption[Transverse and longitudinal abundances]{Transverse (left panels) and longitudinal (right panels) abundances for $\sigma=0$ (upper panels) and $\sigma=-1$ (bottom panels), and a reheating temperature of~\mbox{$T_{\text{rh}}=10^2 \, \text{GeV}$}. Note that $\Omega_{\text{T}}$ includes only one transverse degree of freedom. Figure from~\cite{VectorDM2024}.}
    \label{fig:sf.IndividualAbundances}
\end{figure}

We show the longitudinal and transverse abundances in figure \ref{fig:sf.IndividualAbundances}, for two values of the coupling, $\sigma=0$ and $\sigma=-1$, and fixed reheating temperature $T_{\text{rh}}=10^2 \, \text{GeV}$. As expected from the spectra analysis, the abundance of longitudinal modes is much larger, several orders of magnitude even, depending on the values of $m$ and~$\gamma$. At the same time, a negative value of the coupling $\sigma$ increases both abundances by around one order of magnitude. It is also interesting to note that the transverse abundance changes slower with mass, until $m \sim m_{\phi}$ is reached. More importantly, there is no qualitative change in the behavior of the longitudinal mode abundance in the region close to the conformal point $\gamma=1/6, \sigma=0$, in contrast to the scalar field case.
    
\begin{figure}[t!]
    \centering
    \includegraphics[width=0.8\textwidth]{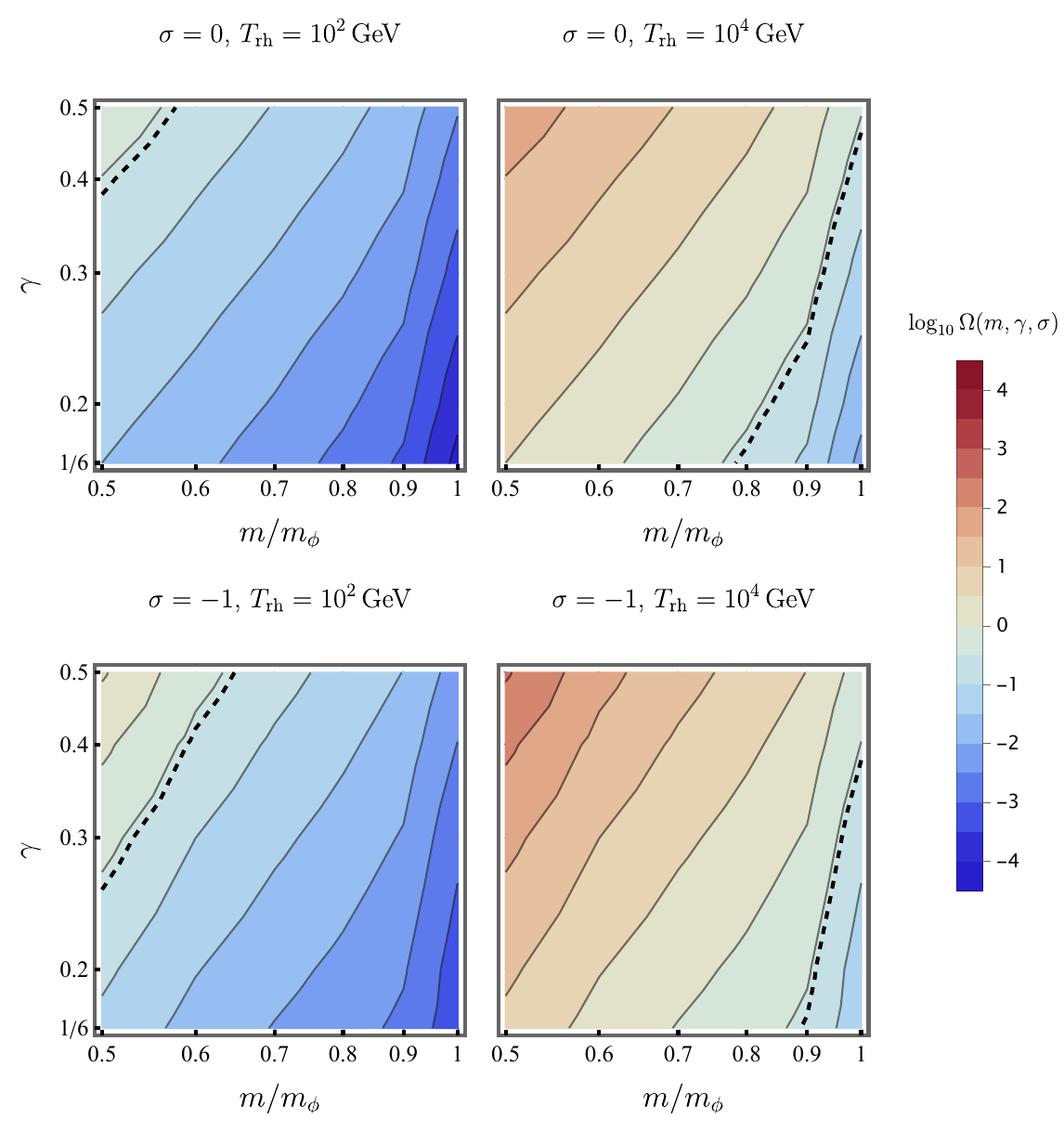}
    \caption[Total vector abundance]{Total abundance for $\sigma=0$ (upper panels) and $\sigma=-1$ (bottom panels), and two different reheating temperatures, $T_{\text{rh}}=10^2 \, \text{GeV}$ (left panels) and~\mbox{$T_{\text{rh}}=10^4 \, \text{GeV}$} (right panels). The dashed line corresponds to the observed dark matter abundance. Figure from \cite{VectorDM2024}.}
    \label{fig:sf.TotalAbundances}
\end{figure}

On the other hand, total abundances are depicted in figure \ref{fig:sf.TotalAbundances}, for~\mbox{$\sigma=0$} (upper panels) and~$\sigma=-1$ (bottom panels) as well as for two different reheating temperatures, $T_{\text{rh}}=10^2 \, \text{GeV}$ and $T_{\text{rh}}=10^4 \, \text{GeV}$. As before, a negative value $\sigma$ increases the total abundance (which is dominated by the longitudinal contribution), and so does increasing the value of the reheating temperature. We observe that gravitational production is able to reproduce observations for masses below the inflaton, and enhancing the production (for example via a negative coupling $\sigma$) shifts the observed abundance toward larger masses. In general, gravitational production of vector fields is much more efficient than that of scalar fields, since one is able to explain observations for a dark matter candidate with a mass of the order of the inflaton mass, in contrast to the case analyzed above. In particular, note that the reheating temperatures in figures \ref{fig:sf.IndividualAbundances} and \ref{fig:sf.TotalAbundances} are much lower than standard values. If we were to reproduce the observed abundance for a typical value of $T_{\text{rh}}$ (for instance,~\mbox{$T_{\text{rh}} \sim 10^9 \, \text{GeV}$}), the mass of the dark matter candidate would have to be orders of magnitude higher than the inflaton mass.

\subsection{Scalar spectra and abundance of dark matter: Starobinsky potential}
\label{subsec:sf.starobinskyabundance}

Let us now show some preliminary results on the production of scalar fields in the Starobinsky-driven inflation. For the numerical computations in the Starobinsky case we have taken a different normalization of the scale factor in order to speed up numerical calculations. In order to compare results from both inflationary models, we show the corresponding spectra at $\eta_{\text{f}}$ in terms of~\mbox{$k^2|\beta_k|^2/a^2(\eta_{\text{f}})$} and $k/a(\eta_{\text{f}})$, since it is in this form that the density enters into the abundance \eqref{eq:sf.Abundance}. Let us remark that the results for the quadratic case in this section have been obtained independently of the ones presented in subsection \ref{subsec:sf.quadraticabundance}, and yet they are consistent with the above.

\begin{figure}
    \centering
    \includegraphics[width=0.85\textwidth]{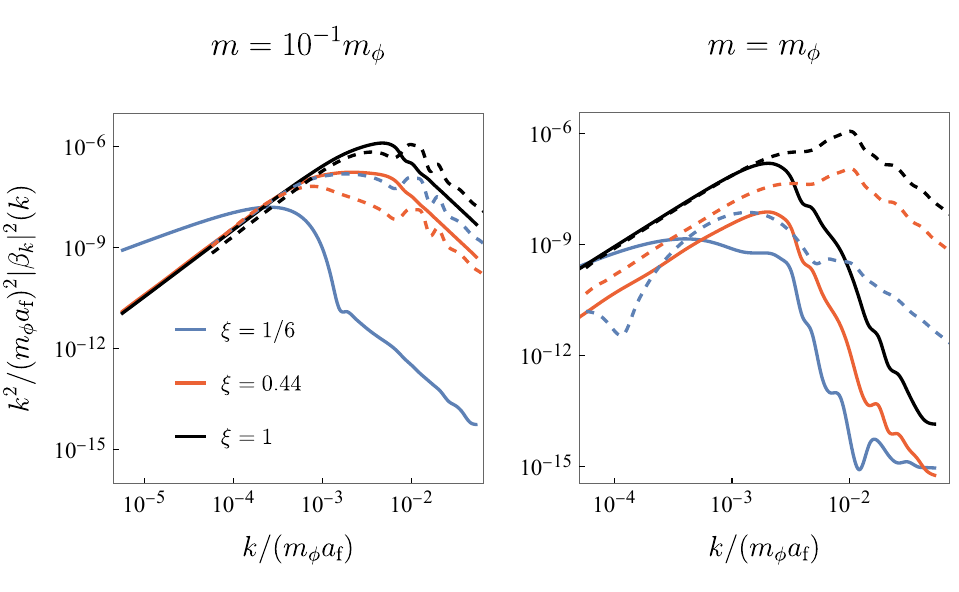}
    \caption[Quadratic and Starobinsky spectra for \mbox{$m=10^{-1}~m_{\phi}$} and~\mbox{$m=m_{\phi}$}, for different values of~$\xi$]{Spectra for $m=10^{-1}~m_{\phi}$ (left) and $m=m_{\phi}$ (right), for different values of $\xi$. Production increases with $\xi$ and is larger for the Starobinsky potential (dashed lines) in comparison to the quadratic case (solid lines).}
    \label{fig:sf.SpectraStarobinskyMasses}
\end{figure}

The spectra for $m=10^{-1}m_{\phi}$ are shown in the left panel of figure~\ref{fig:sf.SpectraStarobinskyMasses}. In the case of Starobinsky potential we can see a bump at a $k$ higher than that of the main peak. This wavenumber $k$ does not depend on the value of the coupling $\xi$. We find that spectra for each potential are qualitatively similar for $\xi>1/6$, while the conformal case $\xi=1/6$, in which the curvature scalar does not contribute to the mode equation, renders qualitatively different spectra for each potential. The rest of the values of $\xi$ result in similar shapes, with production increasing with~$\xi$ and being larger for the Starobinsky potential. We also find that the results for each potential are more similar for low values of $k$ than for higher values. This is unexpected, as the background specific terms in the frequency \cref{eq:qftcs.MasterFrequency} should dominate for lower values of~$k$, and may indicate that long wavelength modes are not sensitive to curvature scalar oscillations. On the other hand, the differences in higher $k$ can be explained by the inflaton field normalization being different in each model.

Now, for $m=m_{\phi}$, shown in the right panel of figure \ref{fig:sf.SpectraStarobinskyMasses}, we again find a different behavior for $\xi=1/6$. In the case of the quadratic potential, a different regime begins at this particular mass, leading to a steeper fall in the ultraviolet, which causes the total production to be much lower than for the Starobinsky case. This is related to the fact that $m_{\phi}$ is the inflaton oscillation frequency at late times, and we had already seen this in subsection \ref{subsec:sf.quadraticabundance}. The Starobinsky potential with~\mbox{$\xi\neq 1/6$} leads to spectra that are similar to their counterparts at~\mbox{$m=10^{-1}m_{\phi}$}, while the case $\xi=1/6$ has a structure with several bumps in the region of low~$k$. The fact that $m=m_{\phi}$ yields a different behavior for the quadratic potential but not for the Starobinsky potential might be explained by $m_{\phi}$ being the scale of the background evolution in quadratic inflation, while it is not in the Starobinsky case. We plan to explore if such a scale exists for the spectra in the Starobinsky inflation at higher mass values, possibly related to the Starobinsky inflaton mass $\bar{m}_{\phi}$.

\begin{figure}
    \centering
    \includegraphics[width=0.45\textwidth]{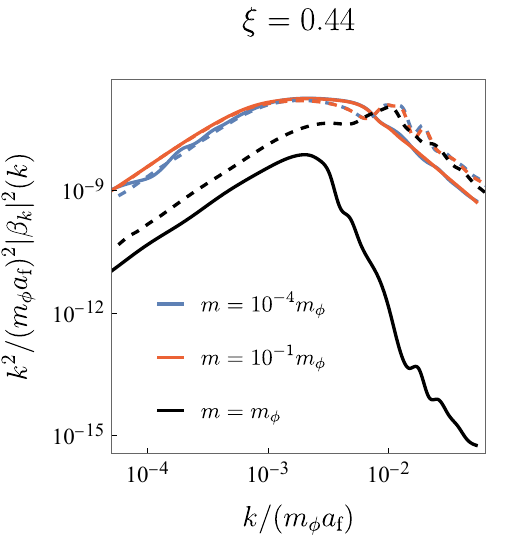}
    \caption[Preliminary quadratic and Starobinsky spectra for $\xi=0.44$ and different $m$]{Spectra for $\xi=0.44$ and varying mass. The spectra are governed by the effect of the coupling for small mass values.}
    \label{fig:sf.SpectraStarobinsy044xi}
\end{figure}

The results for varying mass and fixed $\xi$ are depicted in figure~\ref{fig:sf.SpectraStarobinsy044xi}. We have chosen $\xi=0.44$ because it represents well the behavior for~\mbox{$\xi\neq 1/6$}. We find that for values lower than $m=m_{\phi}$, changing the mass yields only small variations in the spectrum, since it is negligible in the mode equation frequency. Production increases with decreasing mass and, in all cases except for $m=m_{\phi}$ in the quadratic case, the spectra for different masses converge in the ultraviolet. We also see that the $k$ of the bump in the Starobinsky case does not depend on $m$. 

\begin{figure}
    \centering
    \includegraphics[width=0.9\textwidth]{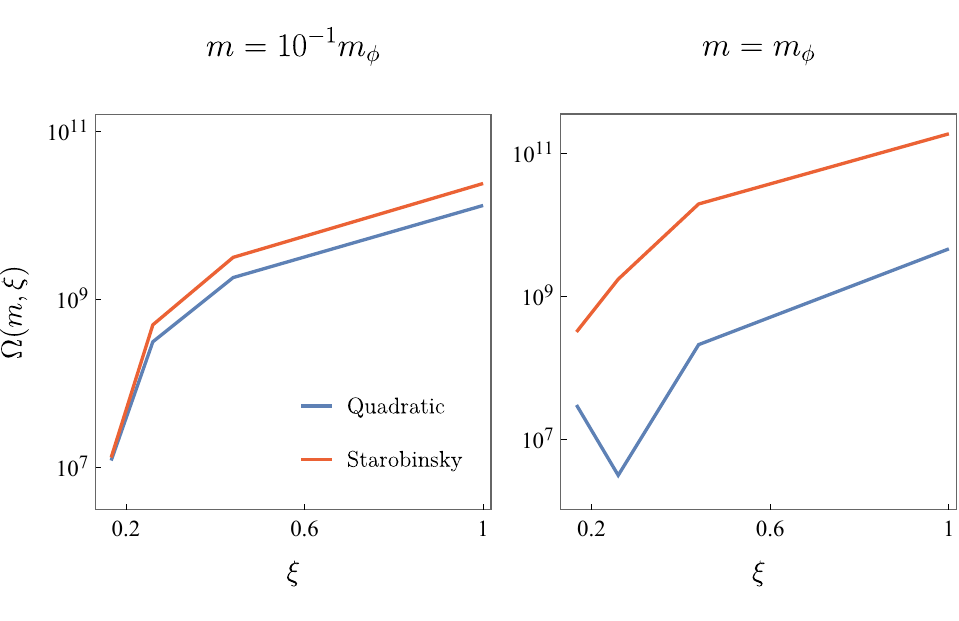}
    \caption[Quadratic and Starobinsky abundances for $m=10^{-1}~m_{\phi}$ and~\mbox{$m=m_{\phi}$}]{Starobinsky and quadratic abundances for $m=10^{-1}~m_{\phi}$ (left) and $m=m_{\phi}$ (right) as a function of the coupling $\xi$.}
    \label{fig:sf.AbundancesStarobinsky}
\end{figure}

We now plot the predicted abundances for different masses as a function of the coupling constant $\xi$. In figure \ref{fig:sf.AbundancesStarobinsky} we show the results for $m=10^{-1}~m_{\phi}$, and we observe that the total production increases with $\xi$ and is slightly higher for Starobinsky potential. Since for masses lower than the ones considered here the spectra, and therefore the number density $n$, are not sensitive to changes in the mass, we can expect the abundances for lower masses to decrease linearly with mass (see \eqref{eq:sf.Abundance}). The case $m=m_{\phi}$ is shown in the right panel of figure \ref{fig:sf.AbundancesStarobinsky}, where we see that production is considerably lower for the quadratic potential due to the change in regime at this mass value. This change also causes the abundance to not always increase with $\xi$, as we see for~\mbox{$\xi=0.26$}. This result is consistent with \cite{ScalarField2023} and subsection \ref{subsec:sf.quadraticabundance}. For the Starobinsky potential, comparing with the $m=10^{-1}~m_{\phi}$  case, we see that production increases with increasing mass. 

Although preliminary, \cref{fig:sf.SpectraStarobinskyMasses,fig:sf.SpectraStarobinsy044xi,fig:sf.AbundancesStarobinsky} already show that the spectra for both potentials are similar for masses lower than the inflaton mass, with slight differences in the ultraviolet. For higher masses, of the order of the quadratic inflaton mass, the production for the quadratic potential behaves differently, while the production in Starobinsky inflation remains in the same regime. This seems to be due to the fact that the inflaton mass is a scale that plays an important role as the frequency of the late reheating oscillations of the Ricci scalar (see \cref{eq:sf.LateRicci}). We have also found that the total abundances are larger for the Starobinsky potential, especially for these higher masses, since the different regime in the quadratic case causes a decrease in the number density of produced particles. Note that the abundances corresponding to the explored parameter space in these preliminary results are much lower than the observations for dark matter. In the following work~\cite{Starobinsky2024}, we will therefore consider a much lighter spectator field, in the line of what we have shown in  subsection \ref{subsec:sf.quadraticabundance}.

\section{Summary}

In this chapter we have studied cosmological production of non-minimally coupled to gravity scalar and vector fields in a $(1+3)$-dimensional FLRW geometry during inflation and reheating. For the scalar field, we have considered both quadratic and Starobinsky's inflationary potentials, and we have shown that for certain values of the dark matter candidate mass and the coupling to the Ricci scalar, we can recover the observed abundance. To perform this analysis, we have derived an approximation to the mode equation valid during slow-roll, we have carefully studied the choices of vacua, and we have paid special attention to the influence of the Ricci scalar oscillations during the reheating era. In the context of the vector field, we have seen that production, under the same background, is enhanced with respect to the scalar field, leading to heavier dark matter particles, and that the longitudinal modes are the main contributors to the total abundance. Importantly, in all scenarios we have found that the Ricci oscillations are very relevant and cannot be ignored.

%% file: Chapters/DMProductionDeSitter.tex


\chapter{Particle production in de Sitter-like inflation} 

\label{ch:deSitter} 



We consider in this chapter a toy model of inflation based on de Sitter spacetime, which can be thought of as a first approximation to the more realistic inflationary models discussed in the previous chapter. The idea of doing so is to obtain analytic solutions to cosmological production rather than having to rely on numerical computations, as we have done above. The exit of the inflationary phase is modeled by a transition to a matter-dominated universe, in which evolution is much more adiabatic than in the case of an exponentially growing scale factor. Therefore, particle production will be negligible after this time. From this point on, the usual Standard Model dynamics hold, and we can employ the fluid description of the thermal Universe. We will discuss two different choices of \textit{out} vacuum. First, the ILES (see \cite{Birrell1982,Mukhanov2007}), which in the conformal coupling case corresponds to assuming that there is a sudden transition to a static universe after inflation. Second, the zeroth-order adiabatic vacuum used in \cref{ch:singlefield}, which, as we will argue below, can be interpreted as assuming that there is a smooth transition to an adiabatic evolution, such as in a matter or radiation dominated universe. We will consider now all possibilities of curvature for the spatial sections of the FLRW metric and study the influence of curvature abundance in the phenomenon of cosmological production, although only considering scalar fields. Similar to what we did in chapter \ref{ch:singlefield}, we obtain the number density of produced particles by solving the corresponding mode equations, and assume a non-interacting spectator field after the production period. This again allows us to compare the resulting abundance with dark matter observations, but also with the predictions of the other models of inflation discussed in \cref{ch:singlefield}.

\section{De Sitter inflation}
\label{sec:ds.deSitter}

The early dynamics of an FLRW universe with spatial curvature $\kappa$ and scale factor~$a$ is described by the Friedmann equation \eqref{eq:cosmo.FriedmannEquation}, with the energetic content of the universe dominated by the inflaton. That is,~$\rho$ is given by \cref{eq:cosmo.InflatonDensityAndPressure}, and therefore the Hubble rate acquires the form~\eqref{eq:cosmo.InflatonFriedmann}. In slow-roll, the latter can be written as
\begin{equation}
    H^2(t) \simeq \frac{8\pi}{3M_P^2}V(\phi) - \frac{\kappa}{a^2(t)}.
\end{equation}
The inflaton potential approaches a constant value $V_0$ in the past, and 
\begin{equation}
    H^2(t) \to \frac{8\pi}{3M_P^2}V_0 - \frac{\kappa}{a^2(t)} = H_0^2 - \frac{\kappa}{a^2(t)}, 
\label{eq:ds.deSitterHubbleRate}
\end{equation}
with $H_0$ denoting the Hubble rate in this limit for $\kappa=0$, as we did in \cref{ch:singlefield}. In other words, one recovers the Friedmann equation for a de Sitter universe with spatial curvature $\kappa$, which is characterized by the energy content corresponding to a cosmological constant~\mbox{($\rho \simeq \text{constant}$)}. This is precisely why the \textit{in} vacuum was chosen as that of Bunch-Davies in the previous chapter. Therefore, de Sitter spacetime is a good approximation to (the early regime of) an inflationary model characterized by a very slow-roll of the inflaton\footnote{Although of course, it does not capture the oscillatory dynamics in reheating.}. In order to perform analytic calculations throughout, we will study the influence of spatial curvature in the production of particles assuming a de Sitter geometry for the whole inflationary stage.

\subsection{Exit of inflation}
\label{subsec:ds.exit}

When the slow-roll regime ends---and therefore the de Sitter description is no longer valid---, the inflaton field~$\phi$ decays and starts oscillating around the minimum of the potential~$V(\phi)$. This is the typical scenario for chaotic, single-field inflation governed by, e.g., a quadratic potential $V(\phi) = \frac{1}{2}m_{\phi}^2\phi^2$, as we have seen in \cref{ch:singlefield}. After that, reheating begins and the energy density of the inflaton is transferred to other particles via their interactions (see \cref{sec:cosmo.reheating}). At this point, the scale factor ceases to behave as a (quasi-)exponential, $a(t) \sim e^{H_0 t}$, and acquires the typical form of the scale factor of a matter-dominated universe, $a(t) \sim t^{2/3}$: Inflation ends and reheating starts. This is what we modeled with the inflationary scenarios in the previous chapter. As we have seen, the rate of expansion of the geometry is much slower during this regime, and production of particles becomes negligible from this point onwards. In terms of cosmological production, one can assume as a first approximation that spacetime becomes static again. We model this transition from inflation to reheating and radiation dominance in two different ways.

\begin{figure}[t]
    \centering
    \includegraphics[width=0.45\textwidth]{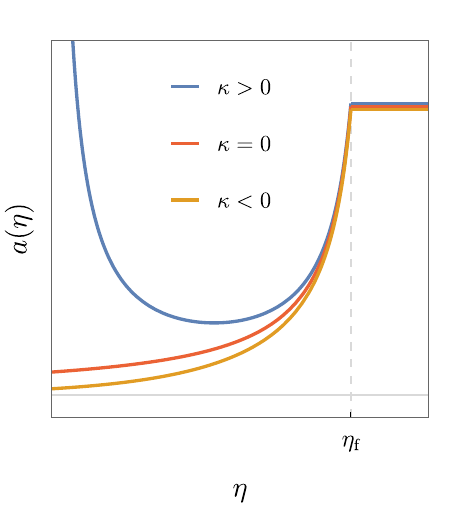}
    \caption[De Sitter scale factors for the three cosmological patches]{Time dependence of the scale factor for the three different spatial curvatures, assuming an instantaneous transition to a static geometry at $\etaf$. The scale factor is normalized to the value it takes at the end of inflation,~\mbox{$a(\eta_{\text{f}}) = 1$}.}
    \label{fig:ds.ScaleFactors}
\end{figure}

As a first approximation, we consider that the de Sitter-like phase suddenly ends at $\eta_{\text{f}}$, where the scale factor becomes constant. The exit of inflation and the transition to reheating and a slowly-expanding geometry is happening at that precise instant. In \cref{fig:ds.ScaleFactors} we show the time dependence of the scale factor as a function of cosmological time. Clearly, an observer living after $\eta_{\text{f}}$ will have a preferred notion of vacuum. Since spacetime is static, quantization follows as in Minkowski spacetime. The frequency of the mode equation \eqref{eq:qftcs.MasterFrequency} becomes constant and the basis defining the Minkowski vacuum is given by plane waves $u_k(\eta)$ of the form \eqref{eq:qftcs.MinkowskiModesGeneral} with positive frequency~\mbox{$\omega_{\text{f}}=\omega(\eta_{\text{f}})$}. One could instead simply choose a different notion of vacuum at $\eta_{\text{f}}$. For example, the ILES (see \cite{Alvarez2023, Ferreiro2023} for comparison with the States of Low Energy). However, since we are in the conformally coupled case,~\mbox{$\xi=1/6$}, the effects of the discontinuities in the scale factor at $\eta_{\text{f}}$ do not affect particle production. As a consequence, choosing the ILES and assuming the expansion of the geometry is suddenly stopped renders the same amount of produced particles (this will become clearer with the discussion in subsection \ref{subsec:bec.InstantaneousSwitch}).

Alternatively, we will also consider the adiabatic notion of vacuum used in \cref{ch:singlefield} for obtaining particle production. There, we argued that this vacuum is the one that an observer living in a sufficiently slowly-expanding universe would define, and we computed the number density of produced particles once the adiabatic regime was reached. One may then ask what is the meaning of choosing the particular solution of the mode equation defined by adiabatic initial conditions~\eqref{eq:sf.AdiabaticVacuum} with de Sitter frequency \eqref{eq:sf.deSitterFrequency} at $\eta_{\text{f}}$, since the expansion is not adiabatic there. We argue that making this choice can be interpreted as assuming a sufficiently slow exit of the de Sitter phase starting at this instant, until an adiabatic regime is reached. Indeed, if this was a stable notion of vacuum, it would mean that the time duration of the exit of inflation is much larger than the characteristic evolution scale in the problem, which is the rate $H_0$, and therefore the change in the mode equation frequency is very slow. In other words, the \textit{switching-off} of the expansion of the geometry is adiabatic. More details on this last concept will come in part~III and IV.

Once the notion of vacuum after inflation is selected, it remains to quantize our field in the de Sitter epoch, to which we devote the next section.

\subsection{Scalar field in a spatially curved FLRW universe}
\label{subsec:ds.scalarfield}

We will particularize to $D=3$ the dynamics discussed in chapter \ref{ch:qftcs} for a non-minimally coupled scalar field $\varphi$ with mass $m$. Let us start from the general line element of the FLRW metric \eqref{eq:cosmo.FLRWLineElementConformal} (where the angular part corresponds now to~\mbox{$D=3$}) and perform the change of radial coordinate 
\begin{equation}
    r = \begin{cases}
    \abs{\kappa}^{-1/2}\sin{\left(\sqrt{\abs{\kappa}}\right)} \quad &\text{with} \quad u \in [0, \pi/\sqrt{\abs{\kappa}}), \quad \text{for} \quad \kappa>0, \\
    u \quad &\text{with} \quad u \in [0, \infty), \quad \text{for} \quad \kappa=0, \\
    \abs{\kappa}^{-1/2}\sinh{\left(\sqrt{\abs{\kappa}}u\right)} \quad &\text{with} \quad u \in [0, \infty), \quad \text{for} \quad \kappa<0.\\
    \end{cases}
\label{eq:ds.RadialCoordTransf}
\end{equation}
In this way, one can write in compact form the line element as
\begin{equation}
    \text{d}s^2 = a^2(\eta)\left[-\text{d}\eta^2 + \left(\text{d}u^2 + r^2(u)\text{d}\theta^2 + r^2(u)\sin^2{\theta}\text{d}\phi^2\right)\right].
\label{eq:ds.Curved3DFLRWLineElement}
\end{equation}
In this case, the Laplace-Beltrami operator $\Delta$ in the equation of motion for the field~\mbox{$\chi(\eta) = a(\eta)\phi(\eta)$} (cf. \cref{eq:qftcs.ChiEOM}) is given by
\begin{equation}
    \Delta = \frac{1}{r^2(u)}\partial_{u}\left(r^2(u)\partial_{u}\right) + \frac{1}{r^2(u)\sin{\theta}}\partial_{\theta}\left(\sin{\theta}\partial_{\theta}\right)+\frac{1}{r^2(u)\sin^2{\theta}}\partial_{\phi}^2.
\label{eq:ds.LaplaceBeltrami}
\end{equation}
The eigenfunctions of this Laplacian are of the form \mbox{$\mathcal{H}_{klq} (u, \theta, \phi)=\mathcal{U}_{k}(u)Y_{lq}(\theta, \phi)$}, where $Y_{lq}$ are the spherical harmonics, and the explicit form of $\mathcal{U}_{k}(u)$ together with the details are given in \cref{sec:ds.eigenfunctions3Dcurved}. The corresponding eigenvalues read
\begin{equation}
    h^2(k) = \begin{cases}
    K(K+2)\abs{\kappa} \quad &\text{for} \quad \kappa>0, \\
    k^2 \quad &\text{for} \quad \kappa = 0, \\
    (K^2 + 1)\abs{\kappa} \quad &\text{for} \quad \kappa<0,
\label{eq:ds.Eigenvalues}
\end{cases}
\end{equation}
with $K = k/\sqrt{\abs{\kappa}}$ and
\begin{align}
    K &= 0, 1, \text{...}, \infty; \quad l = 0, 1, \text{...}, K \quad &&\text{for} \quad \kappa > 0, \\
    k &\in [0, \infty); \hspace{0.575cm}\quad l = 0, 1, \text{...}, \infty \quad &&\text{for} \quad \kappa \leq 0,\\
    q &= -l, -l+1, \text{...}, 0, \text{...}, l-1, l, \quad &&\text{for all values of}\, \kappa.
\label{eq:ds.LabelsDescription}
\end{align}
If one takes the limit $\kappa \to 0$ in the eigenfunctions for the curved cases, one recovers that of the flat geometry (details in \cref{sec:ds.eigenfunctions3Dcurved}). Interestingly, the smallest eigenvalue in the negatively curved case is non-vanishing, acting as an effective mass~gap. 

With this, we can expand the field $\chi$ as
\begin{equation}
\chi(\eta,\vec{u}) = \int \dd \mu_{klq} \left[a_{klq}\mathcal{H}_{klq}(\vec{u})v_k(\eta)
+ a_{klq}^*\mathcal{H}_{klq}^*(\vec{u})v_k^*(\eta)\right],
\label{eq:ds.ScalarFieldExpansion}
\end{equation}
where $\bm{u} = (u, \theta, \phi)$, and the measure in momentum space is given by
\begin{equation}
\int \dd \mu_{klq} = \begin{cases}
\sum_{K=0}^{\infty}\abs{\kappa}^{3/2}(K+1)^2 \sum_{l=0}^K \sum_{q=-l}^l \quad &\text{for} \quad \kappa>0,\\
\int_0^{\infty}\text{d}k \, k^2\sum_{l=0}^{\infty}\sum_{q=-l}^{l} \quad &\text{for} \quad \kappa \leq 0.
\end{cases}
\end{equation}

We can finally write the time-dependent frequency of the mode equation~\eqref{eq:qftcs.ModeEquation},
\begin{equation}
    \omega_k^2(\eta) = h^2(k) + a^2(\eta)\left[m^2+ \left(\xi - 1/6\right)R(\eta)\right],
\label{eq:ds.ScalarFrequency}
\end{equation}
which characterizes the process of particle production for the modes~$v_k$, normalized according to the Wronskian \eqref{eq:qftcs.WronskianModes}.

\subsubsection*{Canonical quantization}

As mentioned in part I, canonical quantization implies promoting the expansion coefficients $a_{klq}, a_{klq}^*$ to creation and annihilation operators, which fulfill the commutation relations
\begin{equation}
\begin{split}
    \big[\hat{a}_{klq}^{\dagger},\hat{a}_{k^{\prime}l^{\prime}q^{\prime}}^{\dagger}\big] &= \left[\hat{a}_{klq},\hat{a}_{k^{\prime}l^{\prime}q^{\prime}}\right]=0,\\
    \big[\hat{a}_{klq},\hat{a}_{k^{\prime}l^{\prime}q^{\prime}}^{\dagger}\big] &= \delta_{ll^{\prime}} \delta_{qq^{\prime}}\times \begin{cases}
    \frac{\delta_{K K^{\prime}}}{\abs{\kappa}^{3/2}(K+1)^2} \quad &\text{for} \quad \kappa > 0,\\
    \frac{\delta(k - k^{\prime})}{k^2} \quad &\text{for} \quad \kappa \leq 0.\\
    \end{cases}
\end{split}
\label{eq:ds.CreationAnnihilationCommutationRelsScalar}
\end{equation}
These are consistent with the commutation relations
\begin{equation}
    \left[\hat{\chi}(t, \vec{u}), \hat{\pi_{\chi}}(t, \vec{u}^{\prime}) \right] =i\delta(\theta - \theta^{\prime})\delta(\phi - \phi^{\prime})\delta(u - u^{\prime})
\label{eq:ds.FieldsCommutationRelsScalar}
\end{equation}
on the field $\chi$ and its conjugate momentum $\pi_{\chi} = \chi^{\prime}$. For the problem at hand, the goal is to extract the number density of produced particles after the evolution of the Universe during inflation and reheating. Let us recall that this is given by~\eqref{eq:qftcs.MeanNumberDensity}, where the corresponding Bogoliubov coefficient relates the natural notions of vacuum for observers living before and after the expansion. We therefore aim to be able to consider particle production in this context as an \textit{in}-\textit{out} process, exactly as we did in \cref{ch:singlefield}. Then, as long as the test particle is not (strongly) interacting, this will be related to the abundance one observer would measure today only by the expansion dilution. The choices of vacua to compare will depend on the particular background in chapter \ref{ch:deSitter} under consideration, and will be discussed further below. 

\section{Quantization in de Sitter spacetime}
\label{sec:desitter}

In the previous subsections we have seen that at the onset of inflation, the geometry approaches that of de Sitter. Moreover, considering an exact de Sitter geometry during the full inflationary phase is a well justified assumption and serves as a first, simple model. Regardless of whether one makes this approximation or not, in order to choose a suitable notion of vacuum at the beginning, one needs to answer the question of how to quantize in de Sitter spacetime. Crucially, the number of symmetries of this geometry is the same as those of Minkowski (cf. subsection~\ref{subsec:qftcs.inout}), which leads to Bunch-Davies as preferred notion of vacuum (for the massive scalar case)~\mbox{\cite{Birrell1982,Allen1985,Mottola1986}}.

In the following, we will show how the three different cosmological cases under study (positive, negative and vanishing $\kappa$) are realized by different choices of coordinates, covering different regions, in a de Sitter geometry.

\subsection{Spatially curved inflationary cosmologies as de Sitter patches}

From equation \eqref{eq:ds.deSitterHubbleRate} we can write a differential equation for the scale factor $a(t)$,
\begin{equation}
    \dot{a}^2(t) = H_0^2 a^2(t) - \kappa,
\end{equation}
whose solutions are
\begin{equation}
    a(t) = \begin{cases}
    \frac{\sqrt{\abs{\kappa}}}{H_0}\cosh{\left[H_0\left(t-t_+\right)\right]} \quad &\text{for} \quad \kappa > 0,\\
    e^{H_0\left(t-t_0\right)} \quad &\text{for} \quad \kappa = 0,\\
    \frac{\sqrt{\abs{\kappa}}}{H_0}\sinh{\left[H_0\left(t-t_-\right)\right]} \quad &\text{for} \quad \kappa < 0,
    \end{cases}
\label{eq:ds.ScaleFactors}    
\end{equation}
where the integration constants $t_{\pm}, t_0$ are taken such that at the point in which the inflationary regime is exited, and particle production is calculated, i.e. $t_{\text{f}}$, the scale factor takes the value that quadratic inflation yields at that same point $\tf$ (see subsection \ref{subsec:sf.quadratic}), so that physical densities can be easily compared there. From that point on, we will assume that the evolution of the dynamics coincides with that studied in \cref{ch:singlefield} (namely the behavior is that of a matter-dominated universe). Consistently, we have taken $H_0$ as the corresponding initial value for~$H(t)$ from the quadratic model. In this way, the~\mbox{$\kappa=0$} case is comparable to the case previously studied in the sense that the inflationary dynamics given by the inflaton equation of motion are exchanged by a de Sitter phase. The other two cases are included for completeness, and to study the influence of spatial curvature in the phenomenon of particle production. Note that in the chosen coordinates, the spatial curvature~$\kappa$ has units of mass squared. 

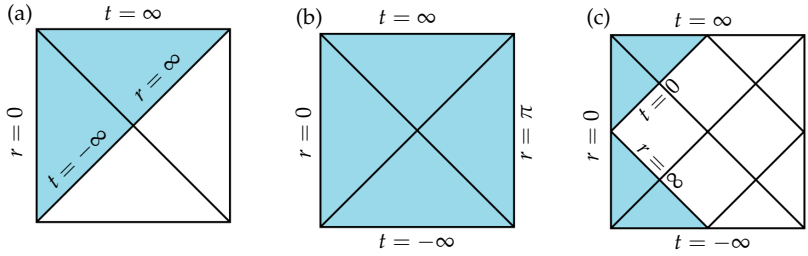
\begin{figure}
    \centering
    \scalebox{0.75}{
        \begin{overpic}[scale=0.45]{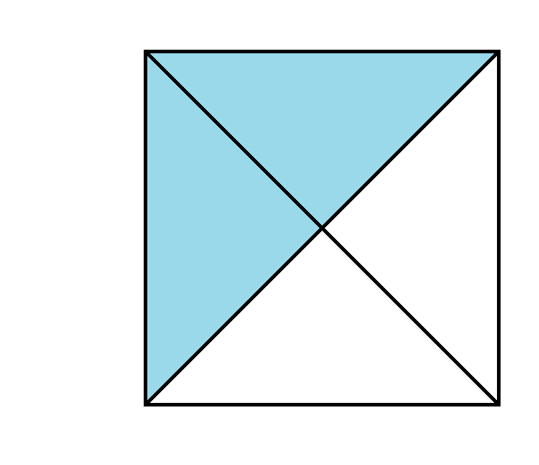}
        \put(17, 80){(a)}
        \put(17, 33){\rotatebox{90}{$r=0$}}
        \put(50, 80){$t=\infty$}
        \put(30, 22){\rotatebox{45}{$t=-\infty$}}
        \put(60, 52){\rotatebox{45}{$r=\infty$}}
        \end{overpic}
        \hspace*{-4mm}
        \begin{overpic}[scale=0.45]{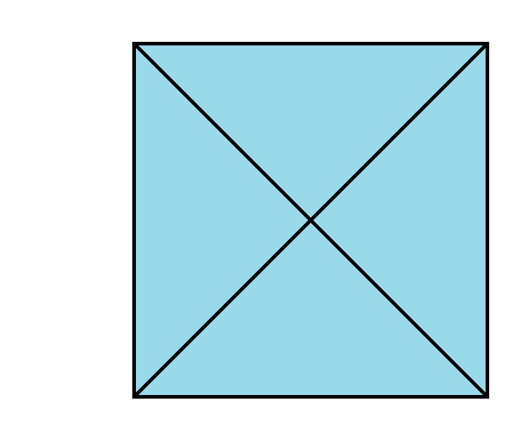}
        \put(17, 81){(b)}
        \put(17, 34){\rotatebox{90}{$r=0$}}
        \put(46, 81){$t=\infty$}
        \put(46, 2){$t=-\infty$}
        \put(96, 33){\rotatebox{90}{$r=\pi$}} 
        \end{overpic}
        \begin{overpic}[scale=0.45]{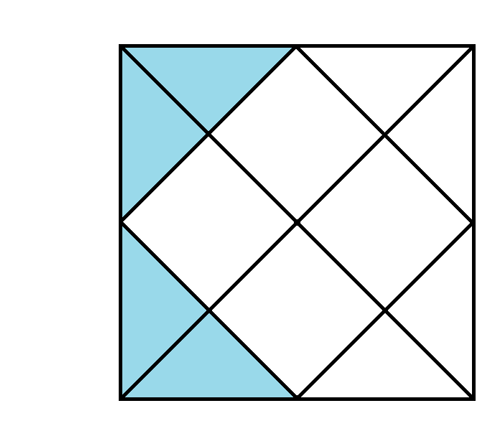}
        \put(15, 84){(c)}
        \put(14, 34){\rotatebox{90}{$r=0$}}
        \put(47, 83){$t=\infty$}
        \put(47, 2){$t=-\infty$}
        \put(33, 48){\rotatebox{45}{$t=0$}}
        \put(33, 38){\rotatebox{-45}{$r=\infty$}}
        \end{overpic}            
        }
\caption[De Sitter Penrose diagrams]{Penrose diagrams for the different coordinate charts in de Sitter: (a) flat, (b) closed, and (c) open coordinates, in terms of cosmological time $t$. The shadowed region corresponds to the covering of the corresponding coordinates, which are denoted with the same symbol in each case.}
\label{fig:ds.PenroseDiagrams}
\end{figure}

The corresponding line element can be written for the different values of~$\kappa$ as in \cref{eq:ds.Curved3DFLRWLineElement}, where the scale factor has to be taken from~\eqref{eq:ds.ScaleFactors}. This is precisely the form that the line element takes in a de Sitter geometry when using closed, flat or open coordinate charts (corresponding to the cases~\mbox{$\kappa>0$}, \mbox{$\kappa=0$} and $\kappa<0$, respectively)~\mbox{\cite{Birrell1982,Mottola1986}}. The Penrose diagrams for the different coordinate charts are displayed in \cref{fig:ds.PenroseDiagrams}.

In terms of conformal time, the scale factors take the form
\begin{equation}
    a(\eta) = \begin{cases}
    \frac{\sqrt{\abs{\kappa}}}{H_0}\sin^{-1}{\left(\sqrt{\abs{\kappa}}\eta\right)} \quad &\text{for} \quad \kappa > 0,\\
    -\frac{1}{H_0\eta} \quad &\text{for} \quad \kappa = 0,\\
    \frac{\sqrt{\abs{\kappa}}}{H_0}\sinh^{-1}{\left(\sqrt{\abs{\kappa}}\eta\right)} \quad &\text{for} \quad \kappa < 0,
    \end{cases}
\end{equation}
where $\eta \in (-\pi/\sqrt{\abs{\kappa}}, 0)$ in the first case and $\eta \in \left(-\infty, 0\right)$ in the rest. On the other hand, the Ricci scalar in this geometry is constant and given by~\mbox{$R = 12H_0^2$}, which obviously does not depend on the patch considered. With this we can explicitly write the mode equations for the three different coordinate choices in de Sitter, or, equivalently, for the three different FLRW cosmologies (closed, flat or open). They are given by\footnote{Note that even though we used the same symbol, the meaning of $\eta$ differs depending on the coordinate chart.}
\begin{equation}
    v_k^{\prime\prime}(\eta) + \omega_k^2(\eta)v_k(\eta)=0,
\label{eq:ds.ModeEquations}
\end{equation}
with
\begin{equation}
    \omega_k^2(\eta)=\begin{cases}
    \left(k+\sqrt{\kappa}\right)^2 + \abs{\kappa}\mu^2/\sin^2{\left(\sqrt{\abs{\kappa}}\eta\right)}, \quad &\text{for} \quad \kappa>0, \\
    k^2 + \mu^2/\eta^2, \quad &\text{for} \quad \kappa=0, \\
    k^2 + \abs{\kappa}\mu^2/\sinh^2{\left(\sqrt{\abs{\kappa}}\eta\right)}, \quad &\text{for} \quad \kappa<0,
    \end{cases}
\label{eq:ds.Frequencies}
\end{equation}
where $\mu$ is the dimensionless, effective mass defined in \cref{eq:sf.deSitterFrequency}.

We have already mentioned that de Sitter spacetime has a preferred notion of vacuum, the Bunch-Davies vacuum, which is associated with creation and annihilation operators that were the coefficients of the expansion of the fields in a particular basis of the solutions of \cref{eq:ds.ModeEquations} before quantization. In the following subsection we discuss which particular solutions of the mode equations lead to an expansion in terms of the operators defining said vacuum, for each of the coordinate charts considered.

\subsection{Bunch-Davies vacuum}

Let us analyze the choice of vacuum for the different de Sitter patches. 

\subsubsection*{Flat patch}
\label{subsubsec:flatBD}

We study first the flat coordinate chart, which is typically considered in the context of Cosmology, since it is often assumed that spatial sections are flat. The general solution to the mode equation \eqref{eq:ds.ModeEquations} can be written in this case as
\begin{equation}
    v_k(\eta) = A_k H_{\nu}^{(1)}(k\abs{\eta})+B_k H_{\nu}^{(2)}(k\abs{\eta}),
\end{equation}
with $\nu = \sqrt{1/4 - \mu^2}$. Note that in the limit $k\abs{\eta} \gg 1$, the frequency of the mode equation \eqref{eq:ds.Frequencies} becomes constant, and therefore one expects that ultraviolet modes behave as Minkowski plane waves. Making use of the asymptotic behavior of Hankel functions, one finds that the coefficients of the linear combination that become positive frequency plane waves for~\mbox{$k \abs{\eta} \gg 1$} are those that satisfy~\mbox{\cite{Birrell1982,Mukhanov2007}}
\begin{equation}
\begin{split}
    A_k &\overset{k\abs{\eta}\gg 1}{\longrightarrow} \frac{1}{2}\sqrt{\frac{\pi}{k}} e^{i\frac{\pi}{2}\left(\nu + \frac{1}{2}\right)},\\
    B_k &\overset{k\abs{\eta}\gg 1}{\longrightarrow} 0.
\end{split}
\end{equation}
For simplicity, we take these values for the constants for all wavenumbers~$k$. The resulting mode takes the form
\begin{equation}
    v_k(\eta) = \frac{\sqrt{\pi \abs{\eta}}}{2} e^{i\frac{\pi}{2}\nu}H_{\nu}^{(1)}(k\abs{\eta}),
\label{eq:ds.flatBD}
\end{equation}
where we neglected the overall phase. Then, the mode \eqref{eq:ds.flatBD} defines the Bunch-Davies vacuum. A natural notion of vacuum for an observer in a flat FLRW spacetime with exponential scale factor will be characterized by an expansion of the field in these modes. This is exactly what we did for choosing the particle notion of the \textit{in} observer in \cref{ch:singlefield}.

\subsubsection*{Closed patch}

In this case, the general solution to the mode equation \eqref{eq:ds.ModeEquations} is given by
\begin{equation}
\begin{split}
    v_k(\eta) = \abs{\kappa}^{-1/4}\sin^{1/2}{\left(\sqrt{\abs{\kappa}}\eta\right)}\Bigg\{&A_k P^{\nu}_{K+1/2}\left[-\cos{\left(\sqrt{\abs{\kappa}}\eta\right)}\right]\\
    &+B_k Q^{\nu}_{K+1/2}\left[-\cos{\left(\sqrt{\abs{\kappa}}\eta\right)}\right]\Bigg\},
\end{split}
\end{equation}
where $P_k^{\nu}$ and $Q_k^{\nu}$ are the associated Legendre polynomials, and $\nu$ is defined as in the flat case. Similar to what we did before, we examine the behavior of the modes in the limit $k \to \infty$, for which the frequency becomes time-independent. This time, we take for the modes that behave as positive frequency plane waves the values of the constants \cite{Birrell1982,Allen1985,Mottola1986}
\begin{equation}
    A_k = \frac{\sqrt{\pi}}{2}\left(K+1\right)^{-\nu}e^{i\left[(K+1)\pi - \frac{\pi}{4} + \frac{\nu\pi}{2}\right]}, \qquad B_k = \frac{2i}{\pi} A_k.
\end{equation}
Finally, the mode reads
\begin{equation}
\begin{split}
    v_k(\eta) = \abs{\kappa}^{-1/4}\sin^{1/2}{\left(\sqrt{\abs{\kappa}}\eta\right)}A_k\Bigg\{&P^{\nu}_{K+1/2}\left[-\cos{\left(\sqrt{\abs{\kappa}}\eta\right)}\right]\\
    &+\frac{2i}{\pi}Q^{\nu}_{K+1/2}\left[-\cos{\left(\sqrt{\abs{\kappa}}\eta\right)}\right]\Bigg\},
\label{eq:ds.closedBD}
\end{split}
\end{equation}
It can be checked that the two-point function calculated in the state defined by eqs.~\eqref{eq:ds.closedBD} and~\eqref{eq:ds.flatBD} yields the same result \cite{Chernikov1968,Tagirov1973, Schomblond1976,Bunch1978}. Therefore, both modes define the same vacuum, i.e. Bunch-Davies.

\subsubsection*{Open patch}

Now, the general solution to the equation \eqref{eq:ds.ModeEquations} is given by
\begin{equation}
    v_k(\eta) = \abs{\kappa}^{-1/4} \left\{A_k P^{iK}_{\nu}\left[\coth{\left(-\sqrt{\abs{\kappa}}\eta\right)}\right] + B_k P^{-iK}_{\nu}\left[\coth{\left(-\sqrt{\abs{\kappa}}\eta\right)}\right] \right\},
\end{equation}
where we again took $\abs{\kappa}=1$. The analysis is more involved in this case. The open coordinates can be used to cover the left or the right open patches (only the left one is highlighted in figure~\ref{fig:ds.PenroseDiagrams}). One can define modes in either patch and extend them analytically to the opposite region. Then, the two following linear combinations of those extended modes (left to the right and right to the left) expand the field in operators annihilating the Bunch-Davies vacuum~\mbox{\cite{Sasaki1994,Maldacena2013,Kanno2014}},
\begin{equation}
    v_{k}^{(\pm)} = \frac{v_k^{(\text{R})}\pm v_k^{(\text{L})}}{\left(1\pm e^{-\pi K+i\left(\nu-1/2\right)\pi}\right)\Gamma(\nu+1/2+iK)},
\end{equation}
where $v_k^{(R)}$ and $v_k^{(L)}$ are given by
\begin{equation}
\begin{split}
    v_k^{(R)} &= \frac{i}{\pi}\left(1+e^{-\pi K+i\left(\nu-1/2\right)\pi}\right)\left(1-e^{\pi K - i\left(\nu-1/2\right)\pi}\right)Q^{iK}_{\nu-1/2}\left[\coth{\left(-\sqrt{\abs{\kappa}}\eta\right)}\right] ,\\
    v_k^{(L)} &= \left(1+e^{-\pi K+i\left(\nu-1/2\right)\pi}\right) P^{iK}_{\nu-1/2}\left[\coth{\left(-\sqrt{\abs{\kappa}}\eta\right)}\right],
\end{split}
\end{equation}
up to a normalization factor. The linear combinations can be rewritten as
\begin{equation}
\begin{split}
    v_{k}^{(\pm)} = \frac{\pm N_k}{2\sinh{\left(\pi K\right)}}\Bigg\{&\frac{e^{\pi K} \mp e^{-i\pi\left(\nu-1/2\right)}}{\Gamma(\nu+1/2+iK)}P^{iK}_{\nu-1/2}\left[\coth{\left(-\sqrt{\abs{\kappa}}\eta\right)}\right] \\ 
    &-\frac{e^{-\pi K} \mp e^{-i\pi\left(\nu-1/2\right)}}{\Gamma(\nu+1/2-iK)}P^{-iK}_{\nu-1/2}\left[\coth{\left(-\sqrt{\abs{\kappa}}\eta\right)}\right]\Bigg\},
\label{eq:ds.openBD}
\end{split}
\end{equation}
where $N_k$ is given by
\begin{equation}
\begin{split}
    &N_k^2=2\pi\sinh{(\pi K)}\Bigg\{\abs{\Gamma(\nu+1/2+iK)}^{-2}\\
    &\hspace{3.5cm}\times\left(e^{2\pi K}+1 \mp e^{\pi K}2\cos{[\pi\left(\nu-1/2\right)]}\right)\\
    &\!-\abs{\Gamma(\nu+1/2-iK)}^{-2}\!\left(e^{-2\pi K}\!+\!1 \mp e^{-\pi K}2\cos{[\pi\left(\nu-1/2\right)]}\right)\!\Bigg\}^{-1}.
\end{split}
\end{equation}
The field expansion \eqref{eq:ds.ScalarFieldExpansion} is written as
\begin{equation}
    \chi(\eta, \vec{u}) = \int \dd \mu_{klq}\sum_{s=\pm} \left[\hat{a}_{klq}^{(s)}\mathcal{H}_{klq}(\vec{u})v_{k}^{(s)}(\eta) + \hat{a}_{klq}^{*,(s)}\mathcal{H}_{klq}^*(\vec{u})v_k^{*, (s)}(\eta)\right].
\end{equation}
The three particular solutions \eqref{eq:ds.flatBD}, \eqref{eq:ds.closedBD} and \eqref{eq:ds.openBD} are associated with an expansion of the field operator corresponding to creation and annihilation operators defining the Bunch-Davies vacuum \cite{Birrell1982,Mottola1986,Sasaki1994}. We will assume that this is the initial state of our system before inflation. In other words, the initial condition for our mode equation will be given by~\cref{eq:ds.flatBD,eq:ds.closedBD,eq:ds.openBD} depending on the case we are studying.

\section{Particle production}
\label{sec:particleproduction}

It only remains to compare the Bunch-Davies modes with those associated with the vacuum notion after the exit of inflation of our choice through~\eqref{eq:qftcs.BogoliubovCoefficientsWronskian}, be it plane waves of the form~\eqref{eq:qftcs.OutModeIC} or zeroth-order adiabatic modes with initial conditions~\eqref{eq:sf.AdiabaticVacuum}, which we evaluate at~\mbox{$\eta=\eta_{\text{f}}$}. In the following, we show some preliminary results for the number density of produced particles in the different cases, as well as the corresponding relic abundance today, which will appear on~\cite{CurvedDM2024}.

\begin{figure}[t!]
    \centering
    \includegraphics[width=0.5\textwidth]{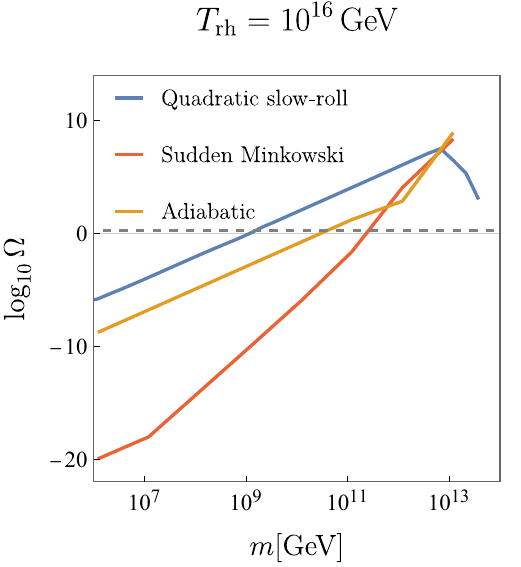}
    \caption[Produced abundance in flat quadratic and de Sitter inflation]{Produced abundance in flat quadratic inflation (blue), de Sitter with ILES (orange), and de Sitter with adiabatic (yellow). The mass of the spectator field is given in GeV. The dashed line denotes the observed dark matter abundance.} 
    \label{fig:ds.TotalDensityFlat}
\end{figure}

Let us first compare production for the quadratic potential (cf. subsection \ref{subsec:sf.quadraticabundance}) with that of the flat ($\kappa=0$) case, for both choices of vacuum after inflation: ILES and adiabatic. Figure \ref{fig:ds.TotalDensityFlat} shows the abundance of produced particles as a function of the spectator field mass $m$, diluted until today, for a reheating temperature of~\mbox{$T_{\text{th}}=10^{16} \, \text{GeV}$}. The horizontal dashed line denotes the observed dark matter abundance.

We can see that production of particles is of the same order in all cases for values of the mass of the field around the inflaton mass $m_{\phi}=1.2 \times 10^{13} \, \text{GeV}$. However, we find large deviations for smaller masses. Interestingly, the adiabatic vacuum prediction evolves with the mass similarly to the quadratic inflation case, whereas it coincides with the ILES prediction in the inflaton mass regime. The latter is due to the fact that the difference between adiabatic and instantaneous modes vanishes for large mass. Indeed, the derivative of the adiabatic mode in \cref{eq:sf.AdiabaticVacuum} contains the term
\begin{equation}
    \frac{\omega_{k,\text{dS}}^{\prime}}{\omega_{k, \text{dS}}^2} = \frac{aa^{\prime}m^2}{\left(k^2+a^2m^2\right)^{3/2}} = \frac{m^2}{H_0^2\left(k^2\eta^2 + m^2/H_0^2\right)^{3/2ç}}.
\label{eq:ds.AdiabaticCoefficientdS}
\end{equation}
For UV modes, $k\gg am = m/(H_0\eta)$, the term in \cref{eq:ds.AdiabaticCoefficientdS} grows with $m$, but it is nevertheless very small.  For wavenumbers for which production is important, those fulfilling $k\lesssim m/(H_0\eta)$, the adiabatic coefficient is of order one. Note that the opposite limit can be reached for sufficiently high masses,~\mbox{$k \ll m/(H_0\eta)$}, and in this situation $\omega_{k,\text{dS}}^{\prime}/\omega_{k, \text{dS}}^2 \sim H_0/m$. This is the reason why the adiabatic and ILES predictions coincide for large masses, and deviate more and more for smaller masses.

One can also extract another lesson from the analysis of \cref{eq:ds.AdiabaticCoefficientdS}. At the beginning of inflation, when the scale factor is very small, the condition~\mbox{$k\lesssim m/(H_0\eta)$} for production is only fulfilled by arbitrarily small wavenumbers (or large wavelengths). Consequently, production of higher $k$ happens after a sufficiently long expansion period.

On the other hand, it seems that the mass dependence of the abundance is not sensitive to the mode evolution during inflation for $m\lesssim H_0 \sim m_{\phi}$, as can be observed when comparing the quadratic and adiabatic curves in figure~\ref{fig:ds.TotalDensityFlat}. Even more, the fact that the ILES prediction disagrees with the other two leads to the idea that this behavior depends mainly on the choice of vacuum.

\begin{figure}[t!]
    \centering
    \includegraphics[width=0.5\textwidth]{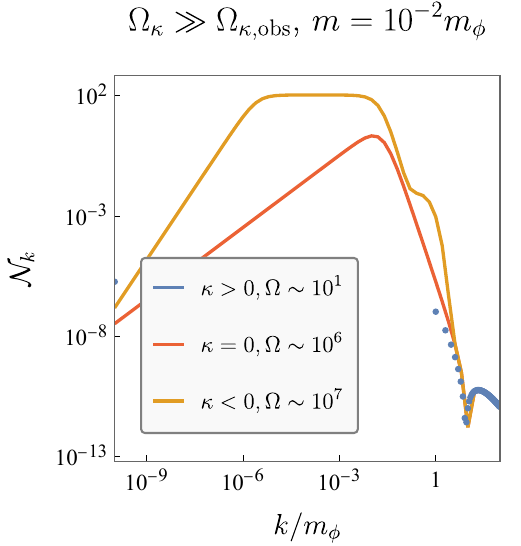}
    \caption[Spectra of produced particles with different spatial curvature in de Sitter inflation, for large values of $\kappa$]{Spectra of produced particles with different spatial curvature in de Sitter inflation using the adiabatic vacuum prescription, for a large value of the curvature abundance $\Omega_{\kappa}$ (corresponding to $\abs{\kappa} = m_{\phi}$) and~\mbox{$m=10^{-2}m_{\phi}$}. Here, $\mathcal{N}_k = (k/m_{\phi})^2\abs{\beta}^2$ for $\kappa \leq 0$, whereas $\mathcal{N}_k = (k+\sqrt{\abs{\kappa}})^2/m_{}\phi^2\abs{\beta_k}^2$ for $\kappa>0$. The abundances $\Omega$ in the legend denote the associated relic abundance obtained by using \cref{eq:sf.Abundance} with $t_{\text{rh}}=10^{15}\, \text{GeV}$. Note that the spectrum for positively curved scenarios is discrete; in particular, the blue point at the far left of the spectrum corresponds to the $k=0$ mode. Positive and negative curvature lead to a smaller and larger abundance than in the flat case, respectively. The differences can be of several orders of magnitude.} 
    \label{fig:ds.SpectraCurvedLargeK}
\end{figure}

\begin{figure}[t!]
    \centering
    \includegraphics[width=0.5\textwidth]{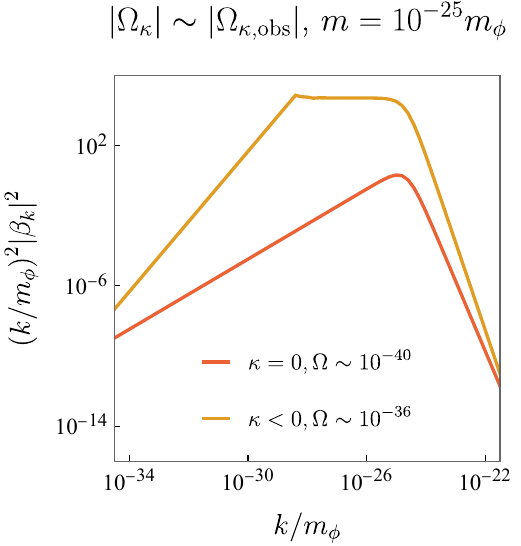}
    \caption[Spectra of produced particles with zero and negative spatial curvature in de Sitter inflation, for small values of $\kappa$]{Spectra of produced particles with zero and negative spatial curvature in de Sitter inflation using the adiabatic vacuum prescription for $\abs{\kappa}=10^{-40} m_{\phi}$ and~\mbox{$m=10^{-25}m_{\phi}$}. It shows that significant deviations in the spectra are obtained for fields in the ultra-light regime even considering curvature values compatible with current observations. Although the total abundance for such light fields represents a subleading contribution to the dark matter density, these results reveal a strong sensitivity of the production mechanism to the global geometry of the universe. The abundance is obtained using~\mbox{$T_{\text{rh}}=10^{15} \, \text{GeV}$}. Note that for such light fields and small spatial curvature, the spectrum of produced particles in the case of positive curvature is negligible compared to the other two cases shown in the Figure.} 
    \label{fig:ds.SpectraCurvedSmallK}
\end{figure}

Let us now consider the scenarios of closed and open universes with spatial curvature $\kappa$. In \cref{fig:ds.SpectraCurvedLargeK} we show the spectra of produced particles for the three curvature cases in the context of de Sitter inflation when choosing adiabatic conditions for the \textit{out} vacuum. We can see that production is enhanced in an open universe, whereas it yields a smaller number density in a closed universe. This is due to the fact that the curvature of the spatial sections affects the evolution of the modes, and therefore the number of particles produced. The effect is more pronounced for small masses, where the curvature of the spatial sections is more relevant and the~\mbox{$\kappa$-dependent} terms in the mode equations \eqref{eq:ds.ModeEquations} become more important. 

From observational constraints, we know that the abundance of curvature in the nucleosynthesis era had to be smaller than \mbox{$\Omega_{\kappa, \text{nuc}} = 10^{-16}$}. Assuming a radiation-dominated evolution backwards in time until the end of reheating, one can deduce
\begin{equation}
    \Omega_{\kappa, \text{rh}} = \Omega_{\kappa, \text{nuc}}\left(\frac{T_{\text{nuc}}}{T_{\text{rh}}}\right)^2 \sim 10^{-16}\left(\frac{T_{\text{nuc}}}{T_{\text{rh}}}\right)^2,
\label{eq:ds.ObservedCurvatureAbundance}
\end{equation}
where $T_{\text{nuc}} \simeq 10^{-3} \, \text{GeV}$ and $T_{\text{rh}}$ are the nucleosynthesis and reheating temperatures, respectively. The latter is treated as a free parameter, as in \cref{ch:singlefield}, and considered to lie somewhere between $T_{\text{nuc}}$ and the Planck scale at $10^{19} \, \text{GeV}$. On the other hand, the abundance of curvature at the end of reheating can be obtained from 
\begin{equation}
    \Omega_{\kappa, \text{rh}} = -\frac{\kappa}{a_{\text{rh}}(T_{\text{rh}})H(T_{\text{rh}})},
\label{eq:ds.CalculatedCurvatureAbundance}
\end{equation}
where both the scale factor and the Hubble rate can be written in terms of the reheating temperature, assuming matter domination until the end of reheating, as we have seen in the previous chapter. From \cref{eq:ds.ObservedCurvatureAbundance,eq:ds.CalculatedCurvatureAbundance} one can obtain the value of $\kappa$ for which the curvature abundance is compatible with observations at each value of $T_{\text{rh}}$ considered. It turns out that compatible values of $\kappa$ are far from the one used in \cref{fig:ds.SpectraCurvedLargeK}. One must therefore go to very low wavenumbers (large enough wavelengths) so that the small spatial curvature is noticed by the corresponding modes. On the other hand, we have seen that the spectra shifts toward lower values of  $k$ with decreasing mass $m$. Therefore, if one aims at observing differences in production due to the curvature, one must consider very low masses, such that production of particles is important at wavenumbers of the order of $\abs{\kappa}$. This is precisely what is shown in \cref{fig:ds.SpectraCurvedSmallK}, where there are differences of four orders of magnitude in the total number density between the different curvatures. However, the associated abundance is very small. We therefore preliminary conclude that, even though the curvature of the spatial sections can have an effect on the production of particles, it is not large enough to be noticeable in sufficiently large abundances when one considers observational compatible values of $\kappa$. This contrasts with some results in the literature, where the effect of spatial curvature is more relevant, for example in the low multipoles of the CMB~\mbox{\cite{Bonga2016,Bonga2017,Hergt2022}}.

\section{Summary}

In this chapter we have studied cosmological production of scalar particles in universes with closed, flat and open spatial sections, assuming an inflationary epoch described by a de Sitter geometry. We have modeled the transition to the hot universe in two ways: Through a sudden transition to Minkowski, and using the adiabatic notion of vacuum to implement a smooth transition. Our preliminary results suggest that spatial curvature may have a large impact on the corresponding abundance, but only in scenarios in which the curvature abundance is very large and not compatible with observations, or for very small spectator field masses, such that the abundances are negligible. These results, however, are sensitive to the specific background dynamics and the choice of vacuum, and we will explore other scenarios in the future.

%% file: Chapters/Weakly-interacting_BECs_as_QFT_in_CS_Simulators.tex

\addtocontents{toc}{\protect\vspace{0.5em}}

\chapter{BECs as cosmological simulators} 

\label{ch:becstheory} 



In this chapter we show how the dynamics of the fluctuations in a BEC can be mapped to that of a scalar quantum field in a curved spacetime, and how the tuning of the background parameters of a $(1+2)$-dimensional BEC, such as a configurable, isotropic trap~\mbox{\cite{SaintJalm2019,Gauthier2021}}, and a time-dependent scattering length, engineered through Feshbach resonances \cite{Stwalley1976,Cornish2000,Chin2010}, can lead to analog particle production in a curved FLRW universe. We achieve this by parametrizing the fluctuations on top of the ground state in terms of the real and imaginary parts of the complex non-relativistic scalar field describing perturbations, as we discussed in \cref{sec:analogs.becs}. 

In \cref{sec:bec.IsotropicTraps}, we show how the resulting acoustic metric (see \cref{eq:analogs.AcousticMetricBEC}) can be engineered to correspond to that of an FLRW universe with possibly curved spatial sections, depending on the spatial dependence of the background density, that is, on the form of the trap. The analog of the scale factor is essentially the time-dependent scattering length, whose tuning allows us to simulate particle production in different types of expanding cosmologies, which is detailed in \cref{sec:bec.ParticleProduction}. The extension of previous works resides in the exact mapping to a more general class of cosmologies (allowing for spatial curvature), and in showing how the effect of particle production is accessible experimentally through the construction of what we will call the rescaled density contrast. We explore particle production for power-law scale factors in \cref{sec:bec.effects}. In \cref{sec:bec.Experiment}, we discuss the implementation of the analogy in a quasi two-dimensional BEC and show experimental results that benchmark this system as a QFTCS simulator. Lastly, in \cref{sec:bec.scattering}, we reinterpret cosmological production in terms of one-dimensional scattering in quantum mechanics, showing an additional analogy that is realized in a BEC, which provides a complementary perspective on our problem.

\section{Acoustic metric in a quasi-2D BEC}
\label{sec:bec.IsotropicTraps}

First, we discuss the acoustic metric and how it can be mapped to curved FLRW metrics for a quasi two-dimensional BEC that is confined in an isotropic trap. We describe the conditions under which the analogy holds, and study the dynamics of the fluctuations in the BEC, governed by such metric.

\subsection{Quasi two-dimensional geometry}
\label{subsec:bec.quasi2D}

As we have discussed in \cref{sec:analogs.becs}, a weakly-interacting BEC is described by a non-relativistic complex scalar field whose dynamics is characterized by the quantum effective action \eqref{eq:analogs.QuantumEffectiveAction}. In this chapter we particularize to a setup consisting on a disk-like geometry in cylindrical coordinates $\vec{r}=(r, \phi, z)$ and a condensate that is tightly confined in the $z$-direction, namely $l_z \ll l_r$, where~$l_z$ and $l_r$ are the condensate extension in the $z$ and longitudinal directions, respectively. This leads to a quasi $(1+2)$-dimensional geometry. Due to the strong confinement, the motion of the atoms in the $z$-direction is essentially frozen, and the mean field $\Phi (t, \vec{r})$ can be separated as $\Phi(t, \vec{r}) = \Psi (t, r, \varphi) \zeta (z)$, where $\zeta(z)$ is the ground state wavefunction along the $z$-direction \cite{Pitaevskii2016}, and the dynamics is encoded in $\Psi(t, r, \varphi)$, which is the field that enters in \cref{eq:analogs.QuantumEffectiveAction}.

We can account for small fluctuations on top of the ground state by means of the parametrization \eqref{eq:analogs.BackgroundSplit}, which was discussed in part~I. Then, the background follows the GP equation \eqref{eq:analogs.GrossPitaevskii} and the imaginary part of the fluctuations on top can be described by the action~\eqref{eq:analogs.QEACurvedSpacetime}, as long as we stay in the acoustic regime and the superfluid velocity vanishes. 

We will consider isotropic trapping potentials of the form
\begin{equation}
    V (t, r) = \frac{m}{2} \omega^2 (t) f(r),
    \label{eq:bec.TrappingPotential}
\end{equation}
where $\omega (t)$ is a time-dependent parameter and $f(r)$ is typically a polynomial in $r$. The potential above has to be understood as defined up to the condensate radius, which we will denote by $R$, point at which a hard wall is hit. We leave this implicit and focus in the following in the region $r < R$. Without loss of generality we can assume $f(0)=0$. For example, $f(r)=r^2$ corresponds to the commonly used harmonic trap, in which case $\omega (t)$ gives the trapping frequency. On the other hand, the interaction strength $\lambda = \lambda(t)$, which we consider time-dependent, can be expressed in terms of the $s$-wave scattering length~$\alpha_s(t)$ within Born's approximation~\cite{Pitaevskii2016} as
\begin{equation}
    \lambda (t) = \sqrt{\frac{ 8 \pi \omega_z \hbar^{3}}{m}} \alpha_s (t),
    \label{eq:bec.BornApproximation}
\end{equation} 
where $\omega_z$ is the trapping frequency in the $z$-direction.

\subsection{Stationary background density profile}
\label{subsec:bec.BackgroundDensity}

As we have discussed in \cref{sec:analogs.becs}, we want to consider a static background, so that background velocity $\vec{v}_0$ vanishes, and background density $n_0$ does not depend on time. In this scenario, the Bernouilli equation \eqref{eq:analogs.EulerBEC} reads
\begin{equation}
    0 = \hbar \partial_t S_0 (t)+ \frac{m}{2} \omega^2 (t) f(r) + \lambda (t) \, n_0 (r),
    \label{eq:bec.EulerEquationStatic}
\end{equation}
where we have neglected the quantum potential. From the above equation, it is straightforward to write the background density profile $n_0$ as
\begin{equation}
    n_0 (r) = -\frac{\hbar}{\lambda(t)}\partial_t S_0(t) - \frac{m\omega^2(t)}{2\lambda(t)}f(r),
\end{equation}
which implies that each term on the right-hand side must be time-independent. That is, we must have
\begin{equation}
    -\frac{\hbar}{\lambda(t)}\partial_t S_0(t) = \bar{n}_0 \quad \text{and} \quad \frac{m\omega^2(t)}{2\lambda(t)} = \frac{\bar{n}_0}{R^2},
\label{eq:bec.ConstantsBernouilli}
\end{equation}
where we have chosen the constants so that the background density profile acquires the form
\begin{equation}
    n_0 (r) = \bar{n}_0 \left(1 - \frac{f(r)}{R^2} \right), \quad \text{for} \quad r < R
    \label{eq:bec.BackgroundDensityProfile}
\end{equation}
with $\bar{n}_0$ denoting the background density at the center of the trap. At the condensate edge $R$, the density drops to zero due to the existence of hard walls in the potential (which are left implicit).

Note that the second equation in \eqref{eq:bec.ConstantsBernouilli} is nothing but a compatibility equation, which ensures that the background density profile is stationary, and can be rewritten~as 
\begin{equation}
    \omega^2(t) = \frac{2 \bar{n}_0}{m R^2} \, \lambda(t).
    \label{eq:bec.TrappingFrequencyTimeDependence}
\end{equation}
The above tells us how the trapping frequency (and by extension, the background phase $S_0$) must be changed when the coupling $\lambda(t)$ varies over time in order to guarantee a stationary density profile of the form~\eqref{eq:bec.BackgroundDensityProfile}. On the other hand, the total particle number $N$ is given by
\begin{equation}
    N = 2 \pi \bar{n}_0 \int_0^R \text{d} r \, r  \left(1 - \frac{f(r)}{R^2} \right).
\label{eq:bec.AtomNumber}
\end{equation}

\begin{figure}[t!]
    \centering
    \includegraphics[width=0.6\textwidth]{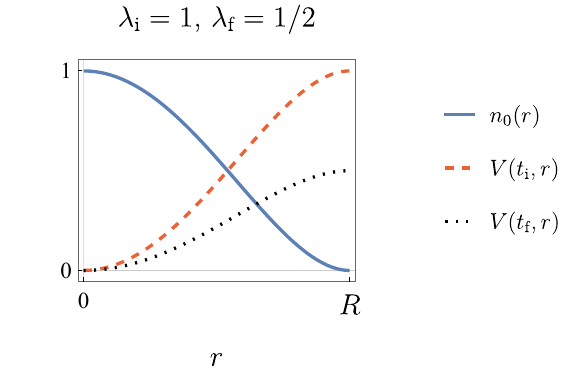}
    \caption[Density profile and time-dependent trapping potential for different values of the interaction strength and $\bm{v}=0$]{The density profile $n_0 (r) = \bar{n}_0 (1 - r^2/R^2)^2$ (blue solid curve) is shown together with the corresponding time-dependent trapping potential~\mbox{$V(t,r)$} at an initial time $\ti$ (red dashed curve), when the interaction strength is $\lambda_{\text{i}}=1$, and at a final time $\tf$ (black dotted line), when $\lambda_{\text{f}}=1/2$. For simplicity, we have set all SI units to $1$. Figure from \cite{BECPaper2022}.} 
    \label{fig:bec.DensityProfile}
\end{figure}

Let us discuss some particular choices for the radial dependence of the trap, encoded in $f(r)$. The simplest function we can implement is $f(r)=0$, which leads to a homogeneous density profile in $r < R$, allowing for a mapping to a flat FLRW cosmology (see e.g. \cite{Weinfurtner2007}), as we will see below. On the other hand, harmonic traps are implemented by
\begin{equation}
    f(r) = \pm r^2 \quad \text{and} \quad n_0 (r) = \bar{n}_0 \left(1 \mp \frac{r^2}{R^2}\right), 
    \label{eq:bec.HarmonicTrapDependence}
\end{equation}
where we allowed also for inverted harmonic traps (recall that hard walls are present at $r=R$, trapping the atoms in the condensate). This choice leads to the well-known Thomas-Fermi density profile, and~$R$ is called in this case the Thomas-Fermi radius. We will see in the following subsection that for these traps positively and negatively curved FLRW universe can be obtained only approximately, in a region around the center of the trap. However, in order to implement curved FLRW universes exactly, one needs to realize trap and density profiles of the form
\begin{equation}
    f(r) = \pm 2 r^2 - \frac{r^4}{R^2} \hspace{0.3cm} \text{and} \hspace{0.3cm} n_0 (r) = \bar{n}_0 \left[1 \mp \frac{r^2}{R^2} \right]^2.
    \label{eq:bec.TrapRadialDependenceExact}
\end{equation}
In \cref{fig:bec.DensityProfile} we illustrate how the potential needs to be adjusted in order to maintain a constant background density profile upon a change in scattering length. Note that decreasing the coupling $\lambda (t)$ implies decreasing also the parameter $\omega (t)$ accordingly, and therefore the confinement condition~\mbox{$\omega(t) \ll \omega_z$} is always fulfilled, provided it holds initially.

\subsection{From the acoustic metric to curved FLRW universes}
\label{subsec:bec.MetricToFLRW}
We showed in \cref{sec:analogs.becs} that restricting ourselves to static density profiles for the background density $n_0 (r)$, with $\vec{v}_0=0$, leads to an acoustic metric of the form \eqref{eq:analogs.AcousticMetricBEC}. For the explicit form of the background density profile \eqref{eq:bec.BackgroundDensityProfile}, the corresponding line element reads 
\begin{equation}
        \text{d} s^2 = - \text{d}t^2 + a^2(t) \left(1 - \frac{f(r)}{R^2}\right)^{-1} (\text{d}r^2 + r^2 \text{d} \phi^2),
\label{eq:bec.BLineElement}
\end{equation}
where we defined a time-dependent, analog scale factor in terms of the speed of sound $c_{\text{s}}$ as
\begin{equation}
    a^2(t) = \frac{1}{c_{\text{s}}^2(t)} = \frac{m \, }{\bar{n}_0 \lambda(t)}.
    \label{eq:bec.ScaleFactorDefinition}
\end{equation}
Crucially, we will work far from the edges of the condensate, and therefore effectively we can regard the radial coordinate $r$ as infinite when studying the dynamics of the fluctuating fields. Let us introduce now the particular trap profiles \eqref{eq:bec.TrapRadialDependenceExact} in the line element \eqref{eq:bec.BLineElement}, which yields
\begin{equation}
    \text{d} s^2 = - \text{d}t^2 + a^2(t) \left(1 \mp \frac{r^2}{R^2}\right)^{-2} (\text{d}r^2 + r^2 \text{d} \phi^2).
    \label{eq:bec.LabCoordinatesLineElement}
\end{equation}
Now, we perform the change of coordinates
\begin{equation}
	\rho(r)=\frac{r}{1 \mp \frac{r^2}{R^2}},
	\label{eq:bec.NewRadialCoordinate}
\end{equation}
where $\rho \in [0, \infty)$ for the negative sign, and $\rho \in [0, R/2]$ for the positive sign. This leads to the relation
\begin{equation}
	\frac{\text{d}r^2}{\left(1 \mp \frac{r^2}{R^2}\right)^2}= \frac{\text{d}\rho^2}{1 \pm 4\frac{\rho^2}{R^2}},
\end{equation}
so that the line element \eqref{eq:bec.LabCoordinatesLineElement} becomes
\begin{equation}
    \text{d} s^2 = - \text{d}t^2 + a^2(t)  \left(\frac{\text{d}\rho^2}{1 - \kappa \rho^2} + \rho^2 \text{d} \phi^2\right),
    \label{eq:bec.FLRWLineElement}
\end{equation}
which corresponds to the line element of $(1+2)$-dimensional FLRW universes \eqref{eq:cosmo.FLRWLineElementSpherical} with positive or negative spatial curvature $\kappa = \mp 4/R^2$. Crucially, the size of the condensate $R$ determines the value of the scalar curvature~$\kappa$, which allows the latter to be engineered in practice. As we mentioned above, decreasing the coupling corresponds to implementing an expansion of spacetime, and vice versa, in light of the relation between the scale factor $a(t)$ and the coupling $\lambda (t)$ (cf. \cref{eq:bec.ScaleFactorDefinition}). Lastly, recall that a flat FLRW universe can be implemented through a homogeneous background density profile~\mbox{$n_0= \text{const.}$}, such that the radial dependent prefactor in the spatial line element \eqref{eq:bec.BLineElement} is absent (and therefore no additional coordinate transformations are needed). This is the situation in a box trap, or in a sufficiently small region around the center of a harmonic trap.

In fact, in the case of harmonic traps with trap and density profiles~\eqref{eq:bec.HarmonicTrapDependence}, the acoustic metric becomes
\begin{equation}
    \text{d} s^2 = - \text{d}t^2 + a^2(t) \left(1 \mp \frac{r^2}{R^2}\right)^{-1} (\text{d}r^2 + r^2 \text{d} \phi^2).   
\end{equation}
If one performs now the coordinate transformation
\begin{equation}
	\rho(r)=\frac{r}{\left(1 \mp \frac{r^2}{R^2}\right)^{1/2}},
\end{equation}
and expands the denominator of the radial differential up to quadratic order in the new radial coordinate $\rho$, one obtains
\begin{equation}
    \frac{\text{d}r^2}{1 \mp \frac{r^2}{R^2}} = \frac{\text{d}\rho^2}{\left(1 \pm \frac{\rho^2}{R^2}\right)^{2}} \approx \frac{\text{d}\rho^2}{1 \pm 2\frac{\rho^2}{R^2}},
\label{eq:bec.HarmonicTrapCoordinateTransf}
\end{equation}
where the approximation in the last step is reasonable in a large region around the center of the trap. Then, the line element acquires the form~\eqref{eq:bec.FLRWLineElement} with $\kappa = \mp 2/R^2$, such that a harmonic trap leads to spatially curved FLRW universes in a macroscopic region around its center (typically up to $r \approx 0.4R$).

Summarizing, it is possible to exactly engineer positive, vanishing and negative spatial curvature through the implementation of
\begin{equation}
    f(r) = \begin{cases} 
    -2 r^2 - r^4/R^2  & \text{for} \hspace{0.3cm} \kappa > 0, \\
    0 & \text{for} \hspace{0.3cm} \kappa = 0, \\
    +2 r^2 - r^4/R^2  & \text{for} \hspace{0.3cm} \kappa < 0,
    \end{cases}
\label{eq:bec.TrapShape}
\end{equation}
whereas harmonic traps approximate these profiles in a sufficiently small region around the center of the condensate. We will give more details on the realization of different trap profiles in \cref{sec:bec.Experiment}.

\section{Particle production}
\label{sec:bec.ParticleProduction}

Now that we have discussed how to realize FLRW universes with different spatial curvature in a quasi two-dimensional BEC, let us study the dynamics of the fluctuations on top of the ground state and analyze the analog of cosmological particle production in this context. In particular, we will show the relation between correlations of density contrast in the BEC and the spectrum of produced particles~$S_k$ (cf.~\eqref{eq:qftcs.SpectrumOffsetAmpPhase}), which allows us to characterize particle production in the experiment.

\subsection{Klein-Gordon equation and mode functions}
\label{subsec:bec.ModeFunctions}
As we have seen in \cref{sec:analogs.becs}, the KG equation for the fluctuation field $\varphi$ in this particular case reads
\begin{equation}
        0 = \ddot{\varphi} + 2\frac{\dot{a}(t)}{a(t)}\dot{\varphi}  - \frac{\Delta}{a^2} \varphi,     
\label{eq:bec.KleinGordonFLRW}
\end{equation}
where the exact form of the Laplace-Beltrami operator $\Delta$ (cf. \cref{eq:qftcs.LaplaceBeltrami}) depends on the spatial curvature (see below).

Following a procedure similar to what was done in \cref{ch:deSitter}, we transform the radial coordinate $\rho$ to $u$ according to \cref{eq:ds.RadialCoordTransf},
\begin{align}
    \rho = \begin{cases}
    \abs{\kappa}^{-1/2}\sin{u} \quad &\text{with} \quad u \in [0, \pi), \quad \hspace{0.04cm}\text{for} \quad \kappa>0, \\
    u \quad &\text{with} \quad u \in [0, \infty), \quad \text{for} \quad \kappa=0, \\
    \abs{\kappa}^{-1/2}\sinh{u} \quad &\text{with} \quad u \in [0, \infty), \quad \hspace{0.008cm}\text{for} \quad \kappa<0.\\
    \end{cases}
    \label{eq:bec.u}
\end{align}
Since we are working in $D=2$, the Laplace-Beltrami operator takes the form~\eqref{eq:ds.LaplaceBeltrami} with the choice $\theta=\pi/2$, that is \cite{Ratra1995,Ratra2017,Argyres1989},
\begin{equation}
    \Delta = \frac{1}{\rho^2}\partial_{\rho}\left(\rho^2\partial_{\rho}\right) +\frac{1}{\rho^2}\partial_{\phi}^2,
    \label{eq:bec.LaplaceOperator}
\end{equation}
which is diagonalized in these geometries through the eigenfunctions~\mbox{$\mathcal{H}_{kq}(u, \phi)$}, the two-dimensional analog of the functions $\mathcal{H}_{klq}(u, \theta, \phi)$ we used in \cref{ch:deSitter} (for details see \cref{sec:bec.eigenfunctions}). The eigenvalues $h^2$ of $-\Delta$ in this case are given~by
\begin{equation}
    h^2(k) = \begin{cases}
    K(K+1)\abs{\kappa} & \text{for}\hspace{0.3cm} \kappa > 0, \\
    k^2 & \text{for}\hspace{0.3cm} \kappa = 0, \\
    \left(K^2 + \frac{1}{4}\right)\abs{\kappa} & \text{for}\hspace{0.3cm} \kappa < 0,
    \end{cases}
    \label{eq:bec.LaplaceEigenvalues}
\end{equation}
with the relation $K=k/\sqrt{\abs{\kappa}}$ as before, and the labels $k$ and $q$ run now over the values
\begin{align}
        K &= 0, 1, \text{...}, \infty; \quad q = -K, -K+1, \text{...}, K-1, K, \quad &&\text{for} \quad \kappa > 0, \\
        k &\in [0, \infty); \hspace{0.55cm} \quad q \in \mathbb{Z} \quad &&\text{for} \quad \kappa \leq 0.
\end{align}

Let us consider now the rescaled field $\chi =a^{1/2}\phi$ and expand it in terms of the corresponding eigenfunctions of the Laplace-Beltrami operator $\mathcal{H}_{kq}$,
\begin{equation}
    \chi(\eta, \bm{u}) = \int \dd{\mu}_{kq} \left[a_{kq} \mathcal{H}_{kq} (\bm{u}) v_k (\eta) + a^*_{kq} \mathcal{H}_{kq}^{*} (\bm{u}) v_k^* (\eta) \right],
\label{eq:bec.ModeExpansion}
\end{equation}
where $\bm{u} = (u, \phi)$, and the measure $\dd\mu_{kq}$ is defined as
\begin{equation}
    \int \dd\mu_{kq}  = \begin{cases}
    \sum_{K=0}^{\infty} \abs{\kappa}\frac{K+1/2}{2\pi} \sum_{q=-K}^{K} &\text{for} \hspace{0.2cm} \kappa > 0, \\
    \int_0^{\infty} \frac{\text{d}k}{2\pi} \, k \sum_{q=-\infty}^{\infty} &\text{for} \hspace{0.2cm} \kappa = 0,\\
    \int \frac{\text{d}K}{2\pi} \,\abs{\kappa} K \tanh(\pi K) \sum_{q=-\infty}^{\infty} &\text{for} \hspace{0.2cm} \kappa < 0,
    \end{cases}
\label{eq:bec.MomentumIntegral}
\end{equation}
for the two-dimensional momentum integral. Canonical quantization follows from promoting the coefficients of the linear expansion to the creation and annihilation operators $\hat{a}^{\dagger}_{kq}$ and $\hat{a}_{kq}$, as we have done before, which fulfill the commutation relations
\begin{equation}
    \begin{split}
        &[\hat{a}_{kq}^{\dagger},\hat{a}_{k^{\prime}q^{\prime}}^{\dagger}] = [\hat{a}_{kq},\hat{a}_{k^{\prime}q^{\prime}}] = 0, \\
        &[\hat{a}_{kq},\hat{a}_{k^{\prime}q^{\prime}}^{\dagger}] = 2\pi \delta_{qq^{\prime}} \begin{cases}
        \frac{\delta_{KK'}}{\abs{\kappa}(K+1/2)} & \text{for}\hspace{0.3cm} \kappa > 0, \\
          \frac{\delta(k-k^{\prime})}{k} & \text{for}\hspace{0.3cm} \kappa = 0,  \\
          \frac{\delta(K-K')}{\abs{\kappa}K \tanh(\pi K)} & \text{for}\hspace{0.3cm} \kappa < 0.
        \end{cases}
    \end{split}
    \label{eq:bec.CommutationRelationsBosonicOperators}
\end{equation}

On the other hand, the mode functions $v_k(\eta)$ are normalized according to the Wronskian\footnote{When restoring units, an extra $\hbar$ factor in $\text{Wr}[v_k, v_k^*] = i\hbar$ has to be included.} \eqref{eq:qftcs.WronskianModes} and follow the mode equation \eqref{eq:qftcs.ModeEquation} with a frequency \eqref{eq:qftcs.MasterFrequency} corresponding to $D=2, m=\xi=0$, namely
\begin{equation}
    \omega_k^2(\eta) = h^2(k) + \frac{1}{4} \left(\frac{a^{\prime}(\eta)}{a(\eta)}\right)^2 - \frac{1}{2}\frac{a^{\prime\prime}(\eta)}{a(\eta)}.
\label{eq:bec.ModeFrequency}
\end{equation}
Note that the influence of the spatial curvature $\kappa$ is fully encoded in $h^2(k)$, because the analog scale factor is determined by the ground state properties (cf. \cref{eq:bec.ScaleFactorDefinition}) and not by the Friedmann equation \eqref{eq:cosmo.FriedmannEquation}. This is a difference with respect to what happens in the de Sitter scenario studied in \cref{ch:deSitter}. There, the background is essentially given by the Friedmann equation, which incorporates spatial curvature. In turn, each $\kappa$ case leads to different scale factors. Here, background equations are not given by General Relativity (as it typically happens in analog models of gravity, where the analogy is at the level of the dynamics of the fields), and the time-dependence of the analog scale factor is engineered differently, independently of the spatial curvature (in this case through a change in the interaction strength). On the scales relevant for typical experiments, i.e. for $R \sim 10^{-5}$ m, the spatial curvature $\kappa$ becomes irrelevant for $k/\sqrt{\bar{n}_0} \gtrsim  0.1$.

\subsection{Implementation of an \textit{in-out} production process}
\label{subsec:bec.BogoliubovTrafo}

In the experiment, we can realize an \textit{in-out} process of the type described in subsection \ref{subsec:qftcs.inout} in an exact manner (meaning that such regions are not reached asymptotically). We will start with the system in the \textit{in} region (which we denote in the following by region I), where the scale factor is constant,~\mbox{$a(\eta)=a_\text{i}$}. At some time $\etai$, the scale factor becomes time-dependent, until it becomes constant again at time $\etaf$, when the \textit{out} region (or region III) is reached. As we have discussed several times already, in regions I and III there exist preferred notions of vacuum. In particular, the choice of mode functions~$v_k(\eta)$ such that (cf. \cref{eq:qftcs.InModeIC})
\begin{equation}
    v_k (\eta) = \frac{1}{\sqrt{2\hbar \omega_{k, \text{i}}}}e^{-i \omega_{k, \text{i}} \eta}, \quad \text{for} \quad \eta \leq \eta_\text{i},
    \label{eq:bec.planewavetime}
\end{equation}
with $\omega_{k, \text{i}} = h(k)$, leads to the \textit{in} vacuum state $\ket{0^a}$, whereas the choice $u_k(\eta)$ that satisfies (see \cref{eq:qftcs.OutModeIC})
\begin{equation}
    u_k (\eta) = \frac{1}{\sqrt{2\hbar \omega_{k, \text{f}}}}e^{-i \omega_{k, \text{f}} \eta}, \quad \text{for} \quad \eta \geq \eta_\text{f},
    \label{eq:bec.uPlaneWaves}
\end{equation}
where $\omega_{k, \text{f}} = \omega_{k, \text{i}} = h(k)$, defines the \textit{out} vacuum state $\ket{0^b}$. Note that the frequencies coincide because $m=0$ (see the general form of the frequency~\eqref{eq:qftcs.MasterFrequency}). We will consider that the system is initially prepared either in the preferred vacuum in region I, namely $\ket{0^a}$, or in a thermal state with temperature~$T$. On the other hand, in the intermediate times, $\eta_\text{i} < \eta < \eta_\text{f}$ (called region II in the following), the notion of vacuum becomes ambiguous. Moreover, particle production due to the spacetime dynamics occurs. Furthermore, let us stress that in region II, the solutions to the mode equation \eqref{eq:qftcs.ModeEquation} are not in general of plane wave form.

In order to extract the number density of produced particles in region~III, one needs to compare the \textit{in} and \textit{out} notions of vacuum through the corresponding Bogoliubov transformation \eqref{eq:qftcs.BogoliubovTransformationModes} between the modes $v_k(\eta)$ and $u_k(\eta)$. In this particular scenario, the relation between the two sets of creation and annihilation operators \eqref{eq:qftcs.BogoliubovTransformationOperators} is given by
\begin{equation}
    \hat{b}_{kq} =  \alpha_k^* \hat{a}_{kq} - \beta_k^* (-1)^q \hat{a}^{\dagger}_{k,-q}, \qquad \hat{a}_{kq} = \alpha_k \hat{b}_{kq} + \beta_k^* (-1)^q \hat{b}^{\dagger}_{k,-q},
\label{eq:bec.BogoliubovTransformationOperators}
\end{equation}
where $\hat{b}_{\bk} = (-1)^q \hat{b}_{k, -q}$ (cf. \cref{ch:qftcs}) for our choice of Laplace-Beltrami eigenfunctions $\mathcal{H}_{kq}$.

As in part II of the thesis, the task is to solve the mode equation~\eqref{eq:qftcs.ModeEquation} in all three regions I, II and III for the \textit{in} and \textit{out} solutions determined by the initial conditions~\eqref{eq:bec.planewavetime} and \eqref{eq:bec.uPlaneWaves}, and to identify the Bogoliubov coefficients $\alpha_k$ and $\beta_k$ through \cref{eq:qftcs.BogoliubovCoefficientsWronskian}, from which all quantities related to particle production can be derived. The conceptual difference in the context of analog experiments lies in the fact that here the \textit{in} and \textit{out} regions are not asymptotic or approximated, but realized in a controlled manner. In particular, we assume that transitions between the regions are instantaneous, such that the scale factor is stationary in regions~I and III, and suddenly becomes time-dependent in region II. This is a reasonable assumption, since the timescale of the expansion is typically much shorter than the timescale of the dynamics of the phonons, which nevertheless requires further theoretical treatment to account for these transitions in the modes. We will discuss this in subsection~\ref{subsec:bec.InstantaneousSwitch}. 

\begin{figure}[t!]
    \centering
    \includegraphics[width=0.5\textwidth]{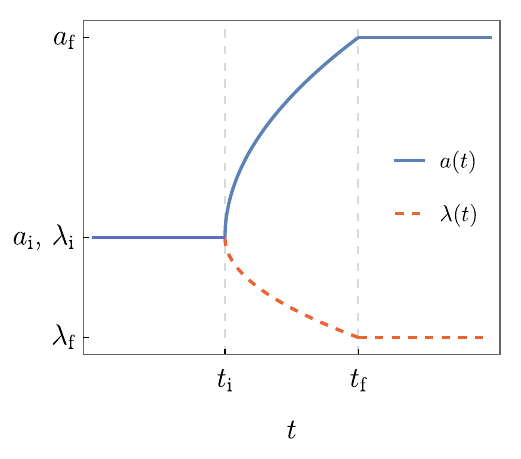}
    \caption[Time dependence of the scale factor and the interaction strength in a typical BEC experiment]{Scale factor $a(t)$ (blue solid line) and interaction strength $\lambda(t)$ (red dashed line) as a function of time in a typical experimental realization. We have set all constants and units to $1$. Figure from \cite{BECPaper2022}.} 
    \label{fig:bec.Expansion}
\end{figure}

In \cref{fig:bec.Expansion} we illustrate the evolution of the scale factor $a(t)$ and the interaction strength $\lambda(t)$---which determines the former through~\eqref{eq:bec.ScaleFactorDefinition}---during an experimental realization of the analog expansion. Note that we used cosmological time $t$ for illustrational purposes, since it corresponds to the laboratory time. 

\subsection{Instantaneous switch-on and -off of the analog expansion}
\label{subsec:bec.InstantaneousSwitch}

We have described the \textit{in-out} process as divided in three different regions. The scale factor is constant in regions I and III, and changes with time in II. However, the transition between regions is instantaneous, which leads to the appearance of Dirac delta contributions in the second derivative of the scale factor, and therefore in the frequency~\eqref{eq:qftcs.MasterFrequency}. Indeed, the derivative of the scale factor can be written as
\begin{equation} 
        a^{\prime}(\eta)=a_{\text{II}}(\eta)\left[\Theta(\eta-\eta_{\text{i}})-\Theta(\eta-\eta_{\text{f}})\right],
\label{eq:bec.DerivativeScaleFactor}
\end{equation}
where $a_{\text{II}}(\eta)$ is the scale factor in region II, and $\Theta$ is the Heaviside function. In order to extend e.g. the \textit{in} solution $v_k$ with initial conditions \eqref{eq:bec.planewavetime} at $\eta=\eta_{\text{i}}$ all the way to~$\eta_{\text{f}}$ in order to compare with $u_k$, we need to treat these discontinuities.

To do so, let us integrate the mode equation \eqref{eq:qftcs.ModeEquation} in an open interval of length~$2\epsilon$ around $\eta_* = \eta_{\text{i}/\text{f}}$, which leads to
\begin{equation}
    v_k^{\prime}(\eta_* + \epsilon) - v_k^{\prime}(\eta_* - \epsilon) = - \int_{\eta_* - \epsilon}^{\eta_* + \epsilon} \text{d}\eta \, \omega_k^2(\eta) v_k(\eta).
\label{eq:bec.JumpDerivative}
\end{equation}
Note that second derivatives of the scale factor will appear in the frequency as long as we are not in the conformally coupled scenario, namely as long as~\mbox{$\xi \neq (D-1)/4D$}. Since in the BEC setup we have $D=2$ and $\xi=0$, the right-hand side of \cref{eq:bec.JumpDerivative} does not vanish, and the mode function $v_k$ is differentiable only at first order. Integrating the above equation yields the jump in the derivative
\begin{equation}
    v_k^{\prime}(\eta_*^-) =  v_k^{\prime}(\eta_*^+) \pm  \frac{1}{2} \frac{a^{\prime}_{\text{II}}(\eta_*)}{a(\eta_*)} v_k(\eta_*),
\label{eq:bec.BoundaryConditionDerivative}
\end{equation}
where the negative sign corresponds to $\eta_{\text{i}}$, the positive sign to $\eta_{\text{f}}$, and $a^{\prime}_{\text{II}}$ is the derivative of the scale factor in region II. This jump is characteristic of non-conformally invariant universes, and significantly enhances particle production, as we will discuss in detail in \cref{ch:switcheffects}. For completeness, we state the boundary conditions for the mode function $v_k$, which is continuous throughout the three regions, namely one has to impose
\begin{equation}
    v_k(\eta_*^-) = v_k(\eta_*^+).
    \label{eq:bec.BoundaryConditionMode}
\end{equation}

Taking into account boundary conditions \eqref{eq:bec.BoundaryConditionDerivative} and \eqref{eq:bec.BoundaryConditionMode} as well as the initial condition \eqref{eq:bec.planewavetime}, one can obtain the \textit{in} mode at all times, in particular at $\eta_{\text{f}}$.

\subsection{Rescaled density contrast, two-point functions and spectrum of fluctuations}
\label{subsec:bec.CorrelationFunctionAndSpectrum}
Let us define the rescaled density contrast
\begin{equation}
    \delta_c (t, \bm{u}) = \sqrt{\frac{n_0 (u)}{\bar{n}_0^3}} \left[n (t, \bm{u}) - n_0 (u)\right],
    \label{eq:bec.RescaledDensityContrast}
\end{equation}
where $n (t,\bm{u}) = \abs{\Psi(t,\bm{u})}^2$ is the full condensate density, and we recall that $n_0$ is the background density, with $\bar n_0$ denoting its value at the center of the trap. As it is defined above, the rescaled density contrast is dimensionless. Importantly, using the relation between the perturbations \eqref{eq:analogs.FluctuationFieldsRelation} and \eqref{eq:analogs.ParametrizationsRelation}, one finds that, to leading order, $\delta_c \sim \partial_t \varphi$. Additionally, the prefactor in \cref{eq:bec.RescaledDensityContrast} is constant only in the case of a flat background density profile. 

From the density contrast \eqref{eq:bec.RescaledDensityContrast} we can build the corresponding equal time two-point correlation function, namely $\braket{\delta_c (t,\bm{u}) \delta_c (t,\bm{u}')}$, which is a typical observable in modern ultracold atom experiments. To leading order in fluctuations, one has that $\braket{n(t, \bm{u}}) = n_0 (u)$, and therefore the density contrast correlator is proportional to the two-point correlation function of the derivative of the phononic field $\varphi$. In particular, for times $t \geq t_{\text{f}}$ when the expansion has ceased, one has that
\begin{equation}
    G_{n n} (t; \bm{u}, \bm{u}') = \braket{\delta_c (t,\bm{u}) \delta_c (t,\bm{u}')} = \frac{\hbar^2 a^4}{m\bar{n}_0} \, \mathcal{G}_{\dot{\varphi} \dot{\varphi}} (t, \bm{u}, \bm{u}'),
    \label{eq:bec.DensityDensityCorrelatorLeadingOrder}
\end{equation}
where
\begin{equation}
   \mathcal{G}_{\dot{\varphi} \dot{\varphi}} (t,\bm{u}, \bm{u}')  = \frac{1}{2}\braket{\{\dot{\varphi}(t,\bm{u}), \dot{\varphi}(t,\bm{u}^{\prime})\}}.
    \label{eq:bec.PhiDotPhiDotCorrelatorDefinition}
\end{equation}
For $t\geq t_{\text{f}}$, the scale factor is constant, and the relation \eqref{eq:qftcs.FieldsRelation} between the fields~$\varphi$ and~$\chi$ simplifies, leading to $\dot{\varphi} = a_{\text{f}}^{-3/2} \chi^{\prime}$ in the case of two spatial dimensions. Therefore, the two-point function \eqref{eq:bec.PhiDotPhiDotCorrelatorDefinition} can be easily written in terms of the correlator~$\mathcal{G}_{\chi^{\prime}\chi^{\prime}}$ defined in \cref{eq:qftcs.ChiDotChiDotCorrelator}, and thus   
\begin{equation}
    G_{n n} (t; \bm{u}, \bm{u}') = \frac{\hbar^2 a_{\text{f}}}{m\bar{n}_0} \, \mathcal{G}_{\chi^{\prime}\chi^{\prime}} (t, \bm{u}, \bm{u}'),
    \label{eq:bec.DensityDensityCorrelatorChi}
\end{equation}
which, as expected from spatial homogeneity of the FLRW universes and discussed in \cref{sec:qftcs.powerspectrum}, does not depend separately on the two spatial positions $\bm{u}$ and $\bm{u}^{\prime}$, but only on the comoving distance $L$ between them. The latter is explicitly given by
\begin{equation}
        L = \begin{cases}
        \frac{1}{\sqrt{\abs{\kappa}}}\cos^{-1}\left(\cos{u}\cos u^{\prime} + \sin u \sin u^{\prime} \cos\Delta\phi\right) &\text{for} \hspace{0.1cm} \kappa > 0, \\
        \left[u^2+u^{\prime 2}-2uu^{\prime}\cos\Delta\phi\right]^{1/2} &\text{for} \hspace{0.1cm} \kappa = 0, \\
        \frac{1}{\sqrt{\abs{\kappa}}} \cosh^{-1}\left(\cosh u \cosh u^{\prime} - \sinh u \sinh u^{\prime} \cos\Delta\phi\right) &\text{for} \hspace{0.1cm} \kappa < 0,
        \end{cases}
        \label{eq:bec.ComovingDistance}
\end{equation}
where $\Delta\phi = \phi - \phi^{\prime}$. Through \eqref{eq:bec.DensityDensityCorrelatorChi}, this observable, defined in \eqref{eq:bec.RescaledDensityContrast}, acquires the symmetries of the geometry, 
\begin{equation}
    G_{n n} (t; \bm{u}, \bar{\bm{u}}') = G_{nn} (t,L).
\label{eq:bec.DensityDensityCorrelatorSymmetry}
\end{equation}
Note that the relation \eqref{eq:bec.DensityDensityCorrelatorChi} is a result of the normalization in the density contrast definition \eqref{eq:bec.RescaledDensityContrast}, chosen as such to ensure that \cref{eq:bec.DensityDensityCorrelatorSymmetry} holds. Indeed, writing the density correlation function \eqref{eq:bec.DensityDensityCorrelatorChi} in terms of the spectrum of fluctuations\footnote{Note that an $\hbar$ seems missing. For simplicity, we have decided to write here $S_k$ without the~$\hbar$ factor that would come from the modes $v_k$ upon unit restoration (see \cref{eq:qftcs.SpectrumDefinition}), such that the spectrum is always dimensionless.} \eqref{eq:qftcs.SpectrumBogoliubov} explicitly shows the $L$ dependence,
\begin{equation}
    G_{n n} (t, L) = \frac{\hbar a_\text{f}}{\bar{n}_0 m} \int \dd \tilde{\mu}_k \, \mathcal{F}_k(L) h(k) S_k (t).
    \label{eq:bec.DensityDensityCorrelatorSpectrum}   
\end{equation}
We introduced here the integration measure $\dd \tilde{\mu}_k$, which excludes the sum over $q$,
\begin{equation}
    \int \dd \tilde{\mu}_k  = \begin{cases}
    \sum_{K=0}^{\infty} \abs{\kappa}\frac{K+1/2}{2\pi} &\text{for} \hspace{0.2cm} \kappa > 0, \\
    \int \frac{\text{d}k}{2\pi} \, k  &\text{for} \hspace{0.2cm} \kappa = 0,\\
    \int \frac{\text{d}K}{2\pi} \,\abs{\kappa} K \tanh(\pi K) &\text{for} \hspace{0.2cm} \kappa < 0,
    \end{cases}
\end{equation}
as well as the functions $\mathcal{F}_k$, defined through
\begin{equation}
    \int \dd \mu_k \mathcal{H}_{\vk}(\bm{u})\mathcal{H}_{\vk}(\bm{u}^{\prime}) = \int \dd \tilde{\mu}_k \mathcal{F}_k(L),
\end{equation}
which allow us to simplify the two-point function by performing the sum over $q$ in~\eqref{eq:qftcs.ChiDotChiDotCorrelator} since the eigenvalue, and thus the mode functions $v_k$, are independent of this wavenumber. Explicitly, one has that
\begin{equation}
    \mathcal{F}_k(L) = \begin{cases}
    P_K\left(\cos{\left(L \sqrt{\abs{\kappa}}\right)}\right) &\text{for} \hspace{0.2cm} \kappa > 0, \\
     J_0\left(kL\right) &\text{for} \hspace{0.2cm} \kappa = 0, \\
    P_{iK - 1/2}\left(\cosh{\left(L \sqrt{\abs{\kappa}}\right)}\right) &\text{for} \hspace{0.2cm} \kappa < 0.
    \label{eq:bec.Kernels}
    \end{cases}
\end{equation}

Lastly, let us recall that an initial state at time $t_\text{i}$ with non-vanishing occupation number, such as a thermal state, would lead to stimulated particle production, as discussed in \cref{sec:qftcs.powerspectrum} (see also \cite{Calzetta2008}). We will consider here thermal initial states characterized by an initial occupation number density of the form 
\begin{equation}
    n^{\text{in}}_k (T) = \frac{1}{e^{\hbar \omega_{k,\text{i}}/(k_\text{B} T)}-1},
    \label{eq:bec.Thermal}
\end{equation}
where $T$ denotes the temperature. 

In the following, to set a temperature scale, we use the critical temperature $T_\text{c}$ of an ideal gas in an anisotropic trap, which in the case of our quasi-2D condensate is given by \cite{Giorgini1996,Dalfovo1999}
\begin{equation}
    T_{\text{c}} = \frac{\hbar \omega }{k_\text{B}}\left(\frac{N}{\zeta(2)} \right)^{1/2},
    \label{eq:bec.CriticalTemperature2D}
\end{equation}
where $N$ is the total number of atoms.

\subsubsection*{Ultraviolet divergence and regularization}

As any two-point function, $G_{nn}$ shows an ultraviolet divergence at the coincidence point \cite{Mukhanov2007}. In order to make predictions for the measurement of the rescaled density contrast correlation function \eqref{eq:bec.DensityDensityCorrelatorSpectrum}, the latter has to be regularized. Importantly, there exist a natural regularization in the context of BECs, coming from the resolution of the measurement apparatus. Additionally, one should keep in mind that the acoustic approximation that we use here is a low momentum effective description, and therefore the analogy does not hold in the ultraviolet, where we would find this problem.
 
In order to implement the above, let us consider smeared-out fields of the form
\begin{equation}
    \Phi(t,\vec{r})=\int \text{d}^2 r^\prime W (\vec{r}-\vec{r}') \phi(t,\vec{r}'), 
\end{equation}
with $W (\vec{r}-\vec{r}')$ denoting a window function, normalized as
\begin{equation}
    \int d^2 r^\prime W(\vec r^\prime) = 1.
\end{equation}
This function acts as an ultraviolet regulator in momentum space, so that the regularized rescaled density contrast correlation function can be written as
\begin{equation}
    \begin{split}
        G_{nn} (t,L) = \frac{\hbar a_{\text{f}}}{\bar{n}_0 m} \int_k \mathcal{F}(k,L)h(k) S_k (t) 
        \tilde{f}_G(k),
    \end{split}
    \label{eq:bec.DensityDensityCorrelatorConvoluted}
\end{equation}
where $\tilde{f}_G(k) = \tilde{W}^* (k)\tilde{W} (k)$ corresponds to the absolute square of the Fourier transformed window function\footnote{In the ultraviolet, spatial curvature can be neglected.}. In the following we work with a window function of Gaussian form with width $w$ in position space (as a function of the comoving distance), such that in the absence of spatial curvature,
\begin{equation}
  \tilde f_G(k) = \tilde W^*(k)\tilde W(k)  = e^{-w^2 k^2}.
   \label{eq:bec.GaussianRegulator}
\end{equation}

\section{Particle production sourced by power-law scale factors}
\label{sec:bec.effects}

Let us now consider cosmological production of particles in our analog system through the study of the spectrum of fluctuations and the rescaled density contrast correlation function in different experimental scenarios. The exact numerical parameters are given in \cref{sec:param.bec}.

We will analyze expansions of different duration $\Delta t$, for several hold times $t_{\text{hold}}$ after the expansion has ceased, the effects of spatial curvature and temperature, and we will consider power-law scale factors of the form
\begin{equation}
    a(t) = Q \, \abs{t - t_0}^{\gamma},
\label{eq:bec.ScaleFactorPolynomial}
\end{equation}
where $Q= a(t=0) \, \abs{t_0}^{- \gamma} > 0$ and $t_0$ are free parameters that can be adjusted in experiments\footnote{The units of $Q$ depend on the power $\gamma$.}. Note that radiation dominated ($\gamma=2/3$) and matter dominated ($\gamma=1$) universes (cf. \cref{ch:cosmo}) are contained in the general form \eqref{eq:bec.ScaleFactorPolynomial}.

In order to extract the spectrum of fluctuations $S_k$, we compare the \textit{in} and \textit{out} modes $v_k$ and $u_k$ at $\eta_{\text{f}}$. Since $u_k(\eta_{\text{f}})$ is already given by the choice of initial conditions \eqref{eq:bec.uPlaneWaves}, the task is to obtain $v_k(\eta_{\text{f}})$. For this, we start with the initial condition \eqref{eq:bec.planewavetime}, and extend it to regions II and III by matching general solutions of the mode equation in these regions to the \textit{in} mode $v_k$ at both $\eta_{\text{i}}$ and $\eta_{\text{f}}$ following the discussion in subsection \ref{subsec:bec.InstantaneousSwitch}. General solutions in region III are given by plane waves with frequency $\omega_{k,\text{f}}$, as we have discussed, whereas the general form of modes in region II can be found in \cite{CosmologyPaper2022} for the different power-law scale factors considered.  

\subsection{Initial vacuum state}

\begin{figure}[t!]
    \centering
    \includegraphics[width=0.85\textwidth]{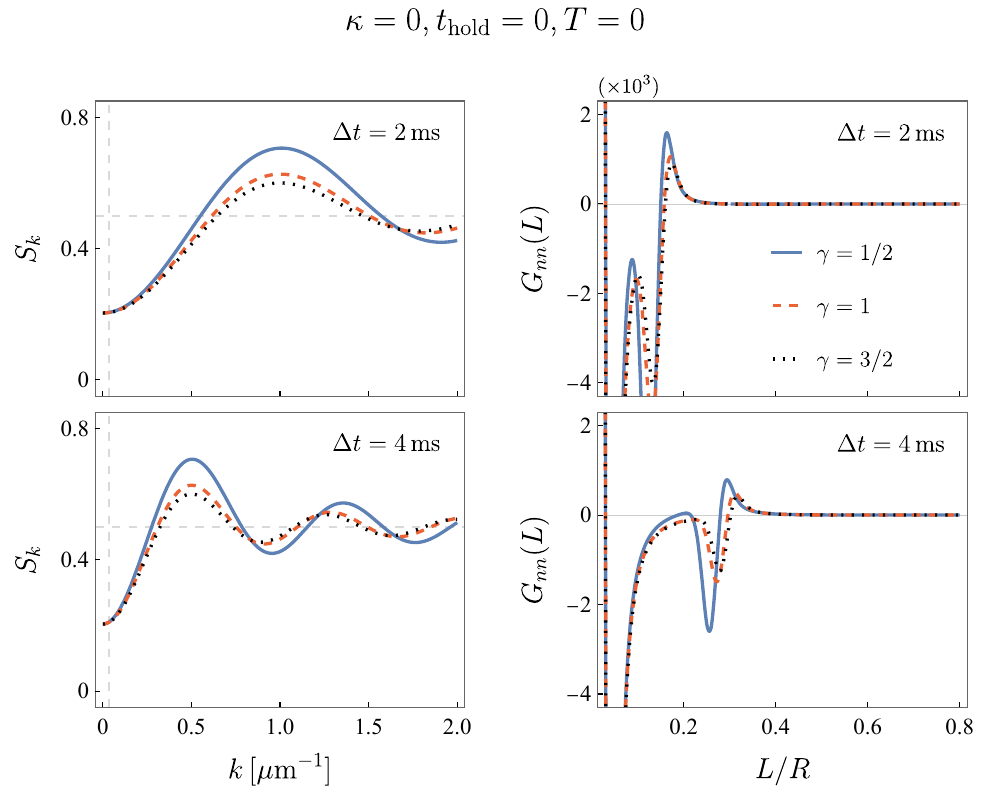}
    \caption[Spectrum of fluctuations as a function of the radial wavenumber together with its corresponding rescaled density contrast correlator as a function of the comoving distance for different types of power-law expansion and zero temperature]{Spectrum of fluctuations $S_k$ as a function of the radial wavenumber $k$ together with its corresponding density contrast correlator $G_{nn}$ as a function of the comoving distance $L$ measured in units of the parameter $R$, for a vacuum initial state, right after the expansion ($\tho = 0$). We consider power-law scale factors with $\gamma=1/2$ (decelerating), $\gamma=1$ (uniform), and $\gamma=3/2$ (accelerating), and an expansion of duration $\Delta t = 2 \, \text{ms}$ (top row) and $\Delta t = 4 \, \text{ms}$ (bottom row). A gray, vertical dashed line in the spectrum plots denotes the resolution in~$k$ as the inverse of the condensate size, whereas a similar horizontal line denotes the value $1/2$. The position space results are obtained after regularization with a Gaussian window function of width $w= 0.5 \,\mu$m. Figure adapted from~\cite{BECPaper2022}.}
    \label{fig:bec.RampDurationGammas}
    \end{figure}

    We first consider that our system is initially prepared in the \textit{in} vacuum. In \cref{fig:bec.RampDurationGammas} we show the spectrum of fluctuations $S_k$ as a function of the wavenumber $k$ together with the density contrast two-point function $G_{nn}$ as a function of $L/R$ in a flat analog FLRW universe, right after the expansion. We have chosen a scale factor ratio of~\mbox{$a_\text{f}/a_\text{i}=\sqrt{6}$} and three examples of power-law expansions: decelerating ($\gamma=1/2$), uniform ($\gamma=1$), and accelerating ($\gamma=3/2$). In the top row of \cref{fig:bec.RampDurationGammas} we show the result of production right after an expansion of duration $\Delta t = 2 \, \text{ms}$, whereas in the bottom row we have taken $\Delta t = 4 \, \text{ms}$. In all panels, we observe very similar results for the three different scale factors. Interestingly, the results for an accelerating universe are closer to the case of uniform expansion than those corresponding to a decelerating universe, which yields larger amplitudes both in the spectrum $S_k$ and the two-point correlator $G_{nn}$. On the other hand, we observe that the characteristic features of the spectrum are shifted toward lower $k$ for longer experiments, since the expansion is more adiabatic. Correspondingly, the anticorrelation-correlation pair in position space appears at larger distances and its amplitude decreases for larger $\Delta t$. As expected, the spectrum converges to $1/2$ for large wavenumbers (denoted by the horizontal dashed line), which do not feel spacetime curvature. Interestingly, for $\gamma=1$, particle production does not occur for certain wavenumbers (i.e. $\beta_k = 0$). We will discuss this further in subsection~\ref{subsec:bec.phase}, as well as in \cref{sec:bec.scattering}. The regularization does not affect the position of correlations, except for the autocorrelation, which moves to $L=0$ for vanishing width $w$. 

    \begin{figure}[t!]
    \centering
    \includegraphics[width=0.85\textwidth]{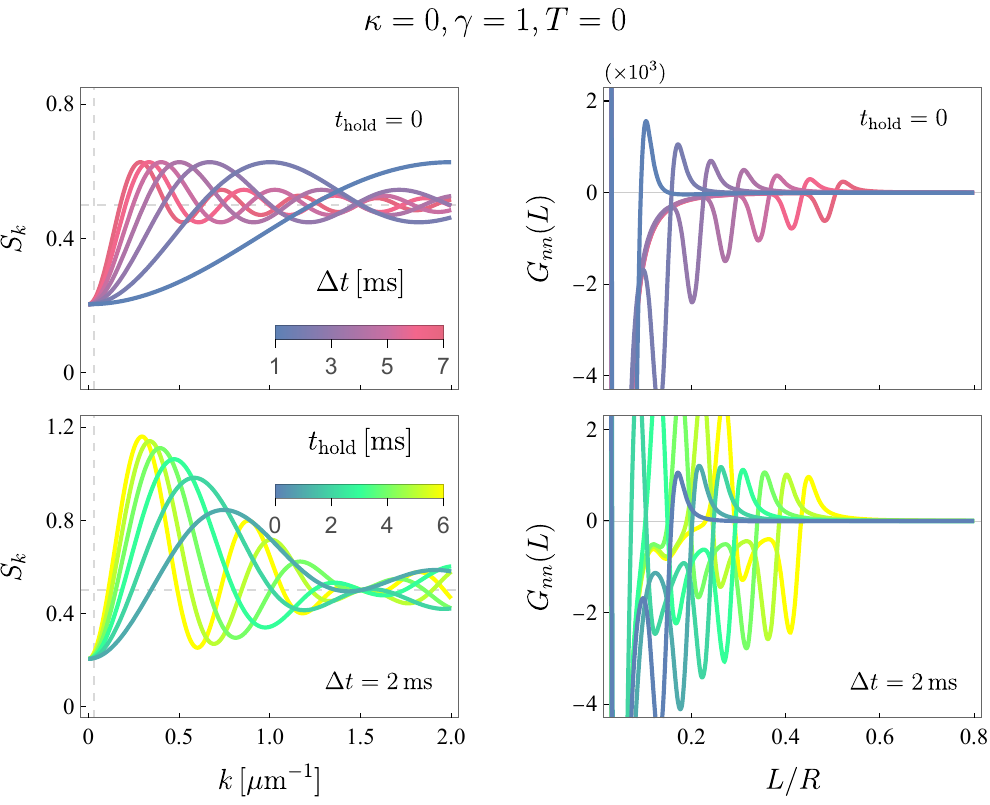}
    \caption[Spectrum of fluctuations as a function of the radial wavenumber together with its corresponding rescaled density contrast correlator as a function of the comoving distance for different expansion durations and hold times]{Spectrum of fluctuations $S_k$ as a function of the radial wavenumber $k$ together with its corresponding density contrast correlator $G_{nn}$ as a function of the comoving distance $L$ measured in units of the parameter $R$, for a vacuum initial state. We fix now $\gamma=1$, and consider the results right after the expansion for different values of $\Delta t$ (upper row), and different hold times $\tho$ (lower row) for fixed $\Delta t =2 \, \text{ms}$. The gray, vertical dashed line in the spectrum plots denotes the resolution in $k$ as the inverse of the condensate size, whereas the horizontal one denotes the value $1/2$. The position space results are obtained after regularization with a Gaussian window function of width~\mbox{$w= 0.5 \,\mu$m}. Figure adapted from \cite{BECPaper2022}.}
    \label{fig:bec.RampAndHoldDependence}
    \end{figure}

    In \cref{fig:bec.RampAndHoldDependence}, we focus on the case $\gamma=1$ and study both the expansion duration dependence and the evolution of the system after the expansion has ceased. The results on the top row are consistent with those of \cref{fig:bec.RampDurationGammas}, showing that upon increasing the duration of the expansion, the maximum of the spectrum stays the same, but the features are shifted toward lower wavenumbers. However, the amplitude of the correlations does decrease (see top right panel of \cref{fig:bec.RampAndHoldDependence}). In the lower row, we fix $\Delta = 2 \, \text{ms}$ and vary the hold time $\tho$. This causes the spectrum to oscillate around the offset $S_{k,0}$ while moving toward the infrared. In position space, correlations propagate with twice the speed of sound at the final scattering length $\asf$, with decreasing amplitude. Note that an additional correlation peak appears after some $t_{\text{hold}}$.

    \subsection{Initial thermal state}
\label{sec:bec.ThermalInitialState}

\begin{figure}[t!]
    \centering
    \includegraphics[width=0.85\textwidth]{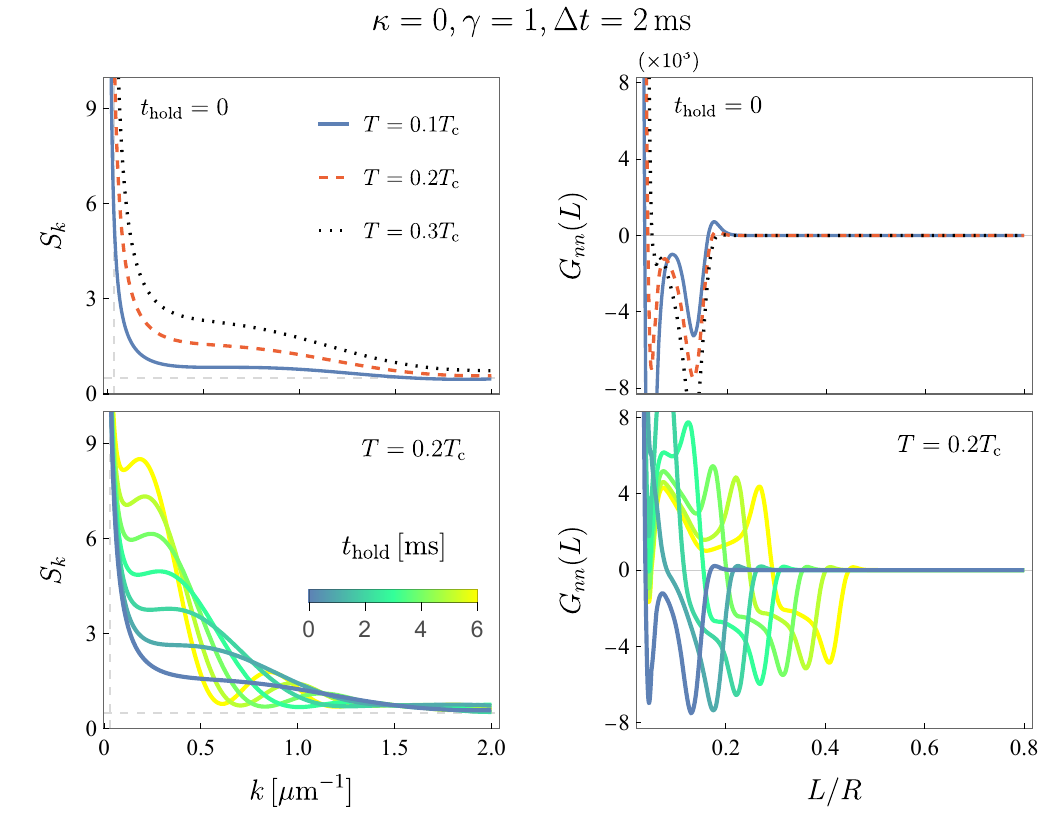}
    \caption[Spectrum of fluctuations and density contrast correlation for thermal initial states with different temperatures]{Spectrum of fluctuations $S_k$ as a function of the wavenumber $k$ together with its corresponding density contrast correlation function $G_{nn}(L)$ as a function of the comoving distance $L$ measured in units of $R$. We consider a uniform expansion ($\gamma=1$). We depict the results for initial thermal states with $T=0.1 \, T_{\text{c}}$ (red),~\mbox{$T=0.2 \, T_{\text{c}}$} (blue), and $T=0.3 \, T_{\text{c}}$ (black) in the upper row (right after the expansion), as well as for different hold times $\tho$, for a fixed temperature~\mbox{$T=0.2T_{\text{c}}$} (bottom row). As before, the gray, vertical dashed line in the momentum plots indicates the low $k$ limit at inverse condensate size, whereas the horizontal one denotes $1/2$. The position space results given here correspond to a Gaussian convolution of  $w= 0.5 \,\mu$m standard deviation. Figure adapted from \cite{BECPaper2022}.}
\label{fig:bec.ThermalState}
\end{figure}

Let us consider now that the initial state is thermally occupied according to \eqref{eq:bec.Thermal}, which is expected in realistic experimental scenarios, and will lead to stimulated particle production. In the upper row of \cref{fig:bec.ThermalState} we again show the spectrum $S_k$ and the two-point density contrast $G_{nn}$, now for three different initial temperatures in units of $T_{\text{c}}$ (upper row), fixed $\Delta t = 2 \, \text{ms}$ and $\gamma=1$. On the other hand, in the bottom row, we fix $T=0.2T_{\text{c}}$ and consider different values of hold time $t_{\text{hold}}$. The important features stay the same, as the thermal state acts as a ($k$-dependent) factor in the spectrum, but particle production is enhanced. Note that the spectrum diverges as~$k$ goes to zero, which is a consequence of the Bose-Einstein distribution~\eqref{eq:bec.Thermal}, in contrast to the case of an initial vacuum state. There are similar shifts toward lower wavenumbers in the spectra and larger distances in the two-point functions with increased hold time $t_{\text{hold}}$ after the expansion, and the spectrum oscillates around~\mbox{$S_{k, 0}=(1+2n_k^{\text{in}})(1/2 + \abs{\beta_k}^2)$}. 

\subsection{Spatial curvature}

\begin{figure}[t!]
    \centering
    \includegraphics[width=0.75\textwidth]{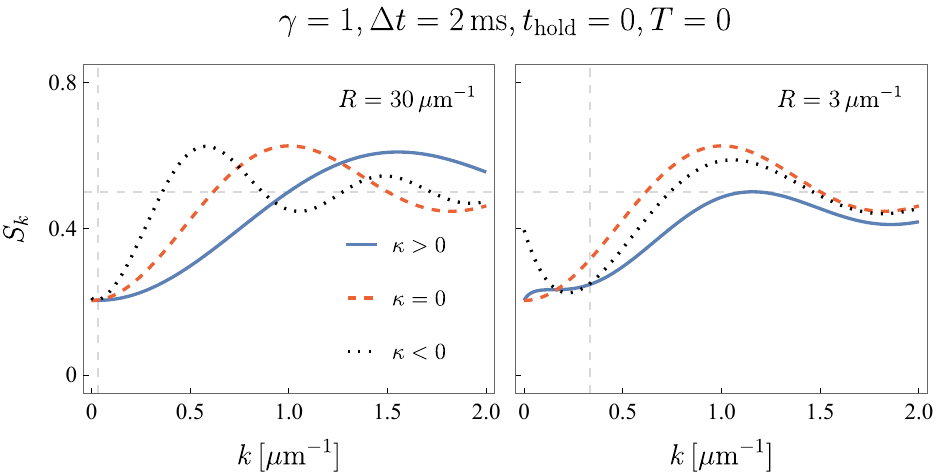}
    \caption[Spectrum of fluctuations for different spatial curvature]{Spectrum of fluctuations after a uniform expansion ($\gamma=1$) in universes with positive (red), vanishing (blue) and negative (black) curvature. In the left panel we consider a fixed atom number, so that the density at the center of the condensate depends on the shape of the trap (cf. \eqref{eq:bec.AtomNumber} and \eqref{eq:bec.TrapShape}). In this case, there are visible differences between the different curvatures. For the right panel, we take the density at the center to be the same for all curvatures. As a consequence, the shape of the spectrum $S_k$ is only affected by spatial curvature at low $k$, provided that $R$ is sufficiently small. The vertical and horizontal gray dashed lines denote the resolution in $k$ and the value $1/2$, respectively. Figure adapted from \cite{BECPaper2022}.}
    \label{fig:bec.Kappa}
\end{figure}

Let us now study how spatial curvature affects particle production. Recall that FLRW universes with positive and negative spatial curvature can be engineered through the radial dependencies of the background density given in \cref{eq:bec.TrapRadialDependenceExact}. The effects of spatial curvature in the frequency of the mode equation \eqref{eq:bec.ModeFrequency} are reflected in eigenvalues $-h^2$ of the Laplace-Beltrami operator. 
In \cref{fig:bec.Kappa} we show the spectrum for a uniform expansion ($\gamma=1$), assuming an initial vacuum state. The influence of curvature is typically negligible for fixed background density $n_0$ (see right panel of \cref{fig:bec.Kappa}), unless one decreases the condensate radius, which determines the value of the spatial curvature $\kappa$ (see subsection \ref{subsec:bec.MetricToFLRW}).

\subsection{Phase of the spectrum of fluctuations}
\label{subsec:bec.phase}

\begin{figure}[t!]
    \centering
    \includegraphics[width=0.75\textwidth]{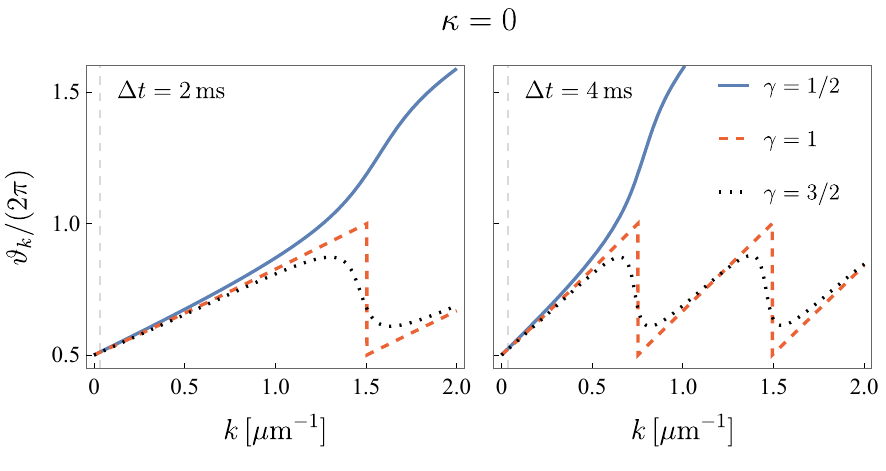}
    \caption[Phase after the expansion for three different scale factors]{Phase $\vartheta_k$ as a function of $k$ after the expansion for three different scale factors. There are qualitative differences depending on $\gamma$. For $\gamma = 1$ and $\gamma = 3/2$ the phase remains in a range of $(\pi,2\pi)$. In particular, in the $\gamma = 1$ case, the phase is a linear function of $k$, and importantly exhibits a $\pi$ jump (for wavenumbers $k$ fulfilling $\abs{\beta_k}^2 = 0$). In contrast, for $\gamma = 1/2$ the phase increases continuously in the range $(\pi,3\pi)$. The gray dashed line denotes the resolution in $k$. Figure from \cite{BECPaper2022}.}
    \label{fig:bec.Phases}
\end{figure}

As we have seen above, the qualitative differences coming from different exponents~$\gamma$ in the scale factor could be difficult to appreciate experimentally, at least when looking into the full spectrum and rescaled density contrast. Nevertheless, there is one particular observable that is particularly sensitive to the different expansion histories. This is the phase of the spectrum of fluctuations $\vartheta_k$ that each mode acquires after expansion, defined in \cref{eq:qftcs.OffsAmpPhaseDef}, which we consider next in fig.\ \ref{fig:bec.Phases}. There, we see that the phases of each wavenumber~$k$ strongly depend on whether we realize a decelerated, uniform, or accelerated expansion. In the case of uniform expansion there are phase jumps appearing at each wavenumber $k$ for which $\beta_k$ turns out to be zero. Indeed, we find that the wavenumbers for which this phase jump happens are given by the analytical solution
\begin{equation}
 k_n = \frac{\af - \ai}{\Delta t} \left[ \left(\frac{n \pi}{ \ln{(\af/\ai)}} \right)^2 + \frac{1}{4} \right]^{1/2},
\label{eq:bec.PhaseJumps}
\end{equation}
where $n$ is an integer counting the number of recurrences. It is worthwhile to note that phases as a function of $k$ also give an insight into the expansion duration $\Delta t$. More details on the behavior of the phase are given in \cref{sec:bec.scattering}.

\section{Experimental realization}
\label{sec:bec.Experiment}

After having studied the theory describing the analogy between fluctuations on top of the ground state of a weakly-interacting BEC and the motion of a massless, minimally coupled to gravity scalar field in an FLRW universe, we now turn to the experimental realization of this analog system. First, we show the emergence of FLRW spacetimes with positive and negative spatial curvature by wave packet propagation in the BEC. Once the analogy is established, we proceed with the implementation of cosmological production out of a thermal initial state in power-law expansions, as discussed above, which we characterize through the amplitude and phase of the spectrum of fluctuations obtained via density contrast measurements (see \cref{eq:bec.DensityDensityCorrelatorSpectrum}).

\subsection{Experimental setup}

The experimental system consists of a quasi two-dimensional condensate of approximately $23.000$ potassium-39 atoms with tunable background density distribution~$n_0(r)$ and interaction strength $\lambda(t)$, which can be adjusted independently. This allows us to implement an acoustic metric of the form \eqref{eq:bec.FLRWLineElement} for the phononic field. We recall that the spatial curvature $\kappa$ depends on the density profile, whereas the analog scale factor $a(t)$ is engineered through the interaction strength $\lambda(t)$. The BEC is tightly confined in the $z$-direction with a trap frequency of $\omega_z = 2\pi \times 1.6\,\mathrm{kHz}$, hence the effective geometry.

\begin{figure}[t!]
    \centering
    \includegraphics[width = \textwidth]{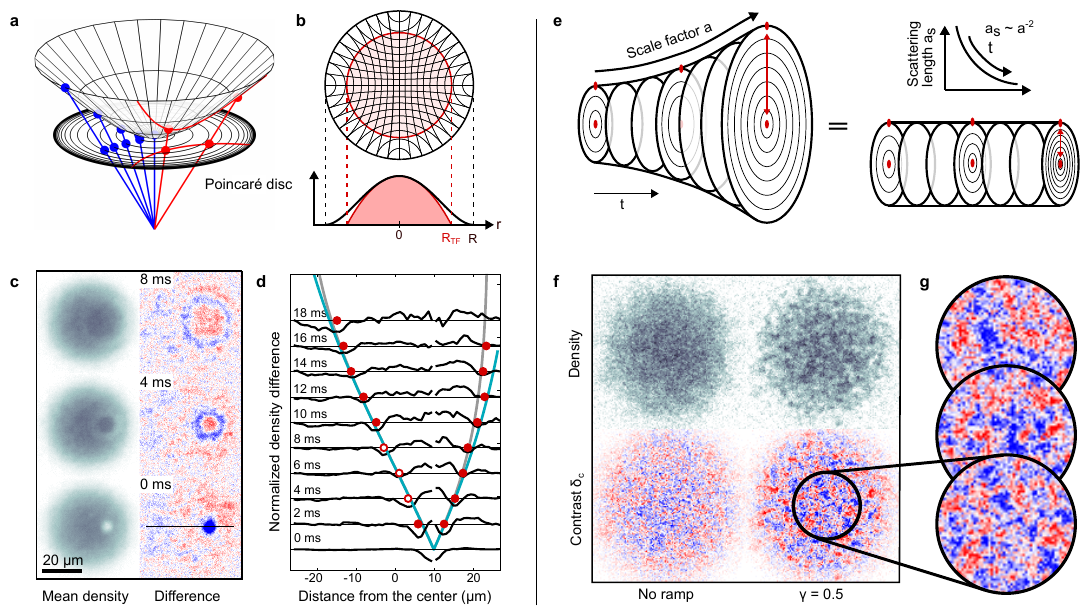}
     \caption[Summary of the realization of the cosmological analogy in a quasi-$2$D BEC]{Summary of the realization of the cosmological analogy in a quasi-$2$D BEC. (a) Negatively curved, infinite $2$D spatial section mapped onto the finite-sized Poincaré disc. (b) Density profiles corresponding to a hyperbolic geometry (black) and a harmonic trap (red), which approximates the former around the center. (c) Propagation of a phononic wave packet averaged over $\sim 100$ realizations (left) and difference to the unperturbed condensate (right). (d) Theoretical and experimental propagation along the geodesic depicted in the lower right panel of (c) (black line). The red dots mark the position of the wave packet at each time. The blue line is the theory prediction for the hyperbolic space and the gray line that of the acoustic metric of a harmonic trap. (e) Equivalence between an expanding spacetime and a stationary BEC with time-dependent scattering length $\alpha_s$. (f) Density and density contrast $\delta_c$ of a single realization before and after an expansion with scale factor $a(t) \propto t^{\gamma}$. The emergent fluctuations are associated with particle production. (g) Different unique realizations of the analog expansion. Figure from \cite{Experiment2022}.}
    \label{fig:bec.Fig1}
\end{figure}

In the following, we describe two types of experiments. On the one hand, we probe spatial curvature by engineering two background density profiles corresponding to negative and positively curved FLRW universes. For the former, we implement a harmonic potential, which approximates a hyperbolic geometry in a region around the center of the condensate (see \cref{eq:bec.HarmonicTrapCoordinateTransf}). In the case of positive curvature, we implement the corresponding density profile in \eqref{eq:bec.TrapRadialDependenceExact}, so that the effective geometry is realized in an exact manner. On the other hand, we realize cosmological particle production in a flat FLRW cosmology by restricting ourselves to a region around the center of a harmonically trapped BEC. Expansion of spacetime is realized by the engineering of the scattering length $\alpha_s(t)$ via Feshbach resonances~\mbox{\cite{Stwalley1976,Cornish2000,Chin2010}}, which allows us to implement power-law scale factors such as the ones analyzed in \cref{sec:bec.effects}.

\subsection{Probing spatial curvature}

As we mentioned, a harmonic trap with a density profile
\begin{equation}
    n_{\mathrm{TF}}(r) = \bar n_0[1-r^2 /R_\mathrm{{TF}}^2] \quad \text{for} \quad r \leq R_{\text{TF}}
\end{equation}
approximates the hyperbolic density profile ($\kappa<0$) in \cref{eq:bec.TrapRadialDependenceExact} around the center of the condensate for $R = \sqrt 2 R_\text{TF}$ and $\kappa = -2/R_\text{TF}^2$ (cf. \cref{eq:bec.HarmonicTrapCoordinateTransf}), where the Thomas-Fermi radius is measured to be \mbox{$R_\text{TF} = 25\,\mu\text{m}$}. This leads to a hyperbolic, two-dimensional spatial geometry with a radial coordinate of infinite range for the phononic excitations (provided we stay far from its edges). One can project the infinite spatial hypersurface onto a finite (Poincaré) disk (more details in \cref{sec:bec.Geometries}), as depicted in \cref{fig:bec.Fig1} (a). Note that equidistant lines in this metric become denser toward the disk edges. Figure \ref{fig:bec.Fig1} (b) shows the exact density profile for the implementation of a hyperbolic metric (black line, see \cref{eq:bec.TrapRadialDependenceExact}), which is approximated in the central region of a harmonically trapped BEC (red curve, see \cref{eq:bec.HarmonicTrapCoordinateTransf}). 

To demonstrate the implementation of hyperbolic space with our harmonically trapped condensate, we observe the propagation of a phononic wave packet that follows geodesics, that is, $\dd s ^2 = 0$, as corresponds to massless excitations. We prepare the wave packet by focusing a laser beam onto the atomic cloud, which then propagates in the hyperbolic geometry once the laser is switched off. Figure~\mbox{\ref{fig:bec.Fig1} (c)} shows this propagation averaged over $\sim 100$ realizations, as well as the density difference to the unperturbed system. For each time step, the profile of the density is extracted along the geodesic connecting the initial perturbation with the center of the condensate, i.e. along the black line in the lower right panel of~\mbox{\cref{fig:bec.Fig1} (c)}. Figure~\ref{fig:bec.Fig1}~(d) shows the normalized density profiles from which the positions of the minima are extracted (red points). We use the three points marked with open symbols to fit the speed of sound at the center of the condensate, yielding~\mbox{$c_{\text{s}} = 1.5\,\mu \text{m}/\text{ms}$}. The solid gray line in figure \ref{fig:bec.Fig1} (d) shows the prediction for the harmonically trapped condensate, and the blue line the prediction for a hyperbolic space (more details in \cref{sec:bec.PhononTrajectories}). Note that the distance $L$ between two points is given by the~\mbox{$\kappa<0$} scenario in \cref{eq:bec.ComovingDistance}. For the actual prediction in~\mbox{\cref{fig:bec.Fig1} (d)} we use the latter equation divided by the speed of sound at the center of the condensate with $\Delta\phi=0$. This serves as a quantitative demonstration that a BEC in a harmonic trap approximates a hyperbolic geometry, corresponding to negative curvature. 

\begin{figure}[t!]
    \centering
    \includegraphics[width = 0.75\textwidth]{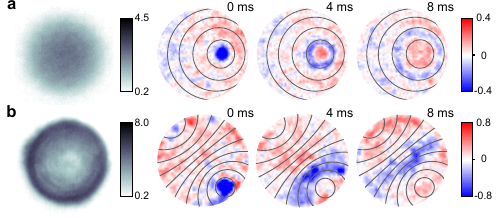}
    \caption[Density profiles for analog FLRW cosmologies with spatial curvature together with the propagation of phononic wavepackets in the emergent geometries]{Density profiles corresponding to analog FLRW cosmologies with negative (top) and positive (bottom) spatial curvature, together with the propagation of phononic wavepackets in the emergent geometries (underdensities in blue, overdensities in red). Black lines indicate wavefront propagation as predicted by the acoustic metric in each case. Figure from \cite{Experiment2022}.}
    \label{fig:bec.FigX}
\end{figure}

These ideas can be extended to different curvatures by choosing the appropriate density profile (cf. \cref{eq:bec.TrapRadialDependenceExact}). For spacetime geometries beyond hyperbolic, we use a digital micromirror device (DMD) \cite{SaintJalm2019,Gauthier2021} to configure arbitrary spatial curvatures. In particular, we do so for positively curved geometries. Figure \ref{fig:bec.FigX} contrasts the wave packet propagation on background densities corresponding to (approximately) hyperbolic and spherical metrics. We observe fundamentally different evolution in agreement with the expected dynamics.

\subsection{Measurement of cosmological production from a thermal state} 

Now that we have shown the implementation of the cosmological analogy, let us turn to the experimental realization of cosmological production due to power-law scale factors as theoretically discussed in \cref{sec:bec.effects}. We will focus on the flat scenario, by implementing a harmonic trap and restricting ourselves to a region around the center of the condensate, such that $n_0 \simeq \bar{n}_0 = 1.3(2)\times10^9 \, \mathrm{cm^{-2}}$.

In figure \ref{fig:bec.Fig1} (e), we illustrate the analogy between an expanding, flat, FLRW universe and a static BEC with time-dependent $s$-wave scattering length $\alpha_s$, which we recall that essentially plays the role of the scale factor\footnote{Control of the scattering length is achieved by changing an external magnetic field in the vicinity of the Feshbach resonance in potassium at $562.2(1.5)\,\mathrm{G}$ \cite{DErrico2007}.} (cf. \cref{eq:bec.BornApproximation,eq:bec.ScaleFactorDefinition}). The distance covered by a phonon moving at the speed of sound in a unit of time is denoted by the separation between the circles. Instead of increasing the size of the spatial sections, which would correspond to the scenario on the left-hand side, the analogy is implemented by decreasing the speed of sound (cf. \cref{eq:bec.ScaleFactorDefinition}), as the right hand side shows. From the point of view of the phonons, it takes more time for them to cover the same spatial distance as the expansion occurs, because the speed of sound from the laboratory perspective is decreased. 

\begin{figure}[t] 
    \centering
    \includegraphics[width = 0.75\textwidth]{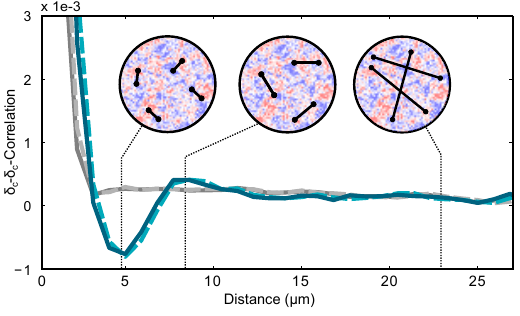}
     \caption[Correlation of fluctuations before and after the ramp]{Correlation function of fluctuations before and after the ramp, extracted from the density contrast $\delta_c$ (see eq. \eqref{eq:bec.RescaledDensityContrast}) as an average over all pairs of points separated by the same distance. The correlation function after a ramp (blue line) shows clear structure in comparison to the correlation function before the ramp (gray line). Distances are evaluated in the flat (solid lines) and in the hyperbolic metric (dashed lines). As expected from the geometry of the Poincaré disc in its central region, these two are very close. Figure from \cite{Experiment2022}.}
    \label{fig:bec.Fig2} 
\end{figure}

We perform accelerated ($\gamma = 3/2$), uniform ($\gamma = 1$), and decelerated ($\gamma = 1/2$) expansions, which are experimentally implemented by a \textit{ramp} of the scattering length~$\alpha_s$ from $400\aB$ to $50\aB$, where $\aB$ is the Bohr radius (this is an increase by a factor~$\sqrt{8}$, namely $\sim 1$ $e$-fold). In \cref{fig:bec.Fig1} (f), we illustrate typical density distributions before and after a specific expansion with $\gamma = 1/2$ (see also~\mbox{\cref{fig:bec.Fig1} (g)} for different realizations). In order to compare with theory, we extract the density contrast \eqref{eq:bec.RescaledDensityContrast} within the central region of the condensate up to half the Thomas-Fermi radius, with the density distribution of the background condensate~\mbox{$n_0(x,y) = \langle n(x,y)\rangle$}, inferred from the mean over all experimental realizations at each pixel position with coordinates $x$ and $y$. The density contrast two-point function \eqref{eq:bec.DensityDensityCorrelatorSpectrum} is then built by evaluating correlations as an average over all pixel-pairs of a given hyperbolic distance in the set of experimental realizations (see figure \ref{fig:bec.Fig2}). After the ramp (that is, in region III), we observe density fluctuations (which have also been observed in very fast ramp experiments, performed in atomic \cite{Westbrook2012, Hung2013,Chen2021} and photonic systems \cite{Steinhauer2022}) in accordance with the oscillations of the spectrum $S_k$ (see \cref{eq:qftcs.SpectrumOffsetAmpPhase}). Figure \ref{fig:bec.Fig2} shows the correlation functions before (gray) and after (blue) an expansion with $\gamma=1/2$ evaluated in a circular region at the center of the condensate with diameter $R_\text{TF}$. While there are only short-range correlations of the initial state over the area of interest, we find a clear anti-correlation signal at a length scale of $5 \, \mu$m after the decelerated expansion, followed by a small correlation peak. We find the same correlation function for an analysis with distances evaluated in the hyperbolic and in the flat metric, as expected for the central region of the Poincaré disc.

\begin{figure}[t]
    \centering
    \includegraphics[width = 0.65\textwidth]{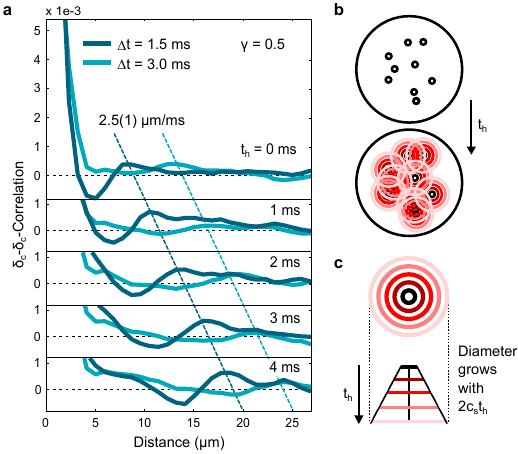}
     \caption[Propagation of correlations after the end of the expansion]{Propagation of correlations after the end of the expansion. (a) The characteristic features in the correlation functions propagate to larger distances, shown for two different ramp durations $\Delta t$. The speed of propagation is twice the speed of sound in both cases. (b) Localized excitations at random initial positions emit waves that propagate at the speed of sound (red circles). (c) A correlation is expected at the diameter of the sound cone, which grows with twice the speed of sound. Figure from \cite{Experiment2022}.}
    \label{fig:bec.Fig3}
\end{figure}

Through the time evolution of the density contrast correlation, we can extract the amplitude $A_k$ and phase $\vartheta_k$ of the spectrum of fluctuations $S_k$. For this, we let the system evolve at constant scale factor for a hold time $t_{\text{hold}}$ after the end of the expansion. Figure \ref{fig:bec.Fig3} (a) shows the correlation functions at different hold times for $\gamma = 0.5$ and two different ramp durations $\Delta t = 1.5 \,\text{ms}$ and $3.0\, \text{ms}$. For the slower ramp, the characteristic correlation feature appears at a larger distance and is less pronounced, as predicted by theory in \cref{sec:bec.effects}. Correlations move to larger distances with a constant speed of $2c_{\text{s}} = 2.5(1) \, \mu \mathrm{m}/\mathrm{ms}$, for both ramp durations. The propagation of the correlation is the result of the time evolution of the spontaneously created excitations, which propagate radially outwards, with each wavefront moving at the speed of sound. In turn, a correlation peak is expected at the diameter of the circular wavefront, which increases with twice the speed of sound (cf. \cref{fig:bec.Fig3} (b) and (c)). 

\begin{figure}[t!]
    \centering
    \includegraphics[width = \textwidth]{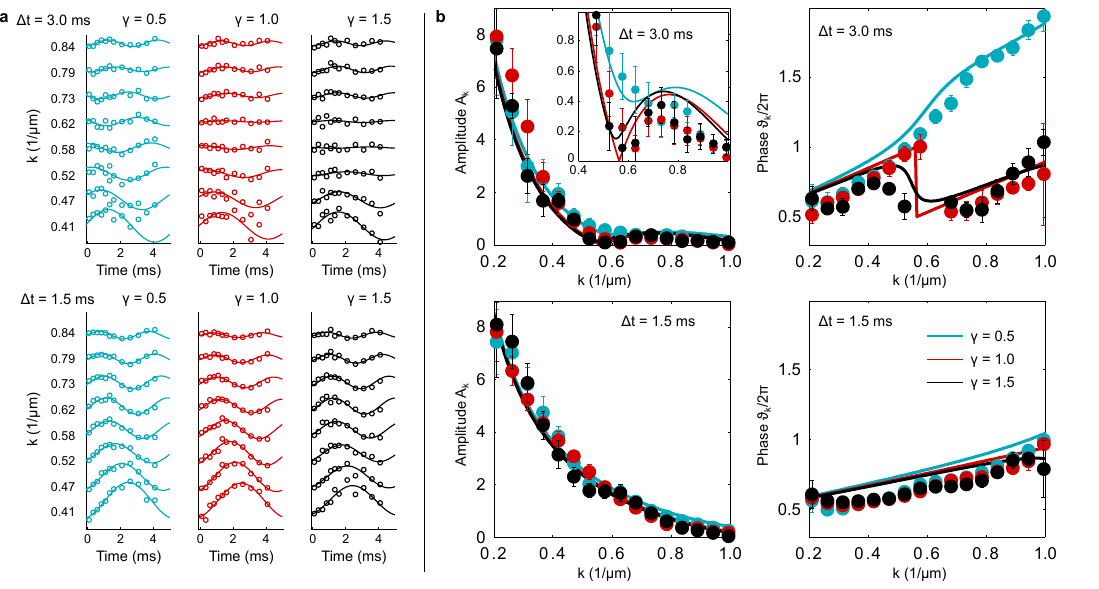}
    \caption[Spectrum of fluctuations and its time evolution after the expansion]{(a) Spectrum oscillations for different wavenumbers $k$ and decelerating (blue), uniform (red) and accelerating (black) expansions, together with cosine fits. (b) Amplitude $A_k$ and phase $\vartheta_k$ together with analytical theory predictions (solid lines). Figure from \cite{Experiment2022}.}
    \label{fig:bec.Fig4}
\end{figure}

By performing a transformation to momentum space, we can obtain the spectrum of fluctuations $S_k$, following \cref{eq:bec.DensityDensityCorrelatorSpectrum}. In figure \ref{fig:bec.Fig4} (a) we show the time evolution of $S_k$ for specific wavenumbers $k$, for the three different power-laws in consideration, and different expansion durations, with solid lines denoting fits to the data points of the form \eqref{eq:qftcs.OffsAmpPhaseDef}. As expected, oscillations happen at twice the frequency~\mbox{$\omega_{k, \text{f}}/\af = c_{\text{s,f}} \, k$} in laboratory time $t$. All observed momentum modes stay in the acoustic regime, corresponding to $k<1/\xi$, with the healing length being $\xi \, \sim 0.5 \, \mu$m in this case. Finally, figure \ref{fig:bec.Fig4} (b) shows amplitude and phase of the oscillations for the two different ramp speeds. The error bars are $1\sigma$ errors from the fit. The solid lines correspond to theory predictions derived from the previous sections of this chapter, using a final speed of sound of ${c_{\text{s}} = 1.2 \, \mu\text{m/ms}}$ at a scattering length of~$50\aB$, an initial scattering length of $350\aB$, and a temperature of $T=40\, \mathrm{n K}$, which are compatible with the experimental values within the errors. Experimentally, the temperature is independently determined to be $T= 60(10)\, \mathrm{n K }$ by fitting the density profile to a thermal spectrum of the form \eqref{eq:bec.Thermal}~\cite{Giorgini1996}.

Crucially, through the time evolution we are able to determine the phase $\vartheta_k$, which carries information about the expansion history, showing qualitative differences for the three type of cosmological expansions probed. Moreover, we confirm the phase jump predicted by theory in the case of a uniform expansion ($\gamma = 1$), for which the amplitude $A_k$ vanishes, i.e. there is no particle production (upper-right panel in figure \ref{fig:bec.Fig4} (b)). For the short ramp, we find $k_1 = 0.56\, \mu\text{m}^{-1}$, while for the fast ramp one has $k_1 = 1.12\, \mu\text{m}^{-1}$, which is not experimentally accessible. For accelerated ($\gamma = 1.5$) and decelerated ($\gamma = 0.5$) expansion, the particle production is not fully suppressed, resulting in a continuous phase evolution. Recall that we theoretically find that the phase is independent of temperature, and can thus serve as a robust indicator for the expansion history of the metric.

\section{Cosmological production as a scattering problem}
\label{sec:bec.scattering}

In this section, we will discuss a different analogy. We will pursue the idea that quantum mechanical scattering in one-dimensional potentials is formally equivalent to cosmological particle production. This other point of view will allow us to gain further insights into QFTCS, at the same time that we demonstrate another application of BECs as quantum simulators. 

Quantum mechanical scattering in one-dimensional potentials is a well-established problem \cite{Flügge1999,Schwabl2007,Boya2008,Landau2013, Griffiths2018, Sakurai2020}. However, as happens with cosmological scenarios, it is not easy to implement experimentally, let alone to be able to design arbitrary forms of the scattering potential within one experimental setup. This is precisely what a BEC offers, as we will see. The idea that the scattering problem and cosmological production share a connection is not new (see e.g. \cite{Mukhanov2007}), although it has not been exploited to the extent that we will do here. 

In the following, we will discuss the theory behind this analogy, and show how one can realize different scattering problems in a BEC, as well as their translation to cosmological scenarios. Further details on the aspects of the scattering analogy can be found in \cite{ScatteringTheory2024}. Additionally, we would like to mention that an experimental realization of this scattering analogy in a BEC is part of the work \cite{ScatteringExp2024}, which is to be published, and thus we will not show any explicit results in this thesis.

\subsection{Description of the scattering analogy}
\label{sec:scatt.CosmPartProd}

The idea behind the scattering analogy is very simple. Take the mode equation~\eqref{eq:qftcs.ModeEquation} for the rescaled field $\chi$, and interpret conformal time $\eta$ as a spatial coordinate. Then, the equation can be written in the form of a stationary Schrödinger equation, namely 
\begin{equation}
    \left[ - \dv[2]{}{\eta} + V(\eta) \right] v_k(\eta) = E_k v_k(\eta),
    \label{eq:scatt.SchrodingerEq}
\end{equation}
with energy eigenvalues $E_k = h^2(k)$, given by the eigenvalues of the Laplace-Beltrami operator $-\Delta$, and a scattering potential of the form 
\begin{equation}
   \! V(\eta)\! = \!- \Bigg\{m^2a^2(\eta) + \frac{1+(4\xi-1)D}{4} \left[(D-3) \left( \frac{a'(\eta)}{a(\eta)} \right)^2\!+2 \frac{a''(\eta)}{a(\eta)} \right]\Bigg\},
\label{eq:scatt.GeneralScatteringPotential}
\end{equation}
from the definition of the frequency \eqref{eq:qftcs.MasterFrequency}. We see that the problem of finding a solution to the mode equation \eqref{eq:qftcs.ModeEquation} is equivalent to solving the above Schrödinger equation, in which conformal time~$\eta$ plays the role of the spatial direction in which the scattering takes place\footnote{Note, however, that the modes $v_k$ are not properly normalized as wavefunctions in quantum mechanics.}. Thus, there is a direct correspondence between the scale factor~$a(\eta)$ characterizing the expansion of spacetime and the scattering potential~$V(\eta)$.

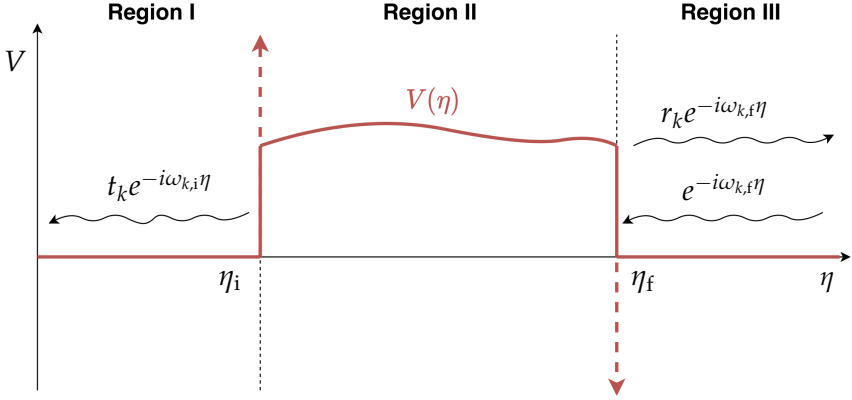
\begin{figure}
    \centering
    \begin{overpic}[width=0.8\columnwidth]{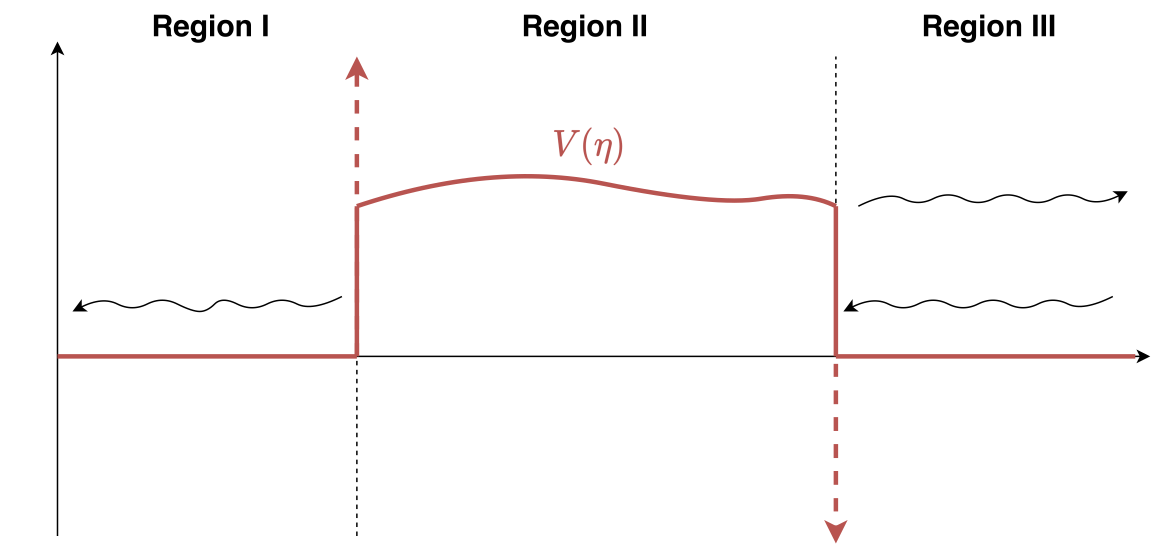}
    \put(13,23.5){$t_k e^{-i \omega_{k, \text{i}}\eta}$}
    \put(80,23){$e^{-i \omega_{k, \text{f}}\eta}$}
    \put(77.5,32){$r_ke^{-i \omega_{k, \text{f}}\eta}$}
    \put(26,13){$\etai$}
    \put(74,13){$\etaf$}
    \put(96,13){\small{$\eta$}}
    \put(1,38){$V$}
    \end{overpic}
    \caption[Representation of the scattering analogy]{Representation of the scattering analogy (for clarity we have taken $a_k=1$). The dashed lines and the arrows indicate Dirac deltas with its sign. Figure from \cite{ScatteringTheory2024}.}
    \label{fig:scatt.ScatteringAnalogy}
\end{figure}

Let us rephrase now the \textit{in-out} formalism in the language of the scattering analogy, which is illustrated in \cref{fig:scatt.ScatteringAnalogy}. We will start from region III ($\eta \geq \eta_{\text{f}}$). In the cosmological picture, expansion has ceased, and the \textit{in} modes are a superposition of positive and negative frequencies~$\omega_{k, \text{f}}$. From the scattering point of view, an incoming wavefunction with amplitude~\mbox{$\sim a_k$} is reflected at the potential boundary~\mbox{$V(\eta_{\text{f}})$}, leading to a reflection wavefunction proportional to $b_k$ (propagation happens from right to left in \cref{fig:scatt.ScatteringAnalogy}). In region~II ($\eta_{\text{i}}<\eta<\eta_{\text{f}}$), the potential $V(\eta)$ is non-vanishing, playing the role of a time-dependent frequency in the cosmological picture. Lastly, in region I ($\eta \leq \eta_{\text{i}}$), the expansion has not yet taken place. The \textit{in} modes are defined as plane waves with positive frequency $\omega_{k,\text{i}}$ in that region (cf. \cref{eq:qftcs.InModeIC}), and in the scattering analogy they correspond to a wave that is transmitted through the potential with amplitude~$c_k$. 

Let us now define the customary reflection and transmission amplitudes $r_k$ and~$t_k$, which in this context are given by
\begin{equation}
    r_k = \frac{b_k}{a_k} \quad \text{ and } \quad t_k = \frac{c_k}{a_k}.
    \label{eq:scatt.ScatteringAmplitudes}
\end{equation}
These amplitudes satisfy 
\begin{equation}
    1=\abs{r_k}^2 + \abs{t_k}^2,  \quad \text{or} \quad \abs{a_k}^2 = \abs{b_k}^2 + \abs{c_k}^2,
    \label{eq:scatt.ProbabilityConservation} 
\end{equation}
which reflects probability conservation in quantum mechanics. Now we can write the Bogoliubov coefficients (see e.g. \eqref{eq:qftcs.BogoliubovTransformationModes}) in terms of the scattering amplitudes as\footnote{This analogy was already used in \cite{Visser1999} to derive general bounds on the process of one-dimensional scattering.}
\begin{equation}
    \begin{aligned}
        \alpha_k &= \frac{a_k^*}{c_k^*} \quad \text{and} \quad \beta_k = - \frac{b_k}{c_k^*}.
    \end{aligned}
\label{eq:scatt.BogoliubovCoeffScattering}
\end{equation} 
Note that the condition \eqref{eq:qftcs.BogoliubovCoefficientsNorm}, necessary to preserve the Wronskian under the Bogoliubov transformation \eqref{eq:qftcs.WronskianModes}, is provided by \cref{eq:scatt.ProbabilityConservation}.  

It is now straightforward to translate the elements of the spectrum of fluctuations~$S_k$ to the scattering language. By making use of \cref{eq:scatt.BogoliubovCoeffScattering}, we find that the mean number density of particles can be written as 
\begin{equation}
    n_k = \abs{\beta_k}^2 = \abs{\frac{b_k}{c_k}}^2 = \abs{\frac{r_k}{t_k}}^2,
\label{eq:scatt.SakharovOffset}
\end{equation}
whereas the amplitude $A_k$ reads
\begin{equation}
    A_k = \left(1+2n_k^{\text{in}}\right)\abs{\alpha_k \beta_k} = \left(1+2n_k^{\text{in}}\right)\abs{\frac{r_k}{t_k^2}},
    \label{eq:scatt.SakharovAmplitude}
\end{equation}
and the phase $\vartheta_k$ can be written as
\begin{equation}
    \vartheta_k = \arg (\alpha_k \beta_k e^{2 i \omega_{k, \text{f}} \eta_{\text{f}}} ) =\arg(-r_k e^{2 i \omega_{k, \text{f}} \eta_{\text{f}}}),
\label{eq:scatt.PhaseShiftInTermsOfROnly}
\end{equation}
where we used that $c_k = 1/\sqrt{2 \omega_{k, \text{f}}}$ (see \cref{eq:qftcs.InModeIC}). The above equation~\eqref{eq:scatt.PhaseShiftInTermsOfROnly} tells us that the phase shift of the spectrum $S_k$ discussed in subsection \ref{subsec:bec.phase} can be understood in terms of a phase shift of the reflected wave relative to the incoming left-mover. Moreover, as we know from \cref{eq:qftcs.ChiDotChiDotCorrelator}, the spectrum $S_k$ can be deduced solely from the absolute value of the (derivative of the) mode function in region~III (cf. \cref{eq:qftcs.SpectrumDefinition}), showing that the time-dependent term in $S_k$ comes from the interference between the incoming and reflected waves in region III. 

\subsection{Realization of hallmark scattering potentials}
\label{sec:scatt.ScatteringPotentialsInTheQuantumSimulation}

In this section we will discuss the form of the scattering potential~$V(\eta)$ that corresponds to different classes of scale factors $a(\eta)$, particularizing \cref{eq:scatt.GeneralScatteringPotential} to the BEC setup described in \cref{ch:becstheory}, namely taking $D=2$ and $m=\xi=0$. Recall that the analog scale factor $a(\eta)$ is determined by the background parameters of the BEC, rather than the Friedmann equation \eqref{eq:cosmo.FriedmannEquation}, as happens in Cosmology. In a similar way as we did previously, we can design the scale factor $a(\eta)$ in order to implement the desired scattering potential. 

In the BEC system studied in \cref{ch:becstheory}, the scattering potential \eqref{eq:scatt.GeneralScatteringPotential} takes the form ($D=2$ and $m=\xi=0$),
\begin{equation}
    V(\eta) = \frac{1}{2} \frac{a^{\prime\prime}(\eta)}{a(\eta)} - \frac{1}{4} \left( \frac{a^{\prime}(\eta)}{a(\eta)} \right)^2,
\label{eq:scatt.ScatteringPotential} 
\end{equation}
and we consider, as before, a situation in which we start with a static geometry in region I ($\eta \leq \eta_\text{i}$) with $a=a_{\text{i}}$, then we have a dynamical expansion in region~II ($\eta_\text{i} \leq \eta \leq \eta_\text{f}$), and finally we reach a static geometry in region III ($\eta \geq \eta_\text{f}$) with~\mbox{$a=a_{\text{f}}$}. The scale factor in region II is given by~$\atwo(\eta)$, and is engineered in the analog system through the interaction strength (cf. \cref{eq:bec.BornApproximation,eq:bec.ScaleFactorDefinition}). As we discussed in subsection \ref{subsec:bec.InstantaneousSwitch}, the derivative of the scale factor will have a discontinuity at the boundaries of region II, if the switch between the static and dynamic phases is considered to be instantaneous. Therefore, one needs to take into account the boundary conditions \eqref{eq:bec.BoundaryConditionDerivative} and \eqref{eq:bec.BoundaryConditionMode} when solving Schrödinger's equation \eqref{eq:scatt.SchrodingerEq} throughout the three regions.

In particular, if the derivative of the scale factor is given by
\begin{equation}
    a^{\prime}(\eta) = a_{\text{II}}^{\prime}(\eta) \Theta(\eta - \eta_{\text{i}}) \Theta(\eta_{\text{f}} - \eta),
\end{equation}
then the second derivative reads
\begin{equation}
    \!a^{\prime\prime}(\eta) = a_{\text{II}}^{\prime\prime}(\eta) \Theta(\eta - \eta_{\text{i}}) \Theta(\eta_{\text{f}} - \eta) + a_{\text{II}}^{\prime}(\eta)\left[\delta(\eta - \eta_{\text{i}}) -\delta(\eta_{\text{f}}-\eta)\right].
\end{equation}
This leads to the appearance of $\delta$ functions in the scattering potential \eqref{eq:scatt.ScatteringPotential}, which reads
\begin{equation}
\begin{split}
    V(\eta) = \Bigg[\frac{1}{2} \frac{a_{\text{II}}^{\prime\prime}(\eta)}{a_{\text{II}}(\eta)} - \frac{1}{4} &\left( \frac{a_{\text{II}}^{\prime}(\eta)}{a_{\text{II}}(\eta)} \right)^2\Bigg]\Theta(\eta - \eta_{\text{i}}) \Theta(\eta_{\text{f}} - \eta) \\
    &\hspace{1cm}+ \frac{1}{2} \frac{a_{\text{II}}^{\prime}(\eta)}{a_{\text{II}}(\eta)}\left[\delta(\eta - \eta_{\text{i}}) - \delta(\eta_{\text{f}} - \eta)\right].
\label{eq:scatt.ScatteringPotentialWithDeltas}
\end{split}
\end{equation}

\subsubsection*{Exemplification}

\begin{figure*}
    \raisebox{-0.31cm}{\subfloat{\begin{overpic}[width=0.33\textwidth]{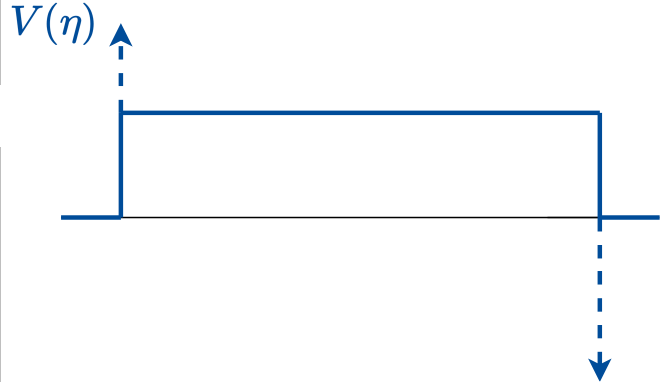}
    \put(13,19){\footnotesize{$\etai$}}
    \put(93,19){\footnotesize{$\etaf$}}
    \put(-1,40){\footnotesize{$H_0^2/4$}}
    \end{overpic}}}
    \subfloat{\begin{overpic}[width=0.33\textwidth]{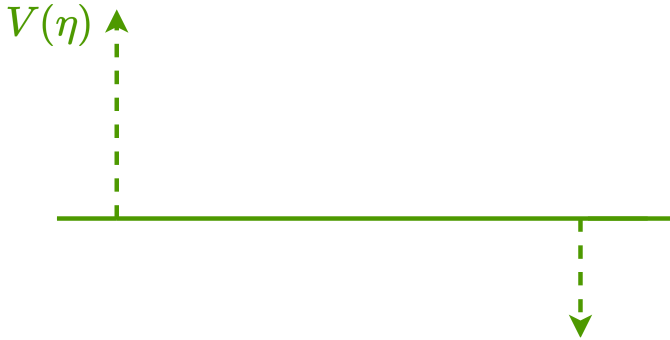}
    \put(13,12){\footnotesize{$\etai$}}
    \put(90,12){\footnotesize{$\etaf$}}
    \end{overpic}}
    \raisebox{-0.04cm}{\subfloat{\begin{overpic}[width=0.33\textwidth]{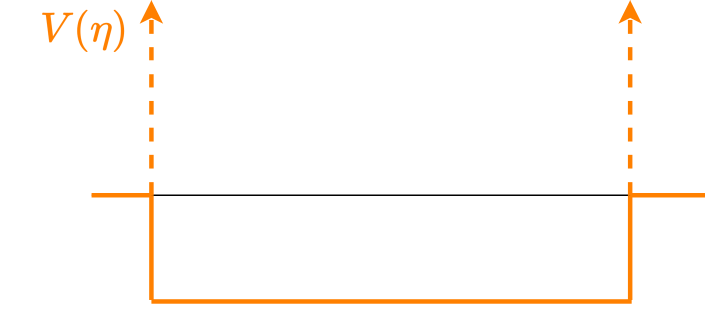}
    \put(13,13){\footnotesize{$\etai$}}
    \put(93,13){\footnotesize{$\etaf$}}
    \put(-3,1){\footnotesize{$-H_0^2/4$}}
    \end{overpic}}}
    \caption[Scattering potentials corresponding to different cosmological expansions in $(1+2)$ dimensions]{Scattering potentials corresponding to cosmological expansions in $(1+2)$ dimensions. We show power-law expansions with~\mbox{$q=0$} (left), $q=1/2$ (center), and antibouncing scale factor~\eqref{eq:scatt.SymmetricExpansionContraction_ScaleFactor}~(right). The solid lines represent the regular part of the potential, while the dashed lines denote Dirac delta contributions. Figure from \cite{ScatteringTheory2024}.}
    \label{fig:scatt.ScatteringPotentialsQuantumSimulator}
\end{figure*}
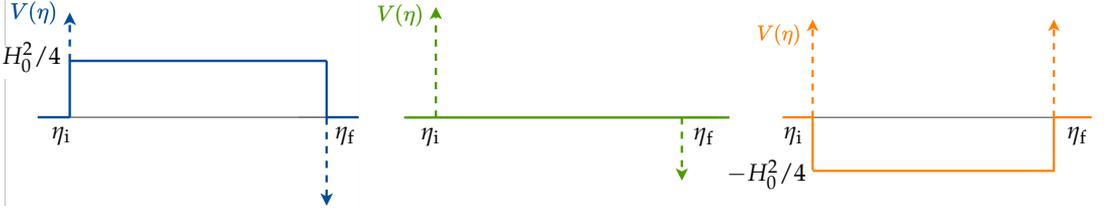

As a first example, let us consider power-law expansions in region II of the form
\begin{equation}
    a_{\text{II}}(t) = \left[1+ (q+1)\tilde{H}_0 t \right]^{1/(q+1)},
\label{eq:scatt.PowerLawScaleFactorII}
\end{equation}
where $q$ is the (constant) deceleration parameter defined in \cref{eq:cosmo.DecelerationParameter} and the parameter $\tilde{H}_0$ is proportional to the Hubble rate $H_0 = \dot{a}/a$. This notation will be more convenient than the one used before in this chapter (see \cref{eq:bec.ScaleFactorPolynomial}) for our purposes here, but one can easily identify
\begin{equation}
    \gamma = \frac{1}{q+1}.
\end{equation}
In conformal time, the family of scale factors \eqref{eq:scatt.PowerLawScaleFactorII} reads
\begin{equation}
    a_{\text{II}}(\eta) = \left(\tilde{H}_0q\eta\right)^{1/q}.
\label{eq:scatt.PowerLawScaleFactorConformalTime}
\end{equation}
Introducing the latter into the scattering potential \eqref{eq:scatt.ScatteringPotential}, one finds for $q \neq 0$ that 
\begin{equation} 
    \begin{split}
    V(\eta)
    = \left( \frac{1}{4q^2} - \frac{1}{2q} \right) &\frac{1}{\eta^2} \Theta(\eta - \eta_{\text{i}}) \Theta(\eta_{\text{f}} - \eta) \\
    &+ \frac{1}{ 2 q\eta}\left[\delta(\eta - \eta_{\text{i}}) - \delta(\eta-\eta_{\text{f}})\right],
    \end{split}
\label{eq:scatt.ScatteringPotential_PowerLawExpansion_ConformalTimeQneq0}
\end{equation}
whereas in the $q=0$ ($\gamma=1$) case the potential reads
\begin{equation} 
    V(\eta) = \frac{\tilde{H}_0^2}{4} \Theta(\eta - \eta_{\text{i}}) \Theta(\eta_{\text{f}} - \eta) + \frac{\tilde{H}_0}{ 2}\left[\delta(\eta - \eta_{\text{i}}) - \delta(\eta-\eta_{\text{f}})\right].
\label{eq:scatt.ScatteringPotential_PowerLawExpansion_ConformalTimeQeq0}
\end{equation}
This corresponds to the typical case of a rectangular barrier scattering potential, or a universe expanding with constant acceleration. Interestingly, the potential for~\mbox{$q=1/2$} ($\gamma=2/3$ in our previous notation) only has contributions from $\delta$ functions (compare the left and central panel of \ref{fig:scatt.ScatteringPotentialsQuantumSimulator}).

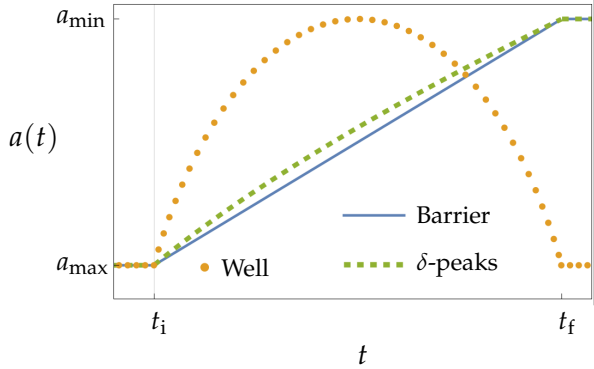
\begin{figure}
    \centering
    \begin{overpic}[width=0.55\columnwidth]{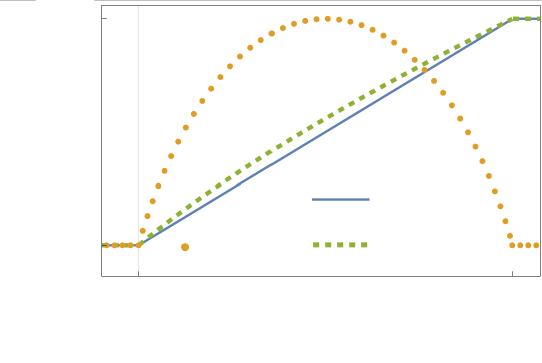}
        \put(25,11){\small{$\ti$}}
        \put(94,11){\small{$\tf$}}
        \put(60,5){\normalsize{$t$}}
        \put(1,42){\normalsize{$a(t)$}}
        \put(9,21){\small{$a_{\text{max}}$}}
        \put(9,63){\small{$a_{\text{min}}$}}
        \put(70,29){\footnotesize{$\text{Barrier}$}}
        \put(37,20){\footnotesize{$\text{Well}$}}
        \put(70,21){\footnotesize{$\delta\text{-peaks}$}}
    \end{overpic}
    \caption[Scale factor as a function of cosmological time for the three different scattering potentials considered]{Scale factor $a(t)$ as a function of cosmological or laboratory time $t$ for the three different scattering potentials considered. Figure from \cite{ScatteringTheory2024}.}
    \label{fig:scatt.ExpansionScenarios}
\end{figure}

Let us now proceed the other way around, that is, starting from the potential and finding the corresponding scale factor. We consider first a rectangular well potential that exactly vanishes at $\etai$ and $\etaf$, namely
\begin{equation}
    V(\eta) = - \frac{\tilde{H}_0^2}{4} \Theta(\eta - \eta_{\text{i}}) \Theta(\eta_{\text{f}} - \eta). 
    \label{eq:scatt.SquarePotentialWell_Definition}
\end{equation}
Introducing this in \eqref{eq:scatt.ScatteringPotentialWithDeltas} and solving for $a_{\text{II}}(\eta)$ one obtains
\begin{equation}
    a_{\text{II}}(\eta) = a_{\text{max}} \cos^2 \left[ \frac{\tilde{H}_0}{2} \left( \eta + \theta \right) \right],
\label{eq:scatt.SquareWellGeneralScaleFactor}
\end{equation}
where $\theta$ is a phase. Importantly, note that this scale factor becomes singular for~\mbox{$\tilde{H}_0\left(\eta + \theta\right) =\pi$}. Thus, let us restrict ourselves to a symmetric anti-bounce, that is, the decelerating half-cycle of \cref{eq:scatt.SquareWellGeneralScaleFactor}, by choosing a suitable phase shift~\mbox{$\theta = -(\etaf + \etai)/2$}. The corresponding scale factor takes the form
\begin{equation}
    a_{\text{II}}(\eta) = \frac{\amax}{2} \left\lbrace 1 + \cos \left[\tilde{H}_0 \left(\eta - \frac{\eta_{\text{f}}+\etai}{2}\right) \right] \right \rbrace,
\label{eq:scatt.SymmetricExpansionContraction_ScaleFactor}
\end{equation}
but we need to take into account the singular derivative at the boundaries of region~II (see \cref{fig:scatt.ExpansionScenarios}). Introducing \cref{eq:scatt.SymmetricExpansionContraction_ScaleFactor} in the general form of the scattering potential~\eqref{eq:scatt.ScatteringPotential}, we find the same result as in \cref{eq:scatt.SquarePotentialWell_Definition}, with additional terms proportional to delta functions,
\begin{equation}
    \!V(\eta)\!= \!- \frac{\tilde{H}_0^2}{4} \Theta(\eta - \eta_{\text{i}}) \Theta(\eta_{\text{f}} - \eta) \! + \frac{\tilde{H}_0}{2} \sqrt{\frac{\amax}{\amin} - 1} \, [\delta(\eta-\eta_{\text{i}}) \!+ \delta(\eta-\eta_{\text{f}})],
\label{eq:scatt.WellSingular}
\end{equation}
where $\amin = a_{\text{II}}(\eta_\text{i})$. 

\section{Summary}

In this chapter we have studied BECs as cosmological simulators. In particular, we have focused on a quasi-$2$D condensate and shown that the dynamics of small, low-energy perturbations on top of its ground state corresponds to that of a massless scalar field minimally coupled to gravity. By properly adjusting the background density profile and the interaction strength, we have demonstrated the implementation of FLRW universes with positive, vanishing and negative spatial curvature, both theoretically and experimentally. Moreover, we have defined a rescaled density contrast function that can be related to the spectrum of fluctuations of the phononic field via the momentum space transform of its two-point correlation functions. In this way, we have realized analog cosmological particle production sourced by power-law scale factors, which manifests as enhanced density fluctuations in the BEC. Importantly, we find quantitative agreement between the experimental observations and the theoretical predictions, confirming the successful implementation of the quantum field simulator. Additionally, we have also discussed how cosmological particle production can be understood as a problem of scattering in quantum mechanics, therefore extending the usefulness of BECs as quantum simulators. We have shown how to interpret the most relevant observables in particle production in the scattering language, and how to model different hallmark potentials. This provides new insights into the dynamics of cosmological production, and opens the door to new experimental possibilities. In fact, we have performed a series of experiments that confirm the theoretical predictions of the scattering framework, although we have not included this analysis here. An important question remains to be analyzed: What is the entanglement of the cosmologically produced particles, and how to measure it in an analog experiment. This is the subject of the next chapter.

%% file: Chapters/Entanglement_between_produced_pairs.tex


\chapter{Entanglement between produced pairs} 

\label{ch:entanglement} 



In this chapter we show how to quantify the entanglement between the produced analog particles in the BEC setup discussed in the previous chapter, providing a guide to optimize experimental configurations aimed at detecting entanglement within current capabilities.

The key element of our proposed procedure is to perform several expansion/contraction cycles as opposed to maximizing the number of $e$-folds in a single expansion ramp (as those discussed and realized in \cref{ch:becstheory}). The repetition of these cycles induces resonances that can be experimentally tuned to enhance entanglement production within the hydrodynamical regime of the condensate, where the cosmological analogy holds. We show that, while entanglement produced in single ramps is not observable within current capabilities, the enhanced resonant production characteristic of expansion/contraction cycles---already suggested by Schr\"odinger~\cite{Schroedinger1939}---produces observable levels of entanglement within attainable experimental configurations.

\section{Experimental setup and state tomography}
\label{sec:ent.ModelExp}

In this section, we explicitly describe the experimental setup, and derive the relations that allow us to perform tomography (that is, to completely determine the quantum state of the system) which is necessary in order to apply the entanglement quantification methods described in \cref{sec:qftcs.entanglement}.

\subsection{Modeling different expansion histories}
\label{subsec:ent.rampmodels}

\begin{figure*}[t!]
    \centering
    \includegraphics[width=\textwidth]{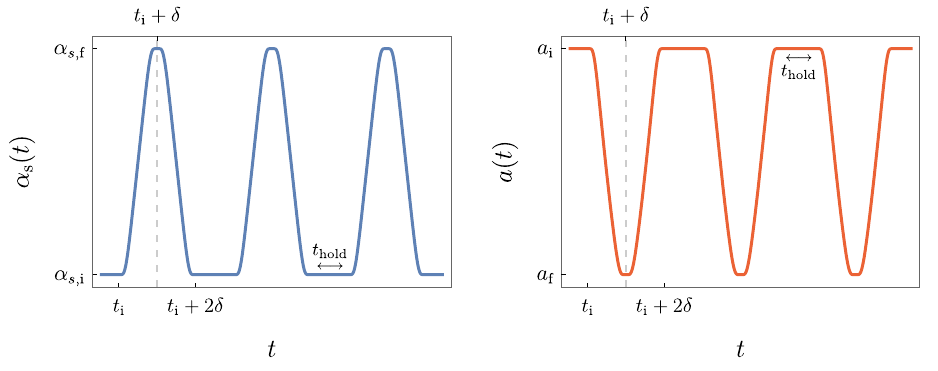}
    \caption[Smooth cusps in the scattering length and the analog scale factor as a function of time]{Scattering length $\alpha_s(t)$ as a function of time, made up of a sequence of symmetric cusps with duration $2\delta$, with $\delta = 0.5 \, \text{ms}$ (left), and corresponding scale factor~$a(t)$ (right). The cusps are separated by an interval $\tho$, during which the scattering length is kept constant and equal to $\asi$. We have highlighted the limits of the static regimes before and after the first cusp, using the labels $\ti$ and~$\ti+2\delta$. Figure from \cite{Entanglement2024}.}  
    \label{fig:ent.ScatteringLength}
\end{figure*}

In the previous chapter, we have discussed power-law scale factors of the form~\mbox{$a(t) \sim t^{\gamma}$}. However, as we will see later, the entanglement produced by such single expansion/contraction ramps is too small to be detected with current experimental capabilities or modest improvements thereof. Instead, we propose using scale factors with repeated expansion and contraction cycles, achievable in current laboratory conditions with the techniques discussed in \cref{ch:becstheory}, and which lead to resonant effects that significantly enhance particle and entanglement production. 

Additionally, we have seen that analog experimental realizations of cosmological expansion are usually modeled by assuming that the derivative of the scale factor (or the scattering length) instantaneously changes from vanishing to a non-vanishing value (see subsection \ref{subsec:bec.InstantaneousSwitch}). Of course, this is not the situation in a real experiment. Instead, here we consider realistic transitions from and to regions of static scale factor, by means of a $C^{\infty}$ step-like function $\Theta_{\sigma}$ of width $\sigma$, which interpolates between~$0$ and~$1$ in the interval $\left(-\sigma/2, \sigma/2\right)$, in such a way that $\sigma=0$ corresponds to the Heaviside function. In particular, we will consider the functional form 
\begin{equation}
    \Theta_{\sigma}(t)=\frac{1}{2}\left(1+\tanh{\left\{\cot{\left[\pi\left(\frac{1}{2}-\frac{t}{\sigma}\right)\right]}\right\}}\right).
\label{eq:ent.SwitchFunction}
\end{equation}
With this, we can model the scattering length as a sequence of \textit{cusps}, joined by periods of staticity of duration $\tho$, as the left panel of figure \ref{fig:ent.ScatteringLength} shows. In particular, each cusp is modeled by the function 
\begin{equation}
    \alpha^{\text{cusp}}_s(t)=\asi+\left(\asf-\asi\right)\left[\Theta_{\delta}(t-\ti-\delta/2)-\Theta_{\delta}(t-\ti-3\delta/2)\right],
\label{eq:eq.OneCuspScatteringLength}
\end{equation}
with $\asi$ and $\asf$ being the smallest and largest values of the scattering length within a cusp, respectively. Note that we consider symmetric cusps in the scattering length of duration $2\delta$, with $\delta$ also controlling the \textit{steepness} of each cusp. Explicitly, the scattering length has the constant value $\asi$ for $t \leq \ti$, and it increases from~$\ti$ to $\ti +\delta$, when it acquires the value~$\asf$. Then, it is similarly decreased from that precise instant until it reaches again $\asi$ at $\ti+2\delta$. After a time $\tho$ while the scattering length is kept constant, this process is repeated. The scattering length for several cycles is depicted in the left panel of figure \ref{fig:ent.ScatteringLength}, whereas the corresponding scale factor is shown in the right panel. 

In summary, the expansion/contraction history we consider is characterized by the initial and final scattering lengths in a single cusp, $\asi$ and $\asf$, respectively, the cusp duration $\delta$, the hold time between cusps $\tho$, and the number of repetitions~$n$. Note that particle production is only sensitive to the ratio $\asf/\asi$, but the absolute value of the scattering length is experimentally relevant, as we will discuss below. In the following, we will fix $\asf=400\aB$.  

We have explored other functional forms for modeling each cusp and found that the predictions for particle production and entanglement are not highly sensitive to the specific form of the function. Instead, the global quasi-periodic structure predominantly determines the magnitude of these effects. The function \eqref{eq:eq.OneCuspScatteringLength} strikes a good balance between accurately modeling experiments and maintaining simplicity. Moreover, the use of a smooth function allows us to avoid the need to treat discontinuities at the edge of the static regions, as it would be needed in the instantaneous case.


In the remaining of the section, we will outline a framework to predict outcomes of the experiments assuming that we have knowledge of the dynamics of the scale factor. We will use this in sections \ref{sec:ent.Entanglement}, and \ref{sec:ent.optimization} to provide concrete predictions for experiments which realize a series of expansion-contraction cycles.

\subsection{State tomography from density contrast and spectrum}
\label{subsec:ent.tomography}

Let us now detail how to perform state tomography from measurements of the time evolution of the condensate density. Recall that we restrict to Gaussian states (cf. \cref{sec:qftcs.gaussian}), which is a reasonable assumption in these experiments. The following method extends in a conceptually straightforward manner to more general states, although reconstruction of higher $n$-point functions from data requires the ability to measure the density of the condensate with increasing time resolution.

In the experiments that we aim to describe, information about the system is gathered through measurements of its density at different times, as we have seen in \cref{ch:becstheory}. Phonons correspond to fluctuations about the condensate's mean value $n_0$, which can be extracted by averaging density measurements over repeated experiments with identical parameter values. Then, measurements of the density at different times in each experiment allow one to compute the density contrast~\mbox{$\delta_c(t, \bm{u})$} (see \cref{eq:bec.RescaledDensityContrast}). As explained in subsection \ref{subsec:bec.CorrelationFunctionAndSpectrum}, the two-point correlation function of the density contrast can be written in terms of the symmetrized two-point function of the derivative of the field $\chi$, which allows one to extract the spectrum of fluctuations~$S_k$ as its momentum-space counterpart (cf. \cref{eq:bec.DensityDensityCorrelatorSpectrum}). Measurements of the density contrast at different times allow for the extraction of the spectrum and its time derivatives via finite-difference methods. Through the relations \eqref{eq:qftcs.CorrelationsSkRelations}, or equivalently \eqref{eq:qftcs.DerivativesPiCorrelator}, this allows us to reconstruct the quantum state of our system---namely the covariance matrix $\bm{\sigma}_{\vk}$ and the mean~$\bm{\mu}_{\vk}$, which vanishes in our case---given certain initial conditions, as long as linear dynamics holds. Assuming that our system is described by a Gaussian state, this provides an algorithm to perform tomography from measurements in experiments such as the one described in \cref{sec:bec.Experiment}.


\subsection{A simple Gaussian loss model}
\label{subsec:ent.losses}

In a real experiment, one has to consider losses. Moreover, since we aim at probing scale factors consisting of subsequent contraction and expansion cycles, we expect efficiency to be a parameter to take into account.

In order to implement possible experimental losses and measurement inefficiencies, we will consider a simple model consisting of a Gaussian loss channel (see e.g.~\cite{Serafini2017}), in which a mode described by some operator $\hat{a}_{\vk}$ is detected with probability $\eta$ and lost to the environment with probability $1-\eta$. Mathematically, this can be implemented by the transformation $\hat{a}_{\vk} \to \sqrt{\eta}\hat{a}_{\vk} + \sqrt{1-\eta} \hat{e}_{\vk}$, where $\hat{e}_{\vk}$ represents the annihilation operator of an environment mode---assumed to be in a Gaussian state so that the loss channel preserves Gaussianity. This translates to the map
\begin{equation}
   \bm{\mu}_{\vk}^{\text{(out)}} \to \sqrt{\eta}\bm{\mu}_{\vk}^{\text{(out)}}, \qquad \bm{\sigma}_{\vk}^{\text{(out)}} \to \eta\bm{\sigma}_{\vk}^{\text{(out)}}+(1-\eta)\bm{\sigma}_{\vk}^{\text{(vac)}}
\label{eq:ent.LossModel}
\end{equation}
for the final state, where $\bm{\mu}_{\vk}$ and $\bm{\sigma}_{\vk}$ are the mean and the covariance matrix defined in \cref{eq:qftcs.MeanAndCov}, and $\bm{\sigma}_{\vk}^{\text{(vac)}}$ denotes the covariance matrix of the vacuum. Recall that the covariance matrix can be written in terms of the offset $S_{k,0}$, the amplitude $A_k$ and the phase $\vartheta$ (cf. \cref{eq:qftcs.CovarianceMatrixOffAmpPhase}). Then, it is straightforward to see that the corresponding implementation of losses implies
\begin{equation}
    S_{k,0} \to \eta S_{k,0} + \frac{1-\eta}{2}, \quad A_k \to \eta A_k, \quad \vartheta_k \to \vartheta_k.
\label{eq:ent.OffAmpLosses}
\end{equation}
Let us stress that the phase $\vartheta_k$ is not only independent of the details of the initial state, such as temperature, as we discussed in \cref{ch:becstheory}, but also of losses. Indeed, it only depends on the dynamics of the system and on the expansion history, as can be deduced from the definition of the spectrum \eqref{eq:qftcs.SpectrumOffsetAmpPhase}. However, all information about entanglement is contained in the other two observables, the offset~$S_{k,0}$ and the amplitude~$A_k$. The fact that detecting entanglement relies on the knowledge of quantities that are sensitive to losses and temperature is crucial, and makes its detection challenging. 

We will study below the generation of entanglement in experiments implementing cyclic cusps as defined in section \ref{sec:ent.ModelExp}, and analyze its robustness against temperature and (cycle-dependent) losses, aiming at finding the optimal experimental setup for detecting entanglement from pair production.
   
\section{Entanglement production}
\label{sec:ent.Entanglement}

This section is dedicated to the quantification of the entanglement created in experiments simulating a dynamical universe in a BEC. To that end, we will use logarithmic negativity $\text{LN}_k$ (see \cref{eq:qftcs.LNOffAmp}), which is an entanglement quantifier for our system, to study entanglement in this context for different tunable experimental parameters determining the expansion or contraction of the analog universe. 


\subsection{Effects of temperature and losses in entanglement production}

According to the monotonically decreasing behavior of logarithmic negativity with~$n_k^{\text{in}}$, a higher occupancy in the initial state (i.e. a higher initial temperature) decreases the degree of entanglement for a fixed amount of produced quanta~$|\beta_k|^2$, as can be deduced from \eqref{eq:qftcs.EigenvaluesOffAmp}. This is in agreement with the well-known fact that entanglement degrades very rapidly with temperature, even vanishing (see e.g.~\cite{Agullo2022a}). Indeed, it is one of the main reasons why it is a formidable task to detect it in such an experiment: The initial state of the system must be prepared at a sufficiently low temperature. 

\begin{figure}[t!]
    \centering
    \includegraphics[width=0.65\textwidth]{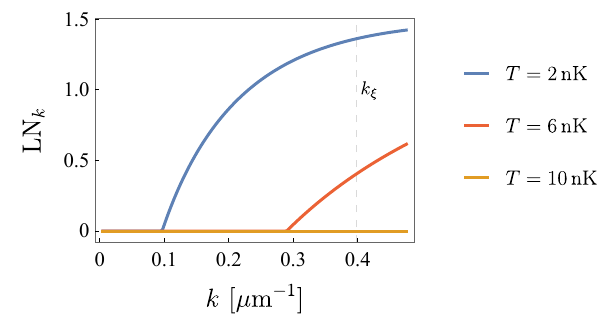}
    \caption[Entanglement as a function of the wavenumber in a typical linear-ramp experiment]{Entanglement as a function of the wavenumber $k$, characterized through $\text{LN}_k$, in a typical linear-ramp experiment ($a(t) \sim t$) of duration $\Delta t = 0.1 \, \text{ms}$. The different curves correspond to initial states of different temperatures $T$. We have used~\mbox{$\alpha_{s,\text{i}} = 400\aB$} and $\alpha_{s,\text{f}} = 50\aB$, which corresponds to approximately one $e$-fold of expansion, i.e. $\log{(a_{\text{f}}/a_{\text{i}})} \simeq 1$. The gray dashed line denotes $k_{\xi}$. This figure indicates that temperatures $\lesssim 10 \, \text{nK}$ are needed to produce entanglement within the hydrodynamic regime. Note that this plot does not account for losses, which would further degrade the entanglement. Figure from \cite{Entanglement2024}.}
    \label{fig:ent.LinearRampLN}
\end{figure}

As an illustration, we show in figure \ref{fig:ent.LinearRampLN} the theoretical prediction for $\text{LN}_k$ in a typical single-ramp experiment such as those described in section \ref{sec:bec.Experiment} and performed in \cite{Experiment2022}. We can see that, in the best-case scenarios, in which ramps are very short and the number of achieved $e$-folds is maximized (in the example of figure~\ref{fig:ent.LinearRampLN} we simulate a ramp of $\Delta t =  0.1 \, \text{ms}$, and initial and final values of the scattering length $\asi = 400\aB$ and $\asf = 50\aB$), entanglement is only present in the hydrodynamic regime---which we denote by $k\lesssim k_{\xi}$, where $k_\xi$ is the value of $k$ for which the corrections to the linear dispersion relation \eqref{eq:analogs.FullBogoliubovDispersion} become of the order of $10\%$---for temperatures $T \lesssim 10 \, \text{nK}$, which lie below the temperatures attained in current experiments ($\sim \SI{12}{\nano\kelvin}$). This tells us that the effects of temperature completely prevent the production of entanglement in experiments with a single expansion ramp such as those in \cref{sec:bec.Experiment}, and provides a case in favor of experiments with several expansion-contraction cycles, where entanglement generation can be enhanced through resonances in production due to the quasi-periodicity of the scale factor in this setup. We will see that these experiments are well positioned to offer the first measurements of entanglement in pair production due to a time-dependent background.

Based on the above, one would be tempted to increase $\abs{\beta_k}^2$ (i.e. produced quanta) as much as possible in order to maximize entanglement. Even though this works well at low production values, blindly focusing on maximizing production can worsen the signal-to-noise ratio, drowning quantum correlations within the errors of the classical signal. This means that maximizing entanglement production might not mean maximizing its detectability, but at the same time too low productions also lead to undetectable entanglement. As mentioned in section \ref{sec:ent.ModelExp}, a clever way of increasing production within current capabilities is to consider periodic ramps with a high number of cycles. However, this requires long experimental times, which increase detection losses, thereby degrading the detectable entanglement. We will see in the next section that a compromise between these two effects can be reached with a relatively low number of cycles. 

The Gaussian loss model introduced in subsection \ref{subsec:ent.losses}, where losses are controlled by the detection efficiency $\eta\in[0,1]$, introduces noise in the detected signal which effectively degrades the detectable entanglement. To see this, note that within the loss model \eqref{eq:ent.LossModel}, the smallest symplectic eigenvalue $\Tilde{\nu}_k^{\text{min}}$ defined in \cref{sec:qftcs.entanglement} becomes
\begin{equation}
    \Tilde{\nu}_{k}^{\text{min}} = 2\eta(S_{k,0}-A_k)+1-\eta.
\label{eq:ent.SymplecticEigenvaluesLosses}
\end{equation}
Therefore, in the regime where there is entanglement ($S_{k,0}-A_k<1/2$), losses increase the value of detectable $\Tilde{\nu}_k^{\text{min}}$ with respect to the lossless value \eqref{eq:qftcs.EigenvaluesOffAmp}. This leads to a decrease in detectable entanglement.

\subsection{Quantification of entanglement produced}

We will now focus on quantifying the production of entanglement and understanding how it depends on the different experimentally controllable parameters. Since we will be exploiting the periodic structure of the scale factor, small details and changes on the specific form of the cusps described in subsection \ref{subsec:ent.rampmodels} will not significantly affect production of particles, dominated by the emergent resonances. Four parameters control the details of the expansion history: (i) The parameter $\delta$, which determines both the sharpness and the duration of each cusp; (ii)~$\tho$, which sets the time interval between consecutive cusps; (iii) the accumulated \mbox{expansion/contraction} at each cusp, $\ai/\af$ (or equivalently $\asi/\asf$); and (iv) the number $n$ of consecutive cusps. Two more parameters control the environment and other aspects of the experiment, namely the ambient temperature $T$ and the efficiency $\eta$. The total efficiency depends on the duration of the experiment; keeping this in mind, we find it convenient to work with the \textit{efficiency per cusp}, $\eta_0$. The total efficiency is then given by $\eta_0^n$. Next, we discuss the impact that varying each of these parameters has on entanglement.

The effects of changing $\eta_0$ and $T$ have already been discussed above; specifically, higher temperatures and lower efficiencies degrade entanglement.

\begin{figure*}[t!]
    \centering
    \includegraphics[width=0.9\textwidth]{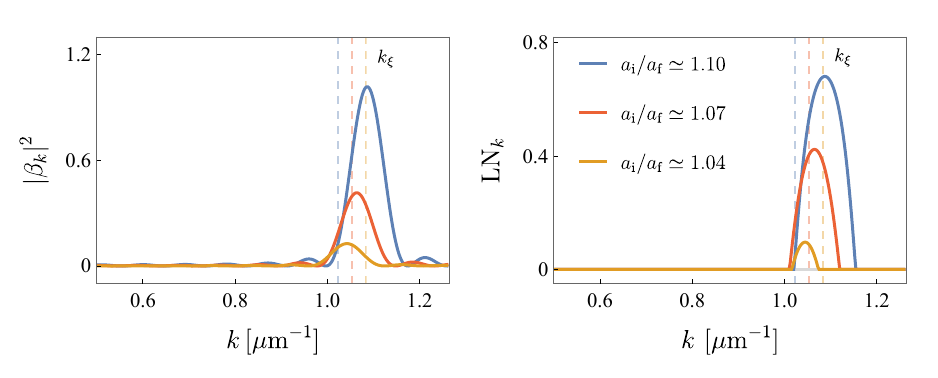}
    \caption[Spectrum of produced particles and entanglement for different values of the initial scattering length]{Number density of produced particles $\abs{\beta_k}^2$ (left) and logarithmic negativity~\mbox{$\text{LN}_k$} (right) as a function of the wavenumber $k$, for three different values of the expansion/contraction accumulated within each cusp, $\ai/\af$. We assumed $T = 12 \, \text{nK}$, $\asf = 400\aB$, $\tho = 0.75 \, \text{ms}$, $\delta = 0.4 \, \text{ms}$, $n = 12$, and $\eta_0 = 0.95$. The three curves correspond to $\asi$ equal to $330$, $350$, and $370$ in units of the Bohr radius $\aB$, which corresponds to $\ai/\af = 1.10$, $1.07$, and $1.04$, respectively. The vertical dashed line denotes the value of $k_{\xi}$ in each case. The figure shows the two effects discussed in the text: An increase in the peak values of $\abs{\beta_k}^2$ and~\mbox{$\text{LN}_{k}$} with $a_{\text i}/a_{\text f}$, and a shift in $k_{\xi}$. For $\ai/\af=1.10$, this shift causes the peak to fall outside the hydrodynamical regime. Figure from \cite{Entanglement2024}.}
    \label{fig:ent.EntanglementSpectraAlpha}
\end{figure*}

Regarding the accumulated expansion per cusp, $\ai/\af$, we find that, as expected, increasing it enhances pair production and the generation of entanglement. However, changing $\ai/\af$ also has a side effect: The wavenumber $k_\xi$, above which the hydrodynamic approximation breaks down, decreases as we increase $\ai$. This is because increasing $\ai$ amounts to decreasing the minimum value of the scattering length, $\asi$. This, in turn, reduces the range of wavevectors $\vk$ of interest, i.e. those within the hydrodynamic approximation. Since we want to discuss analog cosmological production, we will try to maximize entanglement generation within the hydrodynamical regime. Our strategy is to take a relatively high but attainable value of the scattering length at the peak of the cusp, $\asf = 400\aB$, and find the value of the scattering length at the valley of the ramp, $\asi$, that optimizes entanglement production within the hydrodynamic regime. Fig.~\ref{fig:ent.EntanglementSpectraAlpha} shows $\text{LN}_k$ for different accumulated expansions per cusp, highlighting the two effects just mentioned. We also find that the value of $\ai/\af$ which maximizes entanglement generation within the hydrodynamical regime depends slightly on the initial temperature. For the lowest attainable temperatures, it ranges around $\ai/\af\simeq 1.07$, corresponding to $\asf = 400\aB$ and $\asi = 350\aB$. For these values, the hydrodynamical regime corresponds to $k \lesssim k_\xi \approx \SI{1.05}{\micro\meter}^{-1}$.

\begin{figure*}[t!]
    \centering
    \includegraphics[width=0.9\textwidth]{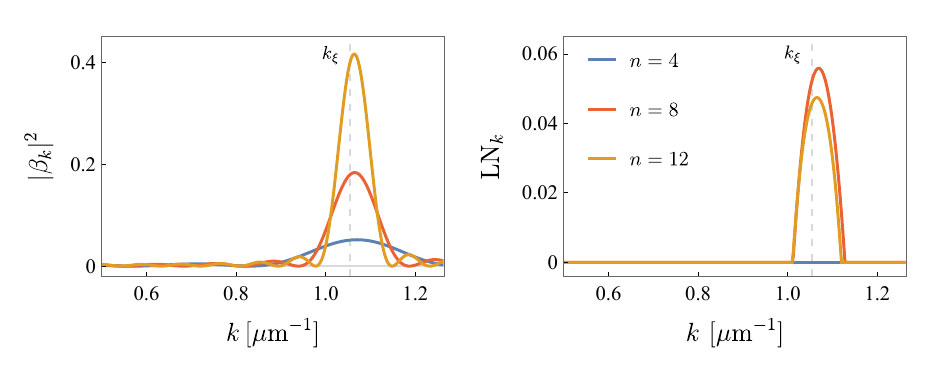}
    \caption[Spectrum of produced particles and entanglement for different numbers of cycles]{Number density of produced particles $\abs{\beta_k}^2$ (left) and logarithmic negativity~$\text{LN}_k$ (right) as a function of the wavenumber $k$, for three different numbers of cusps $n$. We used $T = 12 \, \text{nK}$, $\alpha_{s, \text{i}} = 350\aB$, $\alpha_{s, \text{f}} = 400\aB$, $\tho = 0.75 \, \text{ms}$, $\delta = 0.4 \, \text{ms}$, and $\eta_0 = 0.8$. The vertical gray dashed line denotes $k_{\xi}$. This figure shows the enhancement of peak values of $\abs{\beta_k}^2$ and $\text{LN}_{k}$ for moderate $n$, and the degrading effect that losses have on entanglement for large $n$. Figure from~\cite{Entanglement2024}.}
    \label{fig:ent.ProductionSpectraDifferentN}
\end{figure*}

Modifying the number $n$ of consecutive cusps also has a dual effect. On the one hand, the periodic variation of the scale factor imprints an oscillatory structure on the Bogoliubov coefficients $|\beta_k|^2$. We observe that the value of $|\beta_k|^2$ at the peaks increases with $n$. Thus, in an idealized experiment with no losses, a larger $n$ would lead to more entanglement. However, increasing $n$ also extends the duration of the experiment, thereby increasing losses. As a result, there is an optimal number of cycles, $n_{\text{op}}$, beyond which the degradation due to losses outweighs the enhancement in pair production. The value of $n_{\text{op}}$ depends on the efficiency per cusp, $\eta_0$, which must be characterized experimentally (see fig.~\ref{fig:ent.XmaxCycles} in section \ref{sec:ent.optimization}). A possible approach to achieve this is to measure the spectrum of produced quanta for identically prepared realizations that run over different numbers of cusps. Figure~\ref{fig:ent.ProductionSpectraDifferentN} illustrates the two effects that arise from modifying $n$ (note that we used a relatively low efficiency to exaggerate them).

\begin{figure*}[t!]
    \centering  
    \includegraphics[width=0.9\textwidth]{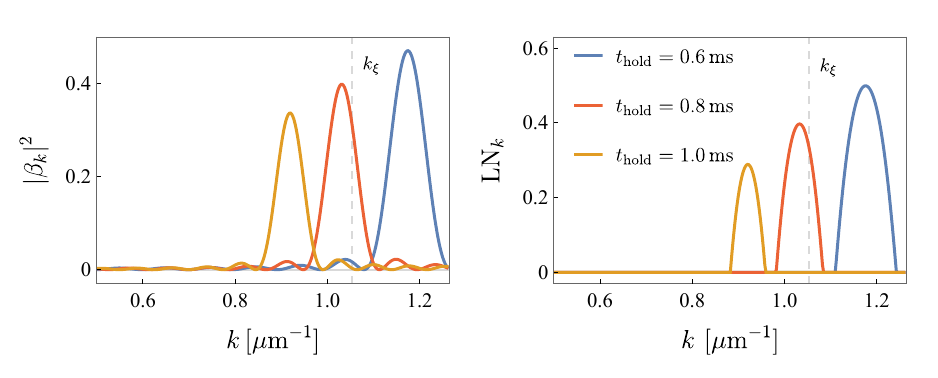}
    \caption[Spectrum of produced particles and entanglement for different hold times]{Particle number density $|\beta_{k}|^2$ (left) and logarithmic negativity $\text{LN}_k$ (right) for three different values of $\tho$ (see fig.~\ref{fig:ent.ScatteringLength} for a graphical definition of this parameter). This plot is obtained using $T = 12 \, \text{nK}$, $\asi = 350\aB$, $\asf = 400 \aB$, $\delta = 0.4 \, \text{ms}$, $n = 12$, and $\eta_0 = 0.95$. The vertical gray dashed line denotes $k_{\xi}$. We observe that $\tho$ controls the position of the resonance, shifting it toward the infrared as $\tho$ increases. Figure from \cite{Entanglement2024}.}
    \label{fig:ent.EntanglementSpectraThold}
\end{figure*}

\begin{figure*}[t!]
    \centering   
    \includegraphics[width=0.9\textwidth]{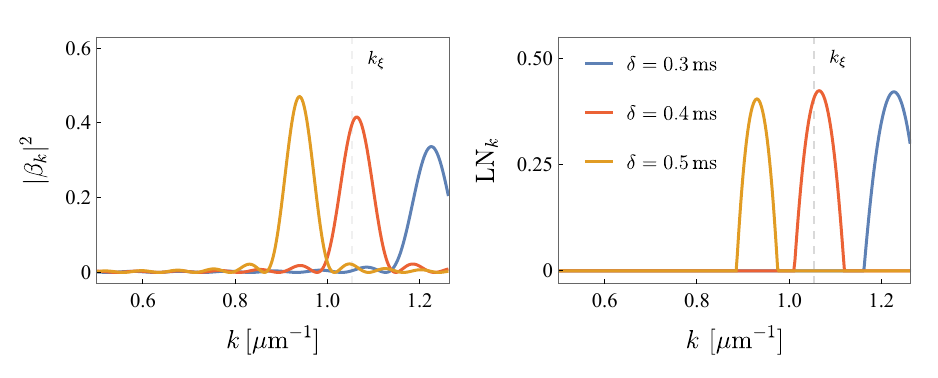}
    \caption[Spectrum of produced particles and entanglement for different cusp durations]{Particle number density $|\beta_{k}|^2$ (left) and logarithmic negativity $\text{LN}_k$ (right) as a function of $k$ for three different values of $\delta$ (see fig.~\ref{fig:ent.ScatteringLength} for the meaning of this parameter). This plot is obtained using $T = 12 \, \text{nK}$, $\asi = 350\aB$, $\asf = 400\aB$, $\tho = 0.75 \, \text{ms}$, $n = 12$, and $\eta_0 = 0.95$. The vertical gray dashed line denotes~$k_{\xi}$. We observe that $\delta$ shifts the position of the resonance. It also has a small effect on the size of $\text{LN}_k$. Figure from \cite{Entanglement2024}.}
\label{fig:ent.EntanglementSpectraDelta}
\end{figure*}

The time lapse $\tho$ between cusps controls the periodic structure of the expansion history $a(t)$. It is therefore expected that $\tho$ determines the location of the resonant peaks observed in the coefficients $|\beta_k|^2$. This is indeed the case, as shown in fig.~\ref{fig:ent.EntanglementSpectraThold}, where we observe that increasing $\tho$ shifts the resonances toward infrared wavenumbers $k$. One takeaway from this figure is that it is important to tune $\tho$ for the resonances in entanglement to fall within the hydrodynamical regime.

Finally, $\delta$ has also an effect in the position of the resonances, since this parameter controls the duration of the cusps (see fig. \ref{fig:ent.EntanglementSpectraDelta}).

\section{Optimizing detectability of entanglement}\label{sec:ent.optimization}
In this section we will use the tools developed above to assess detectability of entanglement in BEC experiments with current capabilities. Since the detection of entanglement would be a benchmark on its own, we will consider in this section the faithful witness $\Delta_k$ (cf. equation \eqref{eq:qftcs.CS}) in order to maximize detectability. Note, however, that a quantifier such as $\text{LN}_k$ would be needed to fully characterize the entanglement produced.

We proceed by assuming a given precision in the values of $S_{k,0}$ and~$A_k$, reconstructed from the measurement of density contrast correlations. Then, we propagate these errors to $\Delta_k$, so that entanglement between phonons with momentum $\hbar\vk$ and~\mbox{$-\hbar\vk$} is detected with statistical significance $X_k$ if
\begin{equation}
\Delta_k>0\quad\text{and}\quad\Delta_k-X_k\sigma_k^{\Delta}=0\,,
\label{eq:ent.OptCondition}
\end{equation}
where $\sigma^{\Delta}_k$ denotes the error in the measurement of $\Delta_k$. Our aim is thus to characterize the regions of experimental parameter space in which the statistical significance of the detection is higher, i.e. we want to maximize $X_k$. 

The optimization procedure will thus iterate over the four relevant parameters that can be experimentally tuned. Recall that these are the initial scattering length~$\asi$, the number of cycles $n$, the waiting time between cycles $\tho$ and the cusps' sharpness $\delta$. To that end, we need to assume some loss per cycle $1-\eta_0$ and a temperature for the initial state $T$. Then, we choose a grid of points in parameter space $(\asi,\tho,n,\delta)$ over which each of the parameters will be optimized. We then quantify the violation of Cauchy-Schwarz inequality within the hydrodynamical regime $k\leq k_\xi$ using $\Delta_k$, and obtain the corresponding error $\sigma^{\Delta}_k$. Lastly, we compute~$X_k$ from the above equation \eqref{eq:ent.OptCondition}, obtaining a \textit{spectrum of statistical significance} as a function of $k$ for each point in the grid of experimental parameters. Finally, we choose the maximum value of $X_k$ for each point in the grid $(\asi,\tho,n,\delta)$, and store it into a function $X_{\text{max}}(\asi,\tho,n,\delta)$. This function has the information of the maximum statistical significance with which entanglement can be detected within the hydrodynamical regime for each combination of parameters and for a given value of initial temperature $T$, loss per cycle $\eta_0$, and assumed precision $\epsilon_{\text{r}}$ of the measurement of  $S_{k,0}$ and $A_k$. 

\begin{figure*}[t!]
    \centering \includegraphics[width=\textwidth]{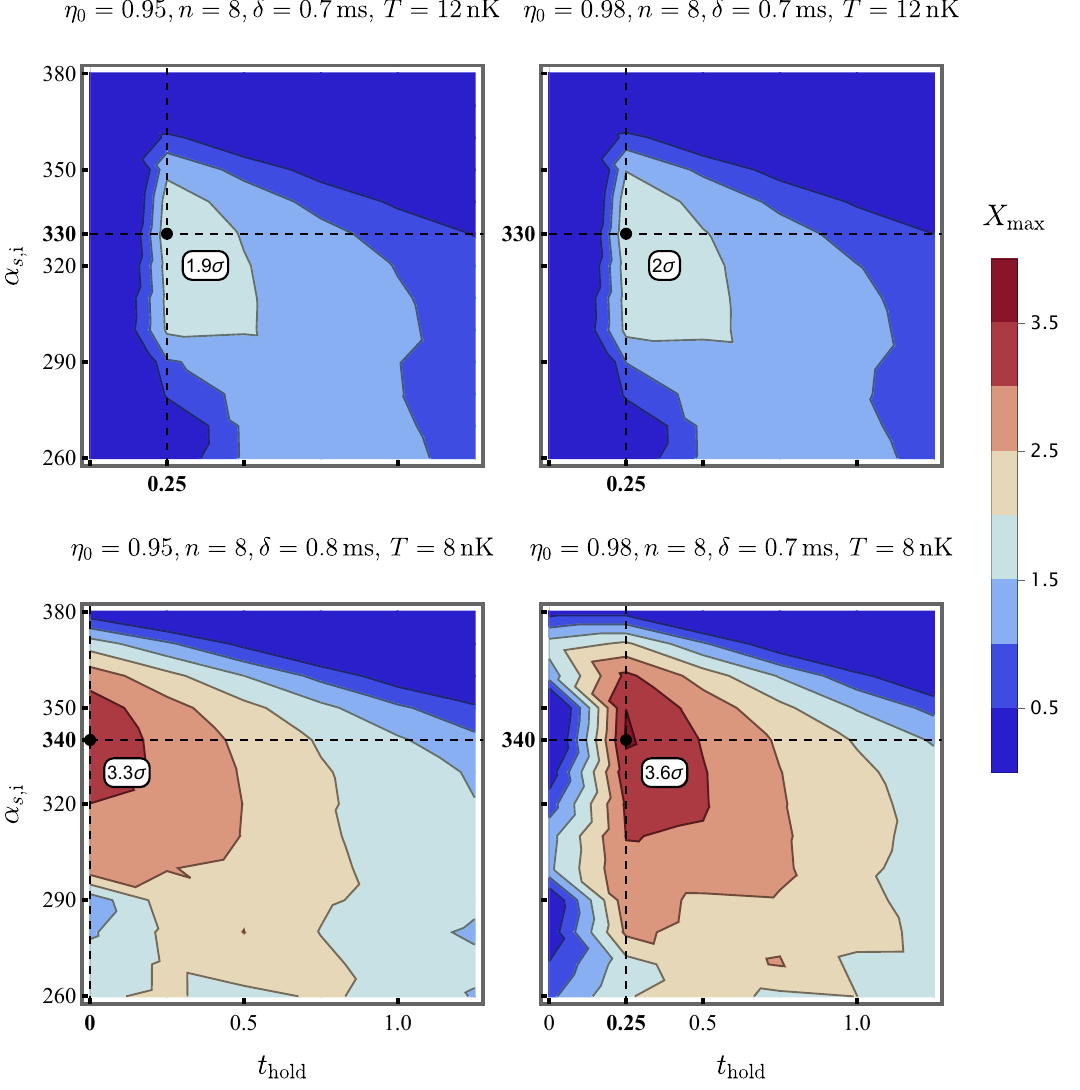}
    \caption[Contour plots in experimental parameter space of the statistical significance of measuring entanglement production by pair creation within the hydrodynamic regime]{Contour plots in experimental parameter space of the maximum significance~$X_{\text{max}}$ of measuring entanglement production due to pair creation in a quasi-2D BEC analog bouncing universe, within the hydrodynamic regime. The axes denote initial scattering length $\asi$ and waiting time between cusps~$\tho$. The optimal configuration for each of the considered initial temperatures~$T$ and loss-per-cycle $\eta_0$ is marked with a black dot. The value of $k$ which maximizes $X_{\text{max}}$ (black dots), and the value of $k_{\xi}$ in the corresponding experimental configuration are $k_{\text{max}}=1.020 \, \mu\text{m}^{-1}, k_{\xi}=1.023\, \mu\text{m}^{-1}$ in the upper panels, and $k_{\text{max}}=1.020 \, \mu\text{m}^{-1}, k_{\xi}=1.038\, \mu\text{m}^{-1}$ in the lower panels. Finally, we have used a relative error of 5\% in the measurement of $S_{k,0}$ and $A_k$. For this plot, we have fixed~$n=8$ in the optimization process. Figure from \cite{Entanglement2024}.}
    \label{fig:ent.DetectabilityContours}
\end{figure*}

To perform the optimization, we have chosen a grid of experimental parameters defined by $\asi\in[260\aB, 380 \aB]$, with step $\Delta\asi=10\aB$; $\tho\in[\SI{0}{\milli\second},\SI{1.5}{\milli\second}]$, with step~\mbox{$\Delta\tho=\SI{0.1}{\milli\second}$}; $n\in[0,12]$ with step $\Delta n=1$; and $\delta \in [0.1 \, \text{ms},1.0 \, \text{ms}]$ with step $\Delta \delta = 0.1 \, \text{ms}$. The only exception is \cref{fig:ent.DetectabilityContours}, where $n=8$ was the only allowed number of cycles. The spectrum of entanglement $\Delta_k$ is computed for each~\mbox{$k\in[\SI{0.01}{\micro\meter},k_{\xi}(\asi)]$} with step $\delta k=\SI{0.01}{\micro\meter}$. We have performed this optimization for $T\in\{8,12,16,20\}$nK, precision $\epsilon_{\text{r}}$ in the determination of $S_{k,0}$ and~$A_k$ in the range of 1\% to 6\%, and values of $\eta_0\in\{0.98,0.95,0.9,0.85\}$, which correspond to losses around 20-85\% for the optimal number of cycles in each case---in experiments such as the one described in section \ref{sec:bec.Experiment}, we expect lower total losses.

The results of this optimization process can be visualized as follows. Choose a value for the number of cycles $n_0$ and the cusp sharpness $\delta_0$, and produce a density plot of $X_{\text{max}}(\asi,\tho,n_0,\delta_0)$, such as those in figure \ref{fig:ent.DetectabilityContours}. Closed contours in this plot, defined by \mbox{$X_{\max}(\asi,\tho,n_0,\delta_0)=X_{\text{c}}$}, enclose regions within which the statistical significance for detecting entanglement is greater than $X_{\text{c}}\sigma$. In figure~\ref{fig:ent.DetectabilityContours} we show the corresponding plots for $T=\SIlist{8;12}{\nano\kelvin}$, relative uncertainty in the measurement of $S_{k,0}$ and $A_k$ of $5\%$, and losses per cycle of $\eta_0=0.95$ and~\mbox{$\eta_0=0.98$}. The values $n_0$ and $\delta_0$ chosen are the ones that maximize detectability in each case. The results show that entanglement can be detected at $\gtrsim 1.9\sigma$ for an initial temperature of $T=12\,$nK and at $\gtrsim3.3\sigma$ for an initial temperature of~\mbox{$T=8\,$nK}.

\begin{figure}[t!]
    \centering
 \includegraphics[width=0.55\textwidth]{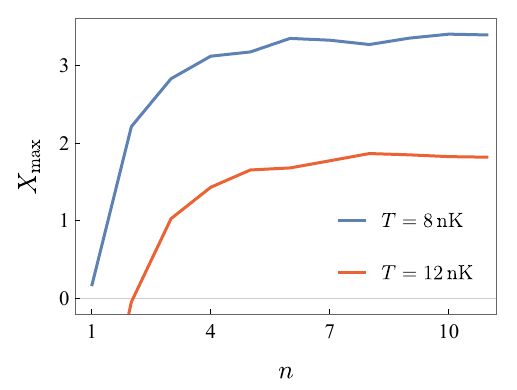}
    \caption[Statistical significance with which entanglement can be detected as a function of the number of cycles]{Statistical significance with which entanglement can be detected~$X_{\text{max}}$ as a function of the number of cycles. The loss model assumed in this plot is $1-\eta_0^n$ with $\eta_0=0.95$, and the relative uncertainty for $S_{k,0}$ and $A_k$ is of $5\%$. Figure from \cite{Entanglement2024}.}
    \label{fig:ent.XmaxCycles}
\end{figure}

An important aspect to be aware of is that losses only influence the produced entanglement through the total loss, not the loss per cycle. For our simple model, this is given by a constant amount of loss for each cycle, so $\eta(n)=1-\eta_0^n$. However, the particular dependence of the total loss in the number of cycles can change the value of~$n$ for which $X_{\text{max}}$ is maximized. Hence, we stress that in any experiment aiming at detection of entanglement from several expansion-contraction cycles, this should be characterized beforehand. We can, however, provide some qualitative information on how statistical significance will depend on the number of cycles. First, we note that $X_{\text{max}}$ grows rapidly with $n$ at low number of cycles, but eventually the growth stops, and afterward increasing $n$ leads to less detectability. This feature, exemplified in figure \ref{fig:ent.XmaxCycles}, is quite robust against reasonable modifications to the loss models.

\begin{figure*}[t!]
    \centering   
    \includegraphics[width=\textwidth]{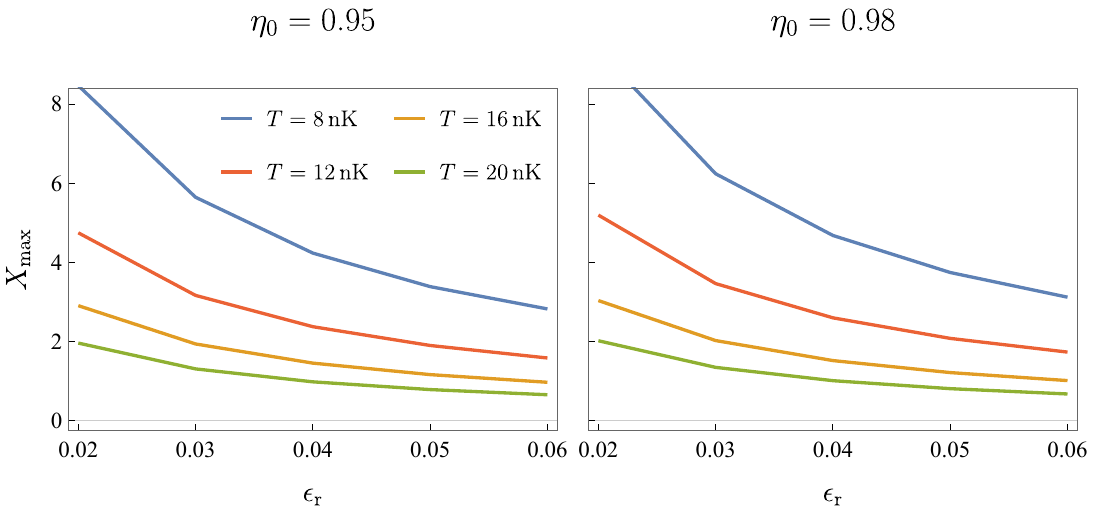}
    \caption[Statistical significance with which entanglement can be detected as a function of precision in the determination of $S_{k,0}$ and $A_k$ for optimal parameters~$\asi$,~$n$,~$\tho$ and ~$\delta$]{Significance at which entanglement can be detected, as a function of relative error $\epsilon_{\text{r}}$ in the experimental determination of $S_{k,0}$ and $A_k$ for optimal parameters~$\asi$,~$n$,~$\tho$ and~$\delta$. The same precision is assumed to be achievable for both~$S_{k,0}$ and~$A_k$. This figure shows that with currently available precision of~\mbox{$5-6\%$} and initial temperatures of $\sim12\,$nK, entanglement can be detected at around $2\sigma$. Mild improvements in precision or temperature rise the statistical significance up to $\sim 4\sigma$. Figure from \cite{Entanglement2024}.}
    \label{fig:ent.MaxSigmas}
\end{figure*}

To study the impact of the assumed precision, in figure \ref{fig:ent.MaxSigmas} we plot the maximum value of the statistical significance at which entanglement is detected for several initial temperatures as a function of the assumed precision in the measurement of amplitude and offset. We see how a small improvement in precision and temperature from currently attainable values ($5\%$ and $12\,$nK, respectively) raises the significance in the detection of entanglement to around $\sim4\,\sigma$. 

\begin{figure*}[t!]
    \centering  
    \includegraphics[width=\textwidth]{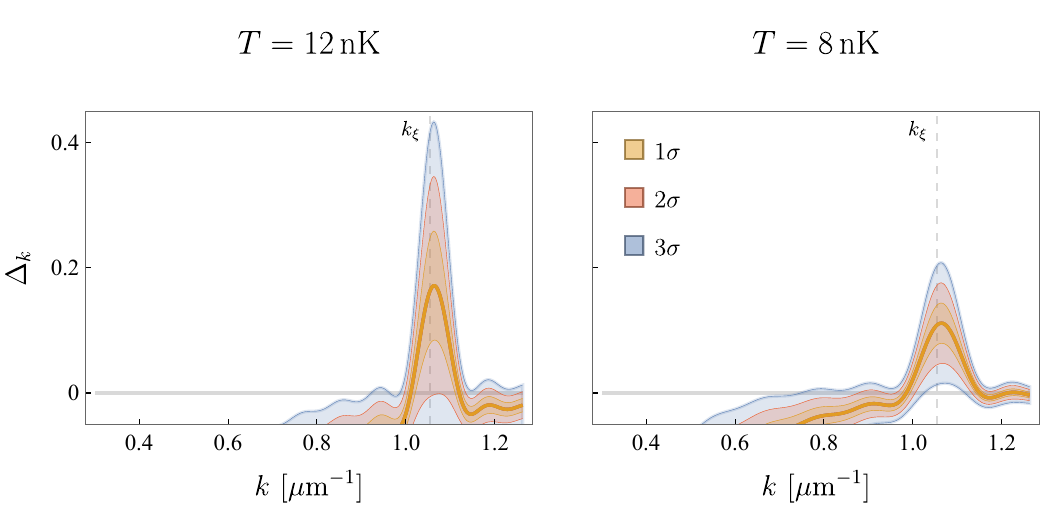}
    \caption[Spectrum of entanglement extracted with $1\sigma$,~$2\sigma$ and~$3\sigma$ bands obtained from propagating the uncertainty of $S_{k,0}$ and~$A_k$]{Spectrum of entanglement as given by $\Delta_k$, with bands indicating $1\sigma$ (orange), $2\sigma$ (red) and $3\sigma$ (blue), using a relative error for $S_{k,0}$ and $A_k$ of 5\%. We take~\mbox{$\eta_0=0.95$}, together with $(\alpha_{s}=350\aB, \tho=0.75 \, \text{ms}, \delta=0.4 \, \text{ms})$, and~\mbox{$n=12$} for the left panel as well as $n=9$ for the right panel, which are the optimal values in each case. The gray dashed lines denote $k_{\xi}$. Figure from~\cite{Entanglement2024}.}
    \label{fig:ent.EntanglementSpectrum}
\end{figure*}

Finally, we plot the spectrum of produced entanglement between each pair $(\vk,-\vk)$ via $\Delta_k$ for the values of experimental parameters that optimize detection at the assumed temperature and precision in figure \ref{fig:ent.EntanglementSpectrum}, including error bands. We see how the detectability of entanglement is always best near the limits of the hydrodynamical regime. This is because optimization of $\asi$ and $\tho$ to maximize statistical significance for detection of entanglement within the hydrodynamical regime tends to select the value of $\asi$ such that there is a resonance in production precisely at the border of the hydrodynamic regime~$k\approx k_\xi$.

\section{Summary}

In this chapter, we have studied the production of entanglement in analog cosmological experiments in BECs such as the one discussed in \cref{sec:bec.Experiment}. For this, we have considered logarithmic negativity as a quantifier, whereas better results regarding detectability can be obtained with the Cauchy-Schwarz inequality, which is a faithful witness in our system. We have seen that considering several expansion-contraction cycles leads to much greater detectability than ordinary single ramps. Moreover, we have found that with currently available techniques, and upon optimization of the relevant experimental parameters, it is possible to detect entanglement at around $2\sigma$ in the acoustic regime, while mild improvements in the temperature or the measurement precision would increase the significance into the range of $3\sigma$ to $5\sigma$, respectively. Further improvements could come, for example, from squeezing the condensate to have a single-mode squeezed thermal state for each perturbation mode as the input for the expansion-contraction cycles~\cite{Agullo2021}, which we leave for future work.

%% file: Chapters/QVA.tex

\addtocontents{toc}{\protect\vspace{0.5em}}

\chapter{Operational realization of quantization ambiguities} 

\label{ch:qva} 




We have seen throughout the thesis that the dynamics of spacetime can lead to particle production of quantum fields. In general, the presence of any time-dependent, external field can lead to particle creation, and quantization follows similarly as for QFTCS. As anticipated in \cref{sec:qftcs.schwinger}, this is the case of the Schwinger effect as well, where the external, classical field is a potential associated with an electric field. The latter introduces energy in the system, leading to pair production even from vacuum. Moreover, in the same way as the time-dependence of spacetime geometry, this external agent breaks time-translational invariance. As a consequence, the notions of vacuum and particle become ambiguous, since the number of symmetries of the classical theory is not large enough to completely determine a preferred quantization (see \cref{ch:qftcs}). 

If the electric field is off at early times, one may define a preferred \textit{in} vacuum (cf. \cref{subsec:qftcs.inout}). However, in order to compute the number density of particles $n_k(\tau)$ produced at some time $\tau$ after the field has been switched-on, it is necessary to make a choice for the vacuum of an observer living at said time. In the analog BEC scenarios in part III, we have found ourselves in situations in which expansion of (analog) spacetime had stopped in the region in which production was computed. Thus, we were able to properly define an \textit{out} vacuum. However, as long as the external field is on at $\tau$ and behaves non-adiabatically (in other words, while Poincaré symmetry is still broken), the number density of particles $n_k(\tau)$ entails quantization ambiguities, which poses questions about the physical interpretation of this quantity \cite{Ilderton2022,Alvarez2022,Dabrowski2014,Dabrowski2016,Yamada2021,Domcke2022,Diez2023}. This is the typical situation in Cosmology, since spacetime is always expanding\footnote{Although we used in part II the fact that in certain regions the expansion of spacetime is adiabatic, there were still several convenient quantization choices (cf. \cref{eq:sf.AdiabaticVacuum,eq:sf.AveragedVacuum}, where we discussed the adiabatic and the averaged vacuum, or \cref{subsec:ds.exit} for a brief comment on the ILES vacuum).}.

In this chapter, we will provide a way of understanding the physical meaning of the possible definitions of $n_k(\tau)$ motivated by the analog experiments discussed in part III, in which the number density of particles is measured after switching off the external agent. For simplicity, we will consider a $(1+1)$-dimensional scenario in which an electric field is smoothly switched on from zero, so that there is a preferred \textit{in} vacuum, and we wonder about the amount of produced particles at a time $\tau$ after the switch-on. We will show how the different theoretical vacuum prescriptions at~$\tau$ can be associated with \textit{out} vacua resulting from a measurement process in which the electric field is switched off at that time $\tau$, identifying both notions of quanta. In this way, we put together some ideas presented in parts II and~III of the thesis, benefiting from the insights each analysis in its context (universe and laboratory) has provided. Note that the discussion in this chapter applies to any QFT in the presence of a time-dependent external agent.

\section{Schwinger effect in one spatial dimension}

The dynamics of a charged scalar field $\psi(t, x)$ in $(1+1)$-dimensional Minkowski spacetime in the presence of a spatially homogeneous, time-dependent, classical electric field is governed by the action \eqref{eq:qftcs.SchwingerAction}, leading to a mode equation of the form~\eqref{eq:qftcs.ModeEquation} for its corresponding Fourier modes, where the time-dependent frequency is given, in contrast to the cosmological case, by \cref{eq:qftcs.SchwingerFrequency}. Note that since we are working in $D=1$, the angle $\theta$ is set to $\pi/2$, so that the frequency $\Omega_k$ reads in this particular case
\begin{equation}
    \Omega_k^2(t) = k^2 + 2qA(t)k + q^2A^2(t) + m^2.
\label{eq:qva.Frequency}
\end{equation}

Quantization of the field $\psi$ follows in the exact same way as described in \cref{ch:qftcs}, with the exception that now the field is complex, and therefore we will have different operators for particles and antiparticles. Indeed, the expansion of the field in terms of a basis of solutions of the mode equation $\{v_k,v_k^*\}$, namely (see \cref{eq:qftcs.ChiFieldExpansion})
\begin{equation}
\label{eq:qva.hatphi}
    \tilde \psi_k(t) = {a}_k v_k(t)+ {b}^{*}_k v_k^*(t),
\end{equation}
determines the corresponding Fock quantization, provided that the expansion coefficients are promoted to annihilation and creation operators $\hat{a}_k$ and~$\hat{b}^{\dagger}_k$, respectively. As usual, homogeneity and isotropy imply that modes with different wave numbers~$k$ are decoupled from one another.

\begin{figure}[t!]
    \centering
    \includegraphics[width=0.55\textwidth]{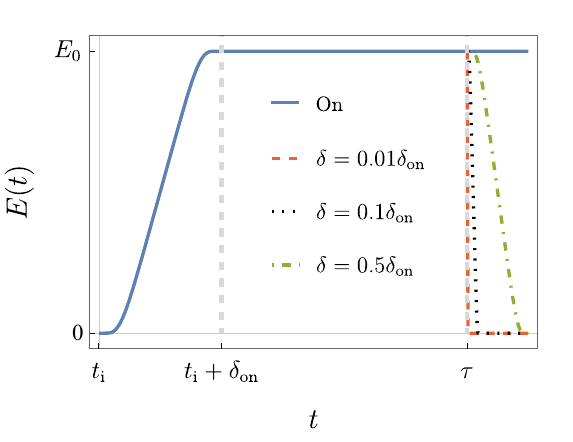}
    \caption[Time evolution of the electric field with different switch-off profiles]{Time evolution of the electric field (solid line) with different switch-off profiles (dashed/dotted lines) corresponding to different values of $\delta$, starting at $\tau$. Figure from~\cite{QVA2023}.}
    \label{fig:qva.E(t)}
\end{figure}

In a similar way as we did in \cref{ch:entanglement}, we consider non-instantaneous switch-ons and -offs of the electric field (see fig.~\ref{fig:qva.E(t)}). The latter is switched on at a time $t_{\text{i}}$, and smoothly reaches the constant value $E_0$ after a time $\delta_{\text{on}}$. Experimentally, in order to access the number of particles created at a time $\tau$, we would start switching it off at that time, until the field vanishes after a time $\delta$. We will describe the finite duration of the switch-on and -off processes by the step function $\Theta_{\sigma}(t)$ of width~$\sigma$ (i.e., $\delta_{\text{on}}$ and $\delta$, respectively, in this case) defined in \cref{eq:ent.SwitchFunction}. Nevertheless, the specific form of~$\Theta_{\sigma}(t)$ does not qualitatively affect the results. In all figures we fix~$m=1$ and~\mbox{$q=1$}, so that the electric field reaches the critical Schwinger limit\footnote{Recall that below this limit, production is negligible (see \cref{sec:qftcs.schwinger}).}~\mbox{$m^2/q=E_0=1$}~\mbox{\cite{Schwinger1951,Yakimenko2019}}. In addition, we set $t_{\text{i}}=0$, $A(t_{\text{i}})=0$, and times are given in units of the switch-on duration $\delta_{\text{on}}$, while we parametrize different switch-offs by varying $\delta$.

\section{Measured particle number}

Given a particular experimental setting, we can compute the asymptotic number of created particles $n^{\text{exp}}_{k,\tau}$ in the mode $k$ that would be measured by our detector when we start the switch-off at the time~$\tau$ (that is, the number of particles in the \textit{out} region). Initially, when the electric field is not on yet, the matter field is in the Minkowski vacuum. As we already know, this state is determined by the solution~$v_k$ to the mode equation which behaves as a positive-frequency plane wave before~$t_{\text{i}}$ (cf. \cref{eq:qftcs.InModeIC}). Similarly, the \textit{out} vacuum is associated with solutions which become positive-frequency plane waves after the electric field is completely switched off (see \cref{eq:qftcs.OutModeIC}), namely at $t_{\text{f}}=\tau+\delta$. We can then calculate the asymptotic number of created particles in the mode $k$ according to the Bogoliubov formalism as (cf. \cref{eq:qftcs.MeanNumberDensity})
\begin{equation}
    n_{k,\tau}^{\text{exp}}=\left|v_k(t_{\text{f}})\dot{u}_k(t_{\text{f}})-\dot{v}_k(t_{\text{f}})u_k(t_{\text{f}}) \right|^2.
    \label{eq:qva.numas}
\end{equation}

\begin{figure}[t!]
    \centering
    \includegraphics[width=0.75\textwidth]{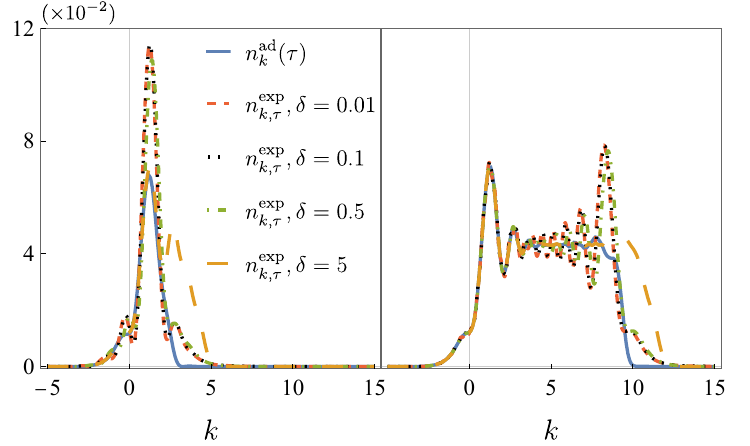}
    \caption[Spectra of the asymptotic number of created particles for different switch-offs]{Spectra of the asymptotic number of created particles~$n_{k,\tau}^{\text{exp}}$ for different switch-offs, starting at times $\tau=3$ (left) and \mbox{$\tau=10$} (right), with different durations~$\delta$ (dashed lines). We also represent the theoretical number density of created particles $n_k^\text{ad}(\tau)$ in the adiabatic vacuum at those times (solid line). Figure from~\cite{QVA2023}.}
    \label{fig:qva.numk}
\end{figure}

Let us stress that for a different measurement process starting at the same $\tau$, here characterized by another duration of the switch-off~$\delta$, the outcome $n_{k,\tau}^{\text{exp}}$ will change. In fig.~\ref{fig:qva.numk}, we show the spectra of asymptotically produced particles $n_{k,\tau}^{\text{exp}}$ for $\tau=3$ and $\tau=10$ for different switch-off durations $\delta$. Observe that the slower the switch-offs, the longer the electric field can accelerate particles, and thus modes with larger wavenumber $k$ become excited. This behavior is in agreement with that of ref. \cite{Adorno2018}, where the authors thoroughly analyze the role played by $\delta$ and~$\tau$ in particle production for a similar profile of the electric field. Note that there appear the typical oscillations in the spectral distribution that have already been observed in previous chapters.

This notion of particle corresponds to the one used in the \textit{in-out} processes in analog systems discussed in detail in part III.

\section{Theoretical particle number}

On the other hand, one may be interested in theoretically computing the number of particles that have been created from an initial time~$t_{\text{i}}$ to some time $\tau$, without relying on any measurement. This question is, however, much more subtle. One would need to choose a particular solution~$u_k^{\tau}$ to eq.~\eqref{eq:qftcs.ModeEquation} by imposing initial conditions at time~$\tau$, i.e., $(u_k^{\tau}(\tau),\dot{u}_k^{\tau}(\tau))$. Contrary to what happens for the \textit{out} solution $u_k$ in the previous section, there is an ambiguity in the selection of $u_k^{\tau}$. Indeed, when the electric field is on, the frequency \eqref{eq:qva.Frequency} is not constant and the physical criterion of preservation of the classical symmetries in the quantum theory is not strong enough to fix a unique vacuum. The spectral number of created particles between $t_{\text{on}}$ and $\tau$ is defined by
\begin{equation}
    n_k(\tau)=\left| v_k(\tau)\dot{u}_k^{\tau}(\tau)-\dot{u}_k(\tau)u_k^{\tau}(\tau) \right|^2.
    \label{eq:qva.numt}
\end{equation}
In contrast with the measured value $n_{k,\tau}^{\text{exp}}$, which is unique given a particular switch-off of the electric field starting at a certain time $\tau$, the theoretical---in the sense that it is not directly measured---quantity~$n_k(\tau)$ strongly depends on the choice of vacuum at that time. This is precisely the reason why the interpretation of this magnitude is not clear in the literature yet \cite{Ilderton2022,Alvarez2022,Dabrowski2014,Dabrowski2016,Yamada2021}. The situation in Cosmology can be even more complicated, because both the \textit{in} and \textit{out} regimes are reached asymptotically (in the best case scenario) and one must make choices for~$v_k$ as well. 

As a well-known example, the solid line in fig.~\ref{fig:qva.numk} shows the computed number of created particles $n_k^\text{ad}(\tau)$ for the adiabatic vacuum (cf. \cref{eq:sf.AdiabaticVacuum})
\begin{equation}
    u_k^{\text{ad}}(\tau)=\frac{1}{\sqrt{2\Omega_k(\tau)}}, \quad \dot{u}_k^{\text{ad}}(\tau)=-i\sqrt{\frac{\Omega_k(\tau)}{2}}-\frac{\dot{\Omega}_k(\tau)}{2\sqrt{2}\Omega_k(\tau)}.
\label{eq:qva.ad0}
\end{equation}
Note the differences between this curve (blue in \cref{fig:qva.numk}) and the corresponding to very fast switch-offs. We will comment on this later~on.

\section{Relating experimental and theoretical particle numbers}

Each measurement procedure selects a specific vacuum whose associated theoretical particle number $n_k(\tau)$ has a clear and well-defined physical interpretation. Indeed, among the different ways to select a mode~$u_k^{\tau}$ in eq.~\eqref{eq:qva.numt}, there is a particular choice for which the number density at time $\tau$ coincides with the measurement outcome~$n^{\text{exp}}_{k,\tau}$ that a specific device would yield. Moreover, we can select, for each time~$\tau$, the mode $u_k^{\tau}$ as the \textit{out} vacuum $u_k$ corresponding to the switch-off starting at that instant. Note that making the replacement $t_{\text{f}} \to \tau$ in eq.~\eqref{eq:qva.numas} does not affect the resulting $n^{\text{exp}}_{k,\tau}$, since this value is independent of the precise instant in which it is computed. Therefore, the choice $u_k^{\tau}=u_k$ effectively makes eqs.~\eqref{eq:qva.numt} and~\eqref{eq:qva.numas} identical. The set of modes $\{u_k^{\text{exp}, \tau}\}$ and the corresponding vacua $\{\ket{0^{\text{exp},\tau}}\}$, defined as described above for each $\tau$, allow for the construction of $n_k^{\text{exp}}(\tau)$ viewed as a function of the time at which we start switching the electric field off. At each time $\tau$, the \textit{in} vacuum is an excited state relative to the vacuum $\ket{0}^{\text{exp},\tau}$, with its excitations directly corresponding to the particles that our detector would measure. It is important to emphasize that $\tau$ represents the time at which we wish to calculate the particles generated by the electric field, not the starting point of a programmed switch-off. This prescription defines a family of physical vacua: Those that can be associated with a switch-off process leading to the same particle number as the one predicted by the vacua themselves. Furthermore, all these vacua unitarily implement the dynamics, as they lead to a finite number of particles by construction, thereby satisfying Shale's theorem~\cite{Shale1962}.

\begin{figure}[t!]
    \centering
    \includegraphics[width=0.75\textwidth]{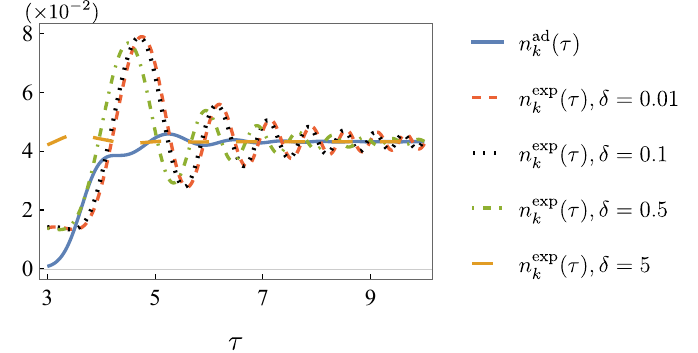}
    \caption[Evolution of the number of created particles for different switch-off durations]{Evolution of the number of created particles $n^{\text{exp}}(\tau)$ with $k=3$ for different switch-off durations~$\delta$. The solid line corresponds to the zeroth-order adiabatic prescription.  Figure from~\cite{QVA2023}.}
    \label{fig:qva.numt}
\end{figure}

In our simplified model, the measurement device is characterized by~$\delta$, and for each value of $\delta$ we obtain a different set of modes $\{u_k^{\text{exp}, \tau}\}$, and thus different interpretations of $n_k^{\text{exp}}(\tau)$. This is shown in fig.~\ref{fig:qva.numt}, where the time evolution of~$n_k^{\text{exp}}(\tau)$ for $k=3$ is depicted, with different durations for the switch-off period $\delta$. For each time $\tau$, we compute the asymptotic particle number $n_{k,\tau}^{\text{exp}}$ when the electric field is switched off at $\tau$. The oscillations observed in $\tau$, previously reported in refs.~\mbox{\cite{Kluger1998,Schmidt1998,Dabrowski2014,Dabrowski2016}}, acquire now a physical meaning through this measurement-based concept of particles. Furthermore, recent works have explored experimental setups that exploit this oscillating behavior to enhance particle production, as discussed in the recent study~\cite{Aleksandrov2022b} and other works on the dynamically assisted Schwinger effect~\cite{Schutzhold2008,Bulanov2010}. In fig. \ref{fig:qva.numt}, we also illustrate the time evolution of the theoretical particle number when the adiabatic vacua are chosen instead at each time $\tau$. The amplitude of these fluctuations decreases as the value of $\delta$ increases. Additionally, as $\tau$ grows, these fluctuations diminish further, making the particle number less dependent on $\delta$. This finding is consistent with~\cite{Adorno2018}, which showed that for a sufficiently large~$\tau$ (beyond those considered here), the effects of switching the field on and off only contribute as next-to-leading order corrections to the constant part of the electric field.

It is worth noting that for each measurement process, the asymptotic result $n_{k,\tau}^{\text{exp}}$ could have been reassigned to a different time than when the switch-off began. However, this would merely result in a simple relabeling of the $\tau$ axis in fig.~\ref{fig:qva.numt}, shifting each curve according to the value of $\delta$.

Lastly, one might ask whether, for a given choice of vacuum, it is possible to find a specific switch-off starting at $\tau$ such that the measured asymptotic particle number~$n_{k,\tau}^{\text{exp}}$ matches the value of $n_k(\tau)$ calculated using that vacuum prescription. This condition does not uniquely determine the time evolution of the \textit{in} mode~$v_k$ after time~$\tau$. Consequently, different functions $v_k$ that satisfy this requirement would lead to different mode equations. Nevertheless, it is noteworthy that a time-dependent frequency $\Omega_k(t)$, as given in eq.~\eqref{eq:qva.Frequency}, satisfies this requirement for all values of $k$.

\subsection{Interpretation of usual vacuum prescriptions}

We now illustrate the application of this operational notion of particles by interpreting two usual notions of vacuum in terms of measurements. First, we consider the ILES, which we discussed in \cref{ch:deSitter}, and is used in many references, especially in those studying the quantum Vlasov equation~\cite{Kluger1998,Schmidt1998,Aleksandrov2022,Mottola2014,Ruffini2010,Roberts2000,Dunne2009,Dumlu2011,Hebenstreit2009}. It minimizes the energy per mode at a particular instant of time~$\tau$ and corresponds to the solution to the mode equation with initial conditions (at $\tau$)
\begin{equation}
\label{eq:qva.ILES}
   u_k^{\text{ILES}}(\tau)=\frac{1}{\sqrt{2\Omega_k(\tau)}}, \quad \dot{u}_k^{\text{ILES}}(\tau)=-i\sqrt{\frac{\Omega_k(\tau)}{2}}.
\end{equation}
In~\cite{Ilderton2022}, it was proved that the theoretical particle number calculated using this vacuum coincides with the asymptotic particle number measured in the case of an instantaneous switch-off of the electric field at $\tau$, i.e. $\delta=0$. This is consistent with the \textit{out} mode in \cref{ch:becstheory}, defined in \cref{eq:bec.uPlaneWaves}. In addition, in agreement with ref.~\cite{Ilderton2022}, the particle number in the instantaneous case $\delta=0$ coincides with the limit $\delta \to 0$. 

As we have seen in cosmological scenarios, another very common vacuum prescription is precisely the adiabatic vacuum (see \cref{eq:sf.AdiabaticVacuum,eq:qva.ad0}). Its corresponding curves in figs.~\ref{fig:qva.numk} and~\ref{fig:qva.numt} deviate from those of an arbitrarily fast switch off, and the amplitude of the oscillations in the spectrum and the time evolution happen to be smaller. Nevertheless, let us mention that a similar analysis in a cosmological context could yield different interpretations for the same vacuum prescriptions, since the frequency of the mode equation would change. This is something that we leave for future work. 

\section{Summary}

We have seen how by using ideas concerning analog experiments we can gain insight into the physical meaning of the different vacuum prescriptions in QFTCS. In particular, we defined a new family of vacua that allows us to compute the number density of produced particles outside a static region, in the line of typical prescriptions such as the ILES or the adiabatic vacuum. The difference is that these vacua are associated with particle notions with a clear physical meaning: They are the particles that would be measured by a specific detector if we started switching off the external agent at the time at which particle density is computed. To each switch-off procedure $\delta$ there is an associated vacuum characterized by the same label. Importantly, this operational procedure can be used to interpret the notion of particle corresponding to usual vacuum prescriptions in terms of the particles measured after certain switch-off with duration $\delta$. In particular, we find that the ILES corresponds to an instantaneous switch-off, whereas the adiabatic vacuum deviates from $\delta=0$.

%% file: Chapters/Switch-on_and_-off_effects_in_particle_production.tex

\chapter{Switch-on and -off effects in particle production} 

\label{ch:switcheffects} 



We will now further investigate the effects of switch-on and -off processes in the context of particle production. These ideas apply to all scenarios studied in this thesis, and we will carry out both general and more specific analyses, focusing on some of the scenarios already discussed.

Although a true switch-on and -off of the expansion or, more generally, of an external time-dependent agent, is something that appears in laboratory situations, one can implement these ideas also in the context of the expansion of the actual Universe. Indeed, in part II we have dealt with the fact that in order to extract the number density of produced particles during inflation and reheating, one must provide suitable notions of vacuum before and after these processes. In practice, one finds asymptotic regions in which the geometry expands adiabatically, and take as vacua the notions corresponding to observers living there. This is what we called asymptotic \textit{in} and \textit{out} regions. Moreover, we discussed in \cref{ch:deSitter} the idea of taking the notion of adiabatic vacuum in order to implement an also adiabatic exit of inflation (of the same degree of the chosen vacuum). Therefore, in practice, one is dealing with a sort of switching-on and -off of inflation in all these scenarios. Thus, the idea of switch-on and -off processes is not only relevant in the context of analog experiments, but also of the early Universe itself.

\section{Instantaneous or smooth switch-ons and -offs}

We will consider that the universe starts expanding at an initial time~$\etai$ and ceases its expansion at a later time $\etaf$, much like in the experimental realizations in part III. Thus, there exist \textit{in} and \textit{out} regions in which there are preferred notions of vacua (see subsection \ref{subsec:qftcs.inout}), those associated to positive frequency modes with~\mbox{$\omega_{k, \text{i}}=\sqrt{k^2+ m^2a^2(\etai)}$} and $\omega_{k, \text{f}}=\sqrt{k^2+m^2a^2(\etaf)}$, respectively. Let us recall that these two notions of particle correspond to the natural choice of vacuum of observers living before and after the expansion of spacetime, and that due to the time-dependence of the geometry, even if the initial state of the system is the vacuum state as understood by the observer at $\eta<\etai$, another observer at $\eta>\etaf$ will measure a non-vanishing occupation in general. 

Instead of assuming an instantaneous switch-on and -off \cite{Barcelo2003}, we keep using the modeling of \cref{ch:entanglement,ch:qva}. We will first implement realistic switch-ons and -offs in a cosmological model. Subsequently, we will analyze their influence on the spectrum of fluctuations, comparing them with the instantaneous limit. Our aim is to understand what is the importance regarding production of the different parts of the expansion process (for example, the importance of the transition from inflation to a thermal universe as compared to inflation \textit{per se}). Regarding analog systems, this study also allows us to quantify up to which point the inevitable switch-on and -off affect the simulation of particle production, which seeks to reproduce the scale factor of the intermediate region. In sections~\ref{sec:switch.BEC} and~\ref{sec:switch.Schwinger}, we will adopt a similar approach for analog experiments in BECs and for the Schwinger effect, emphasizing the similarities and differences with the cosmological case.

First, let us model a flat FLRW universe that undergoes an expansion between \textit{in} and \textit{out} static regions, with a smooth transition between them and the intermediate region II (with scale factor $a_{\text{II}}(\eta)$). We consider that at the initial time $\etai$, the universe starts to expand smoothly so that, after a time~$\delta$, the derivative of the scale factor reaches $a_{\text{II}}^{\prime}(\eta)$. Similarly, at time~$\etaf-\delta$, the universe begins a decelerated expansion lasting~$\delta$, after which it smoothly returns to staticity at~$\etaf$. We will furthermore assume that the switch-on and -off are symmetric, and have the same duration.

\begin{figure}[t!]
    \centering
    \includegraphics[width=\textwidth]{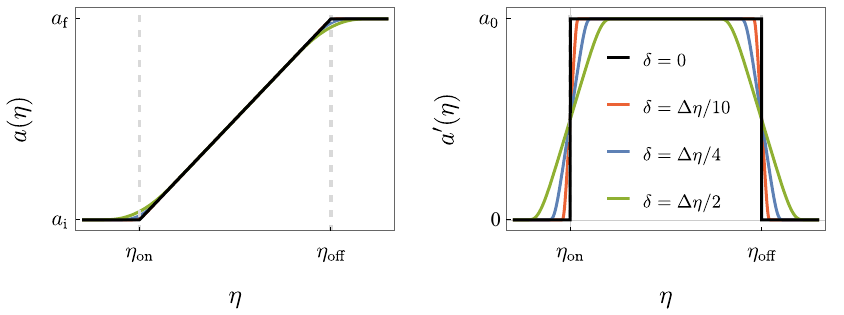}
    \caption[Time evolution of the scale factor and its derivative for different durations of the switch-on and -off]{Time evolution of the scale factor (left panel) and its derivative (right panel) for different durations $\delta$ of the switch-on and -off. We denote by $\etaon$ and $\etaoff$ the points for which $\Theta_{\delta}=1/2$, which coincide for all values of $\delta$.}
    \label{fig:switch.SwitchOnAndOff}
\end{figure}

In order to explicitly implement the switch-on and -off, we will consider a linear expansion in conformal time in region II, i.e. $a_{\text{II}}(\eta)=a_{\text{i}}+(\eta-\etai)a_0$. The switch-on and -off processes will be modeled in the same way as we did in \cref{ch:entanglement,ch:qva}, namely by a smooth step function $\Theta_{\delta}(\eta)$ that interpolates between $0$ and~$1$, remaining constant outside this interval, with width $\delta$, such that for~$\delta=0$ we recover the discontinuous Heaviside step function. Then, the derivative of the scale factor at all times can be written as
\begin{equation} 
    a^{\prime}(\eta)=a_0\left[\Theta_{\delta}(\eta-\etaon)-\Theta_{\delta}(\eta-\etaoff)\right],
\label{eq:switch.DerivativeScale}
\end{equation}
where $\etaon$ and $\etaoff$ are the symmetric points in the middle of the switch-on and -off processes, respectively, which coincide for all values of $\delta$. Explicitly, $\etaon=\etai +\delta/2$, whereas $\etaoff=\etaf-\delta/2$. In the following, we will use the particular set of profiles~$\Theta_{\delta}$ in \cref{eq:ent.SwitchFunction} (with $t \to \eta$). We will then parameterize different switch-ons and -offs only by their duration~$\delta$. While other interpolation profiles could have been chosen, our qualitative results will remain independent of the specific choice. In figure~\ref{fig:switch.SwitchOnAndOff}, we represent both the scale factor~$a(\eta)$ and its derivative $a^{\prime}(\eta)$ for different values of $\delta$, given in terms of the whole expansion duration $\Delta \eta = \etaf -\etai$. 

Recall that a measure of the adiabaticity of particle production is given by the time variation of the frequency $\omega_k(\eta)$, characterized by the adiabatic coefficient~$\mathcal{C}_k$ (c.f. \cref{eq:qftcs.AdiabaticCoefficient}). Therefore, abrupt changes in the scale factor enhance particle production much more in scenarios where time derivatives of the scale factor appear in the frequency~$\omega_k(\eta)$, as we will see below. Depending on the relation between the coupling $\xi$ and the dimension $D$ of the spatial sections, we can distinguish two cases (see the definition of the general frequency \eqref{eq:qftcs.MasterFrequency}): 
\begin{itemize}
    \item[A)] In the conformal coupling case, where $\xi=(D-1)/(4D)$, the frequency has no scale factor derivatives.
    \item[B)] In the non-conformal coupling cases, time-derivatives of the scale factor up to second order appear in the frequency.
\end{itemize}
Each of the scenarios discussed in the chapters above corresponds to particular realizations of the general mode frequency \eqref{eq:qftcs.MasterFrequency}. Depending on the specific values of $D$, $m$ and $\xi$, one can particularize to the different cases that have been studied in the literature. For example, for $D=3$ one would recover the standard mode equation for a massive, non-minimally coupled scalar, which has been extensively analyzed in the context of cosmological production of dark matter during inflation and reheating in \cref{ch:singlefield,ch:deSitter}. Note that in this particular case the contribution of the first derivative of the scale factor to the frequency vanishes. We also discussed in part III of this thesis the cosmological production of the analog field emerging in a $(1+2)$-dimensional BEC (see chapters \cref{ch:becstheory,ch:entanglement}), which corresponds to setting $\xi=m=0$ and $D=2$. We will delve into this specific case in more detail in \cref{sec:switch.BEC}. Although it cannot be realized from \cref{eq:qftcs.MasterFrequency}, the case of Schwinger effect discussed in \cref{ch:qva} can be analyzed similarly (see \cref{sec:switch.Schwinger}).

Regarding the regularity of the solutions $v_k(\eta)$ to the mode equation~\eqref{eq:qftcs.ModeEquation}, they will be infinitely differentiable as long as the switch-on and -off are too. However, with instantaneous changes ($\delta=0$), the second derivative of the scale factor has Dirac delta contributions at~$\eta_{\text{i}}$ and~$\eta_{\text{f}}$, and, as we saw in subsection \ref{subsec:bec.InstantaneousSwitch}, we might need to impose matching conditions at these times. When conformal invariance holds (case A), the continuity of $v_k(\eta)$ implies its differentiability, as the integrand in \cref{eq:bec.JumpDerivative} is bounded and the integral vanishes in the limit $\epsilon\rightarrow 0$. Furthermore, according to the equation of motion~\eqref{eq:qftcs.ModeEquation}, and since in this case~$\omega_k^2(\eta)$ is continuous but not differentiable, $v_k(\eta)$ is differentiable up to third order\footnote{This is why in the conformal case, choosing the ILES as the \textit{out} vacuum yields the same result as a sudden transition to Minkowski, as we mentioned in subsection \ref{subsec:ds.exit}.}. However, the presence of second derivatives of the scale factor in the frequency (case B) reduces the regularity of any continuous solution to its differentiability at first order. This is, for example, the case of the BEC, where we found that the mode function derivative~$v_k^{\prime}$ undergoes a jump of the form \eqref{eq:bec.BoundaryConditionDerivative}. The generalized version of this result for any values of $D, m$ and $\xi$ is given by
\begin{equation}
    v_k^{\prime}(\eta_*^-) =  v_k^{\prime}(\eta_*^+) \mp \frac{1+(4\xi-1)D}{2} \frac{a^{\prime}_0}{a(\eta_*)} v_k(\eta_*),
\end{equation}
where we recall that the negative sign corresponds to~$\eta_*=\etai$ and the positive sign to~$\eta_*=\etaf$.

\section{Relevance of the switch-on and -off processes}
\label{sec:switch.adiabaticity}

We are interested in modeling the expansion of the universe in a potential experiment between two times $\etai$ and $\etaf$. However, complications arise with the introduction of switch-on and switch-off mechanisms, which inevitably influence the spectra of produced particles. The question is: To what degree can we deem this influence negligible? Are there conceivable scenarios where particle production during these transitional phases becomes so pronounced that it obscures the effects of the expansion we aim to simulate? Our study reveals that the answer significantly depends on the coupling parameter $\xi$, which governs the interaction between the matter field and the geometry.

We aim at identifying periods during the entire expansion, including the switch-on and -off processes, when particle production is more intense. To achieve this, we study the adiabatic coefficient $\mathcal{C}_k$ (cf. \cref{eq:qftcs.AdiabaticCoefficient}), which gives us information of the adiabaticity of the entire process over time. When $\abs{\mathcal{C}_k(\eta)} \ll 1$, the expansion of the geometry becomes adiabatic, leading to negligible particle production. Conversely, as suggested by the Quantum Vlasov Equation \cite{Alvarez2022}, when $\abs{\mathcal{C}_k(\eta)} \gtrsim 1$, the time variation in the density of created particles becomes significant. Therefore, analyzing the evolution of $\mathcal{C}_k(\eta)$ over time dictates the primary contributions to the total density $n_k$.

\begin{figure}[t!]
    \centering
    \includegraphics[width=\textwidth]{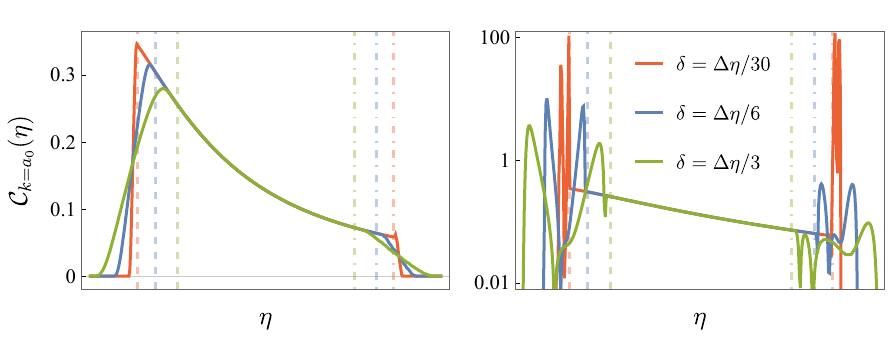}
    \caption[Adiabatic coefficient for a linear cosmological expansion with different duration switch-on and -off]{Adiabatic coefficient $\mathcal{C}_k (\eta)$ for an expansion with $\eta_{\text{off}}=3/a_0$, for different durations $\delta$ of the switch-on and -off processes, fixed $k=a_0$. The left panel illustrates the conformally invariant case ($\xi=1/6$) while the right panel corresponds with the non-conformally invariant case ($\xi = 1$). The dashed lines demarcate region II.}
    \label{fig:switch.AdiabaticityCosmo}
\end{figure}

First, let us analyze the case of the conformal coupling, where $\xi=(D-1)/(4D)$. The frequency~\eqref{eq:qftcs.MasterFrequency} has no scale factor derivatives, and 
\begin{equation}
    \mathcal{C}_k(\eta)=\frac{m^2a(\eta)a^{\prime}(\eta)}{\left[ k^2+m^2a^2(\eta) \right]^{3/2}}.
    \label{eq:switch.adCk16}
\end{equation}
Even in the case of abrupt switch-ons and -offs (i.e. a continuous but not differentiable scale factor), $\mathcal{C}_k(\eta)$ remains bounded, since it only depends on the first derivative of the scale factor. Furthermore, the expansion rate during the switch-on monotonically increases, such that $|a'(\eta)|>|a'(\etai)|$ for $\etai+\delta>\eta>\etai$. Similarly, during the switch-off, the rate decreases, resulting in $|a'(\eta)|>|a'(\etaf)|$ for~\mbox{$\etaf-\delta<\eta<\etaf$}. Moreover, in an expansion the scale factor $a(\eta)$ also increases. Particularly, in the case of the chosen $\Theta_{\sigma}$ function, we see that the adiabatic function~$\mathcal{C}_k(\eta)$ is suppressed during the switch-on and -off processes compared to the main expansion period between~$\etai$ and~$\etaf$. As a consequence, the switch-on and -off mechanisms will have a subdominant effect on the spectra of produced particles. This is illustrated in the left panel of figure~\ref{fig:switch.AdiabaticityCosmo} for the case $D=3, m=a_0$.

The non-conformally invariant case yields markedly different outcomes. In this scenario, the frequency~\eqref{eq:qftcs.MasterFrequency} depends on the derivatives of the scale factor, and
\begin{equation}
\begin{split}
\omega_k(\eta)\omega_k^{\prime}(\eta)=&m^2a(\eta)a^{\prime}(\eta) + \frac{1+(4\xi-1)D}{4}\Bigg[\frac{a^{\prime\prime\prime}(\eta)}{a(\eta)} \\
&+ (D-4)\frac{a^{\prime}(\eta)a^{\prime\prime}(\eta)}{a(\eta)^2} - (D-3)\left( \frac{a^{\prime}(\eta)}{a(\eta)} \right)^3\Bigg].
\end{split}
\end{equation}
For very fast switch-ons and -offs (i.e. $a_0\delta\ll 1$), $a^{\prime\prime\prime}(\eta)$ approaches the derivative of a Dirac delta\footnote{Recall that the derivative of the Dirac delta distribution verifies $x\delta'(x)=-\delta(x)$, so that it has an even stronger divergence around $x=0$ than the Dirac delta.}, which strongly dominates over the terms proportional to $a'(\eta)$ and~$a^{\prime\prime}(\eta)$. On the other hand, the term with $a^{\prime\prime}(\eta)$ dominates in the frequency~\eqref{eq:qftcs.MasterFrequency}. Consequently, the adiabatic coefficient~\eqref{eq:qftcs.AdiabaticCoefficient} quickly diverges during rapid switch-ons and -offs. This is depicted in the right panel of figure~\ref{fig:switch.AdiabaticityCosmo}, where we show the evolution of $\mathcal{C}_k(\eta)$ for exactly the same scenario as before, but with a coupling $\xi=1$. We observe that this situation differs significantly from the conformally invariant case. Therefore, modeling an expanding universe with very fast switch-on and switch-off mechanisms may dramatically impact the spectra of produced particles, masking the contributions from region II to the whole process.

Note that the degree of adiabaticity depends, of course, on the mode~$k$ under study. Thus, let us adopt a spectral point of view in the study of \cref{eq:qftcs.AdiabaticCoefficient}. We will say that a mode $k$ is adiabatic at some time~$\eta$ if $\abs{\mathcal{C}_k(\eta)} \ll 1$. In this case, the mode is not significantly excited, and particle production is negligible. On the other hand, if $\abs{\mathcal{C}_k(\eta)} \gtrsim 1$ for some time $\eta$, the mode is non-adiabatic and its associated number occupation is expected to fluctuate. We define the ultraviolet mode $k_{\text{UV}}$ as the one for which the adiabatic coefficient \eqref{eq:qftcs.AdiabaticCoefficient} equals one, which results in
\begin{equation}
    k_{\text{UV}}(\eta) = \sqrt{\left[M(\eta)M^{\prime}(\eta)\right]^{2/3} - M^2(\eta)},
\label{eq:switch.UVmode}
\end{equation}
where the effective mass $M(\eta)$ is given by
\begin{equation}
  \!  M^2(\eta) = a^2(\eta)m^2 + \frac{1+(4\xi-1)D}{4}\left[(D-3)\left(\frac{a^{\prime}(\eta)}{a(\eta)}\right)^2\!+2\frac{a^{\prime\prime}(\eta)}{a(\eta)}\right].
\end{equation}
Modes with wavenumbers larger than $k_{\text{UV}}(\eta)$ at some instant $\eta$ will be very adiabatic, signalling the beginning of the UV regime. Therefore, the smaller the value of~$k_{\text{UV}}(\eta)$, the broader the UV regime is, and thus the overall total number density of produced particles is smaller.

In the conformal case, \cref{eq:switch.UVmode} takes the form
\begin{equation}
    k_{\text{UV}}(\eta) = \sqrt{\left[m^2a(\eta)a^{\prime}(\eta)\right]^{2/3} - m^2a(\eta)^2}.
\end{equation}
Note that in order to find a real, positive root, one needs
\begin{equation}
    (m^2aa^{\prime})^{2/3}>m^2a^2 \quad \Rightarrow \quad a^{\prime} > ma^2,
\end{equation}
where we used the fact that $a^{\prime}>0$ in an expanding universe. Crucially, this can be rewritten in terms of the Hubble rate as $H > m$, which is not surprising at all: The rate of expansion of the universe must be larger than the mass of the field in order to produce particles. From here one can already see that smaller masses will lead to larger $k_{\text{UV}}$, which means that there will be production in a wider range of modes. This is consistent with the displacement to higher $k$'s of the spectra analyzed in part II of this thesis. Moreover, depending on whether the rate of expansion of the universe $H$ is constant, increasing, or decreasing, the UV regime will stay constant, become larger, or smaller, respectively.

A qualitative analysis in the non-conformal case can be done similarly from \cref{eq:switch.UVmode} by simply using the argument that third derivatives of the scale factor will be introduced in \cref{eq:switch.UVmode} through the derivative of the effective mass, which will dominate the equation. During the switch-on and -off processes, the UV regime---where negligible production is expected---shifts toward larger wavenumbers. This shift becomes more pronounced (i.e. the wavenumbers become higher) as the parameter $\delta$ decreases. In the limit in which~\mbox{$\delta \to 0$}, there are no adiabatic modes (during these transitions)~\cite{Birrell1982}. 

These ideas, which apply to any cosmological production scenario with (asymptotic) \textit{in} and \textit{out} regions (be it analog or not), will be illustrated in the context of two problems that we have already discussed: the Schwinger effect, which is a representative case of the minimally coupled case, and fluctuations on top of a BEC, which corresponds to the non-minimally coupled case.

\section{Switch-on and -off effects in a BEC analog experiment}
\label{sec:switch.BEC}

In order to explicitly study the experimental consequences of finite duration switch-on and -offs, as well as its implications in the context of analogs and its role in the production process, we turn to the problem of analog particle production in a quasi two-dimensional BEC (see \cref{ch:becstheory} and \cite{Experiment2022,BECPaper2022}). We have seen that low-energy fluctuations on top of the ground state of the condensate behave as a massless scalar field in a curved spacetime characterized by the acoustic metric \eqref{eq:analogs.AcousticMetricBEC}, which is determined by the properties of the BEC. The latter can be tuned experimentally so that the corresponding line element acquires the shape of a two-dimensional FLRW metric, and therefore the problem can be mapped to cosmological particle production in $(1+2)$-dimensions of minimally coupled to gravity massless scalar particles. We recall that in the case of flat spatial sections, the frequency of the mode equation in this situation (cf. \cref{eq:bec.ModeFrequency}), which corresponds to taking $D=2$ and~\mbox{$m=\xi=0$} in the frequency \eqref{eq:qftcs.MasterFrequency}, reads
\begin{equation}
    \omega_k^2(\eta) = k^2 + \frac{1}{4}\left( \frac{a'(\eta)}{a(\eta)} \right)^2 - \frac{1}{2} \frac{a''(\eta)}{a(\eta)}.
    \label{eq:switch.BECFrequency}
\end{equation}
Therefore, we are here in the situation in which the coupling is non-conformal, and second derivatives of the scale factor appear in the frequency. Let us recall that in the BEC, the role of the scale factor is played by the scattering length $\alpha_s(t)$, and the expansion is implemented in three regions, with the scattering length being kept constant in regions I and III, which constitute \textit{in} and \textit{out} regimes between which one can calculate particle production. We will treat the switch-on and -off regimes in a manner similar to the treatment in \cref{ch:entanglement}, instead of modeling them as instantaneous. The absence of discontinuities in the derivative will lead to a smoothing of the~\mbox{$\delta$-peaks} in the scattering picture (cf. \cref{sec:bec.scattering}) at~$\eta_{\text{i}}$ and~$\eta_{\text{f}}$. As it is known from the literature~\mbox{\cite{Birrell1982,Visser1999,Glenz2009,Adorno2018,QVA2023}} (also \cref{ch:qva}), we expect the strength of particle production to decrease with an increasing amount of smoothing. The functional form of the scale factor is taken to be exactly the same as in \cref{fig:switch.SwitchOnAndOff}, but now in laboratory time. Perhaps in this context, the question of how the switch-on and -off periods, unavoidable in the context of analog experiments, affect production is even more important, since in principle we only wonder about the intermediate region, where the scale factor acquires the shape we are trying to simulate (linear in $t$ in this example).

\begin{figure*}[t!]
    \centering
    \includegraphics[width=0.9\textwidth]{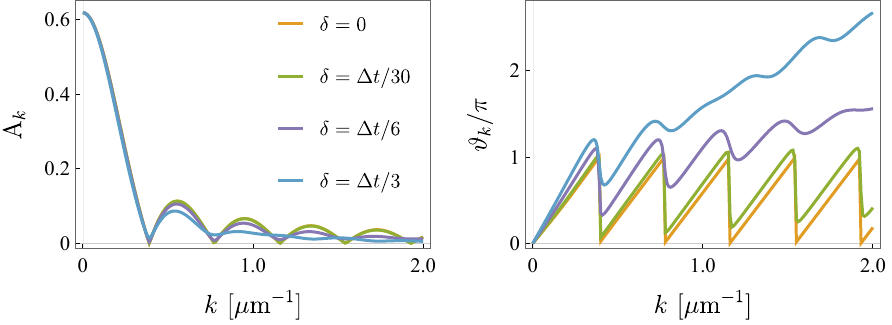}
    \caption[Effect of symmetric, finite duration switch-on and -off in analog particle production in a BEC]{Effect of a finite duration $\delta$, symmetric switch-on and -off in the amplitude $A_k$ and the phase $\vartheta_k$ of the spectrum in a flat, linearly expanding (in laboratory time) FLRW analog. We considered an expansion of $\Delta t = 3 \, \text{ms}$ in which the scattering length is changed from $400 \aB$ to $50 \aB$, and a final speed of sound of approximately $c_s=1.7 \mu\text{m}/ \text{ms}$.}
    \label{fig:switch.SpectraSwitch}
\end{figure*}

We numerically obtain the spectrum of produced particles $S_k$ (cf. \cref{eq:qftcs.SpectrumOffsetAmpPhase}) for~\mbox{$\delta \neq 0$}, and its amplitude $A_k$ as well as its phase $\vartheta_k$ are depicted in \cref{fig:switch.SpectraSwitch}. Here, we observe that the effect of slightly decreasing the duration of an already short switch-on and -off is subtle on $A_k$, which tells us that the limit $\delta\to 0$ is achieved fast. However, the amplitude does not vanish anymore for $\delta \neq 0$, while interestingly the phase jump in~$\vartheta_k$, characteristic of the $\gamma=1$ scale factor (that is, linear in $t$), is smeared out for non-instantaneous switch-ons and -offs. Instead, a behavior corresponding to~\mbox{$\gamma>1$} (accelerated expansion) can be observed for smooth switch-ons and -offs, i.e. oscillations in the phase instead of jumps. Of course, note that since the duration of the process slightly increases with $\delta$, there is an additional constant contributing to the slope of the phase $\vartheta_k$.

\begin{figure}[t!]
    \centering
    \includegraphics[width=\textwidth]{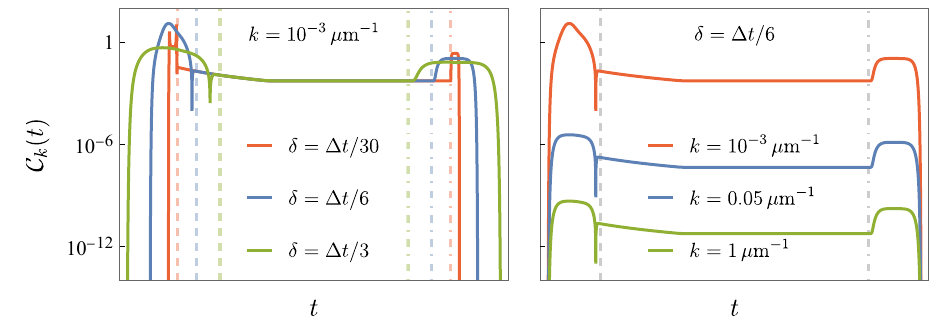}
    \caption[Adiabatic coefficient in typical analog cosmological production in BECs]{Adiabatic coefficient $\mathcal{C}_k (\eta)$ for an experiment of duration \mbox{$\Delta t = 3 \, \text{ms}$}, and scattering length running from $400a_{\text{B}}$ to $50a_{\text{B}}$ in units of the Bohr radius. Different curves show different durations of the symmetric switch-on and -off~$\delta$, fixed $k=10^{-3} \mu\text{m}^{-1}$ (left) and different values of $k$, fixed $\delta=0.5 \, \text{ms}$ (right). The dashed lines demarcate region II.}
    \label{fig:switch.AdiabaticityBEC}
\end{figure}

From the discussion in \cref{sec:switch.adiabaticity}, we expect the switch-on and -off to dominate the production process, since we do not have the freedom to select the value of~$\xi$ in these analog experiments, which is zero in this case. This is illustrated in figure~\ref{fig:switch.AdiabaticityBEC}, which resembles the results of the $D=3$ case in \cref{fig:switch.AdiabaticityCosmo}. The study of adiabaticity tells us that, in fact, most of the resulting density production comes from the switch processes. Interestingly, most of the analyzed range in wavenumber~$k$ behaves adiabatically, which is consistent with the fact that analog production is very subtle in this setup. This highlights the importance of precision measurements in this context, and of the existence of observables such as the phase that allow for a cleaner characterization of the expansion history. 

Of course, this leads to the following question: Are we really probing particle production of a massless scalar field in a FLRW cosmology with scale factor $a(t) \propto t$? Or are we just essentially measuring the effects of the switch-on and -off? During part III of this thesis, we aimed at simulating cosmological particle production for a background characterized by an analog scale factor that is tunable in the laboratory. However, we now realize that switch-on and -off processes, which are unavoidable in the context of such analog experiments, may have a significant impact on the resulting particle production. In fact, they dominate in the context of BECs. This is a crucial point to take into account when interpreting the results of analog experiments. At first, one might think that analog simulators of this type are unable to exactly simulate cosmological production in this sense, but then again, asymptotically static or adiabatic regions are needed in the context of the actual Universe in order to define the notion of particles. Therefore, one may argue that in this sense, experiments are closer to our Universe than one might think.

Lastly, let us mention that analog gravity experiments should be viewed, not only as tests of the theory of quantum fields in curved spacetimes via analogies, but as experiments on the theory of quantum fields in non-trivial backgrounds, which is the one we use to describe BECs in the first place. In this sense, the results of the experiments are not only interesting for the analogies they provide, but also for the insights they give on the theory of quantum fields in non-trivial backgrounds. We will further discuss these ideas in the coda after the conclusions.

\section{Switch-on and -off effects in Schwinger production}
\label{sec:switch.Schwinger}

In the Schwinger effect, the time-dependent frequency of the mode equation is given by \eqref{eq:qftcs.SchwingerFrequency}. Note that one recovers the frequency of a massless, minimally-coupled scalar field for modes perpendicular to the electric potential by making the substitution  $m \to 0$, $q \to m$, and $A(t) \to a(t)$. Nevertheless, the important characteristic of the Schwinger frequency is that there are no derivatives of the potential~$A(t)$. Therefore, the behavior regarding the switch-on and -off of the electric field will be similar to the conformally invariant case in the cosmological scenario. We will, however, leave the details of this analysis for future work\footnote{The details can now be found in the complete, published work \cite{SwitchEffects2024}.}.

\section{Summary}

In this chapter, we have studied the effects of finite duration switch-on and~-off processes in the context of cosmological particle production and its analog realization, together with the importance of these regimes in the whole process. We have seen that switch-on and -off processes can significantly affect particle production as long as the coupling to the geometry is non-conformal. In this case, these regimes become highly non-adiabatic, and dominate particle production. An example of this is the case of analog particle production in a BEC. As a consequence, the interpretation of the results of analog experiments should take into account the effects of these regimes, in the sense that they may not be probing cosmological production due to the engineered scale factor, but rather the effects of the switch-on and -off processes. 

%% file: Chapters/Conclusions.tex

\addtocontents{toc}{\protect\vspace{0.5em}}

\chapter{Conclusions and outlook} 

\label{ch:conclusions} 



We come to the end of this work, at which point we would like to summarize the main results and conclusions of the analyses presented in the previous chapters. We believe that during this thesis, we have thoroughly investigated the phenomenon of cosmological particle production in detail, and from different points of view, ranging from a phenomenological approach involving dark matter production in the early Universe, to a more formal and abstract perspective, that has led us to discuss quantum vacuum ambiguities and the choice of vacuum in the context of~QFTCS. 

In part II of this thesis, we have studied cosmological particle production in the early Universe, during inflation and reheating, for spectator, massive scalar and vector fields. We have analyzed single-field, chaotic inflationary models sourced by both quadratic and Starobinsky potentials (for scalar and vector fields in flat cosmologies), as well as more simple inflation scenarios based on a de Sitter-like expansion of the geometry (for scalar fields in cosmologies of any spatial curvature). The main result is that this mechanism, for certain regions of the parameter space, is able to explain the observed abundance of dark matter, without the need of invoking any additional physics nor having to take backreaction into account. In this sense, it constitutes a simple and natural explanation for the presence of dark matter in the Universe. Another important conclusion is that the presence of tachyonic instabilities can lead to a highly efficient production. This is especially important during the Ricci scalar oscillation phase in the reheating era, when most of the dark matter seems to be produced. Note that these features are not present in the conformally coupled case, since the Ricci scalar is absent from the mode frequency. Thus, oscillations are very important, and they should not be neglected too early in the reheating era, as produced spectra will significantly change. As expected, production depends on the specific inflationary model considered, and the same values of the field mass $m$ and the coupling $\xi$ (in the case of the scalar field), can lead to abundances that vary over a couple of orders of magnitude. We would like to further investigate this effect by making an exhaustive analysis of inflationary models to determine the range of masses and couplings for which observations are reproduced. Additionally, we have seen that production of vector fields is much more efficient than that of scalars due to the unconventional behavior of the longitudinal mode and the possibility of new couplings that cannot be considered in the scalar case. As a consequence, the dark matter produced can be heavier while still reproducing observations. Interestingly, vector fields present ghost instabilities that restrict the region in parameter space in which the theory is correctly defined. Regarding spatial curvature, although it does affect the dynamics of the fields, and can lead to very different produced abundances depending on its value, our preliminary results show that, for curvature abundances compatible with observations, large differences between the curved and the flat scenarios are expected only for very small values of the field mass, which would eventually lead to negligible abundances. Therefore, no effect of curvature is expected for abundances of the order of the observed dark matter abundance. Let us stress that this analysis has been performed in a toy inflationary model, in the conformally coupled case. Therefore, effects of curvature could be more important in more realistic models (especially when including Ricci oscillations), and we would like to investigate this in the future.    

We have devoted part III to establishing a correspondence between phonons on top of the ground state of a $(1+2)$-dimensional Bose-Einstein condensate and a massless, minimally coupled scalar field in FLRW universes with positive, vanishing or negative spatial curvature, engineered through different radially symmetric trapping potentials for the condensate atoms. We have defined a rescaled version of the usual density contrast correlation function typically measured in these ultracold atom experiments, which allows us to fully characterize the spectrum of fluctuations, and thus the whole particle creation process. By changing the effective interaction between atoms in the ground state, we can engineer highly tunable analog scale factors. Importantly, we have experimentally shown that, by taking advantage of the phase features of the spectrum $S_k$, we can deduce the expansion history (i.e. the type of scale factor) that created the particles. We have also provided a reinterpretation of the cosmological particle production process in terms of quantum-mechanical one-dimensional scattering, a language in which the different scale factors and the mode frequency become scattering potentials. Crucially, a certification of the quantum origin of these produced pairs in the BEC is missing, and quantum correlations (i.e., entanglement) are key to verify this. We showed that the current technical status could already, or at least with minor improvements, allow for the significant detection of entanglement in these analog experiments, given that the measurements of the spectrum of fluctuations allow for full state tomography. Moreover, we showed that in this kind of cosmological scenarios, Cauchy-Schwarz inequality, the commonly used entanglement witness, behaves faithfully (although it is not a quantifier). We expect that these ideas will soon be put to work in subsequent experiments in order to measure entanglement due to pair creation for the first time. For future work, we would like to further explore analogs of massive fields, such that scenarios closer to what we discussed in part II can be realized. Beyond the analogy, the ultraviolet behavior of the quantum field theory describing BECs is worth studying, since early estimations of theoretical production very well match the recent experimental results that will be presented in \cite{ScatteringExp2024}.

Finally, part IV served as an explicit demonstration of the usefulness of analog gravity experiments in the study of QFTCS, as it combines ideas coming from both the actual cosmological scenario and the analog experiments. We have discussed the problem of quantum vacuum ambiguities from a different perspective, involving insights from analog experiments. In particular, we have shown that one may define a whole family of well-behaved vacua for QFTCS which are associated with the number of particles that would be measured by a simple measurement process labeled and characterized by the parameter $\delta$. As expected, the notion of particles defined by a vacuum associated with a very fast switch-on coincides with that of the ILES, although we did not find a clear correspondence for very adiabatic switch-ons. However, we would like to remind the reader that we applied these tools in the context of Schwinger effect. As mentioned in \cref{ch:switcheffects} of the thesis, Schwinger effect is partially analogous to cosmological production in the case of a conformal coupling to gravity. This means that the discontinuities introduced by instantaneous switch-ons and -offs do not have any impact. Moreover, the anisotropy characteristic of the Schwinger effect production may be the responsible for the impossibility of relating the adiabatic vacuum used in Cosmology to a very adiabatic switch-off. This is indeed an idea that was used in \cref{ch:deSitter}, and we would like to pursue it further by applying all the concepts in \cref{ch:qva} to cosmological scenarios. After working with switch-ons and switch-offs, it was natural to study what is the influence of these regimes in the total amount of produced particles, and its importance when compared with the rest of the process. This is what we have studied in \cref{ch:switcheffects}, where we have found that, on the one hand, the total amount of particles produced is not significantly changed by shortening the duration of switch-on and -off processes, as long as they are fast enough not to perturb the process in between (i.e., the limit $\delta \to 0$ is achieved fast), whereas on the other hand, and more importantly, as long as derivatives of the scale factors are present in the frequency (that is, as long as the coupling is non-conformal), the switch-on and -off processes are much less adiabatic, in general, than the expansion itself. This means that the particle number is much more subject to change during these periods, and the measured production in the BEC comes mainly from the switch-on and -off. Of course, this does not imply that the discussions in part III are wrong, merely that instead of simulating production in a, say, cosmology with scale factor $a\sim t$, we are actually simulating production in a geometry in which there are transitions from and to \textit{in} and \textit{out} regions, respectively, where most of the production happens, with a very adiabatic process in between corresponding to a scale factor of the form~\mbox{$a\sim t$}. We have also demonstrated in this last part the benefits of analog systems in the study of QFTCS, as they provide new perspectives on these problems. We have also shown how the insights from Cosmology and analog systems can be shared, and how this has provided a more complete view of the problem.

In the future, we would like to explore other related topics such as Higgs inflation and production sourced by it, study in more depth the effects of spatial curvature in early cosmological production, now in more realistic inflationary models, or study other analog systems that could provide new insights into the problems we have addressed. Perhaps entanglement, which is a very hot topic nowadays, deserves special attention. We have considered entanglement between cosmologically produced pairs in a controlled, analog scenario, and pursuing its experimental measurement is obviously a very interesting goal. But at the same time, we have witnessed its interesting and surprising properties and behaviors, and we would like to study them in more detail. In particular, we would like to mention the analysis of entanglement between spatially separated regions, and its realization in analog systems.

We believe that the ideas presented in this thesis can be further developed and expanded, and we hope that they will be useful for future research in the field of cosmological particle production and Quantum Field Theory in Curved Spacetimes.

%% file: Chapters/What_we_can_learn_from_analog_experiments.tex


\chapter{Coda: What we can learn from analog experiments} 

\label{ch:whatwecanlearn} 



Throughout the development of the works that constitute this thesis, the perspective of reproducing the results predicted by QFTCS in analog systems has been a constant, and we feel it has modeled the way in which we understand and have approached the problem. In this sense, we have been inspired by ideas coming from the field of analog gravity, and the way in which we conceive quantum vacuum ambiguities or cosmological particle production would not have been the same without the insights provided by these experiments. In this coda, we would like to reflect on what we have learned from analog experiments, and what is it that we cannot learn from them.

Analog gravity is a field that was born from Unruh's idea of replicating, maybe partially, the physics of BHs in a system that could be built in the laboratory, and therefore accessible. However, one can realize gravitational analogs for probing other scenarios, as we have seen for example in the case of cosmological particle production. One of the key differences with respect to Hawking radiation, as we will discuss below, is that there are plenty of cosmological observations that support QFTCS as correctly describing the dynamics of fields in the early Universe. From the epistemological point of view, an immediate question arises: What can we learn from these experiments? Can we really test Hawking radiation by recreating its analog counterpart in the laboratory? What do they say about cosmological particle production, which has been the main subject in this thesis? This has been a question that has been debated for many years, and has led to a rich discussion in the field of Philosophy of Science. Nevertheless, as we explained in \cref{sec:analogs.history}, the usefulness of analog experiments goes beyond the confirmation (if such thing is possible) of the predictions of the theory.

In this coda, we will elaborate on the principles behind analogies, and on whether analog experiments can be considered as a valid test of the physical theories with which we describe the inaccessible phenomena they aim at simulating. This has been discussed in many philosophical works, such as references \cite{Thebault2019,Crowther2021,Bartha2022}, from the point of view of BH analogs and Hawking radiation. In our case, we will also try to answer the same questions from the point of view of Cosmology. We will then argue that analog experiments do not need confirmatory power to be useful, as well as reflect on the insights that analog experiments have provided to the study of QFTCS, and how they have shaped the way in which we have approached the problems we have addressed in this thesis.

\section{Analog reasoning}

Analog experiments rely on the principle of analogical reasoning to establish a relationship between a source (the system being used as a model) and a target system (the one that is being studied). According to \cite{Bartha2022}, an argument by analogy must have the following structure. Let us consider that certain source system is similar to the target in some aspects, and that the source has some property $P$. Then, it is reasonable to assume that the target system must have the same property $P$ (or at least some similar property $P^*$). Let us remark that the argument is inductive, in the sense that the conclusion is not guaranteed to be true---it is simply reasonable to assume that it is. The idea is that the stronger the analogy (that is, the more similar source and target systems are regarding property $P$), the more likely the conclusion is to hold.

In the case of gravity analogs, the similar aspects between the source and target systems we refer to are the mathematical structure\footnote{To be precise, Bartha asks that the physical laws of both systems are the same \cite{Bartha2022}. This condition can be weakened to include within the analogy systems that have the same equations of motion, which is exactly the case in analog gravity. This is what Dardashti \textit{et al.} call \textit{syntactic isomorphism} \cite{Dardashti2017}.} of the equations that describe them, such that only the interpretation of the mathematical symbols determines to which system the equations are referring to \cite{Bartha2010}. Recall that, in the explicit case of cosmological analogs in BECs, the dynamics of phononic fluctuations on top of the background are described by an equation of motion that, when properly rewritten and interpreted, corresponds to the one followed by a massless scalar field in an FLRW geometry.

In this context and language, one can formulate the conditions under which the source can be considered an analog of the target system (see also \cite{Dardashti2017}). First, one has to be able to describe the source (BECs) with certain theory (GP equation describing the ground state and QFT in non-trivial backgrounds describing small fluctuations on top), and the same for the target system (scalar QFT in an FLRW universe or a BH). Second, there must exist a mathematical equivalence between the two systems, which in our case is realized since the equations of motion of the fields are the same (perhaps in certain regime). Then, if an analogy can be established, one can perform experiments in the source, in some way that is not possible in the target, and infer that the properties of the former will be present in the latter.

Note that there are two conflicting points here. First, one may ask whether the theory that describes the target is indeed accurate, which is going to depend on the tests one is able to perform. Second, we must be sure that the differences between the two systems are not relevant to the property we are interested in, that is, that the analogy indeed holds. If these two conditions are met, then we can safely extrapolate the results from the source to the target system. We will discuss these points in the following section for the case of Hawking radiation and cosmological particle production analogs. 

\section{Is analog confirmation possible?}

There have been contradictory opinions on whether analog experiments of Hawking radiation can be considered as confirmatory. In~\cite{Dardashti2017}, the authors argue in favor, since under certain assumptions on trans-Planckian physics, this effect does not depend on the details of the unknown theory at high energies \cite{Schuetzhold2005}, and at the same time, there is evidence that analog Hawking radiation emerges in other systems~\mbox{\cite{Philbin2008,Weinfurtner2011,Euve2016,MunozDeNova2019,Drori2019,Shi2023}} which are described by the same theory, namely QFTCS. This would support the analogy between source and target. However, in \cite{Crowther2021} it is argued that precisely the validity of the theory of QFTCS in BHs is what is at stake. Since in this case, the target system is inaccessible\footnote{In the epistemology literature, target systems are considered inaccessible by definition \cite{Dardashti2017,Crowther2021}. We do not believe this is the case for QFT in cosmological backgrounds, as we will discuss below.}, one cannot test the theory describing it, and therefore the conditions outlined in last section do not hold. Moreover, if one were able to ensure the validity of QFTCS in BHs, then Hawking radiation would be simply a consequence. In this sense, it is argued that these type of analog experiments cannot provide confirmation.

Let us now discuss the case of cosmological analogs, which we believe is somewhat different. Indeed, we do have plenty of reasons to believe that QFTCS provides an accurate description of the early Universe, and indirect consequences of their predictions have been tested. We are referring here, for example, to the CMB power spectrum, or the formation of structures in the Universe, which can be traced back to the effects of inflation on the posterior cosmological evolution. Therefore, the difference with respect to BHs is the ability to test the theory that describes the target system. We have discussed in part III several analog cosmology experiments in BECs, whose results can be extrapolated to our Universe precisely because QFTCS is well tested. In this sense, we would argue that cosmological analogs can be considered as confirmatory. Moreover, we can test QFTCS in controlled scenarios, compare the results with the theory, and learn about the behavior of cosmological fields in the early Universe, in situations that may not be possible to reproduce in the cosmos, or which cannot be directly observed. In this case, specific situations are the ones that may be inaccessible, but the system itself is not (which again allows for confirmation of the theory). Nevertheless, analog experiments can be useful independently of their confirmatory power, as we will argue~next.

\section{Usefulness of analog experiments}

Let us recall that confirmation has not been the goal of the analog gravity community for many years. As stated in \cite{Almeida2023}, the field has developed in directions other than its original purpose (gaining some insight on BH radiation). QFT in non-trivial backgrounds, as to include quantum field theories in the presence of any external agent (such as in the case of Schwinger effect), is precisely the theory that describes the dynamics of these analog systems (we are referring to, for example, condensed matter analogs). Therefore, performing these experiments constitutes a test of the theoretical framework itself, together with its predictions, and allows one to improve the understanding of these physical systems, regardless of the analogies one is able to establish. Beyond this, analog gravity experiments have gone way past the original idea of realizing Hawking effect. Indeed, in our case we have focused on cosmological particle creation scenarios, but also systems testing different phenomena have proliferated, as discussed in \cref{sec:analogs.history}, e.g. superradiance, dynamical horizons or Casimir effect.

Besides, we believe that we have demonstrated in part IV how an experimental point of view can provide fresh ideas and insights that can improve our understanding of QFTCS. The idea presented in \cref{ch:qva} of defining a quantum vacuum, not from the usual requirements of minimization of energy or oscillations of the particle number, but through hypothetical measurements of particle production---leading to an interpretation of what a \textit{particle} is in this case (the counts after a measurement characterized by $\delta$)---is a clear example of this. Of course, the idea of switching-on and -off the expansion studied in \cref{ch:switcheffects} comes directly from experimental scenarios, although it can be implemented in the context of the early Universe as well, as we discussed. Crucially, the analysis of the importance of these stages in particle production triggered our urge for revisiting the limitations of gravity analogs. After the reflection carried out in these last pages, we are even more convinced of the utility of these experiments, perhaps most interesting when analogies are pushed to their limit. Lastly, let us mention that also the formalism benefited from the sharing of insights from Cosmology to analog systems and the other way around: Chapter~\ref{ch:qftcs} has been written so that it accommodates ideas from the two realms, and we believe this has provided a more complete view of the problem. 

We close this coda by hoping that the reader has enjoyed the journey through the different aspects of Quantum Field Theory in Curved Spacetimes and cosmological particle creation, including its viability as a dark matter production mechanism, and that the insights provided by analog experiments have been useful to understand the problems we have addressed.

%% file: Chapters/AppNumericalParameters.tex

\addtocontents{toc}{\protect\vspace{0.5em}}

\chapter{Numerical parameters} 

\label{ch:app.parameters} 



In this appendix we collect the numerical values of the parameters used in the different computations and figures of the thesis. 

\section{Particle production sourced by single-field inflation}
\label{sec:param.sf}

In most of the figures we have left all the quantities expressed in terms of the mass of the inflaton, $m_{\phi}$, which sets up the scale of the problem. When it has been necessary to assume a numerical value for such a mass, we have taken $m_{\phi} = 1.2 \times 10^{13} \, \text{GeV}$. Accordingly, the Planck mass has the value $\mpl = 1.02 \times 10^6 m_{\phi}$. 

For the quadratic inflationary potential, the initial value of the inflaton field, under the slow-roll assumption, is taken to be~$\phi_{\text{SR}}(\ti) = \phi_{\text{i}} = 3\mpl$. When inflation ends, at~\mbox{$t = 0$}, the field value is $\phi_{\text{SR}}(t=0)= \phi_0 = 0.5\mpl$. The slow-roll approximation can then be used to extract $\ti \simeq -15.35/m_{\phi}$ as the time when inflation starts. The equation of motion~\eqref{eq:cosmo.InflatonEoM} can also be solved numerically taking as initial conditions the same as for slow-roll, $\phi(\ti)=\phi_{\text{i}}$, and the derivative of the approximate solution at this point, $\phi^{\prime}(\ti)=\phi_{\text{SR}}^{\prime}(\ti)$. Both solutions will be very close up to $t_*$, where the slow-roll approximation starts to break down. Then, $\phi(t=0)$ slightly deviates from~$\phi_0$. The scale factor is chosen such that $a(t=0)=a_0=1$. Slow-roll is assumed to be a good approximation until $\eta_*=-500/m_{\phi}$. For the Starobinsky potential, the correct number of $e$-folds are obtained for~\mbox{$L = 6.32 \times 10^{-4} \mpl$} and an initial value for the inflaton field of $\phi_{\text{i}} = 1.088 \mpl$. In this case, normalization of the scale factor is chosen differently to improve numerical performance, although physical quantities do not depend on this. In particular, we have taken $a(t_{\text{f}})=175$.

Unless the contrary is expressly stated, particle production is calculated using the averaged vacuum prescription at $\bar{\eta} = 16.33/m_{\phi}$ for the scalar field. The range of masses explored is $10^{-7}m_{\phi}\leq m \leq 10^{0.5}m_{\phi}$, although for obtaining figure \ref{fig:sf.Abundance} it is assumed that production is the same for~$m\leq 10^{-7}m_{\phi}$. On the other hand, the coupling $\xi$ is such that $1/6 \leq \xi \leq 1$.

Vector production is obtained using the adiabatic vacuum at \mbox{$\eta_{\text{f}} = 7.58/m_{\phi}$}, which is within the adiabatic regime for all the parameter space explored, namely~\mbox{$0.5m_{\phi}\leq m$},~\mbox{$1/6\leq\gamma\leq 1$} and $\sigma \leq 0$.

\section{Particle production in de Sitter-like inflation}
\label{sec:param.ds}

In order to compare with the results on scalar production sourced by single-field inflation, we choose the final time $\eta_{\text{f}}$ so that it coincides with the time at which the inflaton enters the reheating regime, and the Hubble rate of the de Sitter geometry is chosen to be the same as the one at the onset of inflation, $H_0 \simeq 6.15m_{\phi}$, where~$m_{\phi}$ is the inflaton mass for the quadratic inflationary potential. The curvature of the spatial sections~$\kappa$ is left as a free parameter, whose associated curvature abundance can be compared to observations. 

\section{BECs as cosmological simulators}
\label{sec:param.bec}

For the theory only plots, above \cref{sec:bec.Experiment}, we set  $t_{\text{i}}=0$, and the following parameters are used. The condensate consists of \mbox{$N= 15 \times 10^3$} potassium atoms of mass $m=6.47008 \times 10^{-26}$ kg and extends up to a radius $R= 30 \times 10^{-6} \text{m}$. The longitudinal trapping frequency is $\omega=2\pi \times 1.75 \times 10^3 \, \text{s}^{-1}$, while the initial and final $s$-wave scattering lengths are $\asi=300\aB$ and $\asf = 50\aB$, respectively, where~$\aB$ denotes the Bohr radius. The $2$D critical temperature for an anisotropic trap is calculated for these values at initial time, and yields $T_{\text{c}}= 82.41 \times 10^{-9}$ K. Moreover, the regularization for the correlation functions is carried out with a Gaussian of inverse standard deviation $w = 5 \times 10^{-7}$ m.

In the case of the theory curves of the experimental plots, a speed of sound of~\mbox{$\cs = 1.2 \mu\text{m/ms}$} is used, along with scattering lengths $\asi=350\aB$ and $\asf = 50\aB$. An initial temperature of $T=40 \, \text{nK}$ is assumed, and the trapping frequency is~\mbox{$\omega=2\pi \times 1.75 \times 10^3 \, \text{s}^{-1}$}.

\section{Entanglement between produced pairs}
\label{sec:param.entanglement}

In the case of the entanglement predictions, we consider the same quasi-$2$D BEC as in \cref{ch:becstheory}, with $N=45 \times 10^3$ atoms, a radius $R = 25 \, \times 10^{-6} \text{m}$, and a longitudinal trapping frequency of~\mbox{$\omega=2\pi \times 1.6 \times 10^3 \, \text{s}^{-1}$}. For the contraction and expansion cycles, we consider a maximum scattering length of $\alpha_{s, \text{f}}=400\aB$, whereas the linear ramp in \cref{fig:ent.LinearRampLN} corresponds to $\asi = 400\aB$ and $\asf = 50\aB$.


%% file: Chapters/AppCosmoDM.tex

\chapter{Laplace-Beltrami eigenfunctions} 

\label{ch:app.eigenfunctions} 



This appendix is devoted to the eigenfunctions of the Laplace-Beltrami operator corresponding to the spatial sections of the FLRW metric in $D=3$ (\cref{sec:ds.eigenfunctions3Dcurved}) and $D=2$ (\cref{sec:bec.eigenfunctions}). 

\section{Eigenfunctions in (1+3) FLRW universes}
\label{sec:ds.eigenfunctions3Dcurved}

The eigenfunctions of the Laplace-Beltrami operator \eqref{eq:ds.LaplaceBeltrami} are given (up to a normalization factor) by  $\mathcal{H}_{klq}(\vec{u}) = \mathcal{U}_k(u)Y_{lq}(\theta, \phi)$, where $Y_{lq}$ are the spherical harmonics and $\vec{u} = (u, \theta, \phi)$. Explicitly, they read
\begin{equation}
\mathcal{H}_{klq}(\vec{u})=\left\{
\begin{aligned}
W_{Klq}(\vec{u}) &= \sqrt{\frac{M_{Kl}}{(K+1)^2\sin{\left(\sqrt{\abs{\kappa}}u\right)}}}P_{K+1/2}^{-l-1/2}\left[\cos{\left(\sqrt{\abs{\kappa}}u\right)}\right] \\
&\hspace{2cm}\times Y_{lq}(\theta, \phi) \quad \text{for} \quad \kappa>0, \\
X_{klq}(\vec{u})&=\sqrt{\frac{2}{\pi}}j_l(ku)\\
&\hspace{2cm}\times Y_{lq}(\theta, \phi) \quad \text{for} \quad \kappa =0,\\
Z_{Klq}(\vec{u}) &=\sqrt{\frac{N_{Kl}}{K^2\sinh{\left(\sqrt{\abs{\kappa}}u\right)}}}P_{iK - 1/2}^{-l-1/2}\left[\cosh{\left(\sqrt{\abs{\kappa}}u\right)}\right]\\
&\hspace{2cm}\times Y_{lq}(\theta, \phi) \quad \text{for} \quad \kappa<0.
\end{aligned}
\right.
\label{eq:ds.Eigenfunctions}
\end{equation}
Here, $Y_{lq}(\theta, \phi)$ are the usual complex spherical harmonics,
\begin{equation}
Y_{lq}(\theta, \phi) = \sqrt{\frac{2l+1}{4\pi}\frac{(l-q)!}{(l+q)!}}P_l^q(\cos{\theta})e^{iq\phi},
\end{equation}
the function $j_l(kr)$ is the spherical Bessel function of order $l$,
\begin{equation}
j_l(x) = \sqrt{\frac{\pi}{2x}}J_{l+1/2}(x),
\end{equation}
and $P_{K}^{l}$ are the associated Legendre polynomials, where the order $iK - 1/2$ corresponds to the conical or Mehler function. At the same time, the normalization constants $M_{Kl}$ and $N_{Kl}$ read
\begin{equation}
M_{Kl}=\prod_{j=0}^{l}\left[(1+K)^2-j^2\right]=\frac{(K+1)(K+l+1)!}{(K-l)!},
\end{equation}
and
\begin{equation}
N_{Kl} =i^{-2l-1}\prod_{j=0}^{l}\left[K^2+j^2\right] = i^{-2l-1}K^2\frac{\abs{\Gamma(iK+l+1)}^2}{\abs{\Gamma(iK+1)}^2},
\end{equation}
and are chosen such that we obtain the desired volume measure in momentum and real space, as we will see shortly. Note that we follow a slightly different convention for the definition of the eigenfunctions~$\mathcal{H}_{klq}$ than the one used in \cite{Birrell1982}, more on the line of \cite{Abbott1986}, where the associated Legendre polynomials substitute the derivative of the cosine with respect to the hyperbolic cosine ($\kappa<0$), as it is derived in said reference. However, the range we consider here for the wavenumbers is that of~\mbox{\cite{Harrison1967, Parker1974}}.

All three eigenfunctions $\mathcal{H}_{klq}$ coincide when $\kappa \to 0$, which can be checked by noting that the asymptotic behavior of the Legendre polynomials as $\kappa \to 0$ is
\begin{equation}
P_{K+1/2}^{-l-1/2}\left[\cos{\left(\sqrt{\abs{\kappa}}u\right)}\right] \sim \sqrt{\frac{2}{\pi}}\left(k/\sqrt{\abs{\kappa}}\right)^{-l} \left(\sqrt{\abs{\kappa}}u\right)^{1/2} j_l(ku)
\end{equation}
and similarly for complex order,
\begin{equation}
P_{iK-1/2}^{-l-1/2}\left[\cosh{\left(\sqrt{\abs{\kappa}}u\right)}\right] \sim i^{l+1/2}\sqrt{\frac{2}{\pi}}\left(k/\sqrt{\abs{\kappa}}\right)^{-l} \left(\sqrt{\abs{\kappa}}u\right)^{1/2} j_l(ku).
\end{equation}
Therefore,
\begin{equation}
\begin{split}
W_{Klq}(\vec{u}) &=\sqrt{\frac{M_{Kl}}{K^2\sin{u}}}P^{-l-1/2}_{K+1/2}\left[\cos{\left(\sqrt{\abs{\kappa}}u\right)}\right]Y_ {lq}(\theta, \phi)\\
&\underset{\kappa \to 0}{\sim}\left(\frac{(K+1)(K+l+1)!}{K^2(K-l)!}\right)^{1/2}\sin{\left(\sqrt{\abs{\kappa}}u\right)}^{-1/2}\\
&\hspace{3cm}\times K^{-l-1/2}\sqrt{\frac{2ku}{\pi}}j_l(ku)Y_{lq}(\theta, \phi)\\
&=\sqrt{\frac{2}{\pi}}j_l(ku)Y_{lq}(\theta, \phi)= X_{klq}(\vec{u}),
\end{split}
\end{equation}
where we used that
\begin{equation}
\frac{(K+l+1)!}{(K-l)!} \underset{\kappa \to 0}{\sim} \frac{K^{l+1}}{K^{-l}} = K^{2l+1}.
\end{equation}
The limit of $Z_{Klq}(\vec{u})$ follows similarly as long as one recalls that
\begin{equation}
\frac{\abs{\Gamma(iL+l+1)}^2}{\abs{\Gamma(iK+1)}^2} \underset{\kappa \to 0}{\sim} \left(\frac{K^{l+1}}{K}\right)^2 = K^{2l}.
\end{equation}
Thus,
\begin{equation}
\begin{split}
Z_{Klq}(\vec{u}) &= \sqrt{\frac{N_{Kl}}{K^2\sinh{\left(\sqrt{\abs{\kappa}}u\right)}}}P^{-l-1/2}_{iK-1/2}\left[\cosh{\left(\sqrt{\abs{\kappa}}u\right)}\right]Y_{lq}(\theta, \phi)\\
&\underset{\kappa \to 0}{\sim}\sqrt{\frac{2}{\pi}}j_l(ku)Y_{lq}(\theta, \phi) = X_{klq}(\vec{u}).
\end{split}
\end{equation}
In the following, we derive the orthogonality and completeness relations for the three eigenfunctions. We start with the flat universe, where the scalar product
\begin{equation}
\begin{split}
\left(X_{klq}, X_{k^{\prime}l^{\prime}q^{\prime}}\right) &= \frac{2}{\pi}\int\text{d}u\,u^2 j_l(ku)j_{l^{\prime}}(k^{\prime}u)\\
&\hspace{2cm}\times\int \text{d}\theta\text{d}\phi \sin{\theta}Y_{lq}(\theta, \phi)Y^*_{l^{\prime}q^{\prime}}(\theta, \phi)\\
&=\frac{\delta(k-k^{\prime})}{k^2}\delta_{ll^{\prime}}\delta_{qq^{\prime}} 
\end{split}
\label{eq:ds.FlatOrthogonality}
\end{equation}
assures the orthogonality of $X_{klq}$. Here we used the orthogonality of the spherical harmonics,
\begin{equation}
\int \text{d}\theta\text{d}\phi \sin{\theta}Y_{lq}(\theta, \phi)Y^*_{l^{\prime}q^{\prime}}(\theta, \phi) = \delta_{ll^{\prime}}\delta_{qq^{\prime}},
\end{equation}
as well as that of the Bessel functions \citep{Abramowitz1964},
\begin{equation}
\int_{0}^{\infty}\text{d}u\,u^2 j_l(ku)j_l(k^{\prime}u)=\frac{\pi}{2k^2}\delta(k-k^{\prime}).
\label{eq:ds.BesselOrthogonality}
\end{equation}
Completeness follows from that of the spherical harmonics,
\begin{equation}
\sum_{l=0}^{\infty}\sum_{q=-l}^l Y_{lq}(\theta, \phi) Y^*_{lq}(\theta^{\prime}\phi^{\prime}) = \delta(\cos{\theta} - \cos{\theta^{\prime}})\delta(\phi - \phi^{\prime}),
\end{equation}
which together with \eqref{eq:ds.BesselOrthogonality} leads to
\begin{equation}
\begin{split}
\frac{2}{\pi}\int_0^{\infty}\text{d}k\, k^2\sum_{l=0}^{\infty}\sum_{q=-l}^l j_l(ku)&j_l(ku^{\prime}) Y_{lq}(\theta, \phi) Y^*_{lq}(\theta^{\prime}, \phi^{\prime}) \\
&= \frac{\delta(u - u^{\prime})}{u^2}\frac{\delta(\theta - \theta^{\prime})}{\sin{\theta}}\delta(\phi - \phi^{\prime}).
\end{split}
\end{equation}

Let us find the orthogonality relation in the case of positive spatial curvature,
\begin{equation}
\begin{split}
\left(W_{Klq},W_{K^{\prime}l^{\prime}q^{\prime}}\right) &= \abs{\kappa}^{-3/2}\int \text{d}u \text{d}\theta \text{d}\phi\sin{u}^2\sin{\theta} Y_{Klq}(\vec{u})Y_{K^{\prime}l^{\prime}q^{\prime}}^*(\vec{u}) \\
&=\abs{\kappa}^{-3/2}\frac{M_{Kl}}{(K+1)^2}\int_0^{\pi}\text{d}u \sin{u}^2 P_{K+1/2}^{-l-1/2}(\cos{u}) P_{K+1/2}^{-l^{\prime}-1/2}(\cos{u})\\
&\hspace{1cm}\times\int \text{d}\theta\text{d}\phi \sin{\theta} Y_{lq}(\theta, \phi) Y_{lq}^*(\theta, \phi) \\
&=\abs{\kappa}^{-3/2} \frac{\delta_{KK^{\prime}}}{(K+1)^2}\delta_{ll^{\prime}}\delta_{qq^{\prime}},
\end{split}
\end{equation}
where we used the orthogonality of the associated Legendre polynomials for fixed~$l$ with $0\leq l \leq K$,
\begin{equation}
\int_{-1}^1 \text{d}u P_{K+1/2}^{-l-1/2}(u) P_{K^{\prime}+1/2}^{-l-1/2}(u) = \frac{(K-l)!}{(K+1)(K+l+1)!}\delta_{KK^{\prime}}.
\end{equation}
Taking the limit $\kappa \to 0$ yields precisely \eqref{eq:ds.FlatOrthogonality}, taking into account that
\begin{equation}
\delta_{KK^{\prime}} \to \delta\left(\frac{k}{\sqrt{\abs{\kappa}}}-\frac{k^{\prime}}{\sqrt{\abs{\kappa}}}\right) = \sqrt{\abs{\kappa}}\delta(k-k^{\prime}).
\end{equation}
These functions are also complete,
\begin{equation}
\begin{split}
\sum_{K=0}^{\infty}\abs{\kappa}^{3/2}(K+1)^2 \sum_{l=0}^K \sum_{q=-l}^l & Y_{Klq}(\vec{u})Y^*_{Klq}(\vec{u}^{\prime}) \\
&= \abs{\kappa}^{3/2}\frac{\delta(u - u^{\prime})}{\sin^2{u}}\frac{\delta(\theta - \theta^{\prime})}{\sin{\theta}}\delta(\phi - \phi^{\prime}),
\end{split}
\end{equation}
as is shown in ref. \citep{Bander1966} (making $u \to -iu$ and $ K+1\to -iK$ allows one to go from positive to negative spatial curvature).

From the same reference we can extract the relations
\begin{equation}
\left(Z_{Klq}, Z_{K^{\prime}l^{\prime}q^{\prime}}\right) = \frac{\delta(k-k^{\prime})}{k^2}\delta_{ll^{\prime}}\delta_{qq^{\prime}}
\end{equation}
and
\begin{equation}
\begin{split}
\int_0^{\infty}\text{d}K \abs{\kappa}^{3/2}K^2\sum_{l=0}^{\infty}\sum_{q=-l}^{l} & Z_{Klq}(\vec{u})Z_{Klq}^*(\vec{u}^{\prime})\\
&=\abs{\kappa}^{3/2} \frac{\delta(u - u^{\prime})}{\sinh^2{u}}\frac{\delta(\theta - \theta^{\prime})}{\sin{\theta}} \delta(\phi - \phi^{\prime})
\end{split}
\end{equation}
for the negatively curved eigenfunctions (note the equivalence between $\sinh{u}^l \left(\frac{d}{d\cosh{u}}\right)^{l+1}\cos{Ku}$ and $\sinh{u}^{-1/2}P_{iK-1/2}^{-l-1/2}$, which is detailed in \cite{Abbott1986}).

\section{Eigenfunctions in (1+2) FLRW universes}
\label{sec:bec.eigenfunctions}

In the case of two-dimensional spatial sections, the eigenfunctions of the Laplace-Beltrami operator \eqref{eq:bec.LaplaceOperator} are given by
\begin{equation}
    \mathcal{H}_{kq}(\vec{u})=\left\{
    \begin{aligned}
    W_{Kq}(\vec{u}) &= \sqrt{\frac{(K-q)!}{(K+q)!}} \, e^{iq\phi} \, P_{Kq}(\cos u) \quad \text{for} \quad \kappa>0, \\
    X_{kq} (\vec{u}) &= e^{i q \phi} \, J_q (k u) \quad \hspace{2.61cm}\text{for} \quad \kappa=0,\\
    Z_{Kq} (\vec{u}) &= (-i)^q \frac{\Gamma(iK+1/2)}{\Gamma(iK+q+1/2)} \\
        &\hspace{0.965cm} \times e^{iq\phi} P^q_{iK-1/2}\left(\cosh u\right), \hspace{0.35cm} \text{for} \quad \kappa<0,
    \end{aligned}
    \right.
\label{eq:bec.Eigenfunctions}
\end{equation}
where $P_{Kq}(\cos u) = (-1)^qP^q_K(\cos u)$, with $P^q_K$ denoting the associated Legendre polynomials discussed in \cref{sec:ds.eigenfunctions3Dcurved}, and $\vec{u}=(u,\phi)$. Note that $W_{Kq}$ are the usual spherical harmonics except for a normalization factor, the functions $X_{kq}$ are polar waves, and $Z_{Kq}$ is built from conical functions (see \cref{sec:ds.eigenfunctions3Dcurved}). Below we discuss the normalization and completeness relations of the eigenfunctions \eqref{eq:bec.Eigenfunctions}. 

In the flat case, we have the product
\begin{equation}
\begin{split}
    \left(X_{kq},X_{k^{\prime}q^{\prime}}\right) &= \int_0^{\infty}\text{d}u \,u \int_0^{2\pi} \text{d}\phi \, X^{*}_{kq}(u,\phi) X_{k^{\prime}q^{\prime}}(u,\phi)\\
    &= \frac{2\pi}{k}\delta(k-k^{\prime})\delta_{qq^{\prime}},
\end{split}
\end{equation}
while the completeness relation is given by
\begin{equation}
    \int_0^{\infty} \frac{\text{d}k}{2\pi} k \sum_{q=-\infty}^{\infty} X_{kq}(u,\phi) X_{kq}^*(u^{\prime},\phi^{\prime})= \frac{\delta(u-u^{\prime})}{u}\delta(\phi-\phi^{\prime}).
\end{equation}

In the case of positive spatial curvature, the orthogonality relation reads
\begin{equation}
    \begin{split}
        &\left(W_{Kq},W_{K^{\prime}q^{\prime}}\right)\\ &= \abs{\kappa}^{-1} \int_0^{\pi}\text{d}u \sin u \int_0^{2\pi} \text{d}\phi \, W^{*}_{Kq}(u,\phi) W_{K^{\prime}q^{\prime}}(u,\phi)\\
        &= \abs{\kappa}^{-1} \frac{2\pi}{K+1/2}\delta_{KK^{\prime}} \delta_{qq^{\prime}}.
    \end{split}
\end{equation}
On the other hand, the completeness relation reads
\begin{equation}
    \begin{split}
        &\sum_{K=0}^{\infty} \abs{\kappa}\frac{K+1/2}{2\pi} \sum_{q=-K}^{K} W_{Kq}(u,\phi) W_{lm}^*(u^{\prime},\phi^{\prime})\\
        &= \frac{\abs{\kappa}}{\sin u} \delta(u-u^{\prime}) \delta(\phi-\phi^{\prime}).
    \end{split}
\end{equation}

Analogously, the orthonormality relation for the eigenfunctions $Z_{kq}$ follows from
\begin{equation}
\begin{split}
       \big(Z_{Kq}, Z_{K^{\prime}q^{\prime}}\big) &= \abs{\kappa}^{-1}\int_0^{\infty}\text{d}u \sinh u \int_0^{2\pi} \text{d}\phi \, Z^{*}_{Kq}(u,\phi) Z_{K^{\prime}q^{\prime}}(u,\phi) \\
        &= \abs{\kappa}^{-1} \frac{2\pi}{K\tanh(\pi K)} \delta(K-K^{\prime}) \delta_{qq^{\prime}},
\end{split}
\end{equation}
and completeness is guaranteed by
\begin{equation}
    \begin{split}
        &\int_0^{\infty} \frac{\text{d}K}{2\pi} \abs{\kappa} K \tanh(\pi K) \sum_{q=-\infty}^{\infty} Z_{Kq}(u,\phi)Z_{Kq}^*(u^{\prime},\phi^{\prime})\\
        &= \frac{\abs{\kappa}}{\sinh u} \delta(u-u^{\prime}) \delta(\phi-\phi^{\prime}).
    \end{split}
\end{equation}

Similarly to what we did in \cref{sec:ds.eigenfunctions3Dcurved}, one can check that the three eigenfunctions coincide in the limit $\kappa \to 0$.

%% file: Chapters/AppBECTheory.tex

\chapter{Additional analyses and derivations of the BEC setup} 

\label{ch:app.becstheory} 



In this appendix we provide additional details on the BEC setup studied in part III of the thesis. In particular, we further discuss the geometry of the $(1+2)$-dimensional emergent FLRW spacetimes and their mapping to the laboratory coordinates in \cref{sec:bec.Geometries}, as well as the trajectories of phonons through these analog universes in \cref{sec:bec.PhononTrajectories}

\section{Models of spatially curved spacetimes}
\label{sec:bec.Geometries}

\begin{figure}[t!]
    \centering
    \includegraphics[width=0.6\textwidth]{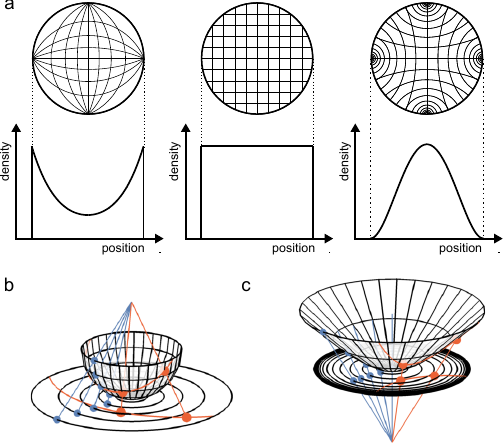}
    \caption[Disk projections of the spatial sections of FLRW universes in two spatial dimensions]{(a) Map between radially symmetric density profiles and spatial sections of FLRW, with the grid denoting the trajectories of null geodesics. (b) Projection of the positively curved half-sphere embedded in three-dimensional Euclidean space onto a polar disk. The projection is performed through the blue lines, and two points connected by a geodesic are depicted in red. (c) Similar projection of the negatively curved hyperboloid embedded in three-dimensional Minkowski spacetime onto the Poincaré disk. Figure from \cite{BECPaper2022}.} 
    \label{fig:bec.FLRWUniverses}
\end{figure}

In this section we will discuss in more detail the geometrical aspects of the spatial sections of the FLRW universes that are realized in the BEC through both a constant density profile and those leading to positively and negatively curved cosmologies (cf. \cref{subsec:bec.BackgroundDensity} and \eqref{eq:bec.TrapRadialDependenceExact}). In \cref{fig:bec.FLRWUniverses} (a), we show the three different spatial geometries together with the corresponding radial density profiles: positive (left), vanishing (middle) and negative (right). Additionally, we illustrate in (b) and (c) the projection of the closed and open two-dimensional spatial surfaces---embedded in a three-dimensional geometry---onto two-dimensional disks, respectively. This helps to intuitively understand the mapping of the FLRW universes to the laboratory coordinates, which we will discuss in the following. 

Let us start with the $\kappa>0$ scenario, and consider a two-sphere of radius $R/2$ embedded in three-dimensional Euclidean space $\mathbb{R}^3$ with Cartesian coordinates $(X,Y,Z)$, given by the equation
\begin{equation}
    \frac{R^2}{4}=X^2+Y^2+Z^2.
\end{equation}
Cartesian coordinates can be parametrized by the angles $u' \in [0,\pi]$ and $\phi' \in [0,2\pi)$, so that one has
\begin{equation}
    (X,Y,Z) = \frac{R}{2} \left(\sin u' \cos \phi', \sin u' \sin \phi', \cos u'\right).
\end{equation}
Then, the induced metric on the two-sphere can be written as
\begin{equation}
    \text{d}s^2 = \frac{R^2}{4}\left( \text{d} u'^2 + \sin^2 u' \, \text{d} \phi'^2 \right).
\label{eq:bec.SphereLineElement}
\end{equation}
By performing a so-called \textit{stereographic projection} from the north pole $(0,0,R/2)$, the two-sphere can be mapped to a disk located at the south pole $(0,0,-R/2)$, with coordinates $(r, \phi)$ (cf. \cref{fig:bec.FLRWUniverses} (b)). The relation between the coordinates is given by
\begin{equation}
    (r, \phi) = \left(R \cot \frac{u'}{2}, \phi' \right),
\end{equation}
and the line element \eqref{eq:bec.SphereLineElement} acquires the form \eqref{eq:bec.LabCoordinatesLineElement}, with positive sign.

Similarly, the negatively curved surface corresponds to a hyperboloid embedded in three-dimensional Minkowski space $\mathbb{M}^3$ \cite{Balazs1986}. In this case, the line element is of the form 
\begin{equation}
    \text{d} s^2 = \text{d} X^2 + \text{d} Y^2 - \text{d} Z^2.
    \label{eq:bec.LineElementMinkwoski}
\end{equation}
The hyperboloid is described by the relation
\begin{equation}
    -R^2/4 = X^2 + Y^2 - Z^2,
    \label{eq:bec.EmbeddingHyperboloid}
\end{equation}
with $Z > 0$, and can be parametrized by a pseudoangle $u' \in [0,\infty)$ and an azimuthal angle ${\phi' \in [0,2\pi)}$. The relation to the Minkowski coordinates is therefore given by
\begin{equation}
    (X,Y,Z) = R/2 \, \left(\sinh u' \cos \phi', \sinh u' \sin \phi', \cosh u' \right),
\end{equation}
leading to an induced metric on the hyperboloid of the form
\begin{equation}
    \text{d}s^2 = R^2/4 \, \left(\text{d} u'^2 + \sinh^2 u' \, \text{d} \phi' \right).
\label{eq:bec.HyperboloidLineElement}
\end{equation}
The projection onto the Poincaré disk located at the south pole, with coordinates $(r, \phi)$, is realized using the apex $(0,0,-R/2)$ of the lower hyperboloid (not shown in fig.\ \ref{fig:bec.FLRWUniverses}) as the base point. Again, the projection is indicated with blue lines in \cref{fig:bec.FLRWUniverses}. Eventually one obtains
\begin{equation}
    (r, \phi) = \left(R \coth \frac{u'}{2}, \phi' \right),
\end{equation}
so that the line element \eqref{eq:bec.HyperboloidLineElement} takes the form \eqref{eq:bec.LabCoordinatesLineElement} (with the minus sign).

\section{Phonon trajectories}
\label{sec:bec.PhononTrajectories}

In this section we will describe the radial motion of phonons in the BEC, which follow null geodesics since they behave as a massless scalar field. We consider radially outmoving phonons from the center of the BEC in the geometry described by the line element \eqref{eq:bec.LabCoordinatesLineElement}. 

In terms of the laboratory radial coordinate $r$, radial geodesics are given by solutions to
\begin{equation}
    \frac{\text{d}t}{a(t)} = \frac{\text{d}r}{\sqrt{1 - f(r)/R^2}},
    \label{eq:bec.TrajectoriesEoM}
\end{equation}
with initial conditions $r(t=0)=0$. For the three different density profiles $f(r)$ in \cref{eq:bec.TrapShape}, the solutions for the different values of the spatial curvature $\kappa$ read
\begin{equation}
    r(t) = R \, \begin{cases}
    \tan z(t)  & \text{for} \hspace{0.3cm} \kappa > 0, \\
    z(t) & \text{for} \hspace{0.3cm} \kappa = 0, \\
    \tanh z(t) & \text{for} \hspace{0.3cm} \kappa < 0,
    \end{cases}
    \label{eq:bec.TrajectoriesSolution}
\end{equation}
where the function $z(t)$ is given by
\begin{equation}
    z(t) = \frac{1}{R} \, \int_0^t \frac{\text{d} t^{\prime}}{a(t^{\prime})}.
\end{equation}
In the case of a quadratic trap profile of the form $f(r)=\pm r^2$, the trigonometric functions in \cref{eq:bec.TrajectoriesSolution} are \textit{sinh} instead of \textit{tan} in the positively curved scenario, and \textit{sin} instead of \textit{tanh} in the case of negative curvature.

\begin{figure}[t!]
    \centering
    \includegraphics[width=0.28\textwidth]{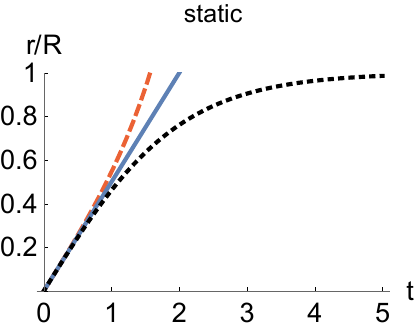}
    \includegraphics[width=0.28\textwidth]{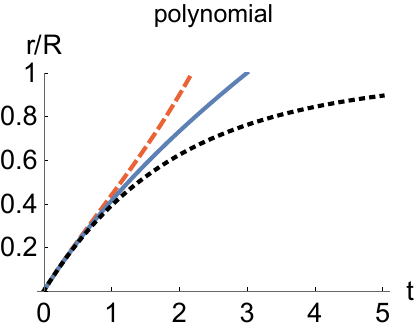}
    \includegraphics[width=0.41\textwidth]{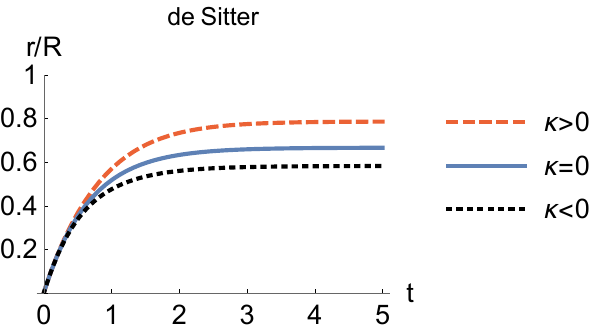}
    \caption[Trajectories of low energy phonons in the BEC]{Trajectories of radially outmoving phonons for the three different spatial curvatures: positive (red dashed line), vanishing (blue solid line) and negative (black dotted line). Three different scale factors are considered: $\gamma=0$ (static, left panel, \mbox{$Q=2, t_0=-1, R=1$}), $\gamma=1$ (middle panel, \mbox{$Q=2, t_0=-1, R=1$}) and exponential (right panel, \mbox{$a_0=1, H=1.5, R=1$}). For illustration purposes, we consider dimensionless parameters. Figure from \cite{BECPaper2022}.}
    \label{fig:bec.PhononTrajectories}
\end{figure}

If we consider power-law scale factors of the form \cref{eq:bec.ScaleFactorPolynomial}, we find, in the case of $\gamma \neq 1$,
\begin{equation}
    z(t) = \frac{1}{Q R} \frac{\text{sgn}(t - t_0)\abs{t - t_0}^{1-\gamma} + \text{sgn}(t_0)\abs{ t_0}^{1-\gamma}}{1-\gamma},
\end{equation}
whereas for $\gamma = 1$ one has
\begin{equation}
    z(t) = \frac{1}{Q R} \ln \frac{\abs{t-t_0}}{\abs{t_0}}.
\end{equation}
Another interesting example is the spatially flat de Sitter universe discussed in \cref{ch:deSitter}, which is characterized by the scale factor
\begin{equation}
    a(t) = a_0 \, e^{H_0 t},
    \label{eq:bec.ScaleFactordeSitter}
\end{equation}
where $H_0$ denotes the constant Hubble parameter and $a_0 = a (t=0)$ is now the initial scale factor. In this case we find
\begin{equation}
    z(t) = \frac{1}{a_0 H R} \left(1 - e^{- H_0 t} \right).
\end{equation}
In \cref{fig:bec.PhononTrajectories} we show the phonon trajectories corresponding to $\gamma=1$, $\gamma=0$ and de Sitter, for the three curvature scenarios. We observe that phonons are slowed down as the curvature becomes more negative.

%% file: FrontBackMatter/Bibliography.tex
\defbibheading{bibintoc}[\bibname]{%
  \phantomsection
  \manualmark
  \markboth{\spacedlowsmallcaps{#1}}{\spacedlowsmallcaps{#1}}%
  \addtocontents{toc}{\protect\vspace{\beforebibskip}}%
  \addcontentsline{toc}{chapter}{\tocEntry{#1}}%
  \chapter*{#1}%
}
\newrefcontext[sorting=none]
\printbibliography[heading=bibintoc,notkeyword=myPapers, resetnumbers=true]




